\documentclass[apj]{emulateapj}
\usepackage{natbib}
\usepackage{threeparttable}
\usepackage{epsfig}
\usepackage{rotating}

\usepackage[linktocpage=true]{hyperref}
\hypersetup{colorlinks=true,citecolor=blue,linkcolor=blue,filecolor=black,runcolor=black}
\usepackage{breakurl} 
\usepackage{amsmath}

\newcommand{\Kkmpers}{\mbox{K~km~s$^{-1}$}}
\newcommand{\xcounits}{\mbox{cm$^{-2}$} \mbox{(K~km~s$^{-1}$)$^{-1}$}}
\newcommand{\ico}{\mbox{$I_\mathrm{CO}$}}
\newcommand{\xco}{\mbox{$X_\mathrm{CO}$}}
\newcommand{\hi}{\mbox{\sc Hi}}
\newcommand{\htwo}{\mbox{H$_{2}$}}
\newcommand{\ebv}{\mbox{E(B$-$V)}}
\newcommand{\av}{\mbox{$A_\mathrm{V}$}}
\newcommand{\icoav}{\mbox{$I_\mathrm{CO} {-} A_\mathrm{V}$}}
\newcommand{\cohtwo}{\mbox{CO-to-H$_2$}}
\newcommand{\mathsc}[1]{{\normalfont\textsc{#1}}}

\begin{document}

\title{The Parsec-Scale Relationship Between $I_{\rm CO}$ and $A_{\rm V}$ in Local Molecular Clouds}

\author{Cheoljong~Lee\altaffilmark{1}}
\author{Adam~K.~Leroy\altaffilmark{2}}
\author{Alberto~D.~Bolatto\altaffilmark{3}}
\author{Simon~C.~O.~Glover\altaffilmark{4}}
\author{Remy~Indebetouw\altaffilmark{5,6}}
\author{Karin~Sandstrom\altaffilmark{7}}
\author{Andreas~Schruba\altaffilmark{8}}
\altaffiltext{1}{Department of Astronomy, University of Virginia, Charlottesville, VA 22904, USA}
\altaffiltext{2}{Department of Astronomy, The Ohio State University, 140 West 18th Avenue, Columbus, OH, 43210, USA}
\altaffiltext{3}{Department of Astronomy, Laboratory for Millimeter-wave Astronomy, and Joint Space Institute, University of Maryland, College Park, Maryland 20742, USA}
\altaffiltext{4}{Institute f\"ur theoretische Astrophysik, Zentrum f\"ur Astronomie der Universit\"at Heidelberg, Albert-Ueberle Str. 2, 69120 Heidelberg, Germany}
\altaffiltext{5}{Department of Astronomy, University of Virginia, Charlottesville, VA 22904, USA}
\altaffiltext{6}{National Radio Astronomy Observatory, 520 Edgemont Rd, Charlottesville, VA 22903, USA}
\altaffiltext{7}{Center for Astrophysics and Space Sciences, Department of Physics, University of California, San Diego, 9500 Gilman Drive, La Jolla, CA 92093, USA}
\altaffiltext{8}{Max-Planck-Institut f\"ur extraterrestrische Physik, Giessenbachstra{\ss}e 1, 85748 Garching, Germany}

\begin{abstract}
We measure the parsec-scale relationship between integrated CO intensity (\ico) and visual extinction (\av) in $24$ local molecular clouds using maps of CO emission and dust optical depth from {\em Planck}. This relationship informs our understanding of CO emission across environments, but clean Milky Way measurements remain scarce. We find uniform \ico\ for a given \av , with the results bracketed by previous studies of the Pipe and Perseus clouds. Our measured \icoav\ relation broadly agrees with the standard Galactic CO-to-H$_2$ conversion factor, the relation found for the Magellanic clouds at coarser resolution, and numerical simulations by \citet{GLOVERCLARK16}. This supports the idea that CO emission primarily depends on shielding, which protects molecules from dissociating radiation. Evidence for CO saturation at high \av\ and a threshold for CO emission at low \av\ varies remains uncertain due to insufficent resolution and ambiguities in background subtraction. Resolution of order $0.1$~pc may be required to measure these features. We use this \icoav\ relation to predict how the \cohtwo\ conversion factor (\xco ) would change if the Solar Neighborhood clouds had different dust-to-gas ratio (metallicity). The calculations highlight the need for improved observations of the CO emission threshold and \hi\ shielding layer depth. They are also sensitive to the shape of the column density distribution. Because local clouds collectively show a self-similar distribution, we predict a shallow metallicity dependence for \xco\ down to a few tenths of solar metallicity. However, our calculations also imply dramatic variations in cloud-to-cloud \xco\ at subsolar metallicity.
\end{abstract}


\section{Introduction}
\label{sec:intro}

CO emission is the main observational tracer of molecular gas in the Milky Way and other galaxies. To use this tracer effectively, we must understand the origin of CO emission in molecular clouds, the relationship between CO emission and H$_2$ mass, and how these vary among different environments. A key aspect of this variation is how the \cohtwo\ conversion factor (\xco; defined as the ratio between the column density of molecular hydrogen, $N_\mathrm{H_2}$, and the integrated CO intensity, \ico) depends on metallicity \citep[see review in][]{BOLATTO13}. 

\citet{LEE15} proposed that a productive way to approach this topic is to consider CO emission from a molecular cloud or an ensemble of clouds as the product of several separable phenomena: (1) the probability distribution function (PDF) of gas column densities within a cloud, (2) the local dust-to-gas ratio which relates a gas column density to a dust column density, (3) the relationship between dust column density and CO emission, and (4) the importance of \hi\ shielding envelopes at low gas column density and low dust abundance \citep[e.g.,][]{LEE12,LEE14}. \citeauthor{LEE15} showed how combining these empirical relationships allow one to predict a scaling for \xco\ as a function of metallicity. 

The major advantage of this approach is that each of these parts is a significant topic of research with its own literature, and that each of these topics can be constrained by observations. For example, the column density PDF of local molecular clouds has been studied by, e.g., \citet{KAIN09,LOMBARDI15,SCHNEIDER15,ABREUVICENTE15}. The dependence of the dust-to-gas ratio on metallicity has been examined in nearby spiral and dwarf galaxies by, e.g., \citet{SANDSTROM13,REMYRUYER14}. The relationship between dust column density and CO emission has been studied by, e.g., \citet{LOMBARDI06,PINEDA08,LEE15}. And the $\hi{-}\htwo$ balance in molecular clouds has been examined by, e.g., \citet[][]{KRUMHOLZ09,LEE12,LEE14,STERNBERG14}.

These individual topics are more tractable to observational studies than a direct estimate of the H$_2$ mass. This is especially true in low metallicity dwarf galaxies where the necessary observations remain difficult to obtain and systematic biases affect all available techniques \citep{BOLATTO13}. Studies targeting these four individual phenomena separately can be constructed more cleanly, and make valuable contributions to our understanding of the H$_2$ content in (low metallicity) galaxies. Thus, the simple, separable approach of \citet{LEE15} represents a practical way to make progress towards understanding the metallicity dependence of \xco\ based on observations. It offers a natural way to fold in our knowledge of ISM structure, dust physics, and PDR structure. As this knowledge improves, so does our understanding of \xco . 

The lynchpin of this approach is the ability to predict CO emission from the line of sight dust extinction (expressed as $V$-band extinction, \av), or dust column density, through a part of a molecular cloud.  Dust is the primary agent shielding CO molecules against dissociating radiation, and thus defines the part of a cloud in which CO represents the dominant form of carbon. Based on this, one can expect that dust shielding is a reasonable tracer of CO emission, at least to first order. To second order, variations in physical conditions such as gas temperature, turbulence, CO opacity, self-shielding, cloud geometry, and the external radiation field may complicate the relationship.

Only two Milky Way clouds---the Perseus and Pipe molecular clouds---have been well-studied in the \icoav\ parameter space \citep{LOMBARDI06,PINEDA08, LEE14}; though many studies related to this topic have been carried out going back to the earliest CO studies. The observational studies by \citet{LOMBARDI06}, \citet{PINEDA08}, and \citet{LEE14} show a clear relationship between \av\ and \ico\ within a molecular cloud. A similar relationship is predicted by analytic models of photon dominated regions \citep[PDRs; e.g.,][]{MB88,VB88,LEQUEUX94,BELL06,WOLFIRE10}, and in simulations that model chemistry and radiative transfer in turbulent clouds \citep{GLOVER11,SHETTY11,GLOVER12B}. 

Motivated by these observational and theoretical works, \citet{LEE15} studied the relationship between CO emission and dust extinction in three Local Group galaxies with different metallicities, aiming to test this simple picture \citep[see also][]{IMARA07,LEROY09}. On the scale of a large part of a molecular cloud, ${\sim}10$~pc, they found that the Large Magellanic Cloud (LMC) and Small Magellanic Cloud (SMC) have similar \ico\ for a given \av\ as nearby Milky Way clouds (with \av\ derived from the far infrared dust emission spectral energy distribution). The agreement in the \icoav\ relationship over a metallicity range of $\sim 0.2{-}1$~$Z_\odot$ suggests a common relationship between CO and dust shielding across galaxies. Although the 10~pc spatial scales analyzed by \citet{LEE15} are coarse, one may expect a similar dependence of CO emission on dust shielding to hold at higher resolution (i.e., on the scale of cloud substructures). This underpins the idea of the simple four-step approach to determine the metallicity dependence of \xco\ as introduced in \citet{LEE15}.

Thus, a better understanding of how to predict CO emission from line of sight extinction in {\em resolved} molecular clouds can help improve our understanding of \xco . The fact that only two Milky Way clouds have been characterized in the \icoav\ relationship at parsec-scale resolution limits our ability to understand CO emission from more extreme environments. Although the local molecular cloud population has been studied extensively, including many studies of \xco\ \citep[e.g.,][]{PINEDA08,PINEDA10}, this specific, very useful measurement---the \icoav\ relationship between line of sight CO intensity and dust extinction---has not been systematically carried out.

In this paper, we take advantage of the new all-sky CO and dust extinction maps from the {\em Planck} mission \citep{PLANCK13_CO,PLANCK13_DUST} to characterize this relationship at ${\sim}1$~pc resolution for 24 local molecular clouds. We present the average \icoav\ relationship, measure the scatter about this relation, and the differences from cloud to cloud. We assess the uncertainty due to foreground and background contamination, and compare our observations with numerical simulations from \citet{GLOVERCLARK16}. Our goal is to create a point of reference for CO and dust studies of other galaxies by analyzing in detail local molecular clouds in the Milky Way.

\begin{table*}
\begin{center}
\caption{Galactic molecular clouds considered in our analysis}
\label{tab:list}
\begin{tabular}{@{}lccccccc@{}}
\hline\hline
Cloud & $l$ & $b$ & width & height & distance & physical resolution & $\langle \Delta A_V \rangle^a$\\
& (degrees) & (degrees) & (degrees) & (degrees) & (parsecs) & (parsecs) &  (mag)\\
\hline
Aquila South & 34.5 & -16.5 & 6.0 & 4.5 & 110.0$^b$ & 0.6 & 0.4\\
California & 161.0 & -9.0 & 5.5 & 5.0 & 410.0$^b$ & 2.1 & 1.3 \\
Camelopardalis & 148.0 & 20.0 & 6.0 & 6.0 & 200.0$^b$ & 1.0 & 0.3 \\
Canis Major & 224.0 & -2.0 & 4.0 & 2.0 & 1150.0$^c$ & 6.0 & 2.8 \\
Cepheus North & 118.0 & 16.2 & 5.0 & 2.0 & 360.0$^b$ & 1.9 & 0.8\\
Cepheus South &	118.0 &	12.7 & 5.0 & 1.5 & 900.0$^b$ & 4.7 & 1.3 \\
Chamaeleon & 299.0 & -15.5 & 7.0 & 8.0 & 150.0$^d$ & 0.8 & 0.5 \\
Gem OB1 & 191.5 & 0.5 & 5.0 & 5.0 & 2000.0$^e$ & 10.5 & 2.2\\
Gum Nebula & 266.0 & -10.0 & 5.0 & 5.0 & 450.0$^f$ & 2.4 &  0.8 \\
Hercules & 44.5 & 9.0 & 3.5 & 3.0 & 200.0$^b$ & 1.0 & 0.8\\
Lacerta & 102.0 & -14.5 & 7.5 & 6.5 & 510.0$^b$ & 2.7 & 0.5 \\
Lupus & 341.0 & 13.5 & 8.0 & 11.0 & 155.0$^g$ & 0.8 & 0.8 \\
Mon OB1 & 201.5 & 2.5 & 5.0 & 3.5 & 890.0$^b$ & 4.7 & 2.1 \\
Mon R2 & 217.0 & -12.0 & 5.0 & 5.0 & 905.0$^c$ & 4.7 & 0.9 \\
Ophiuchus & 358.0 & 16.0 & 10.0 & 8.0 & 125.0$^b$ & 0.7 & 0.8 \\
Orion A & 210.0 & -19.0 & 7.0 & 2.0 & 371.0$^c$ & 2.0 & 0.4 \\
Orion B & 205.5 & -11.0 & 4.5 & 6.0 & 398.0$^c$ & 2.1 & 1.1 \\
Pipe Nebula & 0.0 & 5.5 & 4.0 & 3.0 & 130.7$^g$ & 0.7 & 2.1 \\
Pegasus & 95.0 & -34.0 & 12.0 & 9.0 & 230.0$^b$ & 1.2 & 0.2 \\
Perseus & 160.0 & -20.0 & 5.0 & 5.0 & 240.0$^h$ & 1.3 & 0.5 \\
Polaris Flare & 123.0 & 26.0 & 6.0 & 6.0 & 380.0$^b$ & 2.0 & 0.2 \\
R Coronae Australis & 5.0 & -23.0 & 10.0 & 9.0 & 130.0$^i$  & 0.7 & 0.2 \\
Rosette & 207.0 & -2.0 & 2.0 & 2.0 & 1330.0$^c$ & 7.0 & 2.8 \\
Taurus & 172.5 & -15.0 & 7.5 & 5.0 & 135.0$^b$ & 0.7 & 1.0 \\
\hline\hline
\end{tabular}
\end{center}
$^{a}$ {Mean value of background subtraction across the cloud, estimated from the reference region. 
$^{b}$ \citet{SCHLAFLY14} $^{c}$ \citet{LOMBARDI11} $^{d}$ \citet{BOULANGER98} $^{e}$ \citet{CARPENTER95}
$^{g}$ \citet{LOMBARDI08} $^{h}$ \citet{LOMBARDI10} $^{i}$ \citet{REIPURTH08}}
\end{table*}

\section{Data}
\label{sec:data}

\begin{figure*}[tbh]
\plotone{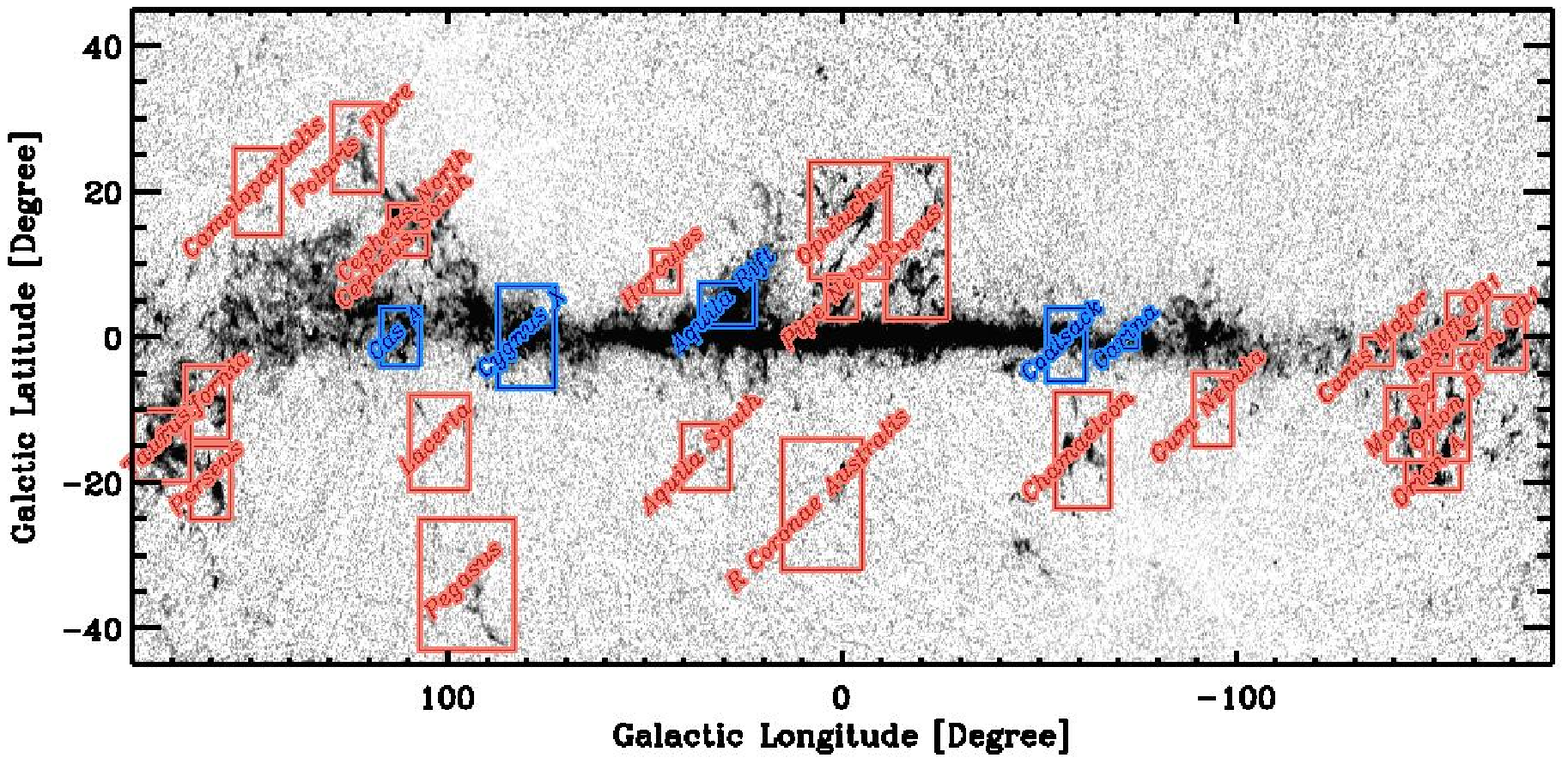}
\caption{\label{fig:coverage} Molecular clouds studied in this paper, highlighted on the {\em Planck} CO map. The rectangles show the regions where we estimate the foreground and background dust column for each molecular cloud. We include only the clouds enclosed by red rectangles in our main analysis. We attempted measurements for clouds marked with blue rectangles, but found them to have too much uncertainty in the foreground and background to yield a reliable dust extinction measurement.}
\end{figure*}

We use the {\em Planck} all-sky CO \citep{PLANCK13_CO} and \ebv\ \citep{PLANCK13_DUST} maps to measure the \icoav\ relationship in local molecular clouds. We consider the clouds listed in Table~1 of \citet{DAME01}, supplemented by a few other well-known nearby regions. Table~\ref{tab:list} lists our targets, and Figure~\ref{fig:coverage} shows their location on the {\em Planck} CO map of the Milky Way.

\subsection{CO Map}
\label{sec:co}

The {\em Planck} team provided three different types of CO maps, all extracted from the HFI broadband photometric data at $100$, $217$, and $353$ GHz \citep{PLANCK13_CO}. In this paper, we use the ``TYPE~1'' CO 1-0 map, which is generated by the single-channel method (see Section \mbox{4.2.1} of their paper for more information). This map has a lower signal-to-noise (S/N) ratio than the other map types, but it suffers less from foreground and background contamination. As we are interested in molecular clouds near the Galactic plane, we expect contamination to be a major issue and choosing the TYPE~1 map significantly improves the fidelity of our results.

The native angular resolution of the {\em Planck} TYPE~1 CO map is $9.65\arcmin$. We convolve this map using a Gaussian kernel to $18\arcmin$ in order to improve the S/N ratio. At the typical distance of ${\sim}200$~pc for our cloud sample, this corresponds to ${\sim}1$~pc spatial resolution. The {\em Planck} TYPE~1 CO map includes a contribution from $^{13}$CO. We correct for this by dividing the provided map by $1.11$, assuming a constant $^{13}$CO/$^{12}$CO ratio following \citet{PLANCK13_CO}. 
After correction, we compared the {\em Planck} CO map with the all-sky CO map by \citet{DAME01} along lines of sight toward the molecular clouds considered in our analysis. The two maps appear almost identical in those regions. \citet{PLANCK13_CO} report the typical $1\sigma$ uncertainty of the CO map to be approximately $1.77$ \Kkmpers\ at $15\arcmin$ resolution, with ${\sim}10$ per cent absolute calibration uncertainty due to $^{13}$CO contamination. We prefer the {\em Planck} CO maps because of their better angular resolution and wider sky coverage, but expect that we would have reached the same result using the \citet{DAME01} map.
 
\subsection{$A_{\rm V}$ Maps}

To estimate a dust extinction, \av , map for each cloud, we use a version of the dust reddening, \ebv , map provided by the {\em Planck} team \citep{PLANCK13_DUST}. We use a simple conversion of their dust optical depth at $850$~\micron\ ($\tau_{850}$), which is the result of a modified blackbody fit to the infrared and sub-millimeter SED. \citet{PLANCK13_DUST} found a strong correlation between \ebv\ from SDSS quasar studies and dust optical depth for diffuse lines of sight at high Galactic latitude ($\ebv < 0.1$, see their Figure~22). We adopt their best fit, $\ebv = 1.49 \times 10^4~\tau_{850}$. We multiply then by $R_\mathrm{V} = 3.1$ to obtain the dust extinction, \av .

We expect this conversion from $\tau_{850}$ to \av\ to appropriate for the diffuse ISM, but to also have some dependence on environment. Studying the Taurus molecular clouds, the {\em Planck} team found the {\em Planck} \ebv\ from $\tau_{850}$ to be ${\sim}25\%$ higher than the \ebv\ derived from the NIR color excess method. We expect that a similar uncertainty in the translation to \av\ may exist in the other molecular clouds. Nevertheless, the correlations between the NIR-based \ebv\ map and the {\em Planck} \ebv\ map are quite strong \citep[see Table~5 in][]{PLANCK13_DUST}. We expect that the qualitative features of our results (e.g.\ the shape of \icoav\ relation) remain reliable, as long as dust optical properties do not change much within a molecular cloud. Similarly, our expressed \av\ may shift if one applies a higher $R_\mathrm{V}$ but only differential changes within a cloud will affect the shape of our results. Still, we caution that the rigorous way to read our results is as a correlation between \ico\ and dust optical depth, $\tau_{850}$, which is closely related to the dust column density.

The angular resolution of the {\em Planck} \ebv\ map is $5\arcmin$. We degrade this resolution to match that of the CO map. The fractional uncertainty in $\tau_{850}$ at its native resolution is about 10 per cent in diffuse regions and typically $2{-}5$ per cent in infrared-bright regions \citep{PLANCK13_DUST}. However, the systematic uncertainty associated with converting $\tau_{850}$ to \av\ and removing the foreground and background contamination is larger.

\begin{figure*}
\plottwo{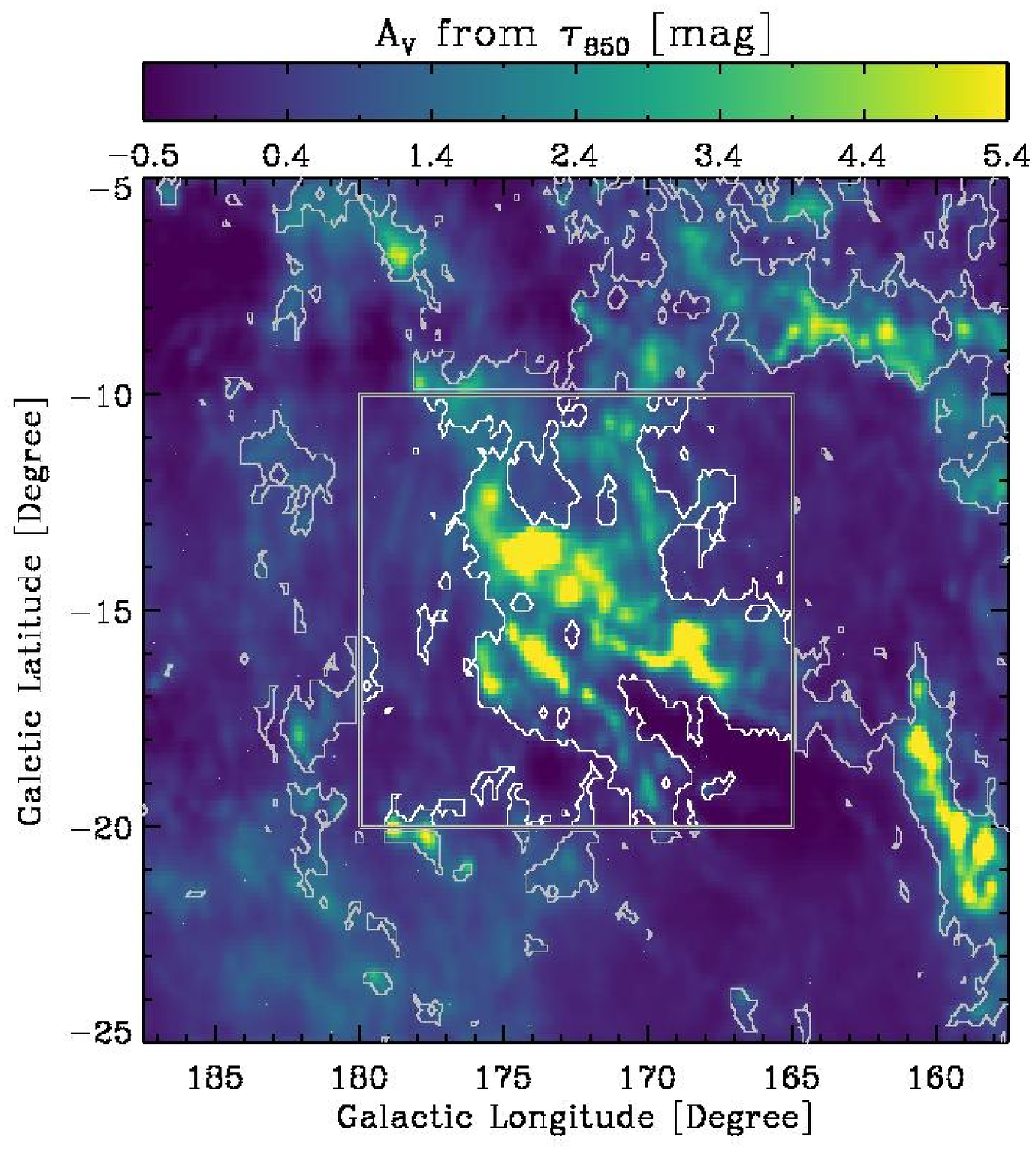}{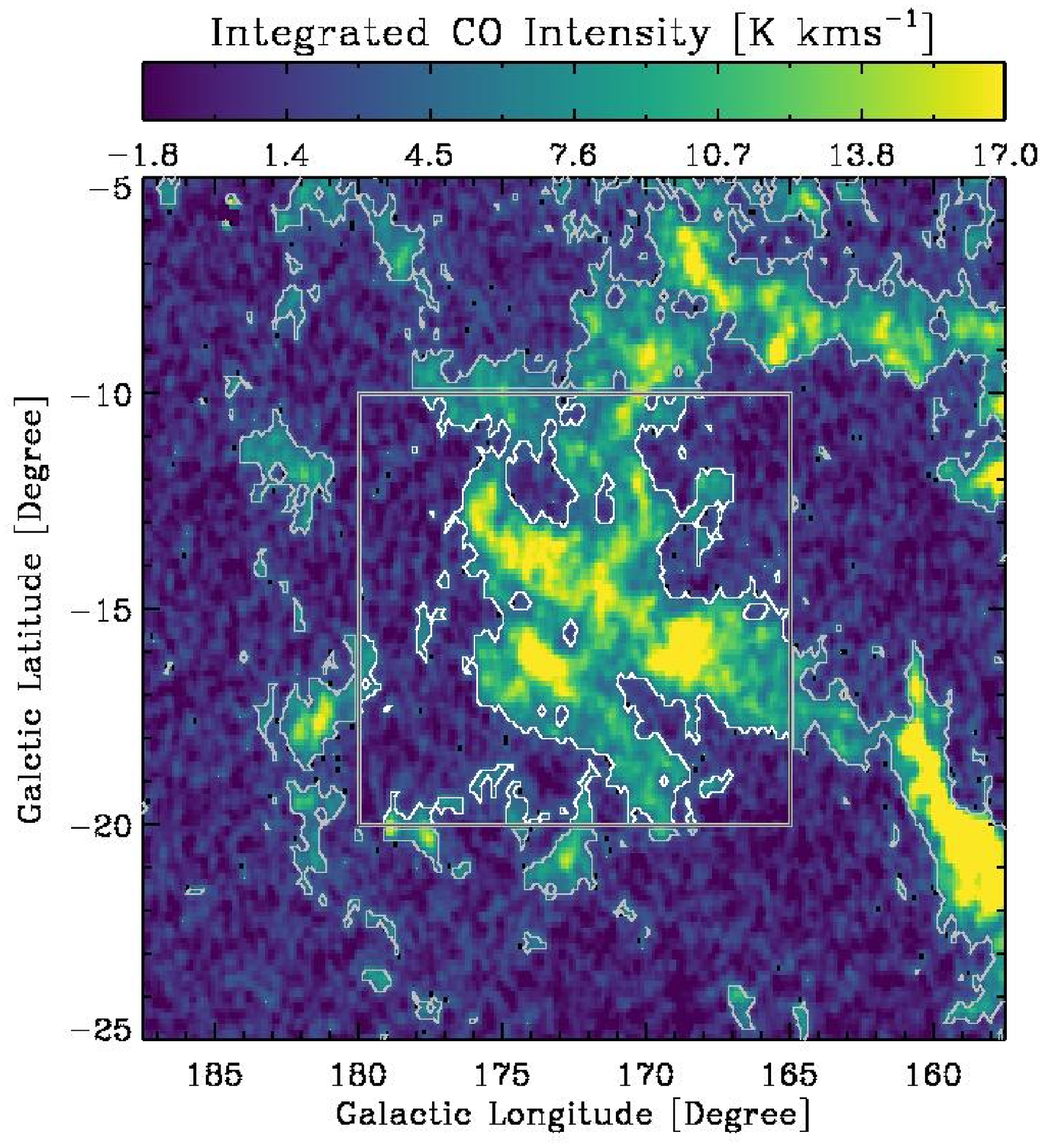}
\plottwo{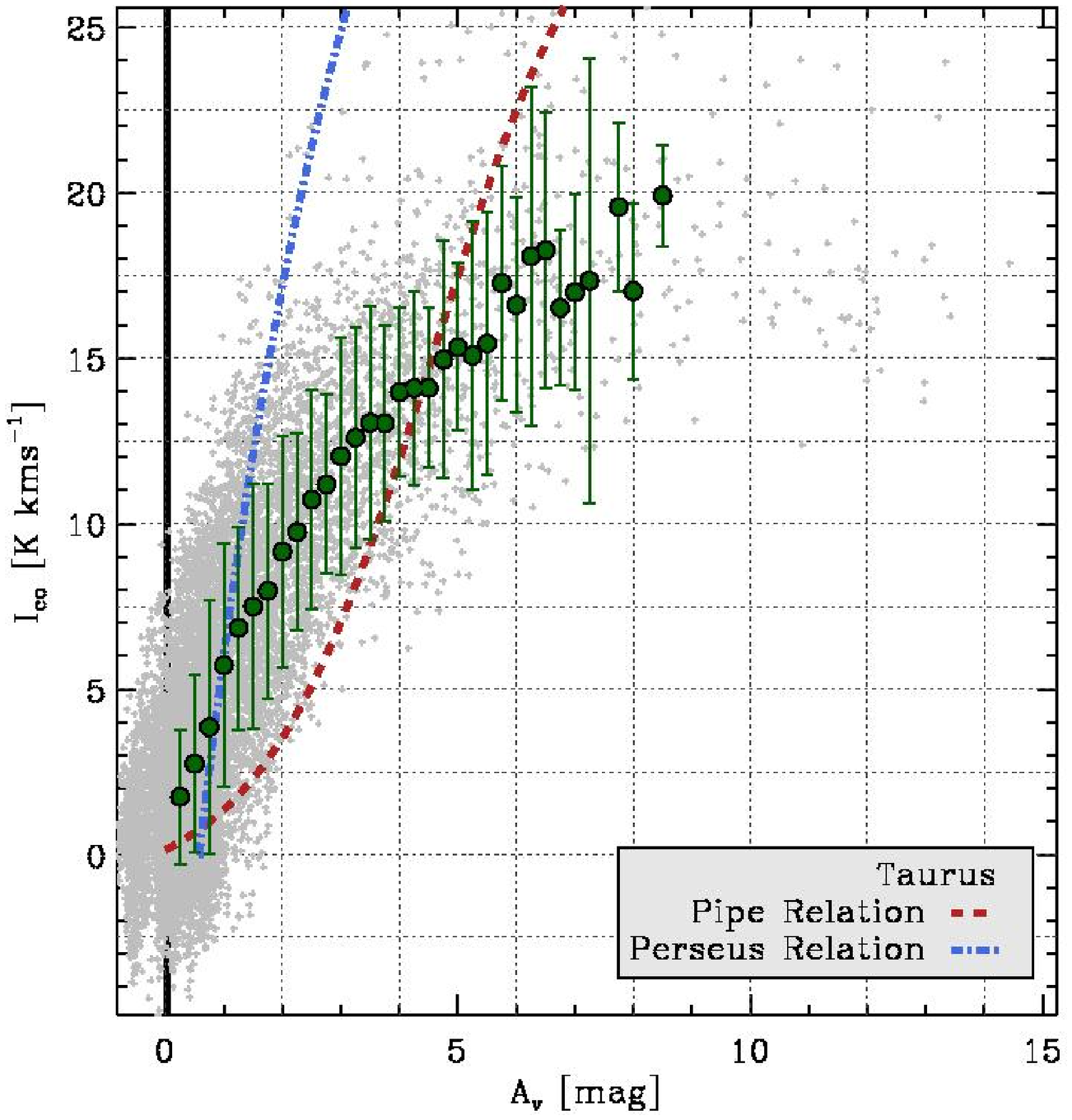}{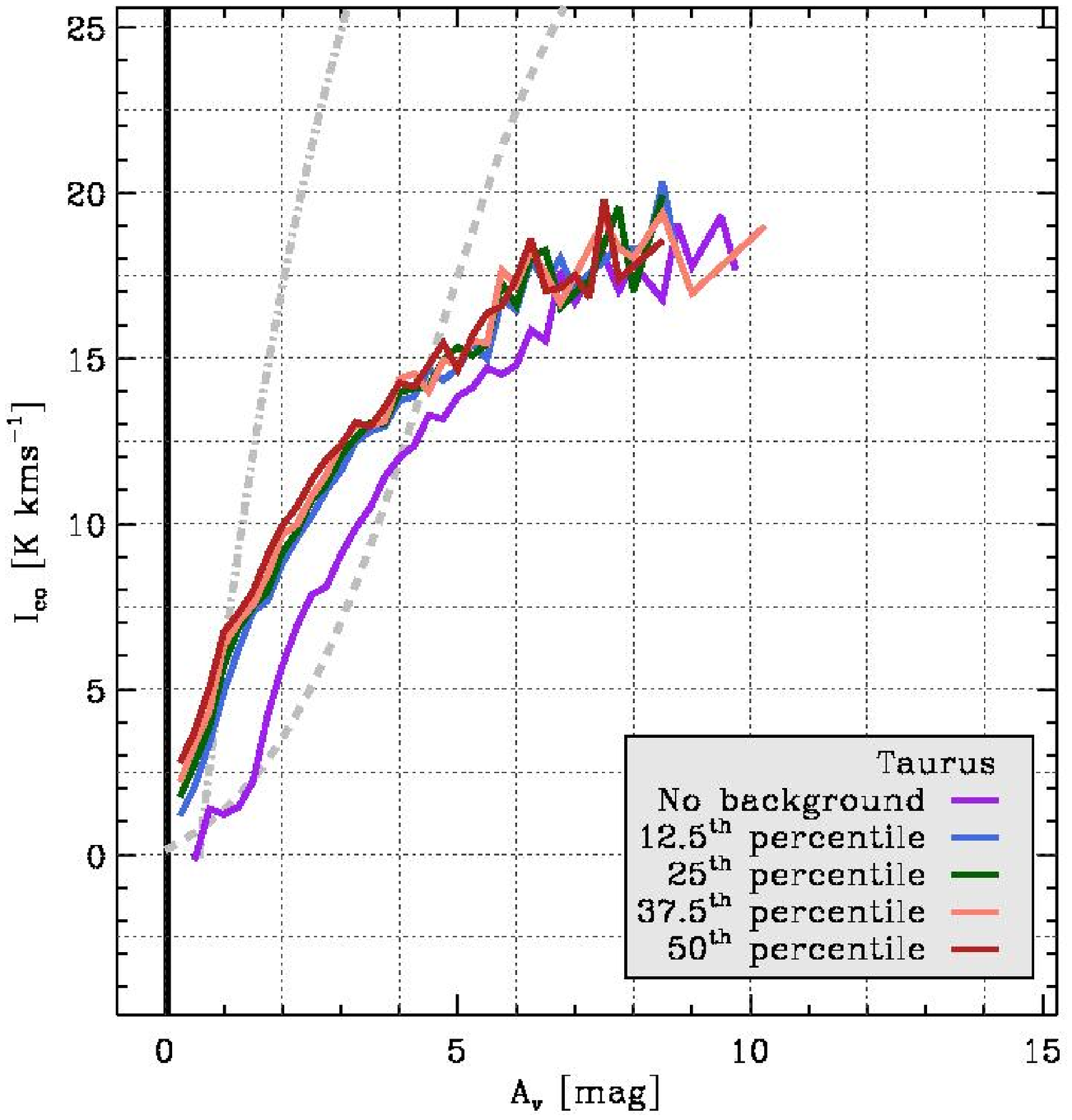}
\caption{\label{fig:taurus} The distribution of dust extinction ({\em top left}), integrated CO intensity ({\em top right}), and the relationship between the two ({\em bottom left and bottom right}) in the Taurus molecular cloud. The rectangle in the upper panels indicates the ``cloud'' region used for the measurement. The white and gray contours mark the region of bright CO emission (${\rm S/N} > 3$). We estimate the contribution of foreground and background dust emission not associated with the cloud outside the marked rectangular region, also excluding the CO-bright regions. The bottom left panel shows the $\icoav$ relationship after background subtraction (Section~\ref{sec:bg}). Here, the gray points are individual lines of sight and the green circles are the binned profile treating \av\ as an independent variable. Error bars show $\pm1\sigma$ scatter about the mean in each bin, and the black line indicates five times the rms \av\ fluctuations in the background region. The bottom right panel shows the same binned profile constructed using different methods for background subtraction: no subtraction (purple) and then varying the level used for the subtraction about our fiducial $25^{\rm th}$ percentile case. In both panels, we plot the sub-parsec \icoav\ relation for the Pipe Nebula \citep{LOMBARDI06} and the Perseus molecular cloud \citep{PINEDA08}.}
\end{figure*}

\subsection{Contamination Subtraction from $A_{\rm V}$ Maps}
\label{sec:bg}

We aim to measure the amount of dust shielding column associated with a molecular cloud along each line of sight and compare it to the CO emission from the same cloud along the same line of sight. Unfortunately, our location within the Galaxy and the lack of velocity information for the dust continuum make it difficult to separate the emission associated with the molecular cloud from background and foreground emission \citep[for a rare exception that proves the rule, see][]{LEE14}. This difficulty also represents one of the major obstacles to accurate measurements of the column density PDF in molecular clouds using dust continuum emission. In that context, the issue has been discussed and solutions have been proposed in several recent papers including \citet{LOMBARDI15} and \citet{SCHNEIDER15}.

Here, we adopt a simple approach to correct for the foreground and background contamination, following \citet{SCHNEIDER15} with slight modifications. The basic idea is to calculate the level of contamination from the sky near the cloud. To define a reference region, we consider an area that extends twice the length of the nominal cloud region (Table~\ref{tab:list}) in Galactic latitude. Next, we mask the cloud region as well as any CO-bright regions outside the cloud (see Figure \ref{fig:taurus}). The remaining unmasked lines of sight represent the sky near the molecular cloud without significant contribution from other molecular clouds. 

\citet{SCHNEIDER15} adopt a singe value of \av, usually the minimum in the reference region, to represent the contamination for a given cloud. We modify this approach to allow the level of contamination to depend on Galactic latitude. This reflects the fact that clouds near the Galactic plane show a strong gradient in their level of contamination with Galactic latitude. This matches the expectation as the path length through the disk increases rapidly as one goes from high Galactic latitude toward the Galactic plane. 

To calculate this latitude-dependent level of contamination, we measure the $25^{\rm th}$ percentile \av\ value in each $0.5$ degree-wide horizontal stripe of the reference region. We take this value as a representative dust contamination level for parts of the cloud with the same Galactic latitude. Then, we subtract this contamination from the {\em Planck} \av\ map to isolate the \av\ that belongs to the targeted cloud. We report the average \av\ contamination for each cloud in Table~\ref{tab:list}.

We tried different statistics (e.g., median, mean, etc.) to measure the level of contamination. As the bottom right panel in Figure \ref{fig:taurus} illustrates, the largest difference is between doing nothing (the purple line) and doing something (the other lines). The choice of statistic plays a secondary role, and taking the $25^{\rm th}$ percentile in the reference region appears to be a good compromise. This avoids the case of widespread negative values of \av\ in the cloud, which can occur using the mean or median ($50^{\rm th}$ percentile). This may be due to imperfect masking, e.g., if the reference region contains some lines of sight associated with the molecular cloud complex even when masking additional CO-bright regions. Or it may stem from other clouds overlapping with the reference region. Overall, the choice of statistic for background subtraction has a minor impact on our results compared to other factors like the choice of reference region \citep[see the discussion in][]{SCHNEIDER15}.

\begin{figure*}
\plottwo{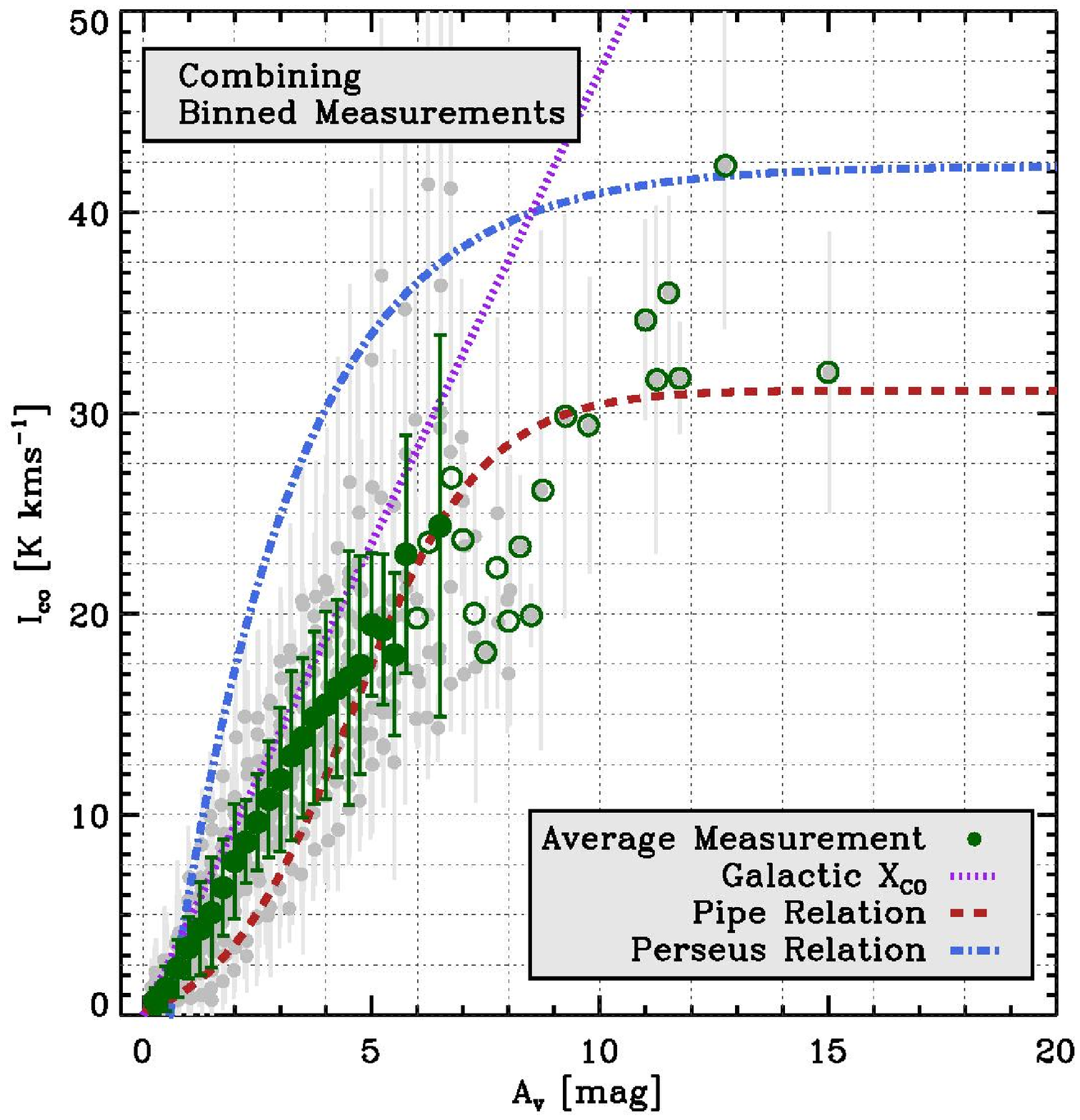}{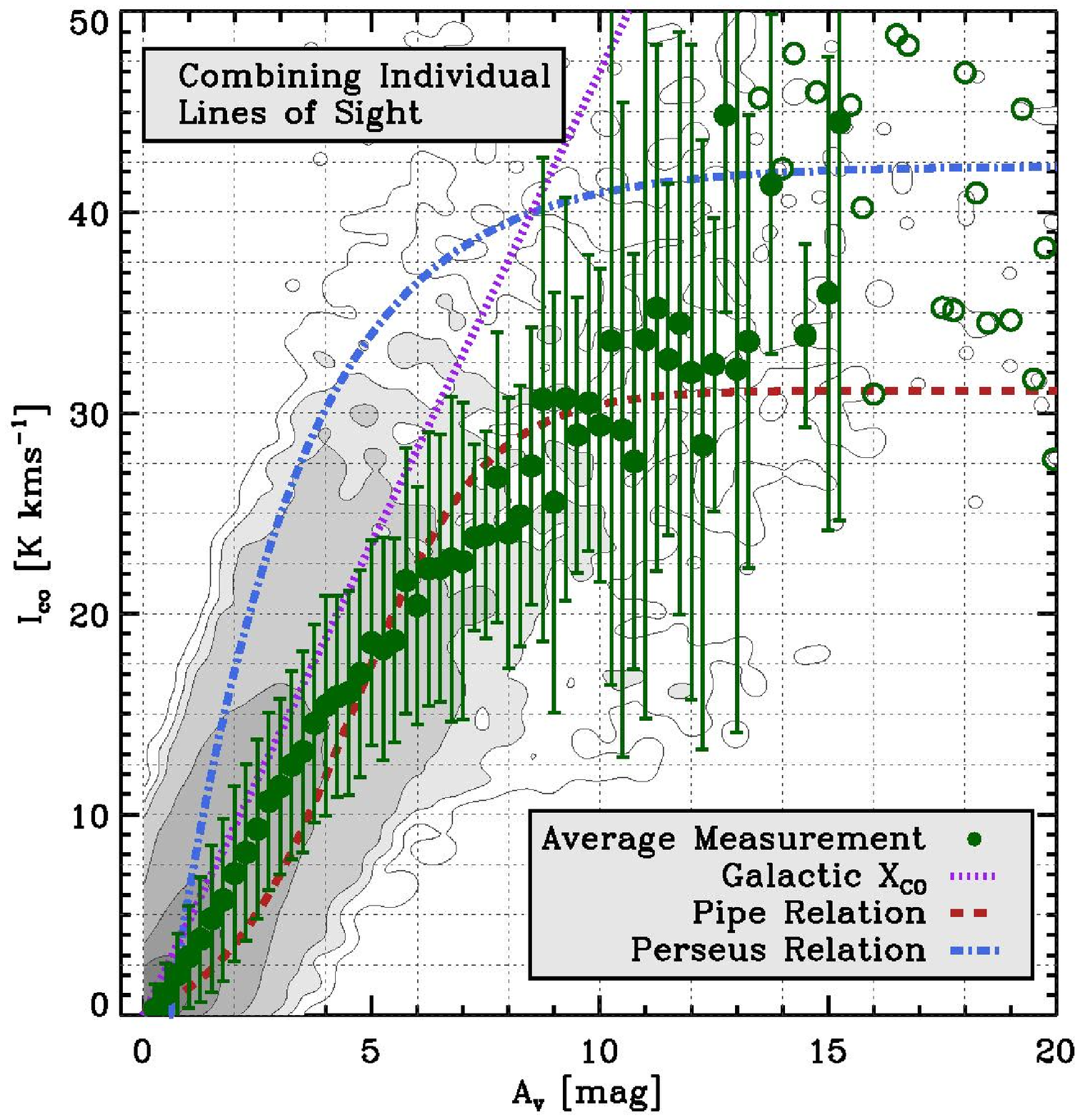}
\caption{\label{fig:synthesis} \ico\ as a function of \av\ combining the 24 local molecular clouds listed in Table~\ref{tab:list}. ({\em left}) The average relation calculated by combining the relations for the individual clouds. Each gray point here shows a bin from an individual cloud; the green profile shows the mean and scatter among the profiles. This approach weights each cloud equally. Unfilled circles mark the average \ico\ in the \av\ bins where we do not have enough statistics to estimate the scatter. ({\em right}) The average relation, now calculated treating each line of sight in the sample equally, so that larger clouds contribute more. The gray contours show data density and are chosen to encompass $99.9\%$, $99.5\%$, $99\%$, $95\%$, $75\%$, and $50\%$ of the data. The green points again show the mean relation and the error bars indicate the $1\sigma$ scatter in the bin. Blue and red lines show the sub-parsec resolution \icoav\ relations measured for the Pipe Nebula \citep{LOMBARDI06} and Perseus molecular cloud \citep{PINEDA08}. }
\end{figure*}

\section{Results}
\label{sec:results}

\subsection{$I_{\rm CO} {-} A_{\rm V}$ Relation for Individual Nearby Clouds}
\label{sec:icoav_synthesis}

\begin{table}
\begin{center}
\caption{\ico\ at a given \av\ in Galactic Molecular Clouds}
\label{tab:result}
\begin{tabular}{llcc}
\hline \hline
Cloud & $A_{\rm V}$ & $I_{\rm CO}$ \\ 
& (mag) & (K km s$^{-1}$) \\ 
\hline \hline
Aquila South  &  0.25& $ 0.2 \pm  1.2$ \\
Aquila South  &  0.50& $ 0.6 \pm  1.4$ \\
Aquila South  &  0.75& $ 1.4 \pm  1.6$ \\
Aquila South  &  1.00& $ 2.2 \pm  1.7$ \\
Aquila South  &  1.25& $ 2.3 \pm  1.0$ \\
Aquila South  &  1.50& $ 0.8 \pm  0.8$ \\
California  &  0.25& $ 0.7 \pm  1.6$ \\
California  &  0.50& $ 1.5 \pm  2.0$ \\
California  &  0.75& $ 2.7 \pm  2.1$ \\
California  &  1.00& $ 3.9 \pm  2.5$ \\
California  &  1.25& $ 5.1 \pm  2.9$ \\
 \nodata & \nodata & \nodata \\
\hline \hline
\end{tabular}

\end{center}
\tablecomments{This is a stub. The full data are available as a machine readable table distributed with the paper. Mean and scatter of \ico\ in $0.25$~mag wide bins of \av\ estimated from $\tau_{850}$. We report results for bins with at least $6$ data points.}
\end{table}

Figure~\ref{fig:coverage} shows the location of the clouds studied in this work, overlaid on the {\em Planck} Galactic CO map. Clouds shown in blue boxes were initially considered but excluded from the sample because of heavy contamination. Other than these, most of the clouds are located well outside of the Galactic plane, making it easier to compare the line of sight dust extinction and integrated CO intensity. Our cloud sample includes the two molecular clouds, the Pipe Nebula and Perseus molecular cloud, in which the $\icoav$ relationship has been measured by previous works \citep{LOMBARDI06, PINEDA08, LEE14}. Although at coarser spatial resolution than these previous studies, we here increase the number of clouds with a parsec-scale measurement of the \icoav\ relationship by an order of magnitude (from two to $24$). Doing so, we attempt to distill a general relationship for local molecular clouds, capture the intrinsic spread, and search for physical variations between clouds. 
Figure~\ref{fig:taurus} shows an example of our analysis for the Taurus molecular cloud. Similar figures for the other clouds can be found in the Appendix. In the top panels, we plot the maps of estimated visual extinction (\av) and integrated CO intensity (\ico). We find general coincidence between the location of bright CO emission and highly shielded lines of sight in the cloud region (marked with a black rectangle). This agrees qualitatively with the theoretical picture that shielding from the dissociating radiation by dust is the main factor in setting the extent of widespread CO emission \citep[e.g,][]{WOLFIRE10,GLOVER11}.

The bottom left panel shows the \icoav\ relationship for Taurus. We plot integrated CO intensity as a function of line of sight dust extinction, with individual ${\sim}1$~pc diameter lines of sight shown as gray points. The ensemble of individual lines of sight show large scatter, far greater than the observational uncertainties. This is somewhat expected not only from the local variation of physical conditions (e.g., temperature, radiation field, turbulence) within a cloud \citep[see][]{PINEDA08}, but also due to complex cloud geometry and projection effects. Physically, we expect that the CO abundance at any given point in a cloud will depend on a weighted average of the dust column in different directions from that point to the edge of the cloud, whereas we measure only the dust column along a single line of sight. Numerical simulations of turbulent molecular clouds find that although these quantities are correlated, there is considerable scatter in this relationship \citep[see e.g.][]{CLARKGLOVER14}, potentially explaining much of the scatter in the observed \icoav\ relationship.

To distill a representative relationship, we calculate the average \icoav\ relation in each cloud by combining many lines of sight. Doing so, we treat \av\ as the independent variable and estimate the median and standard deviation of integrated CO intensity in bins of \av. In Figure~\ref{fig:taurus}, this binned profile appears as the green circles with error bars indicating the rms ($1\sigma$) scatter of the CO emission within a given \av\ bin. For comparison, we also plot the \icoav\ relationships in the Pipe Nebula (dashed line; \citealt{LOMBARDI06}) and Perseus molecular cloud (dash-dotted line; \citealt{PINEDA08}) measured at sub-parsec scales.

In Table~\ref{tab:result}, we report the average integrated CO intensity in bins of \av\ for each cloud. We present results for \av\ bins in which there are enough pixels ($n_\mathrm{pix} > 6$) to calculate the mean \ico\ and the standard deviation about this value. Generally, we find that \ico\ increases with increasing \av\ in our clouds, though there is non-negligible scatter both among clouds and within individual clouds. Qualitatively, some of the clouds show hints of a minimum \av\ threshold for detectable CO emission ($A_\mathrm{V,\,thres}$) or a saturation of CO emission at high \av, but not all clouds exhibit the same features.  We come back to this point in Section~\ref{sec:icoav_shape} where we compare the \icoav\ relationships among clouds, focusing on their various shapes.

Although we treat \av\ as an independent variable, we emphasize that our derived \icoav\ relation remains uncertain along the $x$-axis (\av) for several reasons. Key uncertainties, discussed above, include spatial variations in dust emissivity (which relates dust optical depth, $\tau$, to dust extinction, \av) and uncertainty in the removal of foreground and background contamination. Furthermore, even if we estimate \av\ perfectly, our measured relationship may not perfectly reflect the physical relationship between dust extinction and CO emission. As mentioned, cloud geometry complicates the ability to relate line of sight dust to the true shielding layer. Finally, although we convert dust optical depth to \ebv\ and \av, the opacity to photons that dissociate CO is the real relevant quantity.

\subsection{An Aggregate \icoav\ Relation for Local Clouds }
\label{sec:icoav_synthesis2}

\begin{table}
\begin{center}
\caption{\ico\ at a Given \av\ for the Whole Sample}
\label{tab:synthesis}
\begin{tabular}{llcc}
\hline \hline
$A_{\rm V}$ & $I_{\rm CO}$ (L.o.S.) & $I_{\rm CO}$ (bins) \\ 
(mag) & (K km s$^{-1}$) & (K km s$^{-1}$) \\ 
\hline \hline
 0.25& $ 0.3 \pm  1.2 $& $ 0.7 \pm  0.6 $ \\
 0.50& $ 1.1 \pm  1.5 $& $ 1.3 \pm  1.1 $ \\
 0.75& $ 2.0 \pm  2.0 $& $ 2.3 \pm  1.4 $ \\
 1.00& $ 2.9 \pm  2.5 $& $ 3.4 \pm  1.5 $ \\
 1.25& $ 3.8 \pm  3.1 $& $ 4.3 \pm  2.3 $ \\
 1.50& $ 4.8 \pm  3.7 $& $ 5.1 \pm  2.7 $ \\
 1.75& $ 5.7 \pm  4.1 $& $ 6.4 \pm  2.4 $ \\
 2.00& $ 7.1 \pm  4.4 $& $ 7.7 \pm  2.8 $ \\
 2.25& $ 8.1 \pm  4.4 $& $ 8.6 \pm  2.1 $ \\
 2.50& $ 9.3 \pm  4.5 $& $ 9.6 \pm  2.4 $ \\
 2.75& $10.6 \pm  4.4 $& $10.8 \pm  2.9 $ \\
 3.00& $11.4 \pm  4.4 $& $11.8 \pm  3.6 $ \\
 3.25& $12.5 \pm  4.7 $& $12.9 \pm  4.2 $ \\
 3.50& $13.1 \pm  5.0 $& $13.8 \pm  4.0 $ \\
 3.75& $14.5 \pm  4.9 $& $14.8 \pm  4.3 $ \\
 4.00& $15.4 \pm  5.5 $& $15.4 \pm  4.7 $ \\
 4.25& $15.9 \pm  5.0 $& $16.3 \pm  4.4 $ \\
 4.50& $16.1 \pm  5.1 $& $16.8 \pm  6.3 $ \\
 4.75& $17.0 \pm  5.2 $& $17.4 \pm  5.4 $ \\
 5.00& $18.5 \pm  5.1 $& $19.5 \pm  3.6 $ \\
 5.25& $18.3 \pm  5.5 $& $19.2 \pm  3.8 $ \\
 5.50& $18.7 \pm  5.1 $& $18.0 \pm  4.1 $ \\
 5.75& $21.6 \pm  6.6 $& $23.0 \pm  5.9 $ \\
 6.00& $20.4 \pm  5.9 $& \nodata  \\
 6.25& $22.2 \pm  6.8 $& \nodata  \\
 6.50& $22.3 \pm  6.7 $& $24.4 \pm  9.5 $ \\
 6.75& $22.7 \pm  8.1 $& \nodata  \\
 7.00& $22.6 \pm  7.9 $& \nodata  \\
 7.25& $23.8 \pm  4.6 $& \nodata  \\
 7.50& $23.9 \pm  5.2 $& \nodata  \\
 7.75& $26.8 \pm  7.2 $& \nodata  \\
 8.00& $24.0 \pm  6.8 $& \nodata  \\
 8.25& $24.9 \pm  6.5 $& \nodata  \\
 8.50& $27.4 \pm  6.9 $& \nodata  \\
 8.75& $30.7 \pm 12.0 $& \nodata  \\
 9.00& $25.5 \pm 10.5 $& \nodata  \\
 9.25& $30.7 \pm 10.1 $& \nodata  \\
 9.50& $28.9 \pm  6.9 $& \nodata  \\
 9.75& $30.5 \pm  7.4 $& \nodata  \\
10.00& $29.4 \pm  7.8 $& \nodata  \\
10.25& $33.6 \pm 17.1 $& \nodata  \\
10.50& $29.1 \pm 16.3 $& \nodata  \\
10.75& $27.6 \pm 10.4 $& \nodata  \\
11.00& $33.7 \pm 18.9 $& \nodata  \\
11.25& $35.2 \pm 13.1 $& \nodata  \\
11.50& $32.7 \pm  8.8 $& \nodata  \\
11.75& $34.5 \pm 14.5 $& \nodata  \\
12.00& $32.0 \pm 16.3 $& \nodata  \\
12.25& $28.4 \pm 15.2 $& \nodata  \\
12.50& $32.4 \pm  7.3 $& \nodata  \\
12.75& $44.8 \pm  9.8 $& \nodata  \\
13.00& $32.2 \pm 18.1 $& \nodata  \\
13.25& $33.6 \pm 11.3 $& \nodata  \\
13.50& \nodata & \nodata  \\
13.75& $41.4 \pm  8.5 $& \nodata  \\
14.00& \nodata & \nodata  \\
14.25& \nodata & \nodata  \\
14.50& $33.8 \pm  4.6 $& \nodata  \\
14.75& \nodata & \nodata  \\
15.00& $36.0 \pm 11.8 $& \nodata  \\
\hline \hline
\end{tabular}

\end{center}
\tablecomments{Mean and scatter of \ico\ in $0.25$~mag wide bins of \av\ estimated from $\tau_{850}$. L.o.S.: averaging all lines of sight. Bins: averaging binned profiles for individual clouds. We report results for bins with at least $6$ data points.}
\end{table}

A main goal of this study is to characterize the typical CO intensity at a given \av\ and the associated scatter in a typical Solar Neighborhood molecular cloud. That is, we aim to synthesize a typical \icoav\ relationship. To do this, we combine results for our full sample using two different weighting schemes. First, we equally weight each molecular cloud, then we equally weight each individual line of sight.

Figure~\ref{fig:synthesis} shows this synthesis. The left panel plots \ico\ as a function of \av , combining measurements of the \icoav\ relation for individual clouds.  Here, the cloud-averaged Milky Way \icoav\ relation (shown as green circles with scatter indicated by the error bars) is calculated by averaging the binned \icoav\ profiles of individual clouds. This gives each cloud equal weight and prevents the few clouds with large angular extent from dominating the result. The right panel shows the complementary result. Here, we give all lines of sight across our sample equal weight. In both panels, but especially the right one (equally weighting all lines of sight), the large density of data points in the bottom left corner illustrates that most lines of sight have low \av\ and low \ico .

The two weighting schemes result in similar \ico\ for a given \av\ bin, which lends confidence to the generality of our result. Both methods suggest that averaged over our sample, \ico\ increases close to linearly with increasing \av\ in the low-to-intermediate \av\ regime ($\av \le 4$ mag). As \av\ increases further ($\av \ge 4$ mag), the slope of the \icoav\ relation becomes shallower. This provides some suggestion of the saturation of CO emission due to high optical depth at high \av , though not strong evidence. Evidence for a minimum \av\ threshold for CO emission ($A_\mathrm{V,\,thres}$) is even weaker in the aggregate relations, while this feature is predicted by theoretical PDR models and numerical simulations of molecular clouds \citep{WOLFIRE10, GLOVER11}. We return to these features in detail below.

\section{Discussion}

Our average \icoav\ relationship (Figure~\ref{fig:synthesis}) and those for individual clouds (see Appendix) highlight that \ico\ for a given \av\ is similar across our sample (see also Table~\ref{tab:result}). If we compare our results with previous measurements, we find that the Perseus results of \citet{PINEDA08} form an upper envelope for our data, while our average \icoav\ relation tracks the \citet{LOMBARDI06} result for the Pipe Nebula below $\av \sim 5$~mag. Our results also echo the findings of some of the earliest CO studies, which compared dust extinction to CO emission in nearby molecular clouds \citep[e.g.,][]{DICKMAN78,LISZT82,YS82}. For instance, \citet{YS82} find similar CO intensity for a fixed \av\ in infrared dark clouds and giant molecular clouds in the Galaxy, $\ico / \av \approx 2.35$ $\Kkmpers\ {\rm mag}^{-1}$  (see Table~6 in the Appendix of their paper), using dust extinction and cloud virial mass estimates to compare with CO intensities. Qualitatively, the good match between dust and CO emission reinforces the idea that shielding by dust plays the primary role in defining the location of bright CO emission within molecular clouds \citep{WOLFIRE10,GLOVER11}.

In detail, a number of questions remain: is the normalization of our measurement consistent with expectations? How should we understand the weakness or absence of the expected threshold and saturation features? Are our results consistent with numerical simulations and observations of other galaxies? And if our measured pc-scale relation is universal, what are the implications for the metallicity dependence of the \cohtwo\ conversion factor? In this section we address each of these topics.

\subsection{Recasting the \icoav\ Relation in Terms of the \cohtwo\ Conversion Factor}
\label{sec:xco}

\begin{figure}
\plotone{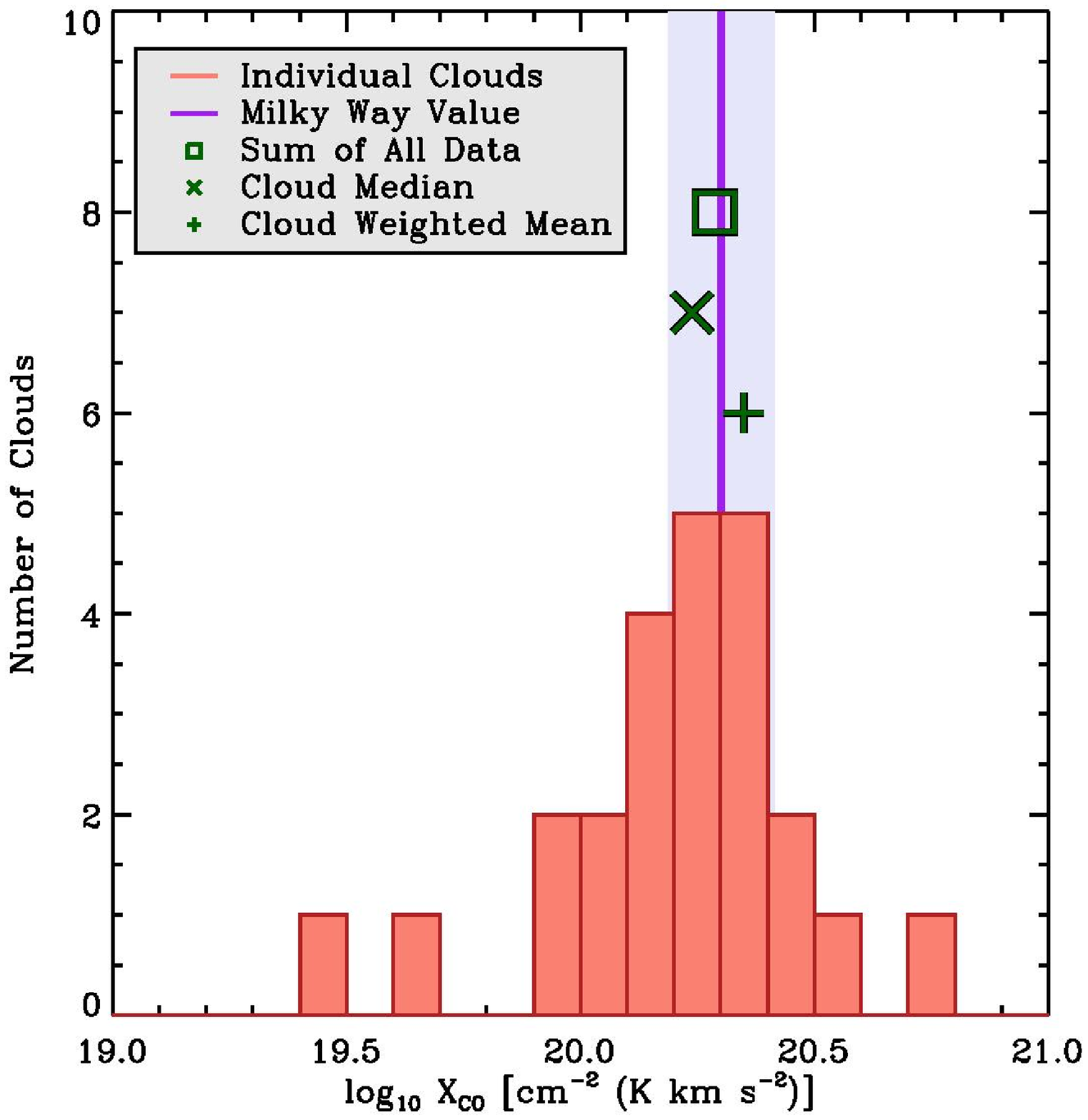}
\caption{\label{xco_histogram} The histogram of implied \cohtwo\ conversion factors in our sample Galactic molecular clouds from the {\em Planck} dust and CO maps. The histogram plots one value for each cloud. The purple line and shaded region show the recommended Milky Way value of $2{\times}10^{20}$~\xcounits\ with a $\pm 30\%$ uncertainty from \citet{BOLATTO13}. The green points show different average values derived from our data assuming a standard dust-to-gas ratio and an \hi\ shielding layer of $0.2$~mag. The median of all cloud values, a mass-weighted mean, and a calculation weighting each line of sight equally all yield values close to the nominal Galactic value. This close agreement between the two offers a sanity check on our measured \icoav\ relation.}
\end{figure}

Dust and gas are often well mixed in the ISM, so that an approach similar to what we present here has been used to study the \cohtwo\ conversion factor directly by comparing the column density of gas implied by dust to the integrated CO intensity. Though we are primarily interested in the actual relationship between \ico\ and \av , recasting our results in terms of the implied \xco\ factor offers a useful check on the normalization of our results. We do so following the equation below:

\begin{equation}
\label{eq:xco}
X_{\rm CO} = \frac{N({\rm H}_2)}{I_{\rm CO}} = \frac{\frac{N_{\rm H}}{A_{\rm V}}(A_{\rm V}-2A_{\rm V}^\mathsc{Hi})}{2I_{\rm CO}}~.
\end{equation} 

\noindent Here $N_{\rm H}$ refers to the total column density of hydrogen atoms in either the atomic or molecular phase. $A_\mathrm{V}^\mathsc{Hi}$ is the visual extinction into a cloud where the dominant gas phase transitions from {\sc Hi} to H$_2$ \citep[see][]{STERNBERG14}. The factor of 2 in front of $A_\mathrm{V}^\mathsc{Hi}$ accounts for the fact that our \av\ from dust emission probes the entire line of sight, and so includes the {\sc Hi} shielding layer on both the front and the back side of the cloud.  The ratio of total hydrogen column density to visual extinction, $N_{\rm H} / \av$, is observed to be $1.87 \times 10^{21}$ ${\rm cm}^{-2}~{\rm mag}^{-1}$ in diffuse Milky Way lines of sight \citep{BOHLIN78}. 

Following Eq.~\ref{eq:xco}, if $A_\mathrm{V}^\mathsc{Hi} = 0$ then a constant \xco\ factor would appear as a straight line in the \icoav\ space. With a finite but small $A_\mathrm{V}^\mathsc{Hi}$, the curve describing a fixed conversion factor has $I_{\rm CO} = 0$ below $A_\mathrm{V}^\mathsc{Hi}$, rises quickly, then asymptotes to a straight line. The purple lines in Figure~\ref{fig:synthesis} shows the standard \xco\ in the Milky Way ($2{\times}10^{20}$ \xcounits , \citealt{BOLATTO13}) for the \citet{BOHLIN78} $N_\mathrm{H} / A_\mathrm{V}$ ratio and neglecting $A_\mathrm{V}^\mathsc{Hi}$. Our data exhibit a slightly shallower slope than this fiducial line (which has a slope of $4.7$ \Kkmpers\ ${\rm mag}^{-1}$), suggesting a slightly higher \xco\ in our sample.

We also directly calculate the implied \xco\ for individual clouds from the maps of \av\ and \ico\ and Eq.~\ref{eq:xco}. Again, we adopt the \citet{BOHLIN78} $N_\mathrm{H} / A_\mathrm{V}$, and now assume that the transition from \hi\ to \htwo\ takes place approximately at $A_\mathrm{V}^\mathsc{Hi} = 0.2$~mag for a fixed radiation field \citep{DRAINE78}, as motivated by \citet{KRUMHOLZ09,STERNBERG14}.

Figure~\ref{xco_histogram} shows the resulting distribution of \xco\ values for our sample of local molecular clouds. The median value treating each cloud as one measurement is $\xco \approx 10^{20.2}$ \xcounits\ and the cloud-to-cloud scatter is ${\sim}0.2$~dex, though with a few significant outliers. If we instead weight each cloud by its mass, to derive a weighted Solar Neighborhood value, we find a mean $\xco \approx 10^{20.3}$ \xcounits , while summing over all of our data to derive a single value, we find $\xco \approx 10^{20.35}$ \xcounits .

All of our estimates agree well with the standard Galactic value, which may not be surprising given that this value is partially based on {\em Planck} and other dust results. Bearing in mind that we expect some departures from the assumed fixed $N_\mathrm{H} / A_\mathrm{V}$ and $A_\mathrm{V}^\mathsc{Hi}$, our \icoav\ results appear consistent with the literature on the \cohtwo\ conversion factor. This provides an important sanity check on our overall measurement.

\subsection{Shape of the \icoav\ Relation}
\label{sec:icoav_shape}

On theoretical grounds, one would expect to observe several features in a highly resolved \icoav\ relation. These should reflect the physics of CO emission from PDRs. 

First, one would expect to see a threshold visual extinction, $A_\mathrm{V,\,thres}$, below which the CO emission drops rapidly due to photodissociation of CO molecules \citep{VB88, VISSER09}. Physically, this reflects the transition of the dominant carbon reservoir from {\sc Cii} in the poorly shielded outskirts of clouds to CO in the well-shielded interiors \citep[e.g.][]{TIELENS85}. Outside this transition, in regions where most carbon is C and {\sc Cii}, gas and dust still exist, but they lack abundant CO molecules. Thus, we expect the emissivity of gas in CO to be much lower in the outer parts of clouds. This {\sc Cii}-to-CO transition is associated with a particular amount of shielding \citep[see][]{WOLFIRE10}. Assuming that the line-of-sight dust column traces the shielding of the gas, then we would expect a drop in $I_{\rm CO}$-per-$A_V$ below some $A_{V,\,thres}$. Recent theoretical models estimate a threshold for bright CO emission to be $A_\mathrm{V,\,thres} \sim 1{-}2$~mag \citep[e.g.][]{WOLFIRE10, GLOVER11, GLOVER12A}.

One does not truly expect to find no CO below this threshold, as UV absorption studies find both \htwo\ and CO down to very low column densities \citep[e.g.,][]{SHEFFER08}. However, at very low columns, the abundance of CO relative to H$_2$ does drop steeply. For example, \citet{SHEFFER08} find $N({\rm CO}) \propto N({\rm H_2})^3$ at modest column densities, demonstrating that indeed much less of the C is included in CO at low column densities in the outskirts of clouds. This phenomenon should manifest as a much steeper slope in the $I_{\rm CO}$-$A_V$ relation below $A_{V,\,thresh}$, leading to very low $I_{\rm CO}-to-A_V$ ratios in this regime. There is evidence for this threshold extinction from CO observations of nearby Galactic molecular clouds \citep{PINEDA08, PINEDA10}, though observations of CO-bright diffuse regions \citep[e.g.,][]{LISZT12} suggest that this simple picture does not capture all of the relevant physics. Beyond the classic PDR models mentioned above, the effect is also evident in high physical resolution simulations, such as the one by \citet{GLOVERCLARK16} that we compare to below.

One also expects a saturation of \ico\ at high \av\ as CO emission becomes optically thick \citep[e.g,][]{SHETTY11} and CO intensity approaches a constant value regardless of gas column. This effect is expected theoretically. For example, the PDFs of integrated CO intensity in turbulent molecular cloud simulations show a `piled-up' feature at some high \ico\ intensity \citep{SHETTY11, GLOVER12B}. It is also observed, with the best-fit relations for the nearby Pipe Nebula and Perseus molecular cloud showing a saturation of \ico\ at $\av \ge 10$~mag (e.g., see Figure~\ref{fig:synthesis}).

The presence or absence of these features have important consequences for the dependence of \xco\ on metallicity. As we will see in Section~\ref{sec:xco_metallicity}, the \av\ threshold plays a crucial role in predicting how \xco\ behaves in low metallicity environments. That can be easily understood in terms of the approach introduced in \citet{LEE15}: for a fixed gas column density PDF, low metallicity (and so a low dust abundance) will shift large amounts of material to have low \av . If CO emission is (almost) totally suppressed in this regime then \xco\ depends strongly on metallicity. The theoretical work by \citet{WOLFIRE10} and \cite{GLOVER11} both show strong \av\ thresholds and consequently strong dependence of \xco\ on metallicity.

\subsubsection{Observed Threshold and Saturation Features}
\label{sec:what_we_see}

Do we see a clear threshold and saturation behavior in our data? As discussed above (Section~\ref{sec:icoav_synthesis}), both the synthesized \icoav\ relationship and the relations of individual clouds show mixed results. There are some cases where we can visually identify these features. Chamaeleon, the Gum Nebula, Hercules, Lupus, Orion~A, and R~Coronae Australis show some evidence for a minimum threshold in \av\ for CO emission. Chamaeleon, Hercules, Lupus, Orion~A, the Pipe Nebula, and Taurus show the saturation behavior. When present, the $A_\mathrm{V,\,thres}$ lies at $0{-}1$~mag, while the CO saturation starts at a large range of \av\ around $2{-}5$~mag with large variations between clouds.

If a cloud shows one of the features, it is likely to have the other feature as well, and most of the clouds with clear features are the closer members of our sample (within $200$~pc), with the exception of the Gum Nebula and Orion~A. This strongly suggests that the spatial resolution plays a major role in our ability to detect these features in the \icoav\ relationship. That is, blending of distinct regions by a large beam appears to remove our ability to cleanly isolate poorly shielded regions, and perhaps also heavily shielded regions. Achieving a physical resolution of $\lesssim\,1$~pc seems to be a necessary condition to be able to visually identify the minimum \av\ threshold for CO emission or the saturation of \ico\ at high \av . Reinforcing this view, we note that we do not find conclusive evidence for these features in the Pipe Nebula and Perseus molecular cloud, where the spatial resolution of the data is $0.7$~pc and $1.3$~pc, respectively. Previous studies of these clouds \citep{LOMBARDI06,PINEDA08} using sub-parsec resolution data did find evidence for a threshold behavior in Perseus and saturation in both clouds. We explore the effect of resolution on the \icoav\ relation in a more quantitative fashion in Section~\ref{sec:resolution}.

Factors other than the spatial resolution also contribute to the observed shape of the \icoav\ relation. Some nearby clouds do not show these features even though the spatial resolution should be good enough to identify them. There are also distant clouds with large \av\ that should show the saturation behavior. Regardless of sub-beam clumping, a beam with average $\av \approx 10$~mag must have a large part of the mass well shielded. However, \ico\ continues to increase as a function of \av\ in those clouds. 

The accuracy of our estimate of the foreground and background contamination also matters critically to the presence or absence of a threshold. As we noted in Section~\ref{sec:bg}, it is notoriously difficult to determine the correct value for the \av\ that is not associated with the clouds. Recent studies of the column density distribution of Galactic molecular clouds \citep{LOMBARDI15, SCHNEIDER15} consider the low end of the column density PDF to be highly unreliable due to the ambiguity in the estimation of material unrelated to the cloud, often citing $\av \lesssim 1$ mag as the limiting regime. Unfortunately, this is exactly the regime where any \av\ threshold for CO emission should emerge. Realistically, because of limited resolution and this uncertainty in the \av\ zero point, our results place only a weak constraint on the presence or absence of an \av\ threshold for CO emission. Even in the case where we find circumstantial evidence for $A_{\rm V,thres}$ by eye, we caution that its value is likely very uncertain.

As for the case of the CO saturation at high \av , the non-detections of such features may arise from a variation of physical conditions inside the molecular clouds such as gas temperature, turbulent line width, CO opacity, and cloud geometry. In Perseus, \citet{PINEDA08} observe a significant variation of parameters that describe \ico\ as a function of \av\ even within this single cloud complex. The clouds missing such features may exhibit similar variations, making the observed CO intensities for high \av\ lines of sight strongly variable. In this case, averaging many lines of sight may not converge to a constant CO intensity; though, we would still expect a shallower slope at high \av .

Finally, we note that not all of our target clouds have enough area that we would expect a large amount of mass at the high \av\ values needed for the CO saturation. In other words, the combination of low cloud mass and coarse resolution means that we may lose the dense parts of the cloud within a few individual pixels. The notable candidates for this effect are Aquila South, Camelopardalis, Lacerta, Pegasus, and the Polaris Flare. 

\subsubsection{Comparison to Theoretical Work}
\label{sec:simulation}

\begin{figure}[tbh]
\plotone{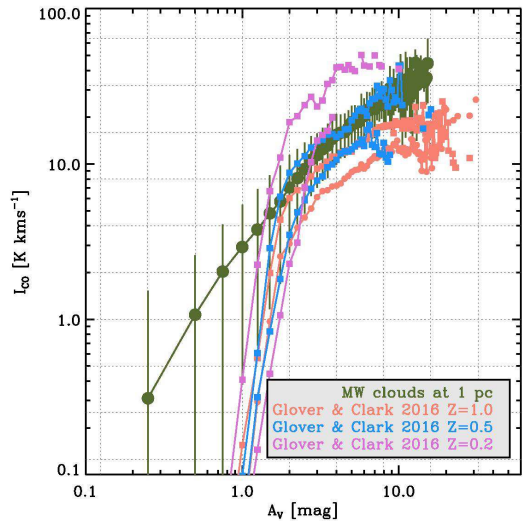}
\caption{\label{fig:compare_simulation}
Comparison of our synthesized Milky Way \icoav\ relationship with theoretical expectations. Our \icoav\ relation weighting all lines of sight equally appears in green with error bars showing the scatter. Points show the \icoav\ relations for three different metallicity clouds (i.e., 1, 0.5, 0.2 solar metallicity) with fixed mass ($10^4$~M$_{\odot}$) and radiation field \citep{DRAINE78} from the simulations by \citet{GLOVERCLARK16}. For each cloud, we plot the simulation at two times: at the onset of star formation (the upper line) and $1$~Myr before this time (the lower line). The theoretical \icoav\ relations show some dependence on metallicity, with higher CO emission at a given \av\ in lower metallicity systems. At intermediate $\av \approx 2{-}5$~mag, our measurements agree well with the star-forming Solar metallicity clouds and both $Z = 0.5$ clouds, and lie between the star-forming low metallicity and inert high metallicity cloud. All of the simulations show clear signatures of a minimum \av\ threshold for CO emission and saturation of the CO line at high \av . Such a features are not obvious in the observations, at least partially due to the lack of resolution.}
\end{figure}

Both photon dominated region models (\citealt{WOLFIRE10}) and numerical simulations \citep[e.g.][]{GLOVER11} identify the dust extinction (\av) as the key parameter for the location of CO in a molecular cloud. Our observational study of the \icoav\ relationship in Milky Way molecular clouds is directly motivated by these previous works. Here, we make a direct comparison between the results of our observations and theoretical predictions.

Specifically, we compare our observational results to the recent simulations by \citet{GLOVERCLARK16}. They used a modified version of the Gadget~2 SPH code \citep{SPRINGEL05} to simulate molecular clouds with different initial conditions. We consider runs with three different metallicities: $1.0$, $0.5$, and $0.2$ solar metallicity ($Z_{\odot}$). They assumed that the dust-to-gas ratio scales linearly with metallicity. All three runs assume the same standard \citet{DRAINE78} UV radiation field, a fixed cosmic ray ionization rate of $10^{-17}$~s$^{-1}$ per hydrogen atom. The cloud mass is $10^4$~M$_\odot$ and the initial volume density is $276$~cm$^{-3}$. The initial turbulent velocity field is the same and chosen such that the cloud is initially marginally gravitationally bound (i.e., the initial kinetic energy is the same as the gravitational binding energy of the cloud). The turbulence is decaying over time, and the simulations are evaluated at two times: at the initial onset of star formation and $1$~Myr before this time. This set of simulations corresponds to placing a low-mass Galactic cloud into environments with metallicities similar to the Milky Way ($1.0~Z_{\odot}$), the LMC ($0.5~Z_{\odot}$), and the SMC ($0.2~Z_{\odot}$).

\citet{GLOVERCLARK16} generate integrated CO intensity maps from the underlying SPH simulations using RADMC-3D\footnote{\href{http://www.ita.uni-heidelberg.de/~dullemond/software/radmc-3d/}{http://www.ita.uni-heidelberg.de/$\sim$dullemond/software/radmc-3d/}}. They used the large velocity gradient (LVG) approximation to calculate the level populations of CO molecules, as described in \citet{SHETTY11}. The size of the final \av\ and CO maps from the simulations is $16.2$~pc per side, and the number of pixels is $256^2$ (making a pixel ${\sim}0.06$~pc big). The spatial resolution of the simulated maps is thus comparable to extinction maps of nearby clouds \citep{LOMBARDI06,PINEDA08} and a factor of few higher than the data that we use in this paper.

Figure~\ref{fig:compare_simulation} compares our observed \icoav\ relation (the green circles and error bars showing the $1\sigma$ scatter) to the simulation results. At intermediate $\av \approx 2{-}5$~mag, the simulation at half solar metallicity most closely resembles our observed relation, though the solar metallicity calculation at the onset of star formation shows only a small offset towards lower \ico\ at fixed \av\ compared to our measurements. We will see below that this intermediate range appears least affected by resolution effects.

The simulations exhibit strong evidence for a minimum \av\ threshold for CO emission. They show a steep drop in CO emission as \av\ approaches ${\sim}1$~mag, reflecting the need for a dust layer to shield CO molecules from dissociating radiation. On the other hand, the observations produce a smooth relationship between CO and \av\ at low \av . Following on the previous section, we highlight uncertainty in the contamination correction and the high spatial resolution of the simulation, ${\sim}0.2$~pc, compared to our $\gtrsim 1$~pc beam. We demonstrate the impact of resolution by blurring the simulation in the next section.

The saturation of CO emission at high \av\ is also apparent in the simulations, consistent with the results found in \citet{SHETTY11}. Our average \icoav\ relation shows some hint of saturation when plotted in linear scale (Figure~\ref{fig:synthesis}), but this plot shows that any such effect remains weak compared to the same behavior at high resolution in the simulations. Again, we expect resolution to play an important role, but it cannot explain the whole difference. Variations among the properties of clouds in our sample must also contribute to the observations, such that any saturation sets in at even higher \av\ for our highest mass clouds.
 
\begin{figure*}[tbh]
\plottwo{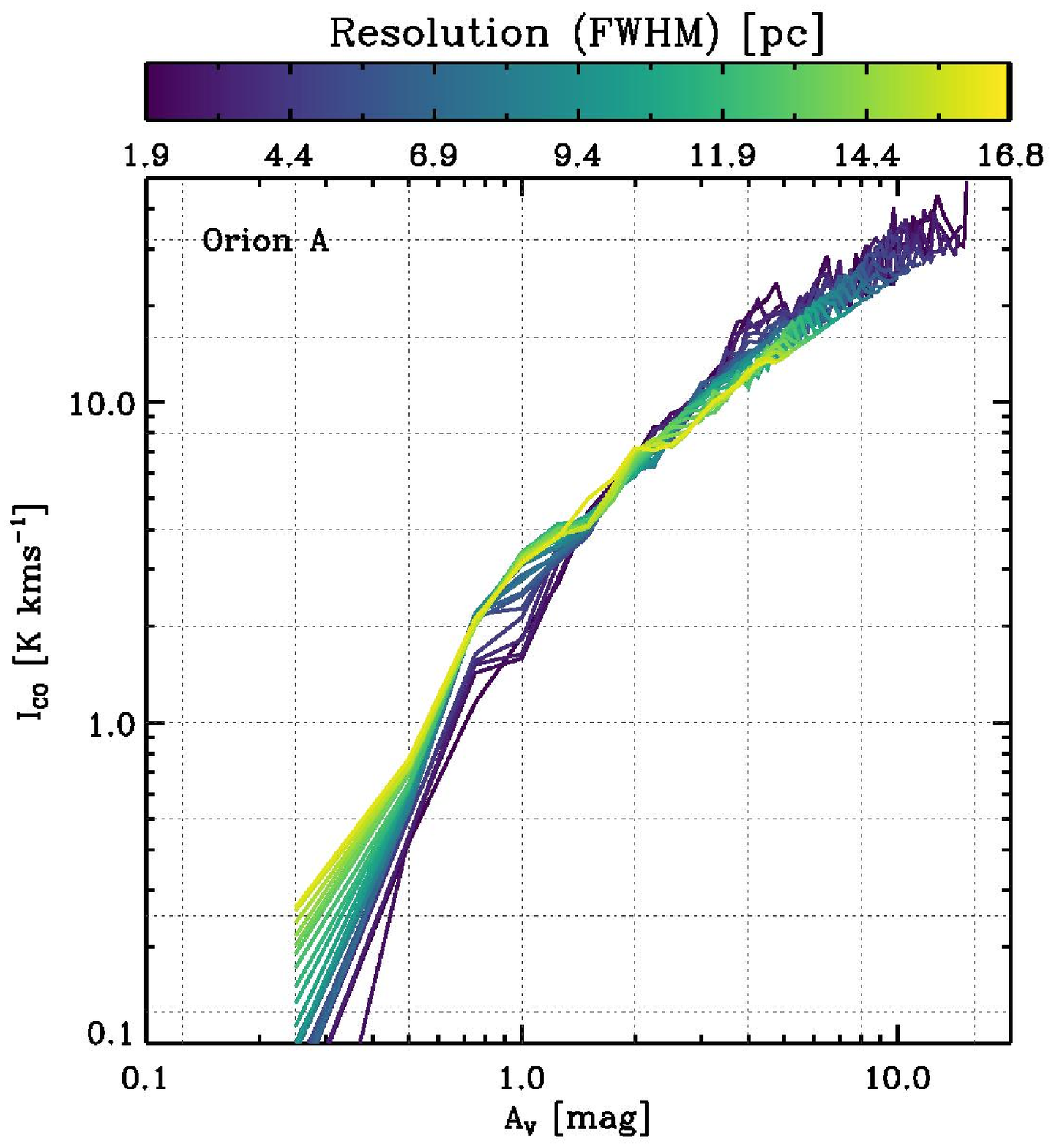}{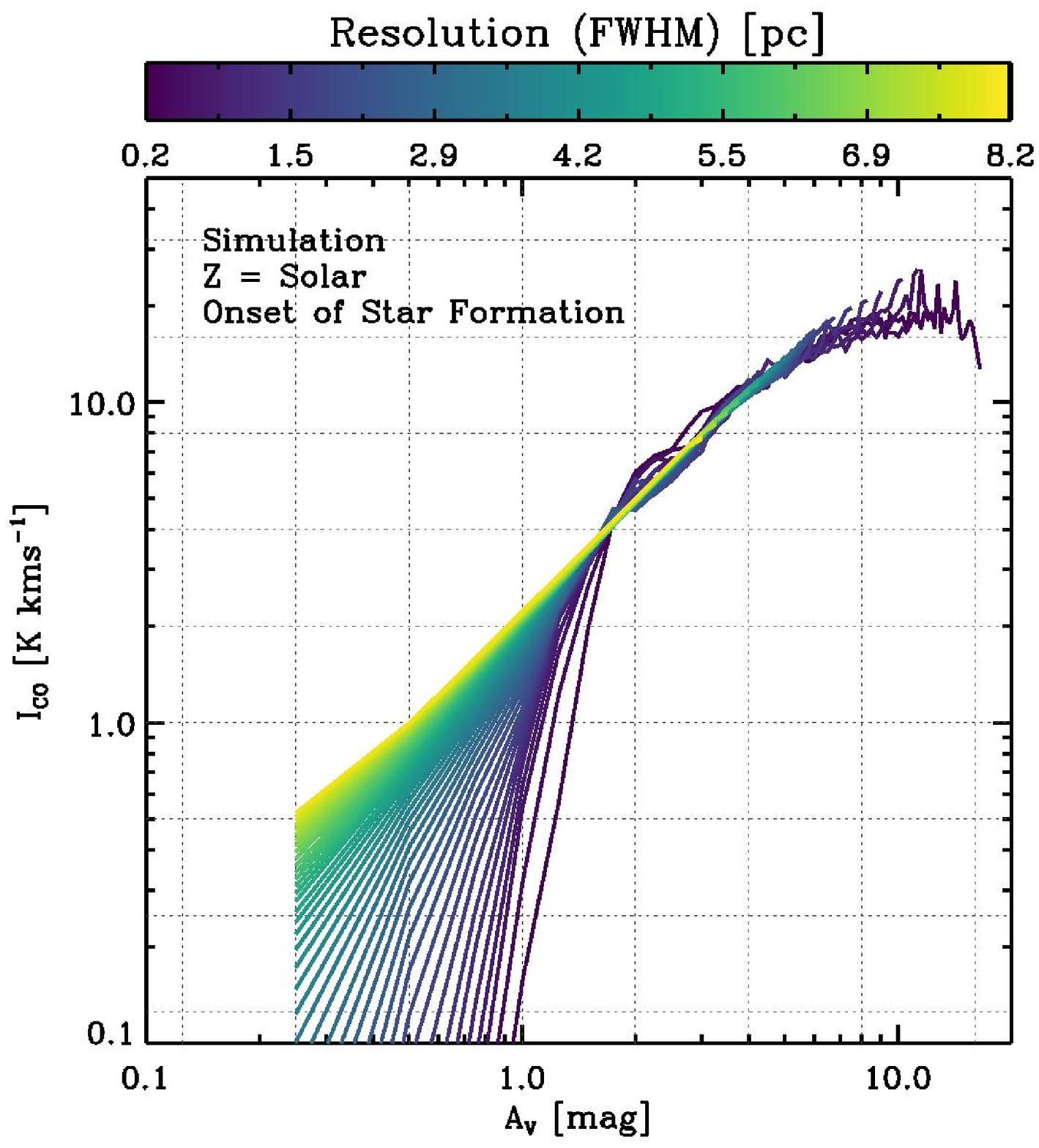}\caption{\label{fig:resolution}
Effect of spatial resolution on the \icoav\ relation. ({\em left}) \ico\ as a function of \av\ in the Orion A molecular cloud measured at different resolutions. We progressively degrade the resolution to simulate observing a more distant cloud and plot the mean relation colored according to the resolution of the data. ({\em right}): A similar exercise using the \citet{GLOVERCLARK16} simulations of a Solar metallicity cloud at the onset of star formation. In both cases, the dynamic range in the \icoav\ relation is reduced as the spatial resolution becomes progressively coarser. As the resolution changes, \ico\ at a given \av\ does not vary much in the intermediate \av\ regime, but at high and low \av\ the situation is different. Any threshold and saturation are washed out as the convolution blurs together intermediate \av\ gas and either high or low \av\ gas. This makes the threshold and saturation effects hard or impossible to measure at coarse spatial resolution. Based on the simulation data, ${\sim}0.1$~pc may be a useful figure of merit.}
\end{figure*}

In Section~\ref{sec:othergal} we compare our results to those found for the Magellanic Clouds by \citet{LEE15}. With that in mind, we highlight the differences among the \citet{GLOVERCLARK16} results for clouds of different metallicities. In the simulations, a molecular cloud with a lower metallicity but otherwise identical initial physical conditions shows stronger integrated CO intensity at a given \av\ compared to a higher metallicity cloud. Plotting the data in this way attempts to control for dust shielding and the simulations fix the strength of impinging radiation field. Therefore, the difference seems most easily attributed to differences in CO excitation. Key to this paper, this sorting by metallicity contrasts with the observation by \citet{LEE15} who find similar CO emission at a given \av\ in three different metallicity galaxies (the Magellanic Clouds and the Milky Way) at $10$~pc resolution.

\subsubsection{Effect of Spatial Resolution}
\label{sec:resolution}

We argue that our ability to detect an \av\ threshold for CO emission or the saturation of CO intensity at high \av\ can be significantly diminished by insufficient spatial resolution. The high resolution of the \citet{GLOVERCLARK16} simulations allows us to test this hypothesis. The simulations show threshold and saturation features at their native resolution of ${\sim}0.1$~pc. To test whether these features could be recovered at coarser spatial resolution, we blur the \ico\ and \av\ maps from their solar metallicity cloud simulation. We convolve these data with progressively larger and larger kernels and then measure the \icoav\ relation at ever coarser resolution. We perform the same test on the data for the Orion~A molecular cloud, which is relatively nearby and shows some hint of an \av\ threshold and \ico\ saturation (see Figure~20).

Figure~\ref{fig:resolution} shows the results of this test. For both the real and simulated cloud, the slope of the \icoav\ relation at low \av\ tends to become shallower as the spatial resolution becomes coarser. At $\av \le 1$~mag the mean CO intensity at a given \av\ become systematically higher for the lower resolution data. The effect is stunningly strong in the simulation, with the threshold all but vanishing by the time we degrade the resolution to a few pc (though we do caution that at these scales, the resolution begins to approach the scale of the whole simulated cloud). The apparent \av\ threshold for CO emission shifts to lower \av\ or vanishes as the resolution of our data becomes worse. At the same time, the dynamic range in \av\ is significantly reduced as we degrade the spatial resolution. As the maximum \av\ becomes smaller, the saturation behavior of \ico\ at high \av\ becomes progressively harder to identify. 

These tests have important implications for our results. First, note that the \ico\ at intermediate \av\ (i.e., a few mag) appears reasonably robust to resolution effects. As a result, this normalization should be viewed as our main, most secure result and, indeed, one of our main goals was to compare this value among clouds and to other galaxies. Second, we will see below that the threshold behavior is crucial for the metallicity dependence of \xco . To constrain this with observations, one needs high (significantly sub-parsec) resolution data, beyond the reach of the {\em Planck} data. The ``blurring'' along the \av\ axis induced by changing resolution will also affect the PDF, so that ideally one also needs to measure or model the \av\ distribution at high resolution. 

\subsection{Comparison to Other Galaxies}
\label{sec:othergal}

\begin{figure}
\plotone{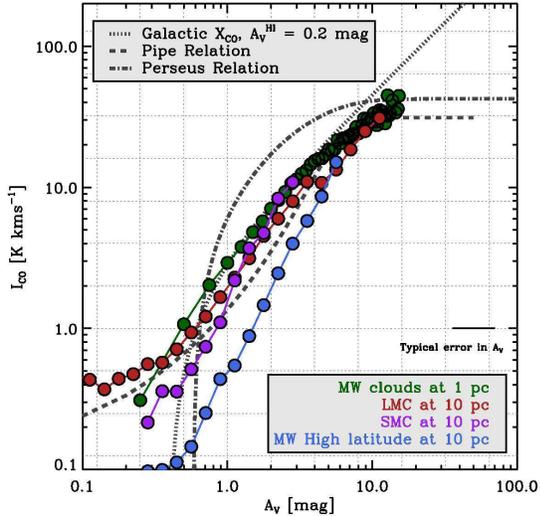}
\caption{\label{fig:compare_local_group} Comparison of \icoav\ relations in Local Group galaxies. The relations for the LMC, SMC, and high latitude Milky Way are taken from \citet{LEE15}, and are measured on $10$ parsec scales. The synthesized Milky Way relation plotted here is calculated by weighting all the lines of sight equally (i.e., the right panel in Figure~\ref{fig:synthesis}) and has resolution of ${\sim}1$ parsec at the median distance of $200$ parsec. The gray dotted line shows the standard Galactic \xco , assuming that all hydrogen is in molecular phase. The gray dashed line shows the same, but for the case of a fixed \hi\ shielding layer of $A_\mathrm{V}^\mathsc{Hi} = 0.2$~mag.}
\end{figure}

A main goal of our paper is to extend the work of \citet{LEE15}, who found a similar amount of CO emission at a given \av\ using matched spatial resolution ($10$~pc) data in the Milky Way, the LMC, and the SMC. This similarity across a wide range of metallicity implies that the amount of dust shielding is the primary factor in determining the extent of CO emission, and this idea agrees with theoretical expectations \citep[e.g.,][]{WOLFIRE10, GLOVER11} and the results of the recent simulations by \citet[][see above]{GLOVERCLARK16}. But in the work by \citet{LEE15}, the available Milky Way data represented the limiting factor allowing only a few cloud-averaged and high latitude lines of sight measurements. Here, we ask how our new results for our large set of local molecular clouds compare to the measurements of the Magellanic Clouds in \citet{LEE15}.

Figure~\ref{fig:compare_local_group} plots this comparison. We show our synthesized Milky Way \icoav\ relation with measurements at lower resolution from \citet{LEE15} for the Large Magellanic Cloud ($0.5~Z_{\odot}$), the Small Magellanic Cloud ($0.2~Z_{\odot}$), and high latitude ($|b|>5$ degrees) sight lines in the Milky Way ($Z_{\odot}$). All these comparison data have spatial resolution of ${\sim}10$~pc, approximately an order of magnitude coarser than our synthesized Milky Way relation. We also plot the curves that are expected for the standard Galactic \cohtwo\ conversion factor (gray dotted line) and an \hi\ shielding layer of $A_\mathrm{V}^\mathsc{Hi} = 0.2$~mag (gray dashed line).

To first order, our synthesized Milky Way, LMC, and SMC relations agree well, especially given the uncertainty in the $x$-axis. These three galaxies span a factor of ${\sim}5$ in metallicity. Their good agreement in this parameter space highlights the central role of dust shielding in determining CO emission. A given amount of dust column on the scale of a cloud predicts the amount of CO emission well, to first order independent of environment. There are fine differences in the \icoav\ relationship from region to region, of course. These are clear from our atlas of clouds, \citet{PINEDA08} showed them in Perseus, and \citet{LEE15} demonstrated differences in the \icoav\ relation between different regions in the LMC when sorted by $T_{\rm dust}$. Despite these differences, Figure~\ref{fig:compare_local_group} offers good support to the idea that as a practical tool, \av\ can be used to predict \ico\ with reasonable accuracy across systems with different metallicity.

Note that the high latitude Milky Way lines of sight stand out most in this plot, especially at low \av . These have the lowest CO intensity per unit visual extinction. This behavior was noted by \citet{LEE15}, who attributed the low CO emission per unit visual extinction to the long path length through the thick Galactic atomic gas disk at high latitudes. Most of the gas, and most of the dust, in most lines of sight at $|b| > 5^\circ$ is physically unassociated material spread our over $\sim$kpc along the line of sight.

Finally, note that with ${\sim}1$~pc resolution, we expect the present measurements to be more accurate than the LMC and SMC relations or the cloud-averages in \citet{LEE15}. As we saw above, the better resolution does a better job of not blurring out cloud structure. The higher resolution also allows us to correct for contamination by material unassociated with the clouds. The coarser resolution in the Magellanic clouds prevented such an operation, and we relied on the external line of sight (and lower dust content of contaminating \hi) in \citet{LEE15}. Future, higher resolution work, especially with ALMA, will improve the quality of mapping of the Magellanic Clouds and will allow a similar approach to what we use in this paper.

\subsection{Implications for \xco\ as a Function of Metallicity}
\label{sec:xco_metallicity}

\begin{figure*}
\plottwo{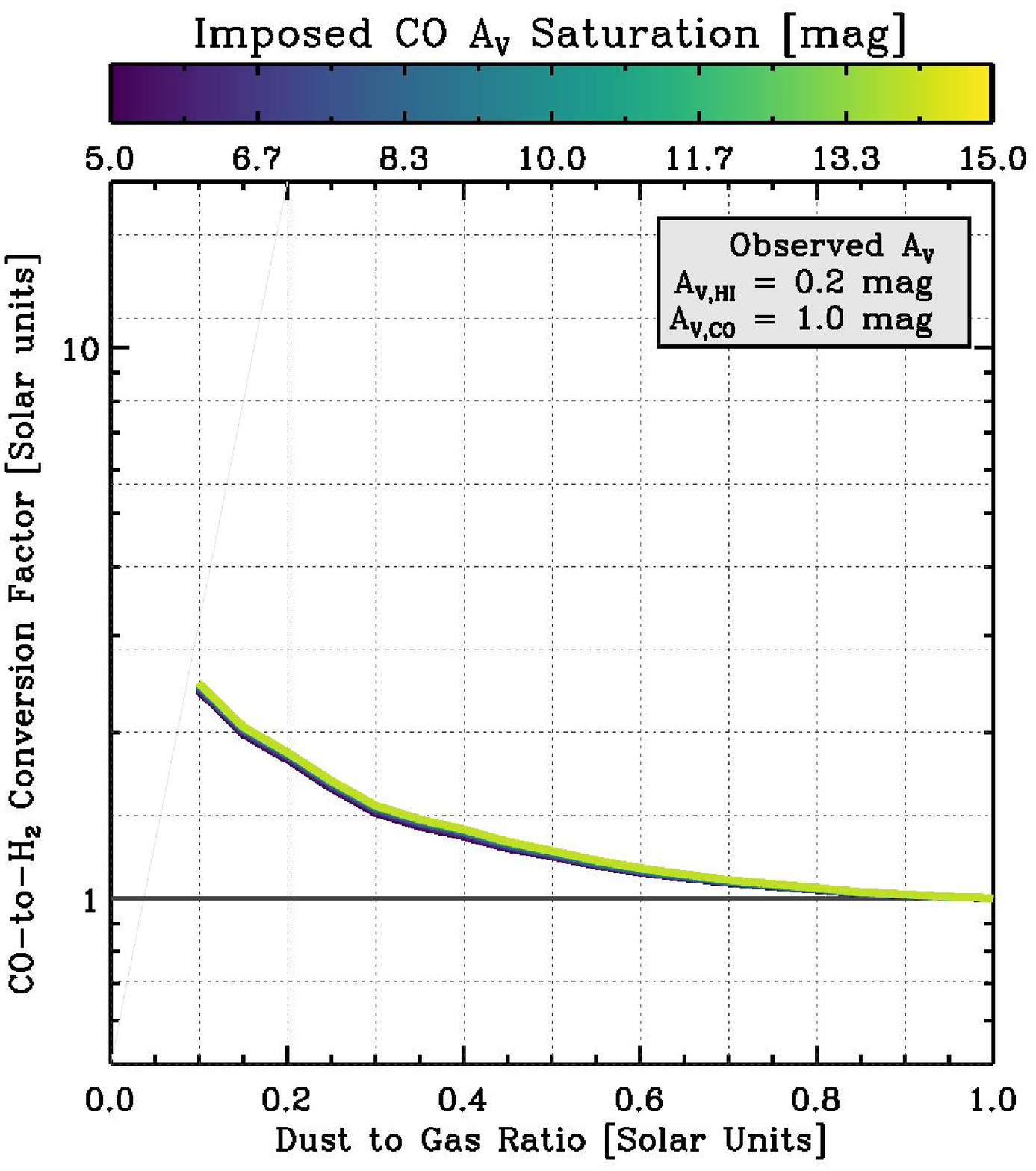}{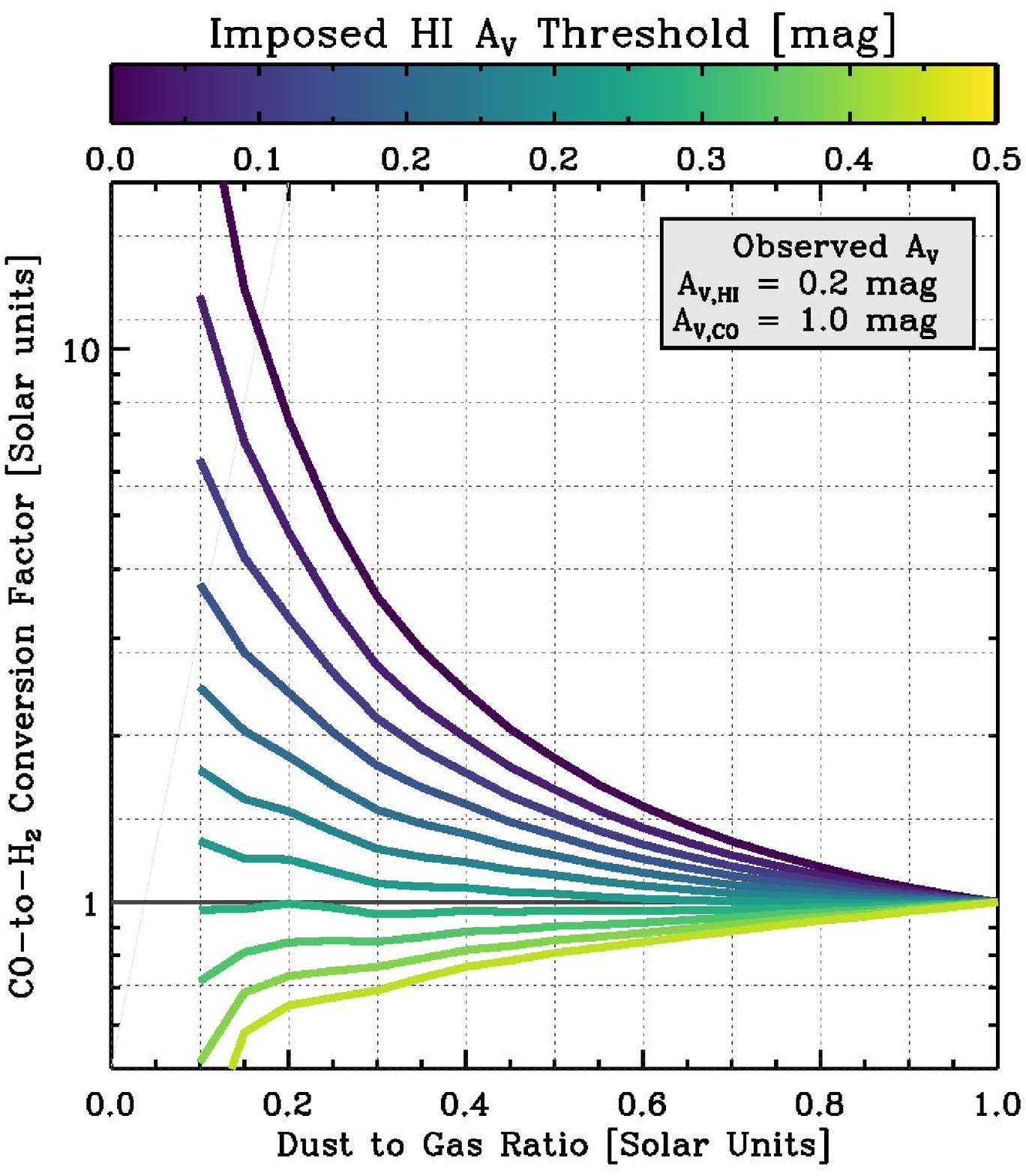}
\plottwo{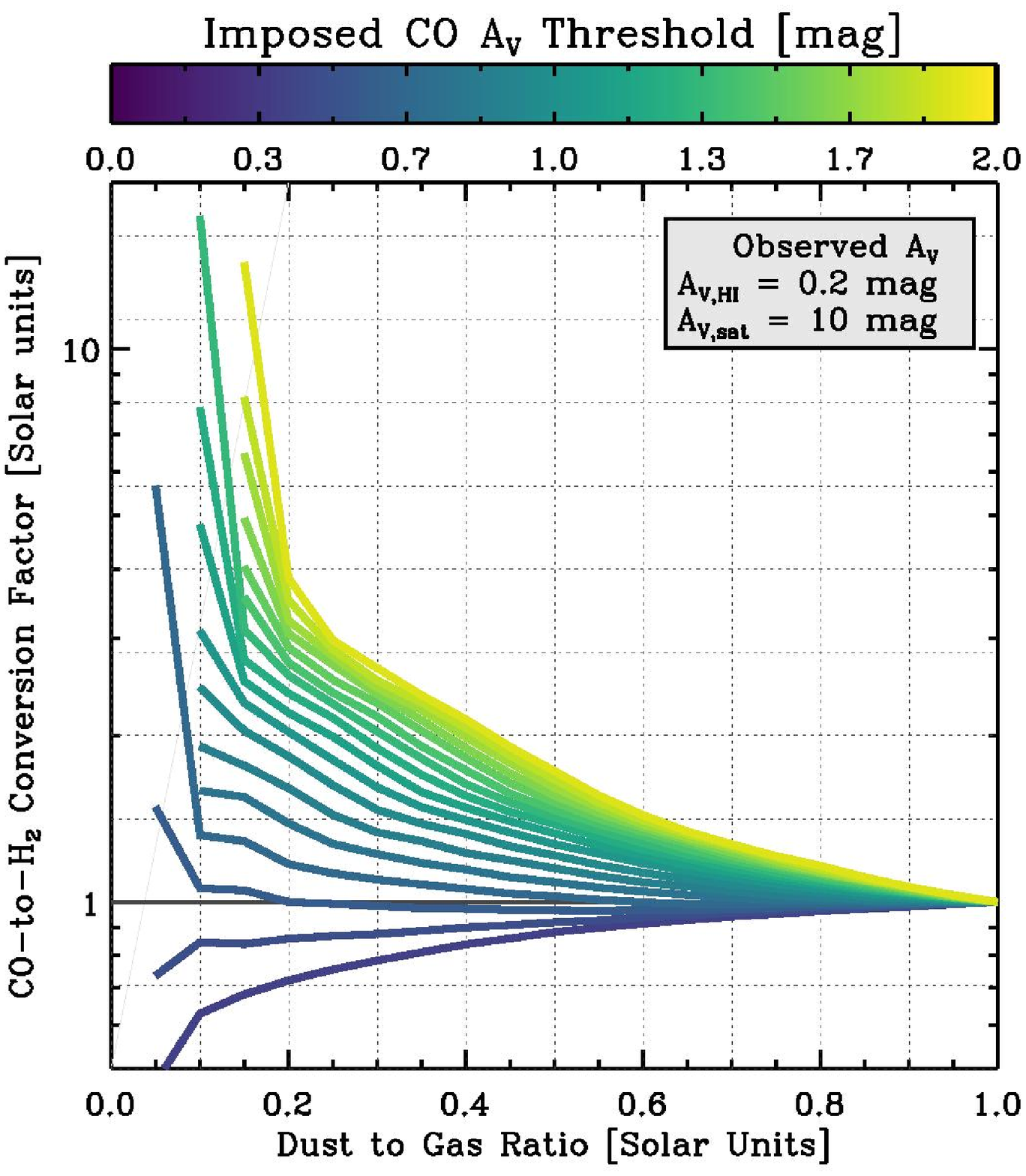}{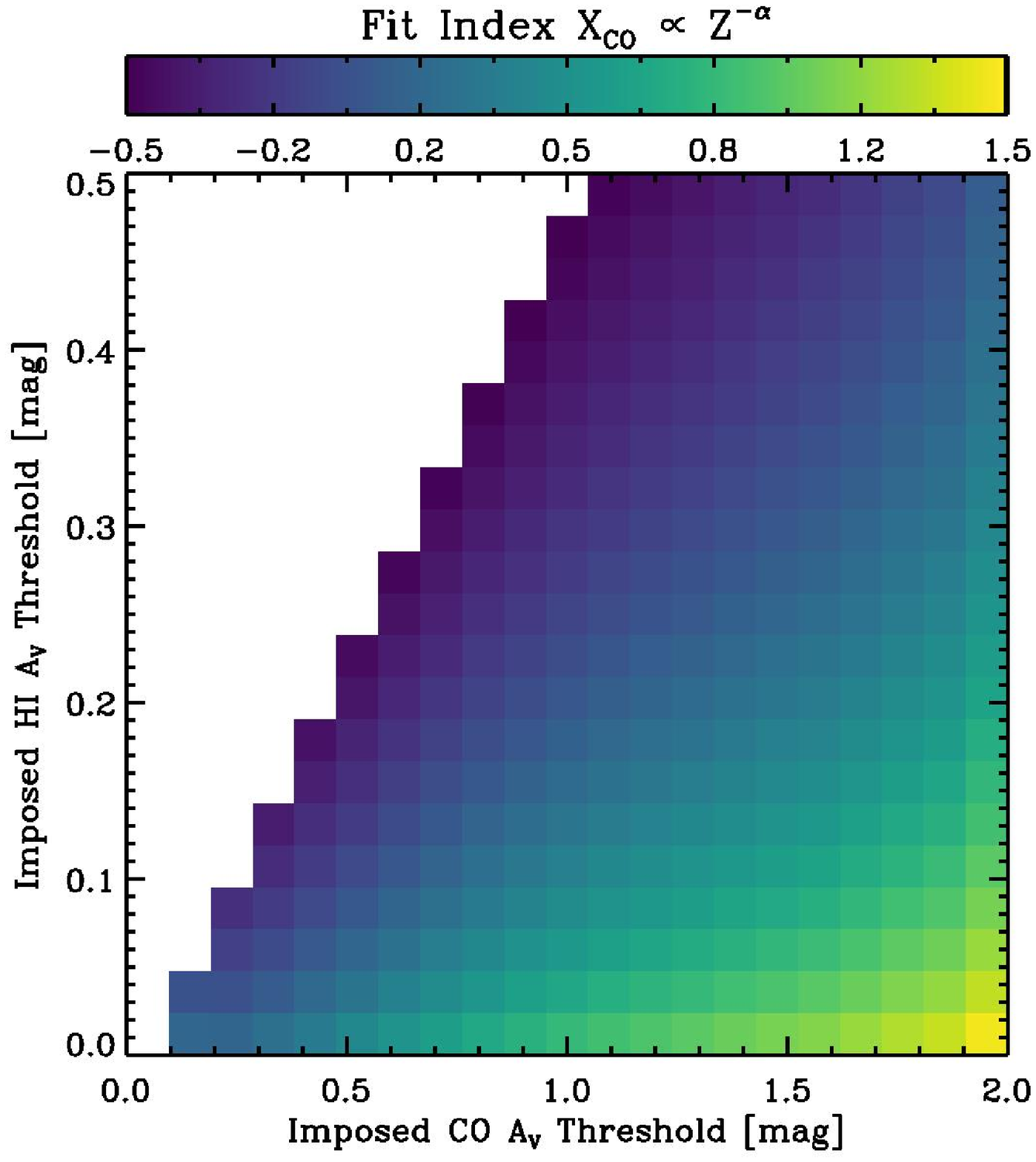}
\caption{\label{fig:xco_metallicity} The \cohtwo\ conversion factor as a function of metallicity, following the four-step approach in \citet[][see that paper for an illustration]{LEE15}. We use a modified version our aggregate \icoav\ measurement for local molecular clouds, along with the measured distribution of \av\ from $\tau_{850}$ to explore how \xco\ would change if we observe an analog of the Solar Neighborhood clouds but at different dust-to-gas ratio. In this approach the dust-to-gas ratio and \xco\ are both normalized to equal $1$ at the value calculated for our maps. The behavior of \xco\ as a function of $DGR$ pivots on features still below the resolution of our data, and we show the sense of this here. We vary the ({\em top left}) CO saturation, ({\em top right}) \av\ threshold of the \cohtwo\ transition, and ({\em bottom left}) \av\ threshold for detectable CO emission. The CO saturation has negligible impact on the metallicity dependence. By contrast, any threshold for CO emission and the thickness of the atomic gas shielding layer both have large impacts. The {\em bottom right} panel shows the interplay of the CO threshold and \hi\ shielding layer. We plot the best fit power law index, $\alpha$, in $\xco \propto Z^{-\alpha}$ as a function of the adopted \cohtwo\ and CO thresholds. We only show results where the CO threshold exceeds twice the \hi\ threshold (in our book-keeping the \cohtwo\ threshold is one sided and the CO threshold is two sided). The steepest metallicity dependence comes from cases with a large CO-dark layer, i.e., high CO threshold and low \cohtwo\ transition.}
\end{figure*}

\begin{figure*}
\plottwo{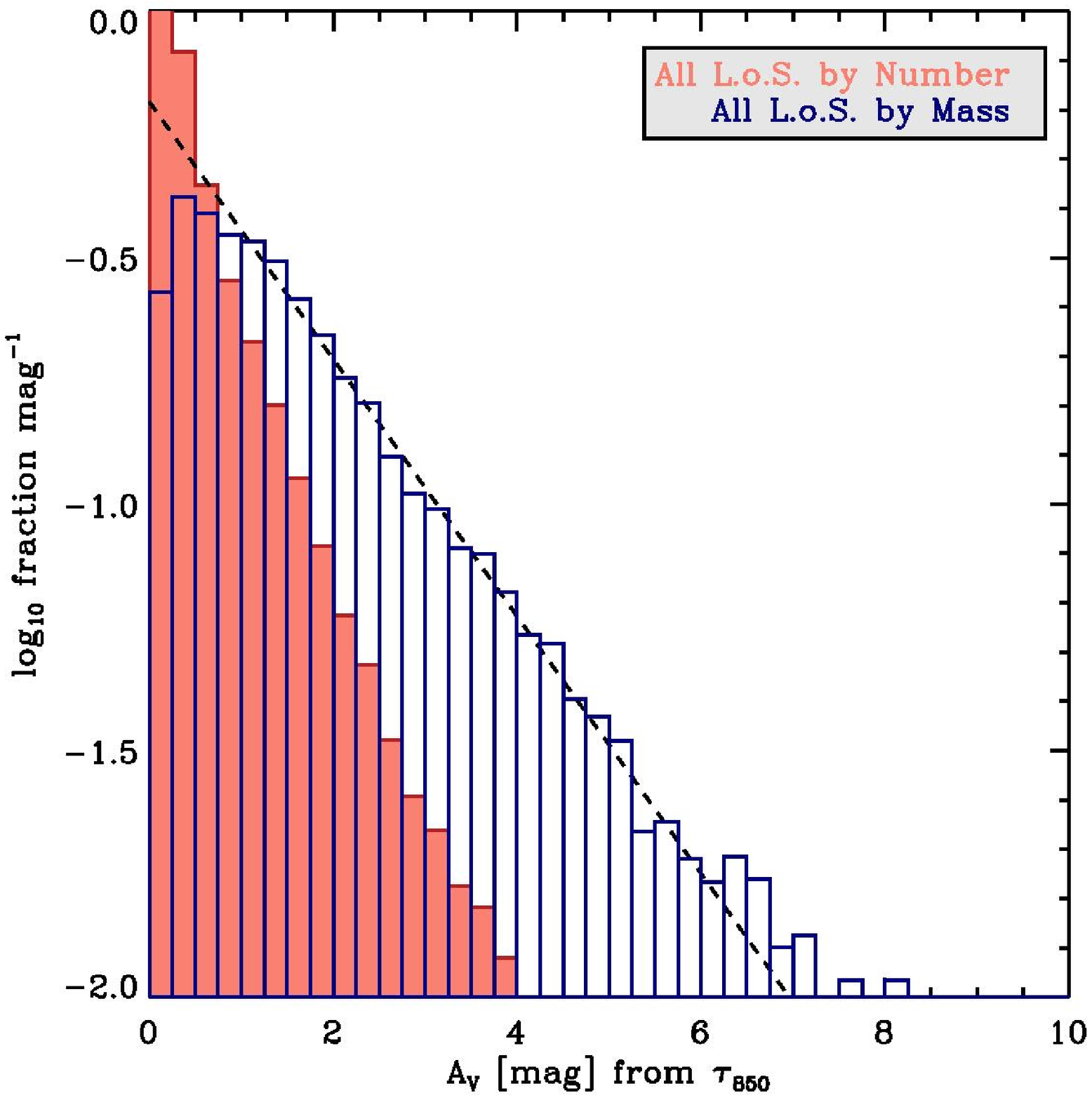}{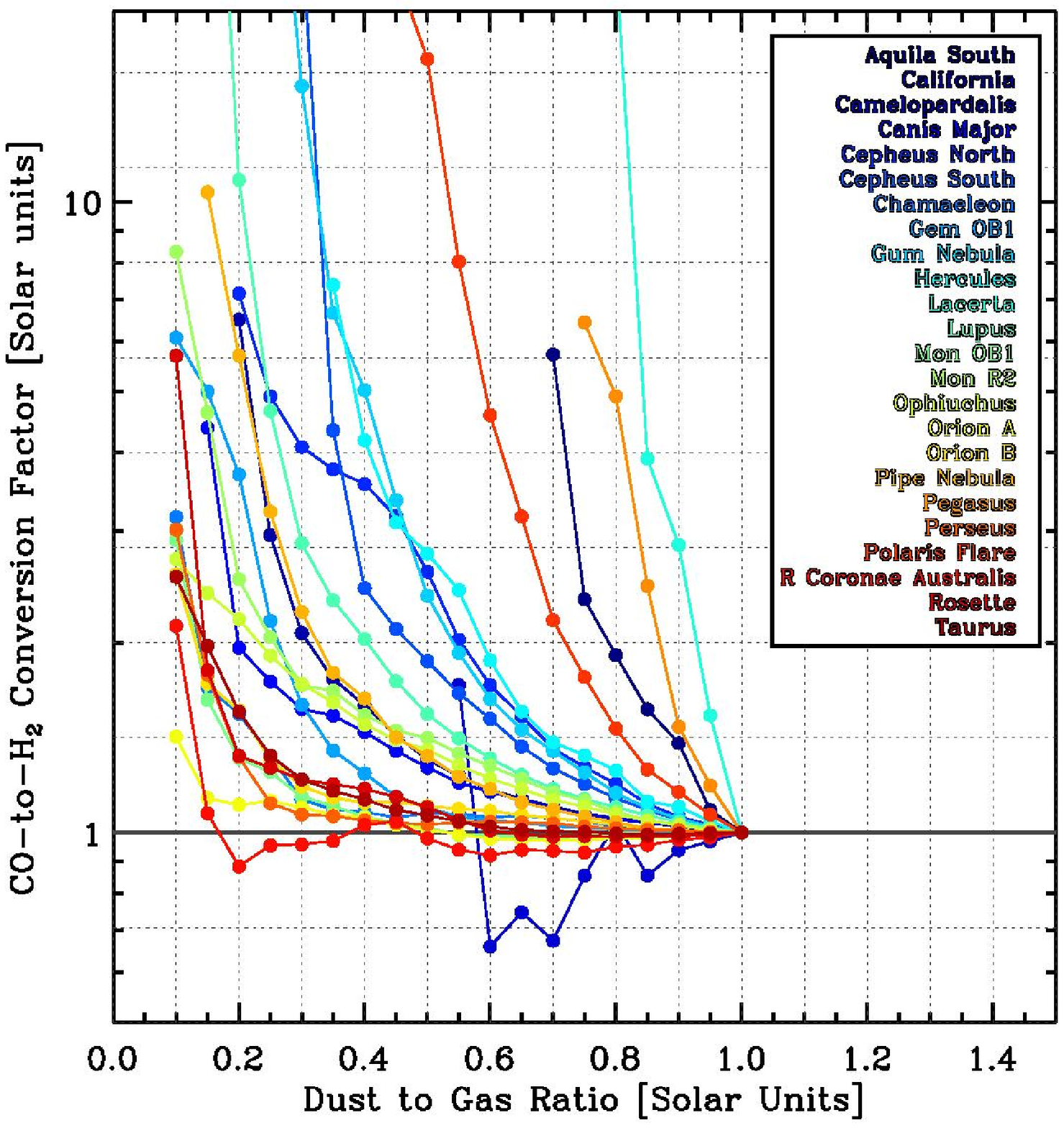}
\caption{\label{fig:xco_clouds} ({\em left}) The distribution of lines of sight (red, filled histogram) and mass ($\propto \av$, blue, open lines) as a function of \av\ for our whole sample. The shape of the combined distribution is self-similar (nearly a power law; the dashed line shows slope $-0.26$.), contributing to the flat behavior of \xco\ in Figure \ref{fig:xco_metallicity}. ({\em right}) \xco\ as a function of dust to gas ratio inferred for individual clouds, assuming a threshold for CO emission of $A_{\rm V,thres} = 1$~mag and an {\sc Hi} shielding layer of $0.2$~mag. We observe a large diversity in behavior from cloud-to-cloud, so that \xco\ can be expected to be a strong function of the cloud column density distribution, properties, and evolutionary state. At any given subsolar metallicity, the plot suggests that we should expect large cloud-to-cloud scatter in \xco .}
\end{figure*}

Based on the observation of similar \ico\ at a given \av\ in the LMC, SMC, and Milky Way, \citet{LEE15} argued that \ico\ may be reasonably predicted from \av\ in an approximately universal way in highly resolved molecular clouds. In this case, they suggest that the dependence of \xco\ can be modeled as a separable problem with four individually tractable parts: 

\begin{enumerate}
\item Clouds have some distribution of gas column densities ({\em the column density PDF}).
\item That distribution of column densities translates to some distribution of dust column densities, expressed as \av . The gas column relates to the dust column via the {\em dust-to-gas} ratio, which is a function of metallicity.
\item Below some \av , most of the gas is atomic. Inside that {\em atomic shielding layer}, the gas is mostly molecular (\htwo).
\item In the molecular gas, the amount of CO emission can be predicted from the line of sight extinction via the {\em \icoav\ relation.}
\end{enumerate}

\noindent Our findings in Section~\ref{sec:icoav_synthesis2} suggest that the fourth step, the prediction of the CO intensity from \av\ may be valid to good approximation. Thus this further motivates the empirical approach. Here we revisit and extend the calculations of \citet{LEE15}, also noting future areas for improvement in the calculations and needed observations.

We refer the readers to \citet{LEE15} for more details on the basic calculation, but summarize the approach here. The starting point of the calculation is a realistic distribution of gas column densities ($N_{\rm H}$). In \citet{LEE15}, this is computed from a library of cloud column density PDFs observed in the Milky Way \citep{KAIN09}. In this work, we try a more direct approach instead and consider the observed \av\ distribution from the {\em Planck} map to infer an alternative baseline gas column density PDF for the solar metallicity clouds.

{\em \xco\ and DGR for the whole sample:} First, we work with the full set of lines of sight across our sample and the average \icoav\ relation derived from them. This renders this exercise approximate, because the clouds are at different distances and in different physical states. But the exercise is still illuminating. We suggest thinking of it as asking how CO emission from an ensemble of Solar Neighborhood clouds would change if this region were moved to a low metallicity galaxy.

We translate the \av\ distribution for the local clouds into a distribution of line of sight dust column densities at a different metallicity by multiplying the gas column density by the dust-to-gas ratio. Following \citet{LEE15} we assume that the dust-to-gas ratio is directly proportional to metallicity. This calculation provides an estimate of the dust extinction, \av , distribution for a given molecular cloud PDF at any metallicity.

Next, we label the parts of the \av\ PDF below the $\hi{-}\htwo$ threshold as \hi\ and then remove these from the rest of the calculation. Formally, we subtract $2 A_\mathrm{V}^\mathsc{Hi}$ from all of our predicted \av\ values. This accounts for the shielding layer along the line of sight and identifies the well-shielded molecular gas in the clouds. As in Section~\ref{sec:xco}, our fiducial \hi\ shielding layer is $A_\mathrm{V}^\mathsc{Hi} = 0.2$~mag \citep{KRUMHOLZ09,WOLFIRE10,STERNBERG14}, but we also test the effect of varying $A_\mathrm{V}^\mathsc{Hi}$. 

Finally, from the \av\ distribution associated only with molecular gas, we predict the amount of CO emission using an adopted \icoav\ relation. We use our measured relation with two modifications: we explore the effect of adding a threshold for CO emission, $A_{\rm V,thres}$ even though we show above that our data have too coarse a resolution to find such a feature. We also explore the effect of implementing a sharp saturation, an \av\ above which \ico\ remains fixed at some value. 

By summing the CO emission and the molecular gas column in the model, we calculate a \cohtwo\ conversion factor for that model cloud. We change the dust-to-gas ratio from ${\sim}0.1~Z_\odot$ to  $1~Z_\odot$ for each model. Repeating the calculation for many values of the dust-to-gas ratio, we can predict the behavior of \xco\ versus metallicity ($Z$) given an input PDF (here the whole Solar Neighborhood ensemble) and the assumptions derived above.

{\em Central Role of the CO Threshold and $\hi{-}\htwo$ Transition in \xco\ vs $DGR$:} Figure \ref{fig:xco_metallicity} shows that our adopted $A_\mathrm{V}^\mathsc{Hi}$ and $A_{\rm V,thres}$ essentially determine the behavior of \xco . Each panel plots the calculated behavior of \xco\ (i.e., the ratio of \htwo\ mass to CO emission) as a function of metallicity for many models. In the top left panel we vary the saturation level for CO while holding the other parameters fixed. This has negligible effect for the metallicity dependence, though these physics can be highly relevant in other regimes \citep{SHETTY11}.

The top right and bottom left panels show \xco\ as a function of metallicity as we vary the \av\ threshold for the atomic-to-molecular transition. The bottom left shows the results of varying the threshold for CO emission, below which we take $\ico = 0$. Both of these quantities dramatically affect the behavior of the conversion factor as a function of metallicity. Over the plausible range of values that we explore the behavior shifts from almost no dependence of \xco\ on $Z$ to very steep values. 

The most basic conclusion from this exercise should be that better statistical constraints on both the threshold for CO emission and the threshold for \hi\ shielding will help this field. Certainly a great deal is already known from absorption work \citep[e.g.,][]{SHEFFER08} and theory \citep[e.g.,][]{WOLFIRE10,STERNBERG14}, but we still emphasize that these quantities are pivotal to the behavior of \xco . 

In detail, Figure \ref{fig:xco_metallicity} shows that one needs to know both $A_\mathrm{V}^\mathsc{Hi}$ and $A_{\rm V, thres}$. The strongest dependence of \xco\ on $DGR$ occurs in regions with a large threshold for CO emission but a low $A_\mathrm{V}^\mathsc{Hi}$, i.e., in regions with a large ``CO-dark'' component expressed in units of $A_V$. This agrees with the detailed models of \citet[see][]{WOLFIRE10}. 

A plane fit to the parts of the diagram where $A_{\rm V,thres} < 2 A_\mathrm{V}^\mathsc{Hi}$ yields $\alpha \approx 0.3 + 0.725 A_{\rm V, thres} - 3.0 A_\mathrm{V}^\mathsc{Hi}$ where $\xco \propto DGR^{-\alpha}$ over the range $DGR = 0.1{-}1.0$. A fit using $\Delta A_{\rm V} = A_{\rm V, thres} - A_\mathrm{V}^\mathsc{Hi}$ can also predict $\alpha$, but with less precision.

{\em Divergence at very Low DGR and Shallow Metallicity Dependence Above:} We see the divergence at metallicity $Z < 0.2~Z_\odot$ also noted by \citet{LEE15}. This breakdown occurs over a fairly wide range of assumptions. It reflects that in the Solar Neighborhood, most lines of sight and most mass in molecular gas lies at comparatively low $\av \lesssim 5$~mag. We show this distribution in the left panel of Figure \ref{fig:xco_clouds}. If we imagine keeping the gas column density distribution but scaling down the $DGR$, then at $Z \lesssim 0.2$~Z$_\odot$ all of this gas lies at $\av \lesssim 1$~mag and its behavior becomes highly sensitive to our adopted assumptions. As a result, at even moderately low $DGR$, one moves into the situation where only a small fraction of the cloud emits strongly in CO \citep[see][among many others]{GLOVER11,LEE15}.

At intermediate metallicities ($Z > 0.2~Z_\odot$, we find a relatively weak metallicity dependence for \xco . The exact number depends strongly on our assumptions, but often in the range $\alpha \sim 0.5{-}1.5$ in $\xco \propto Z^{-\alpha}$. This range of $\alpha$ is shallower than many previous studies, including \citet{ISRAEL97, LEROY11, SCHRUBA12, LEE15}, and more similar to the shallower slopes ($\alpha > -1$) often found from analyses based on CO virial masses \citep{WILSON95, ROSOLOWSKY03, BOLATTO08}. However, those measurements have also been interpreted to apply only to the CO-bright parts of clouds, so we do not expect the comparison to be rigorous.

Again the distribution of \av\ and mass explains the relatively weak dependence for \xco\ in our exercise. As we see from the left panel in Figure \ref{fig:xco_clouds}, our ensemble of measurements shows an approximately power law distribution over the range $\av \approx 2{-}6$~mag. This does not necessarily bear on the appropriate general shape of the column density PDF, which has been explored in detail by \citet{KAIN09,ABREUVICENTE15,LOMBARDI15}. We know that power law tails certainly do exist in some of our clouds \citep{KAIN09}, and these are enough to give the ensemble distribution a self-similar shape. As a result, the ratio of gas above any pair of thresholds will remain relatively fixed. The shape of the \icoav\ relation and deviations from self-similarity will create the behavior that we observe. By contrast, in a curving distribution like a log-normal, the shape of the PDF can contribute strongly to a steep \xco\ vs. metallicity dependence.

{\em Strong Cloud-to-Cloud Variations:} We also carry out this exercise for each individual cloud. We use our fiducial $A_\mathrm{V}^\mathsc{Hi} = 0.2$~mag and a CO threshold $A_{\rm V,thres} = 1$~mag. We plot the results in the right panel of Figure~\ref{fig:xco_clouds}. This exercise highlights that we should expect dramatic conversion factor variations from cloud-to-cloud in a low metallicity system. Taking this result at face value, if we shifted the Solar Neighborhood clouds to the metallicity of the SMC, some would appear approximately Galactic in nature, while others would disappear almost entirely, showing little or no CO emission. Here we continue to use our average \icoav\ relation, so this plot mainly reflects variations in the underlying \av\ PDFs of the clouds.

Some evidence of such effects have recently been reported in \citet{SCHRUBA17}, who found suggestions of strong field-to-field conversion factor variations in the Local Group low metallicity galaxy NGC~6822. They suggest that these may relate to either temporal or structural changes related to the star-forming state of a cloud. Simulations of molecular clouds at these metallicities by \citet{GLOVERCLARK16} also find evidence for strong temporal variations in the conversion factor.

\section{Conclusions}

We measure the parsec scale relationship between \ico\ and \av\ in $24$ local Milky Way molecular clouds using {\em Planck} dust and CO maps \citep{PLANCK13_CO, PLANCK13_DUST}. After correcting for contamination along the line of sight, we present measurements for each cloud individually and a combined relation derived from our ensemble of local clouds. These local clouds occupy a common region in the \icoav\ parameter space, sweeping out a relation that closely resembles that found for the Magellanic Clouds by \citet{LEE15} at coarser ($10$~pc) resolution. This agreement across a factor of five in metallicity reinforces the idea that dust shielding (\av) is the primary determinant of the location of CO emission. Moreover, the observed line of sight dust column can be used to estimate CO emission (\ico) with reasonably good accuracy across environment. We also show that our measured relation agrees well with numerical simulations of CO emission from molecular clouds by \citet{GLOVERCLARK16}.

Although our measured synthesized relation shows a declining slope, expected sharp features such as a minimum \av\ threshold for CO emission or the saturation of \ico\ at high \av\ are weak or absent in the aggregate \icoav\ relationship and the individual \icoav\ relation for many clouds. The clouds that do show such features tend to be the closest in our sample. By convolving high resolution observations and simulations, we show that degrading sub-parsec resolution data to coarser values tends to wash out such features and lower the dynamic range in \av . This, combined with the significant difficulty in estimating line of sight contamination, renders direct observations of the \av\ threshold for CO emission challenging. But we also emphasize that quantitative measurements of the value and variation in the \av\ threshold for CO emission (and the closely related threshold for the molecular-to-atomic transition) remain absolutely crucial to an accurate estimate of \xco\ as a function metallicity.

Using the \icoav\ relations that we observe for Milky Way clouds, we explore the implied metallicity dependence of the \cohtwo\ conversion factor. Our calculations, based on the empirical approach from \citet{LEE15} leverage the observed \av\ distribution and \icoav\ relation. A main result of these calculations is that, as one might expect, the dependence of \xco\ on metallicity pivots on the adopted threshold for CO emission and \hi-to-\htwo\ shielding layer; any CO saturation plays only a small role. These two quantities interact with one another, so that strong variation in \xco\ as a function of metallicity arise in cases with a large ``CO dark'' layer. Again, improved observational constraints are key and appear to require spatial resolutions of order $0.1$~pc.

Our calculations imply strong cloud-to-cloud variations in how \xco\ depends on metallicity, so that one should expect strong conversion factor variations across a highly resolved data set studying a low metallicity galaxy. Treating our whole sample together, we tend to find divergence in \xco\ below $Z \sim 0.2~Z_\odot$, as only a small part of a Solar Neighborhood cloud remains well-shielded at these metallcities. Above this, we find relatively weak dependence of \xco\ on $DGR$. In large part, this reflects the self-similar shape of the \av\ distribution in our sample. Power law tails in the column density distribution \citep[e.g.,][]{KAIN09, SCHNEIDER15, ABREUVICENTE15} combine to give our overall sample a power law shape, which minimizes the impact of the PDF on \xco\ (compared to, say, a pure lognormal shape).

In the near future, we expect similar ($\sim$pc) resolution observations of Local Group dwarf galaxies. The results here are intended to serve as a point of comparison for such studies.

\acknowledgments We thank the referee of the paper, Paul Goldsmith, for constructive and encouraging reports that helped improve the paper. We thank the {\em Planck} team for the public release of the dust and CO maps that form the core of this paper. These are a product of ESA/Planck, the Planck Collaboration, and the papers cited in the main text. During preparation of this paper, support for the work of C. Lee was partially provided by NASA for program HST-GO-12055.027-A through a grant from the Space Telescope Science Institute, which is operated by the Association of Universities for Research in Astronomy, Inc., under NASA contract NAS 5-2655. The work of AKL is partially supported by the National Science Foundation under Grants No. 1615105, 1615109, and 1653300. ADB acknowledges partial support from NSF-AST1412419. SCOG acknowledges financial support from the Deutsche Forschungsgemeinschaft via SFB 881 ``The Milky Way System'' (sub-projects B1, B2, B8) and SPP 1573 ``Physics of the Interstellar Medium'' (grant number \mbox{GL668/2-1}), and by the European Research Council under the European Community's Seventh Framework Programme (FP7/2007-2013) via the ERC Advanced Grant STARLIGHT (project number 339177).

\appendix
\section{$\ico{-}\av$ Relation for Individual Molecular Clouds}
\label{sec:appendix}

\begin{figure*}
\plottwo{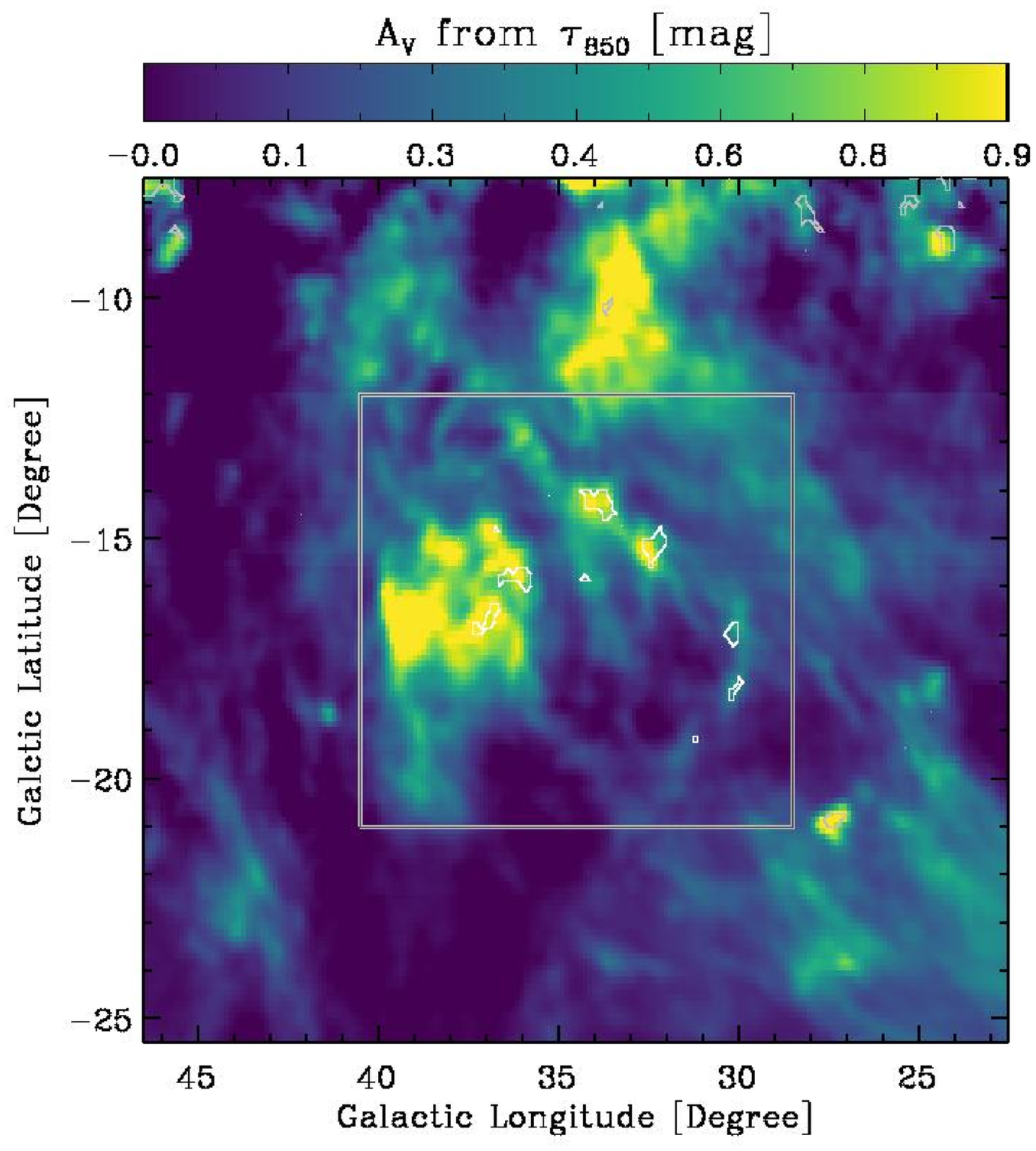}{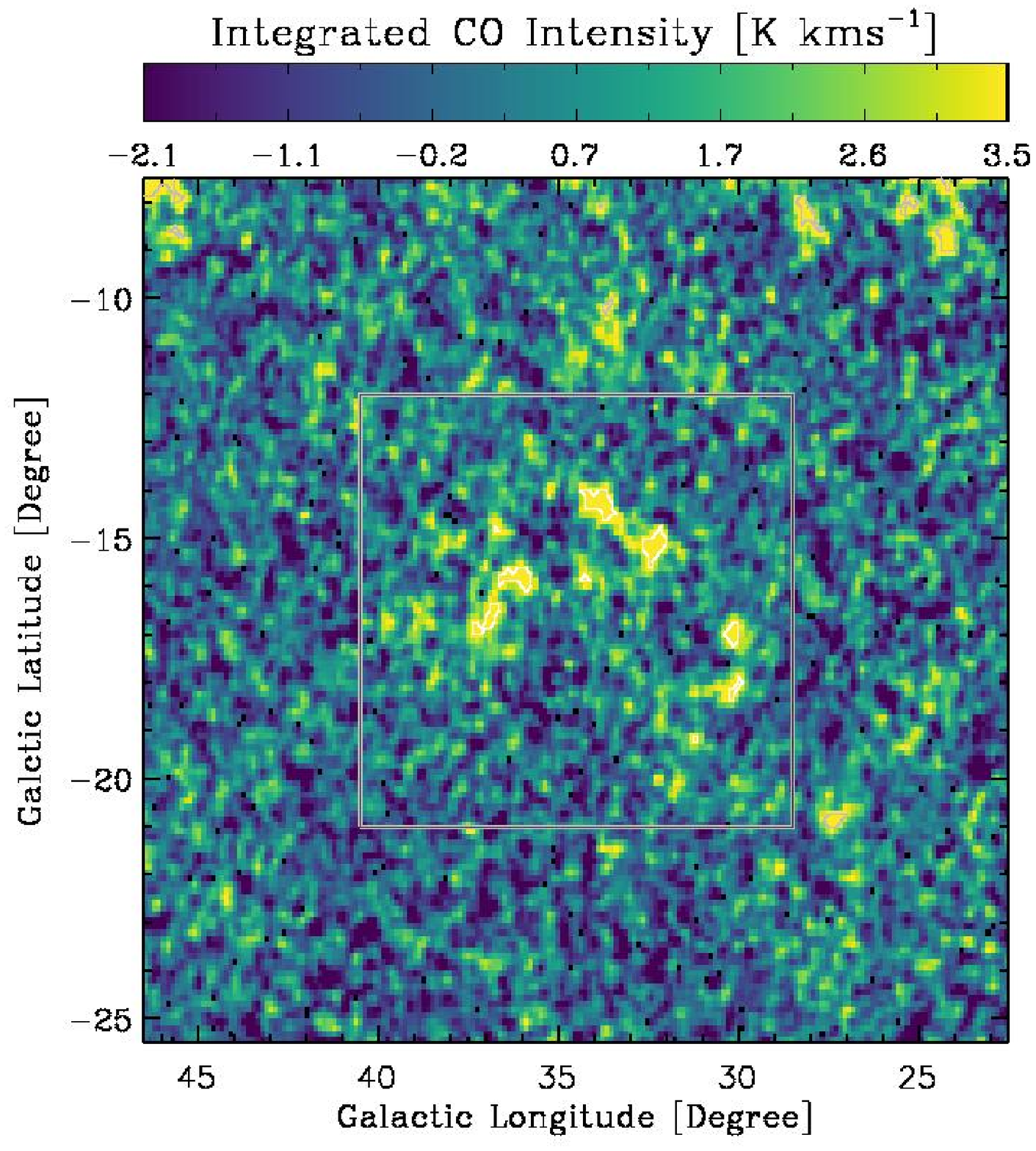}
\plottwo{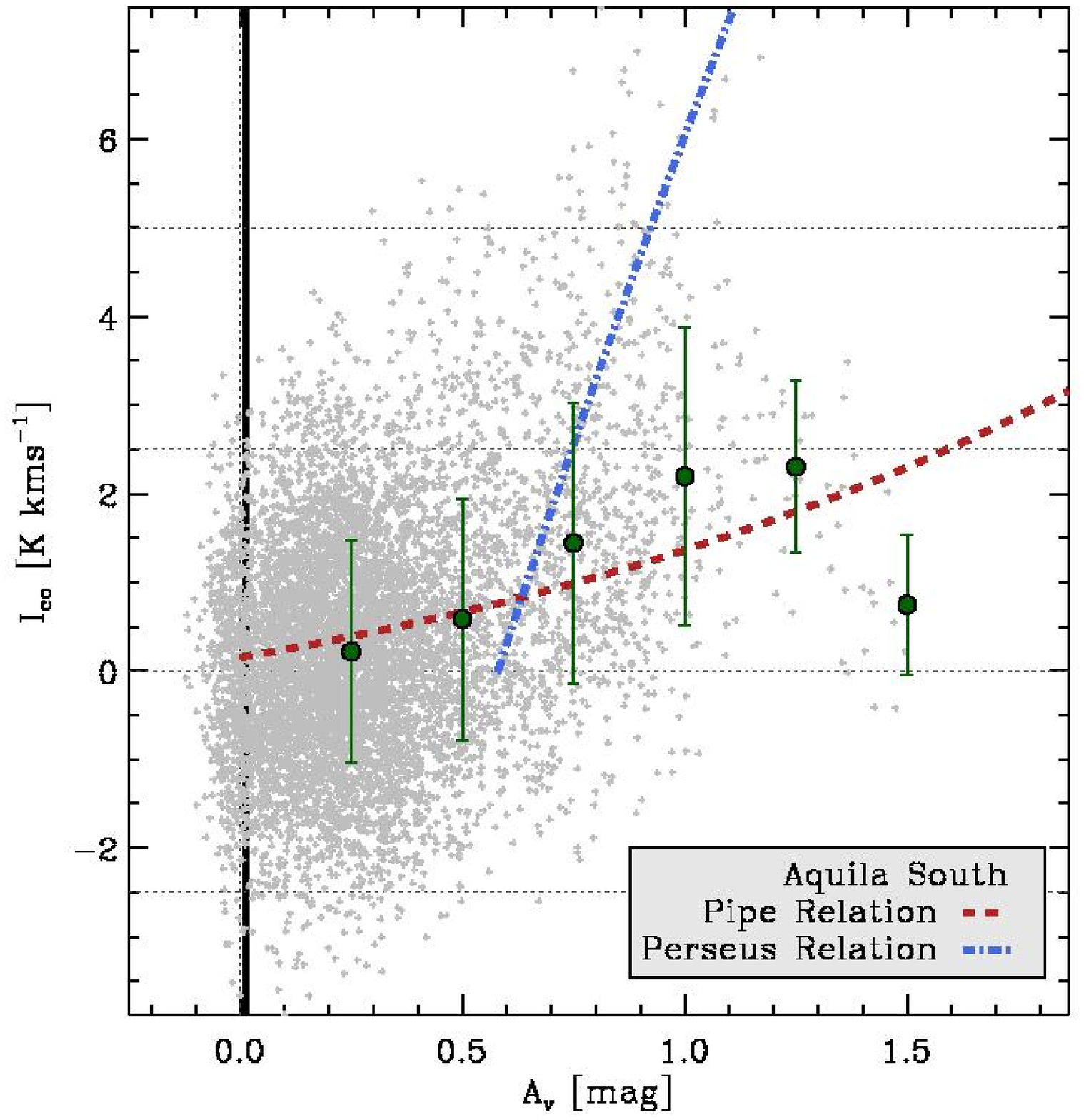}{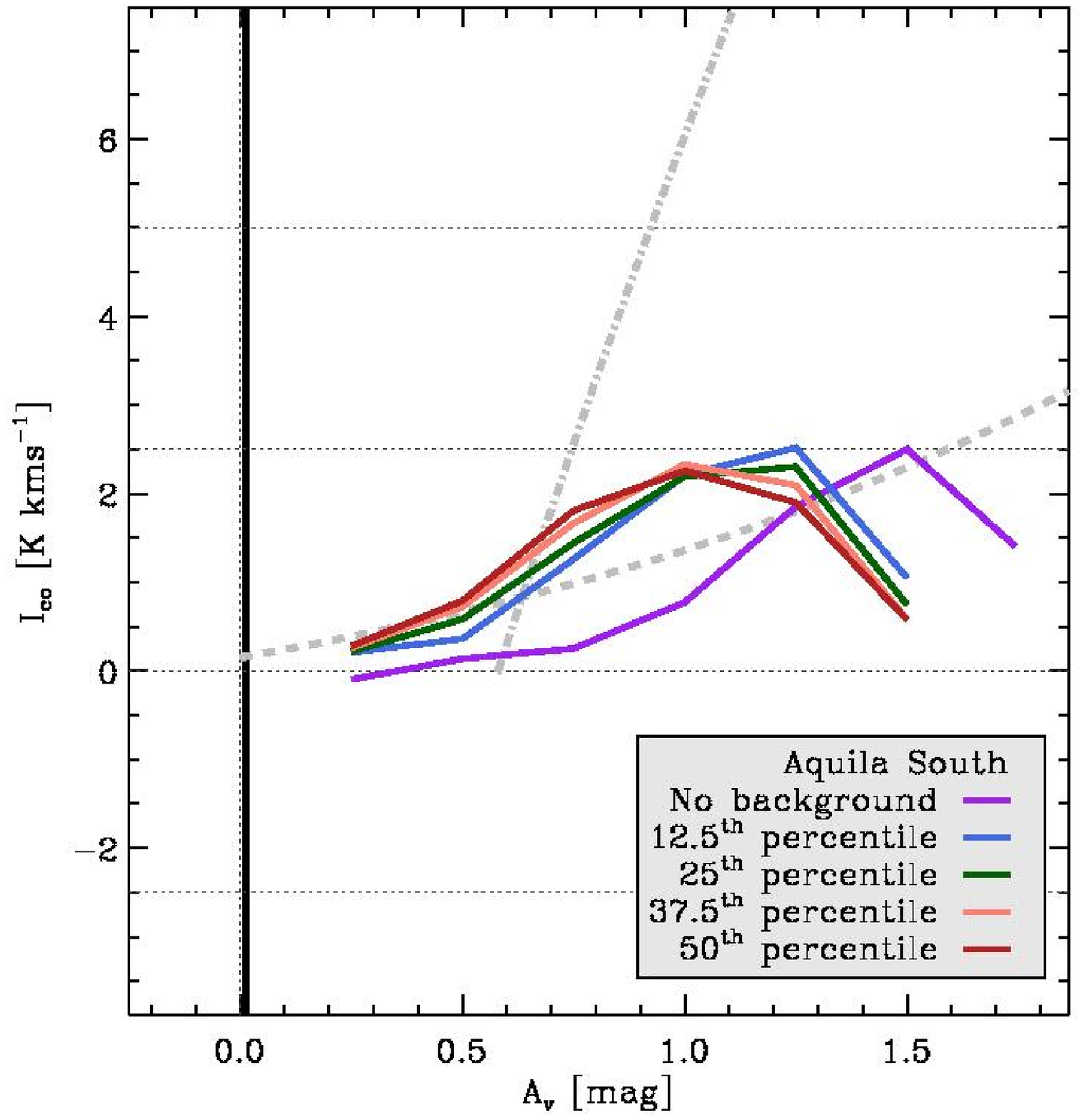}
\caption{Same as Figure~\ref{fig:taurus}, for the case of Aquila South.}
\end{figure*}

\begin{figure*}
\plottwo{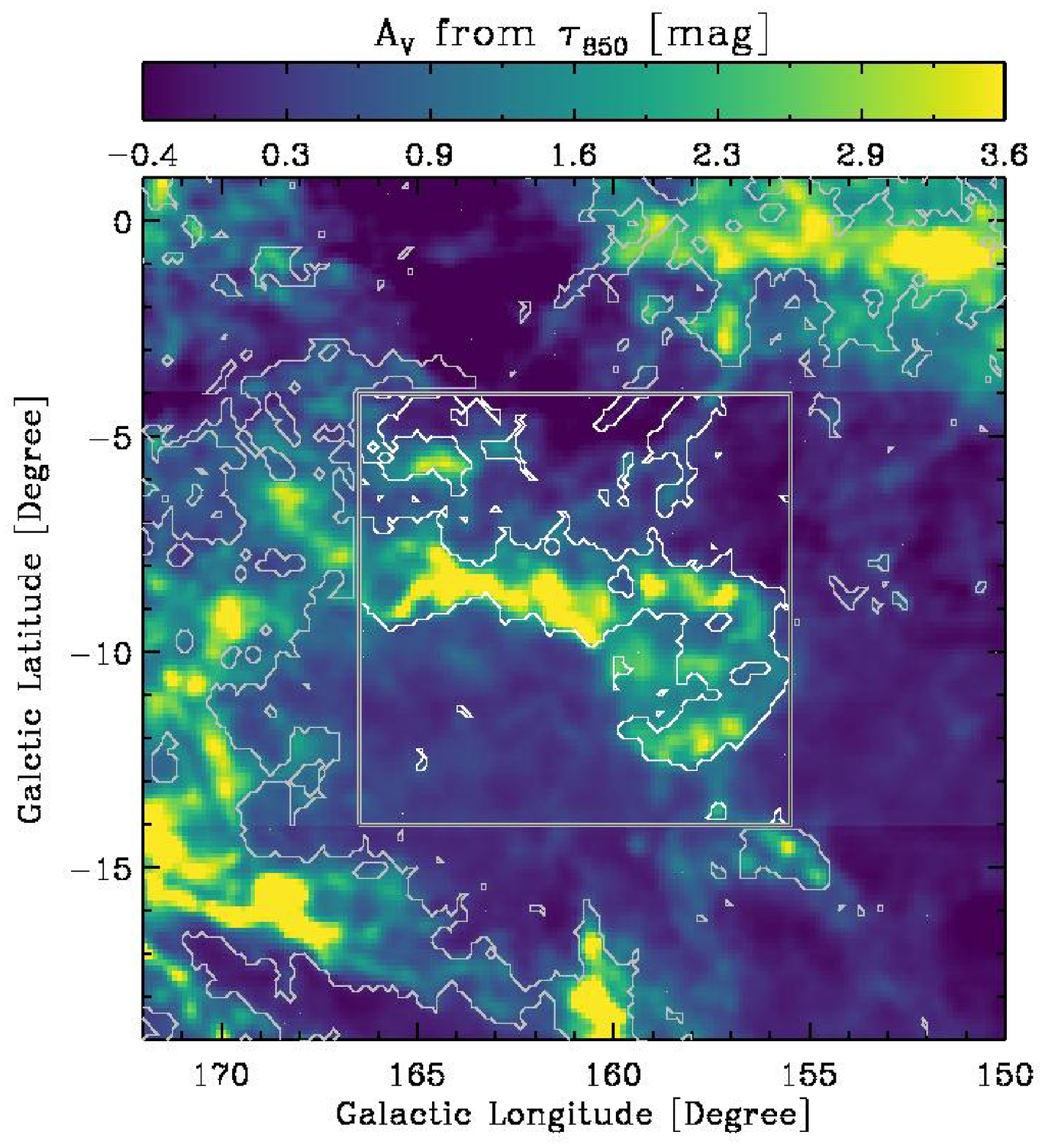}{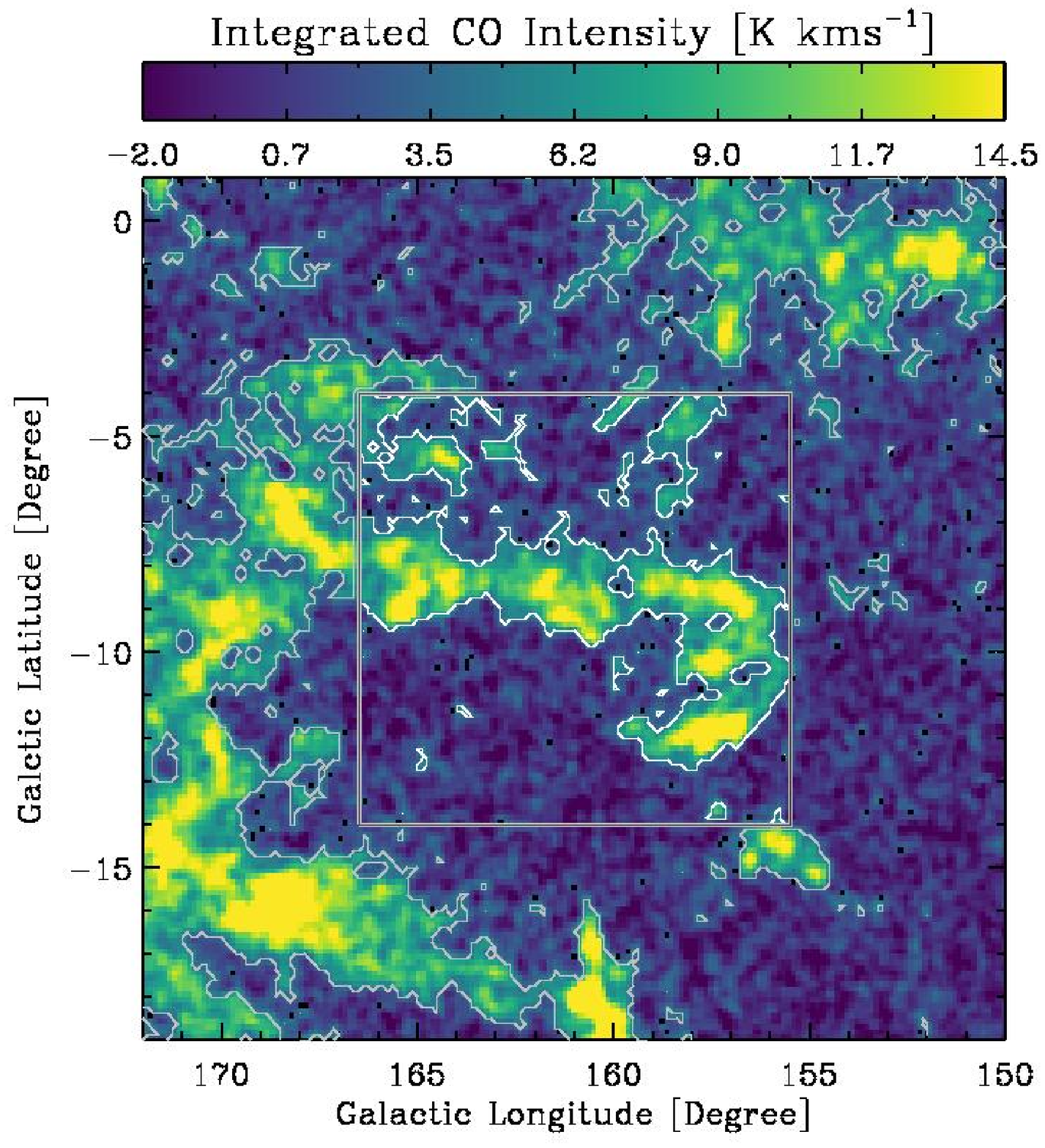}
\plottwo{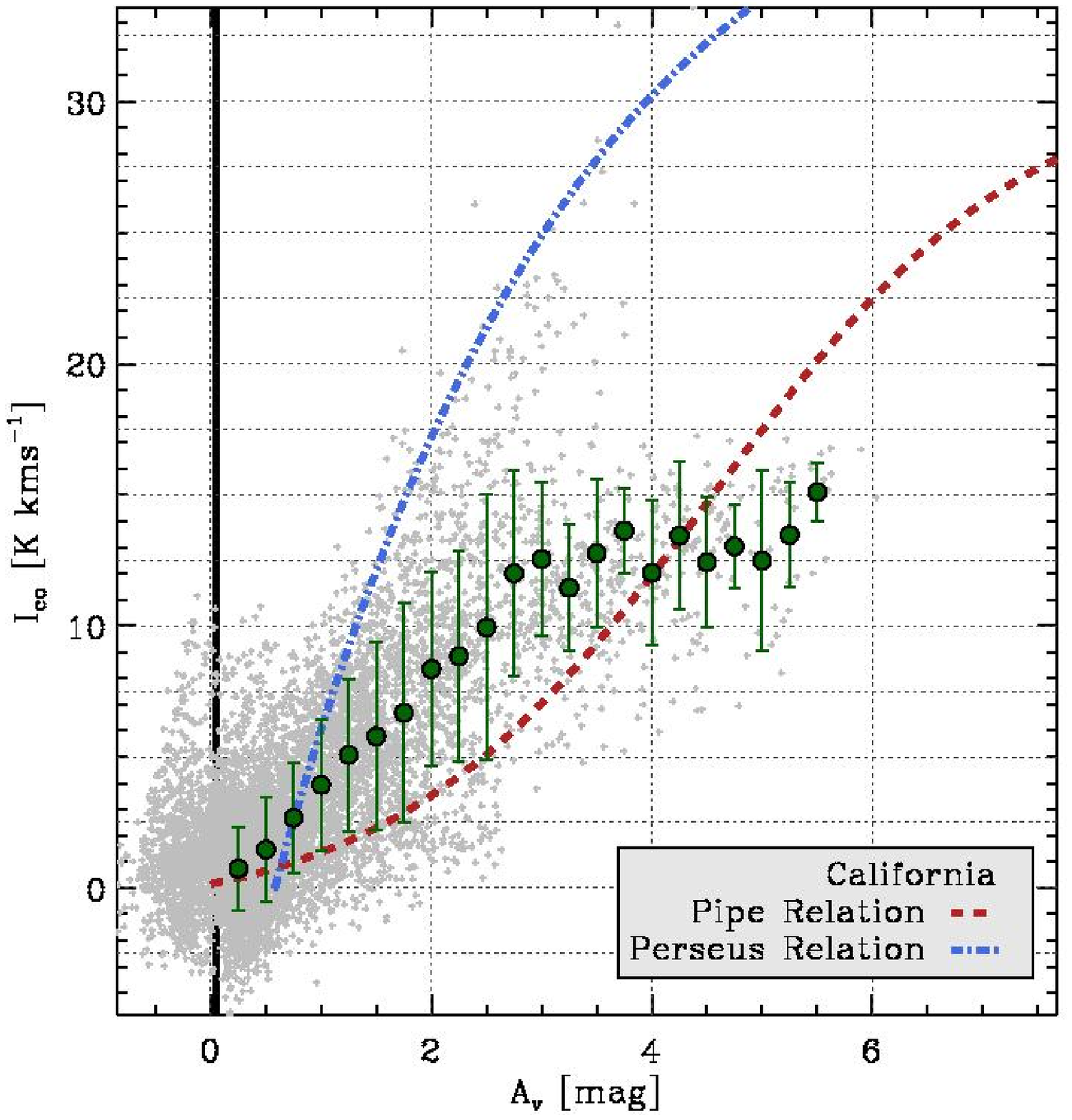}{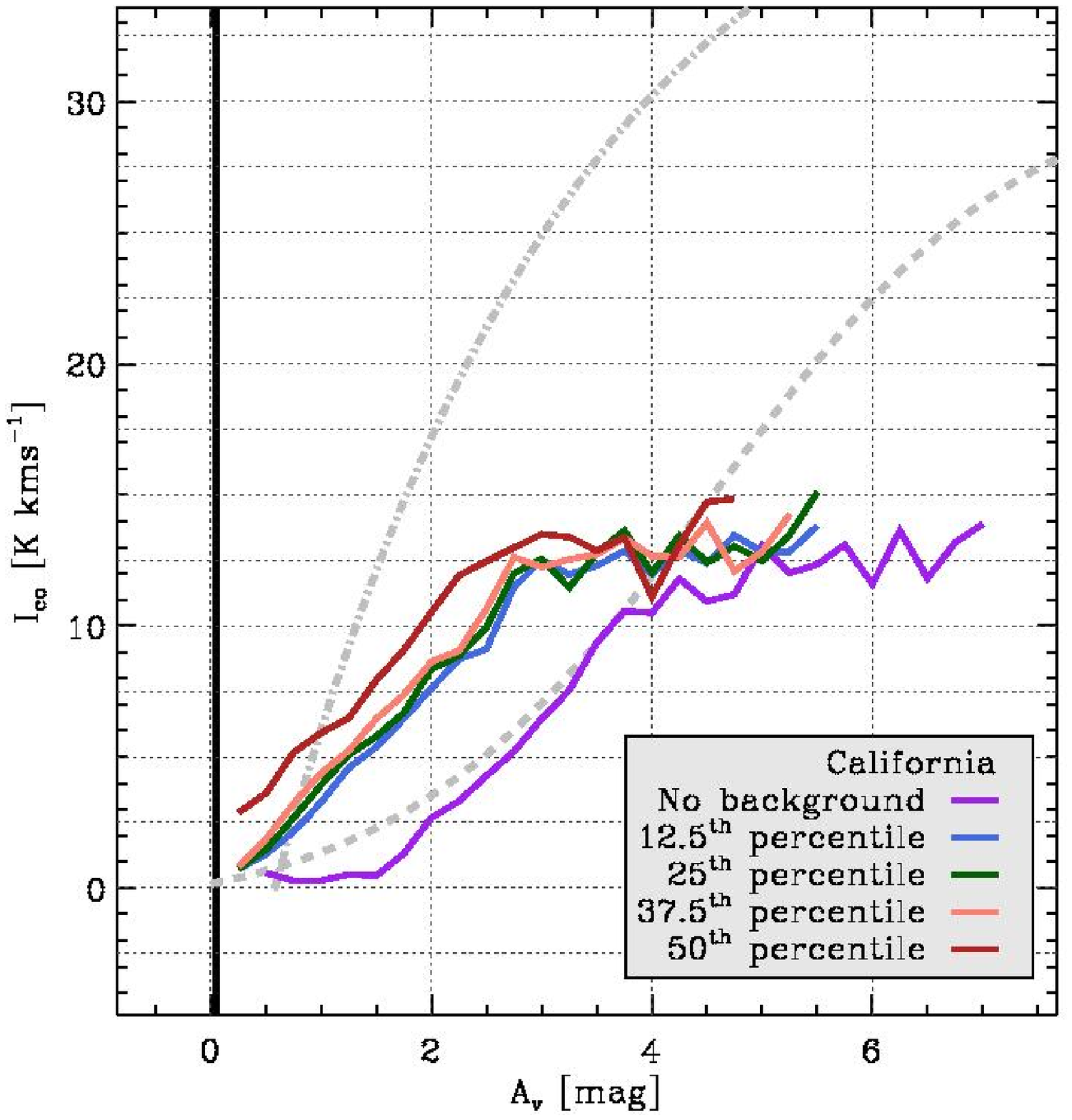}
\caption{Same as Figure~\ref{fig:taurus}, for the case of California.}
\end{figure*}

\begin{figure*}
\plottwo{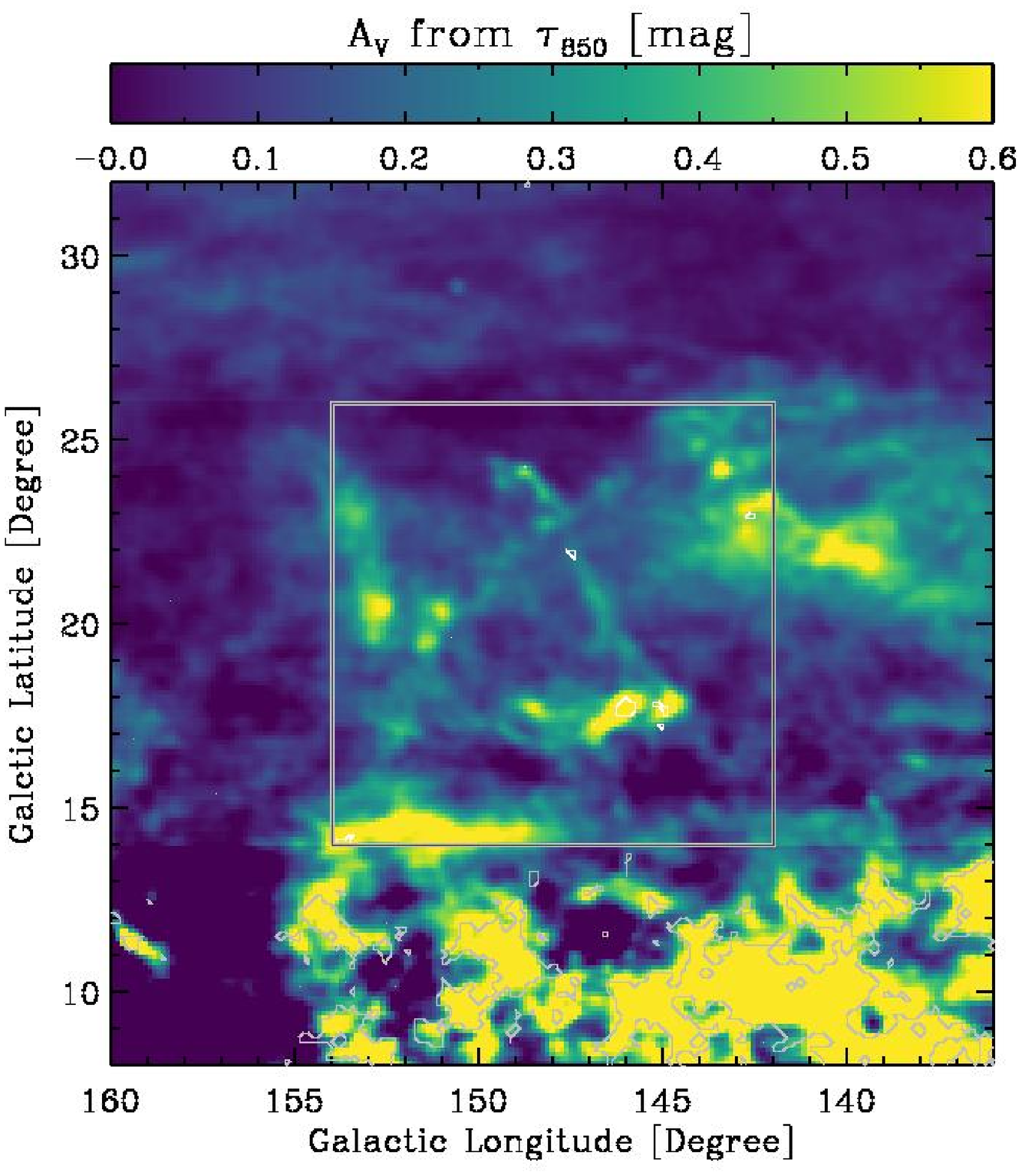}{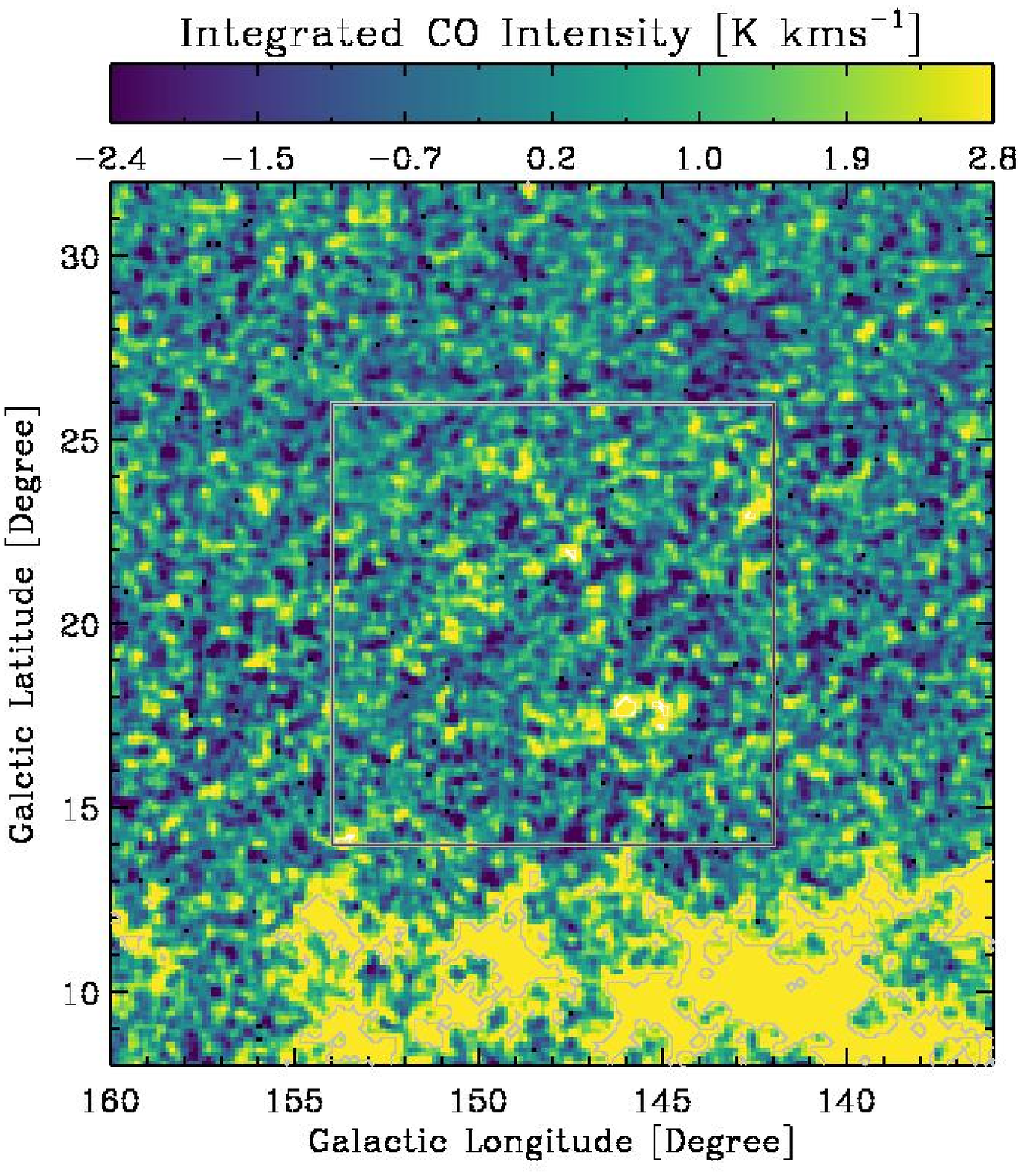}
\plottwo{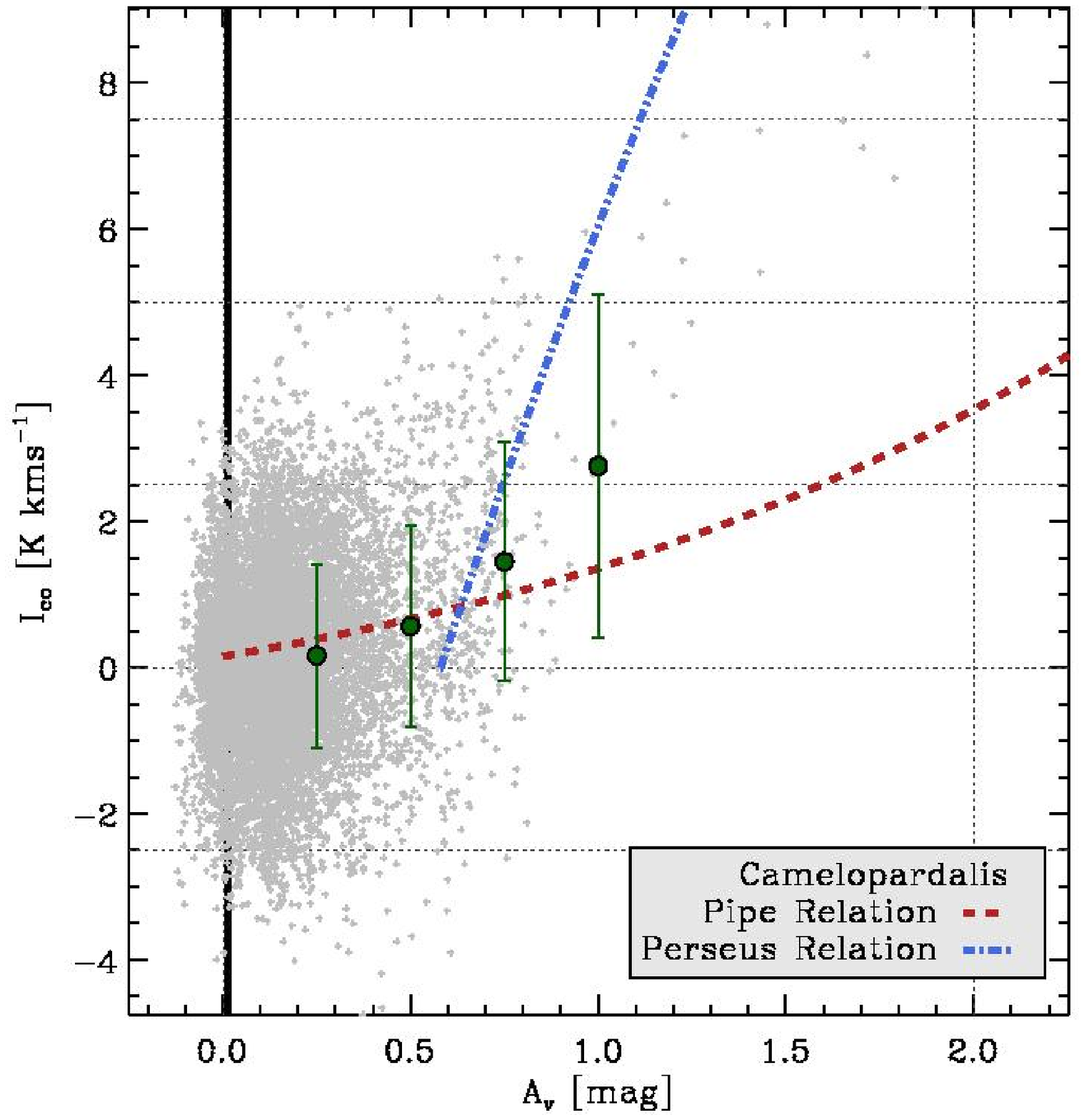}{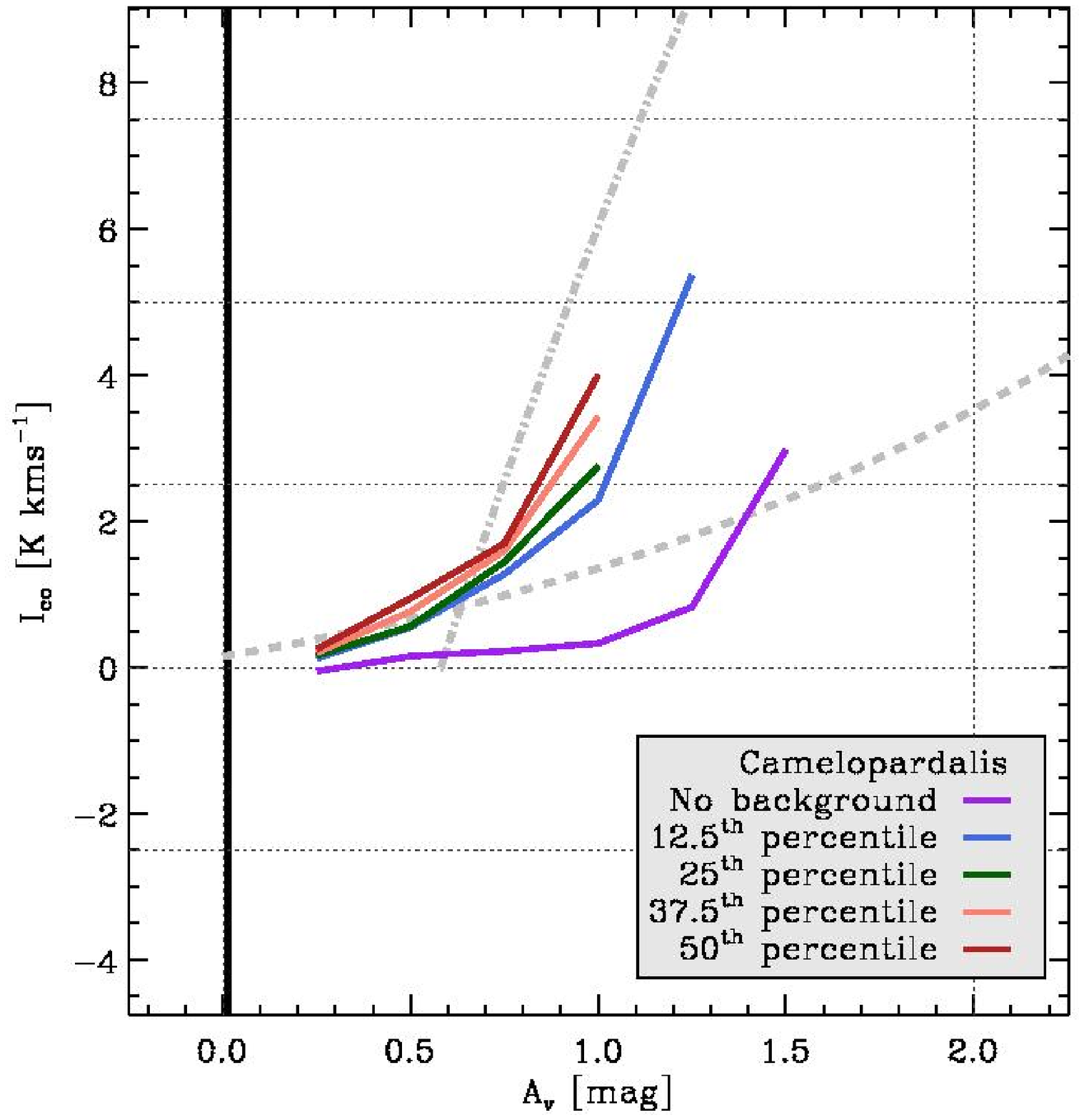}
\caption{Same as Figure~\ref{fig:taurus}, for the case of Camelopardalis.}
\end{figure*}

\begin{figure*}
\plottwo{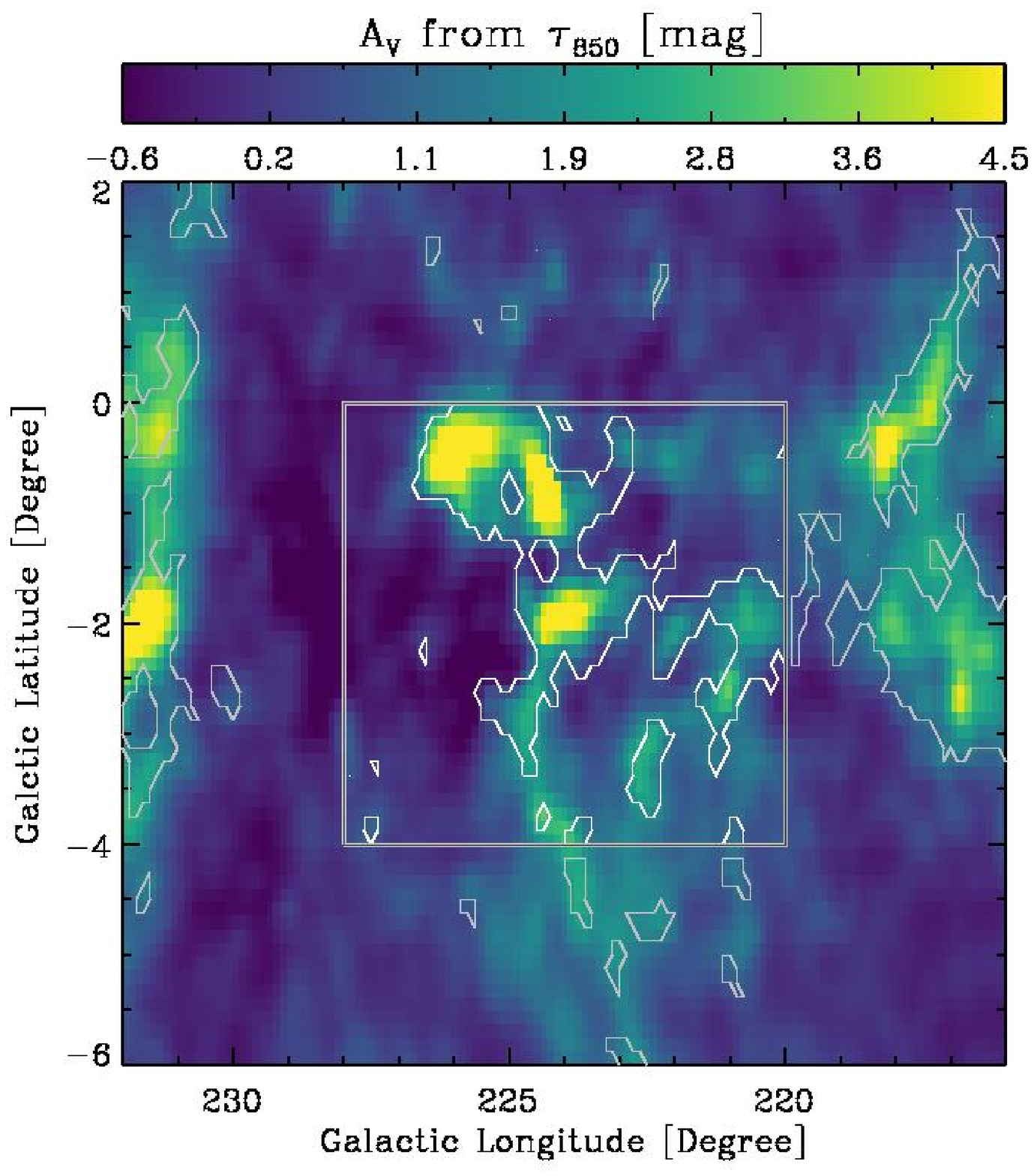}{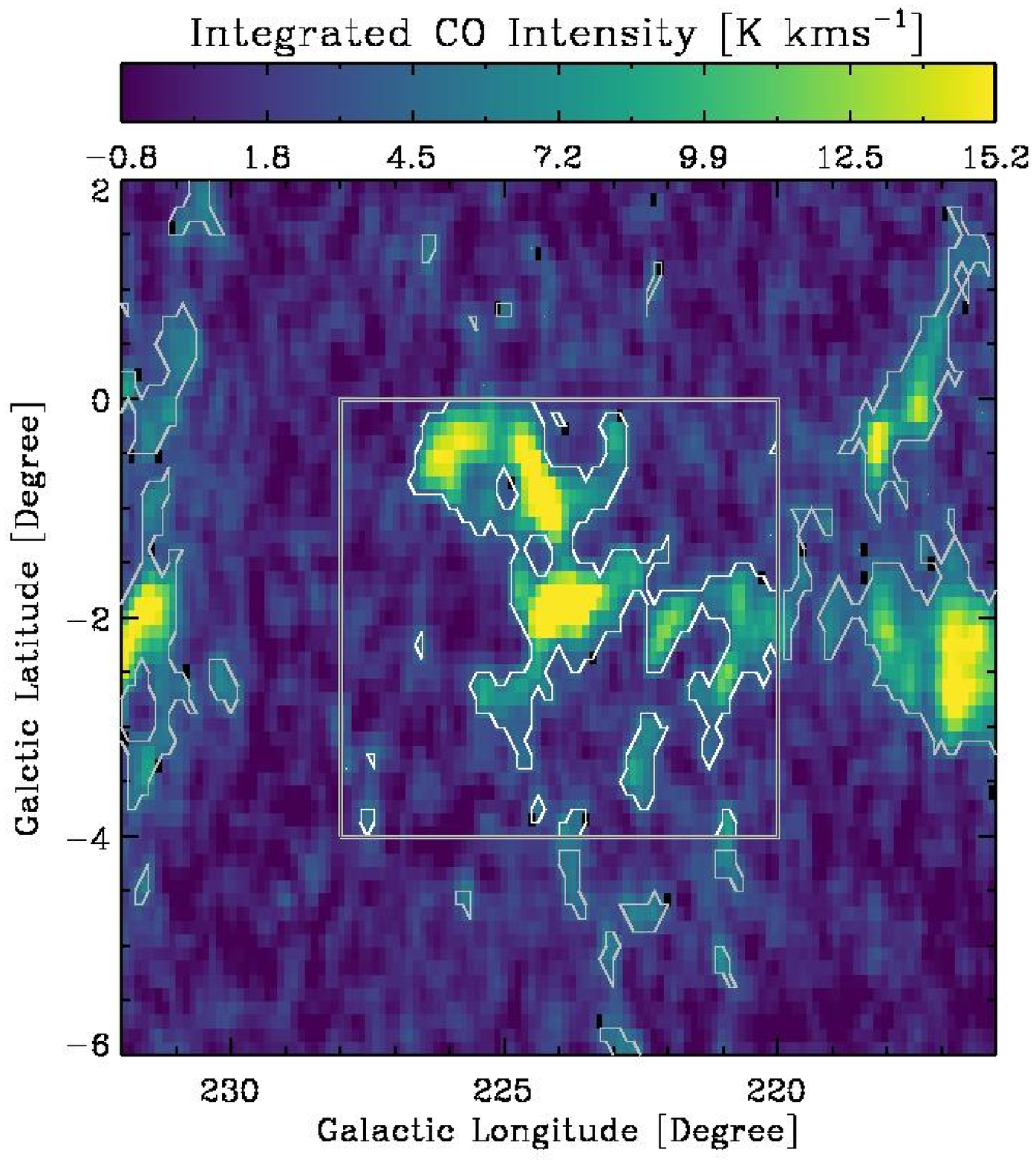}
\plottwo{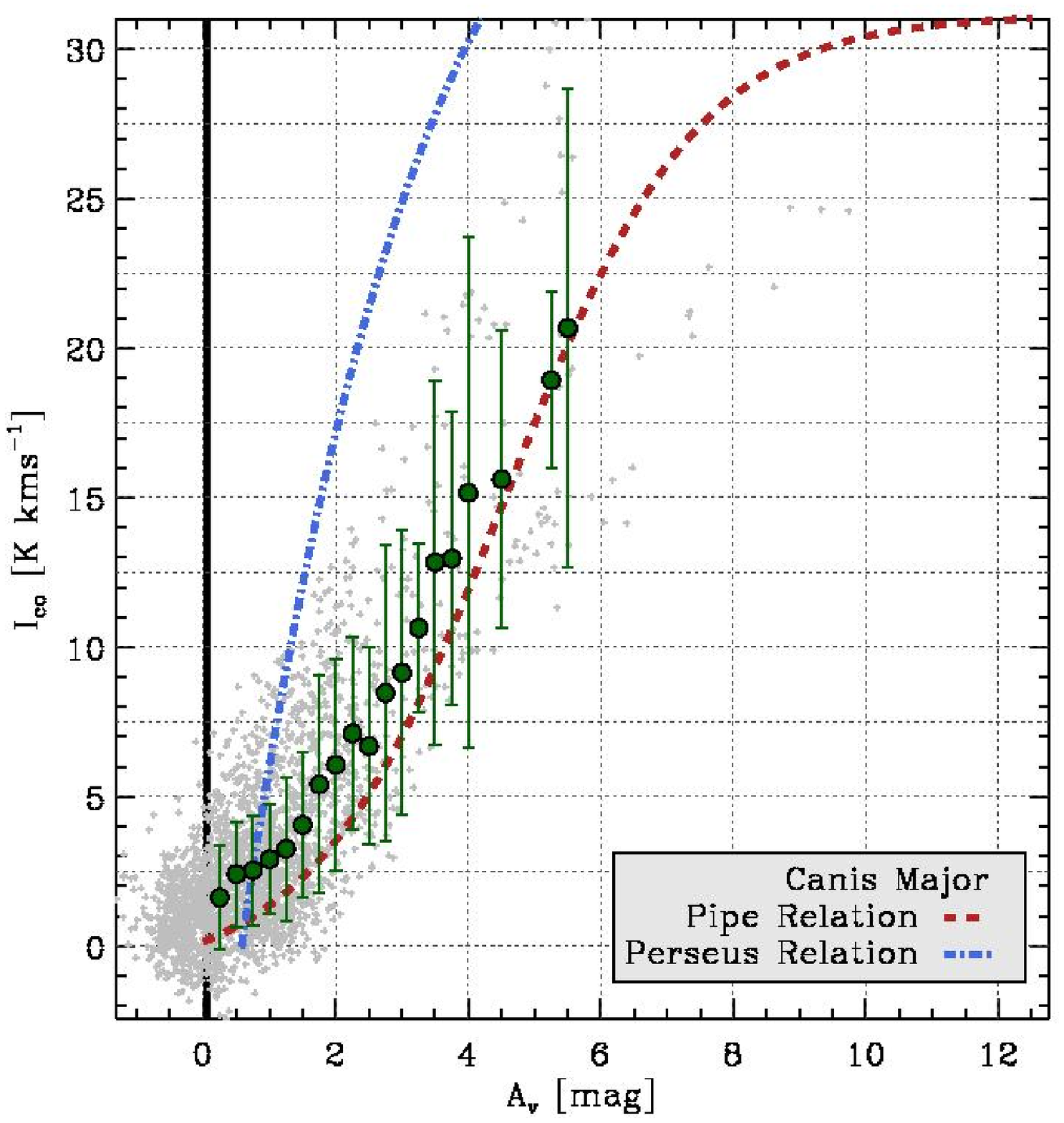}{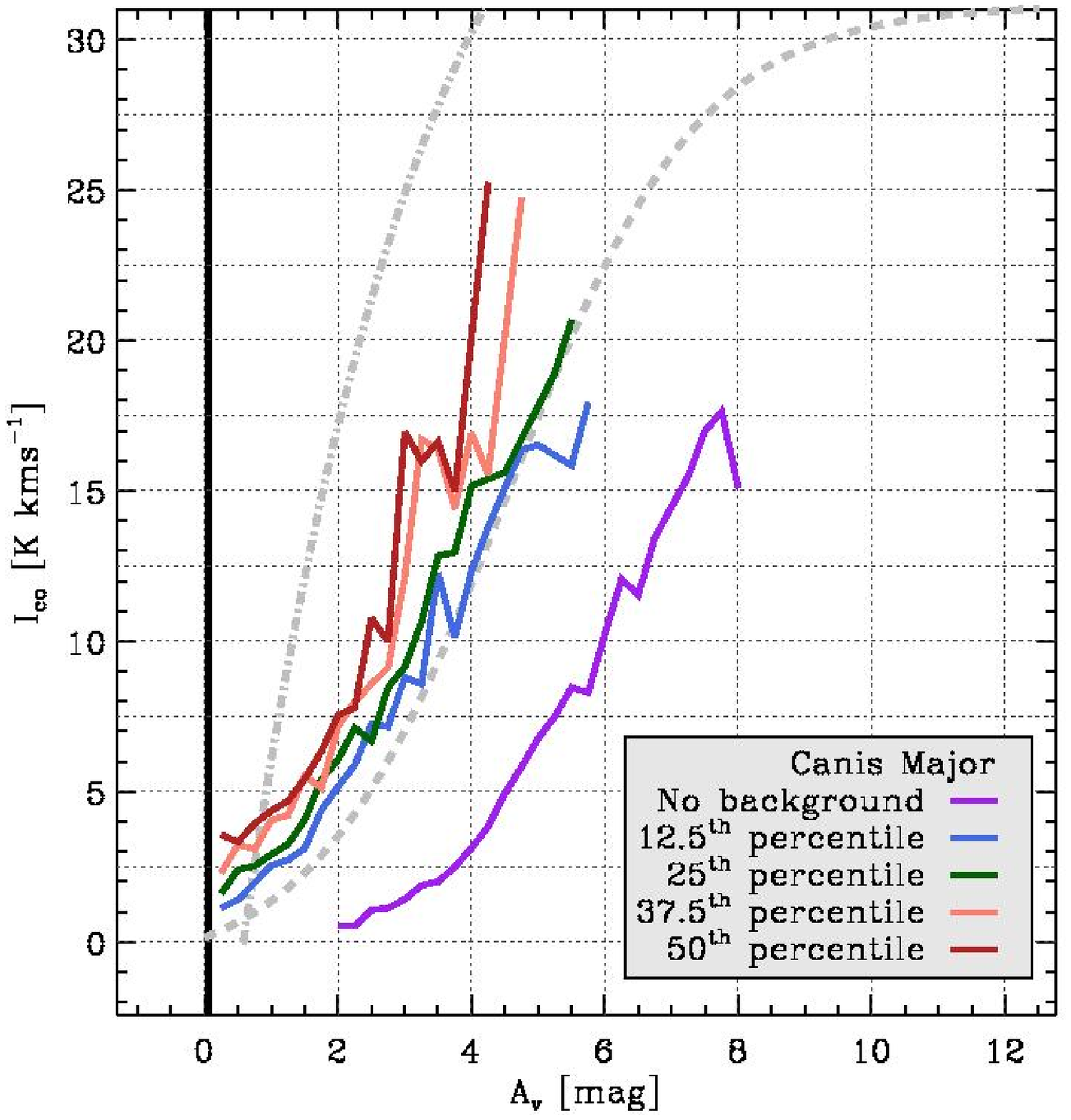}
\caption{Same as Figure~\ref{fig:taurus}, for the case of Canis Major.}
\end{figure*}

\begin{figure*}
\plottwo{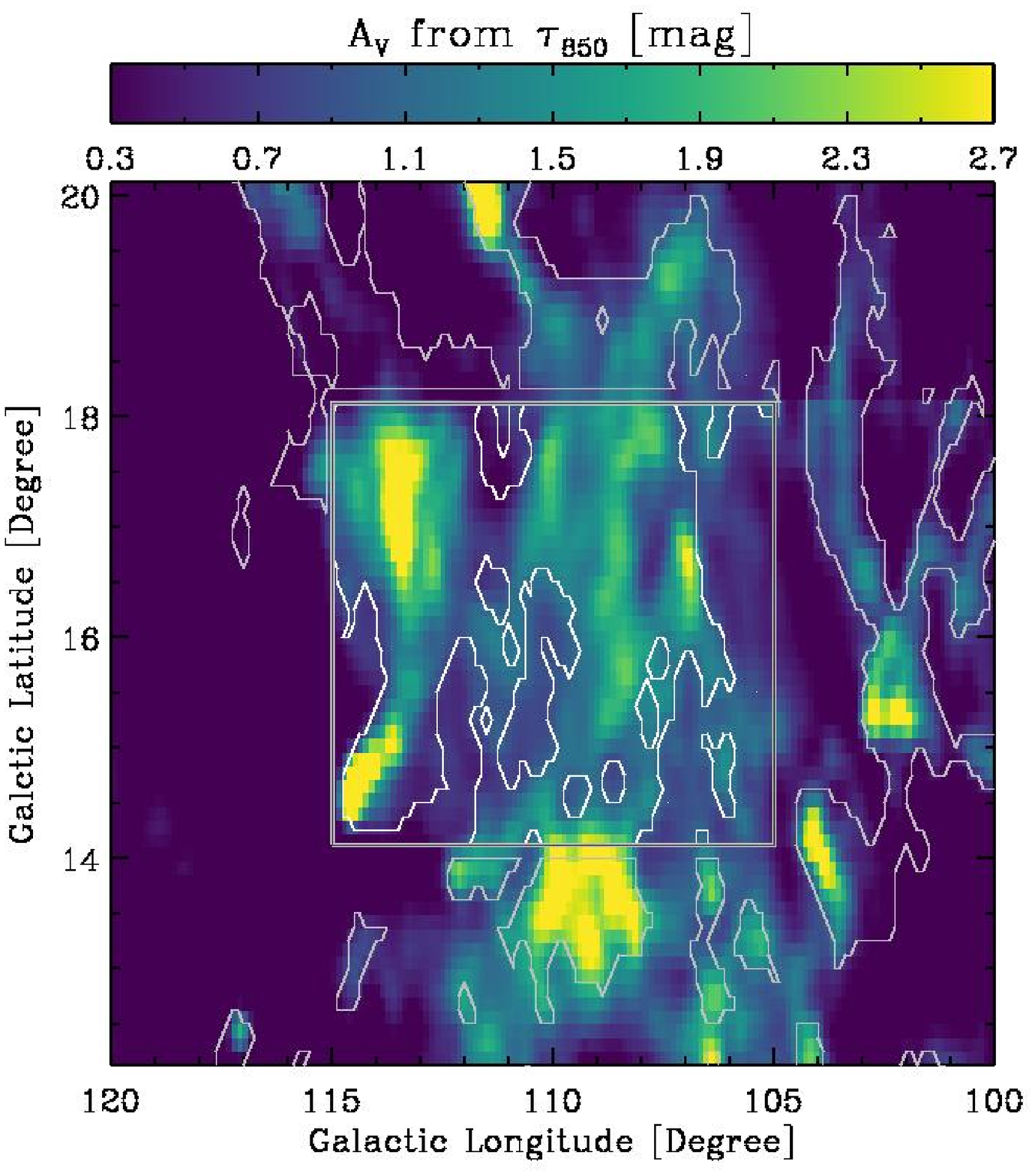}{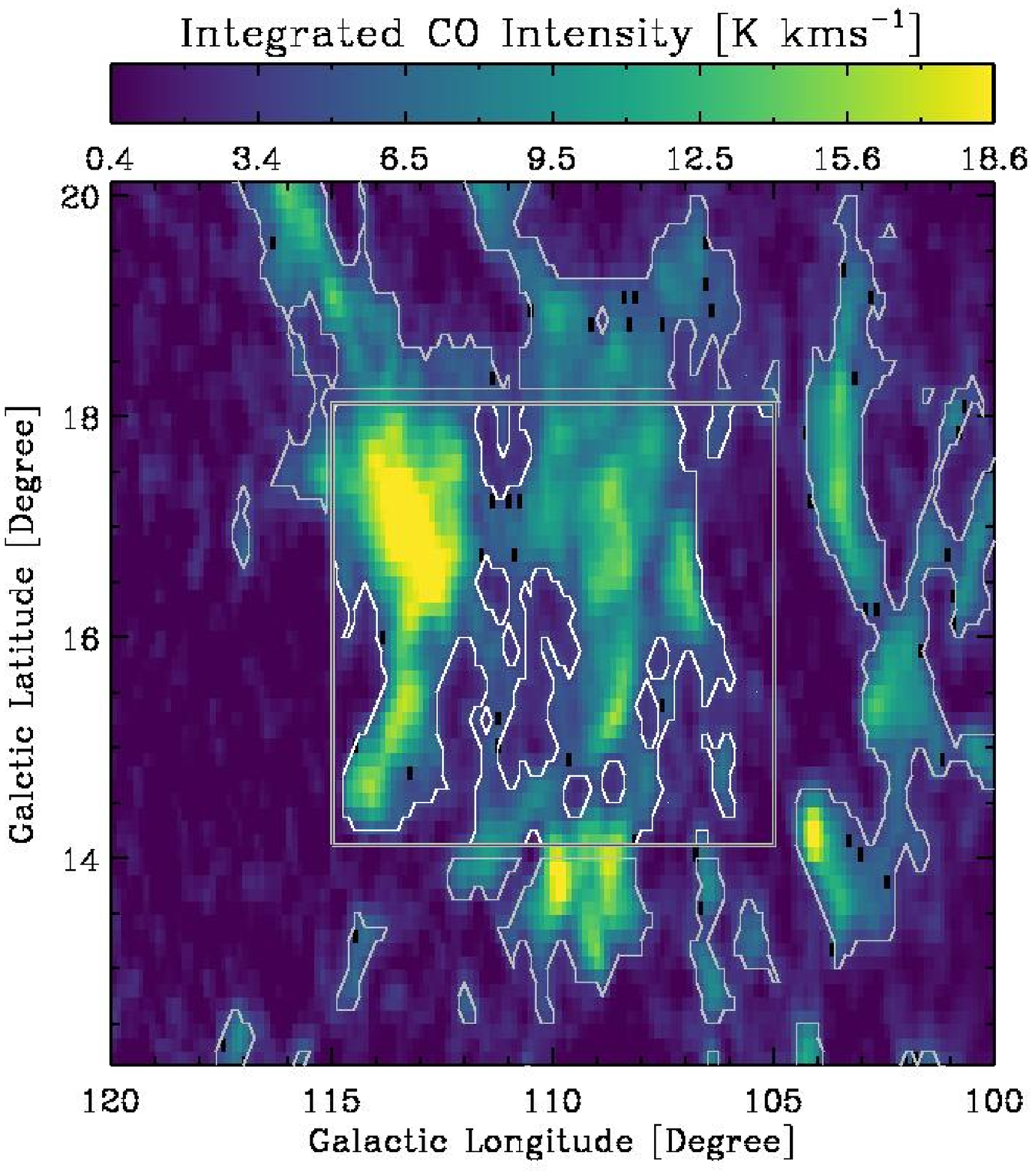}
\plottwo{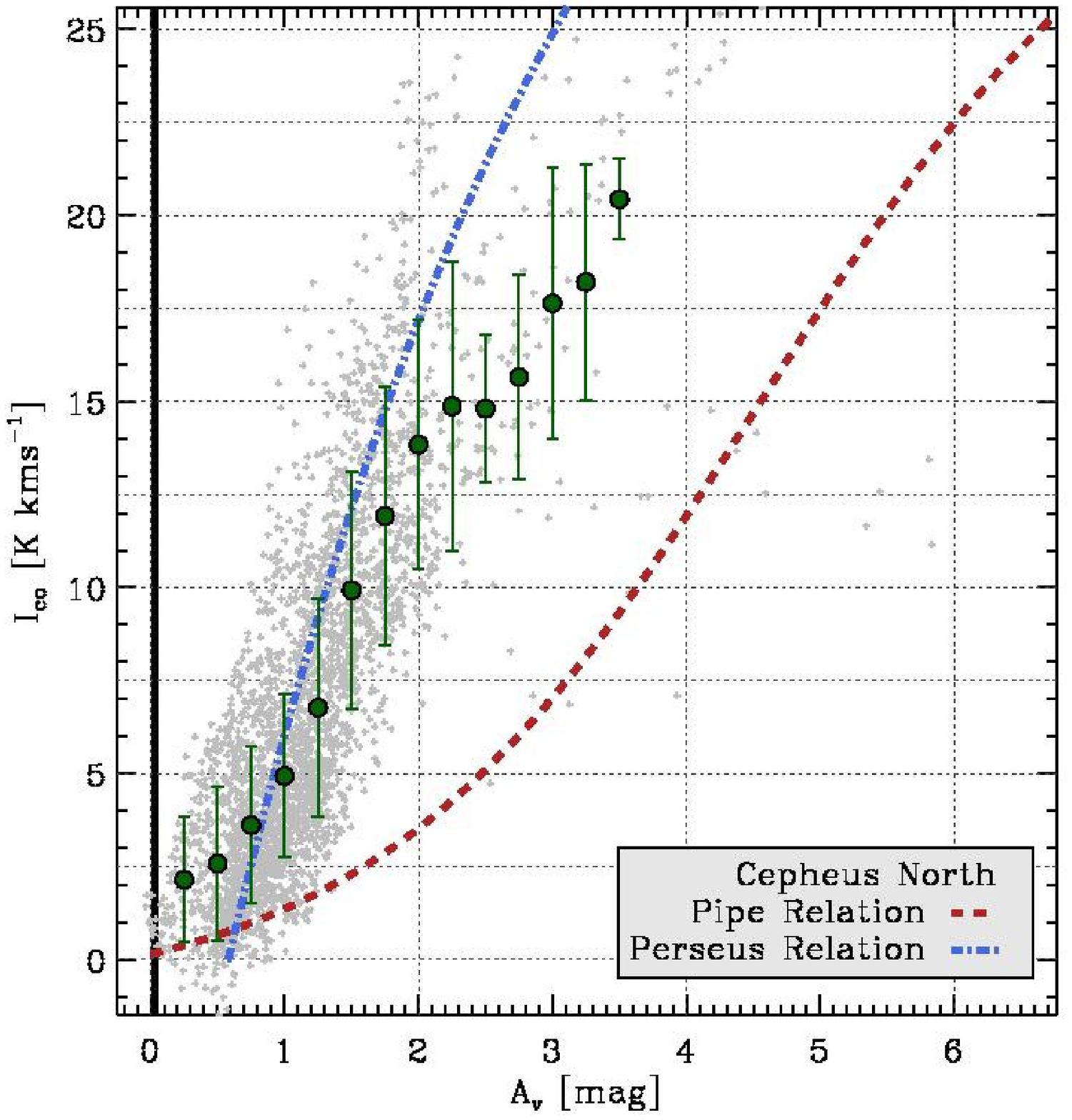}{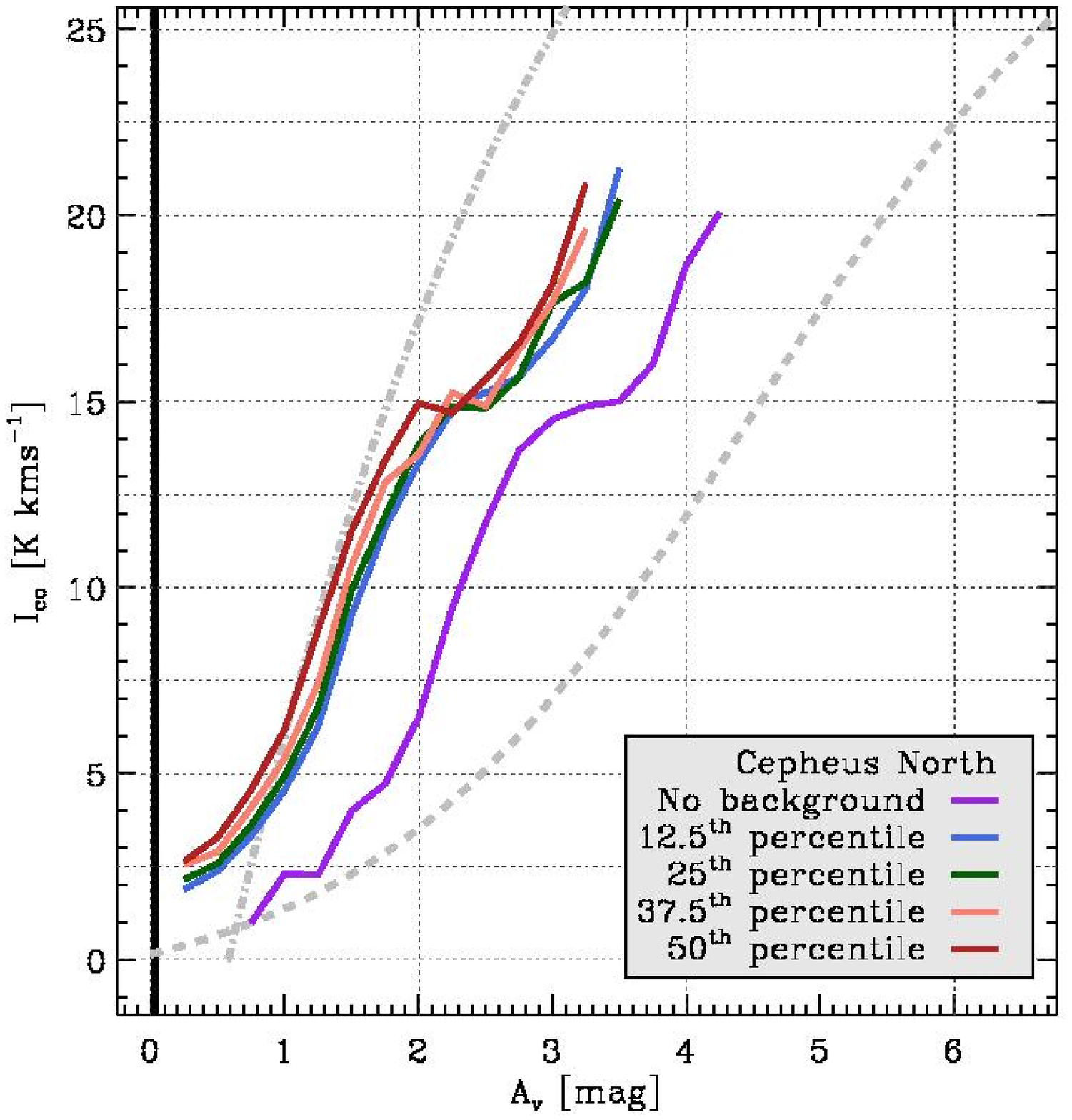}
\caption{Same as Figure~\ref{fig:taurus}, for the case of Cepheus North.}
\end{figure*}

\begin{figure*}
\plottwo{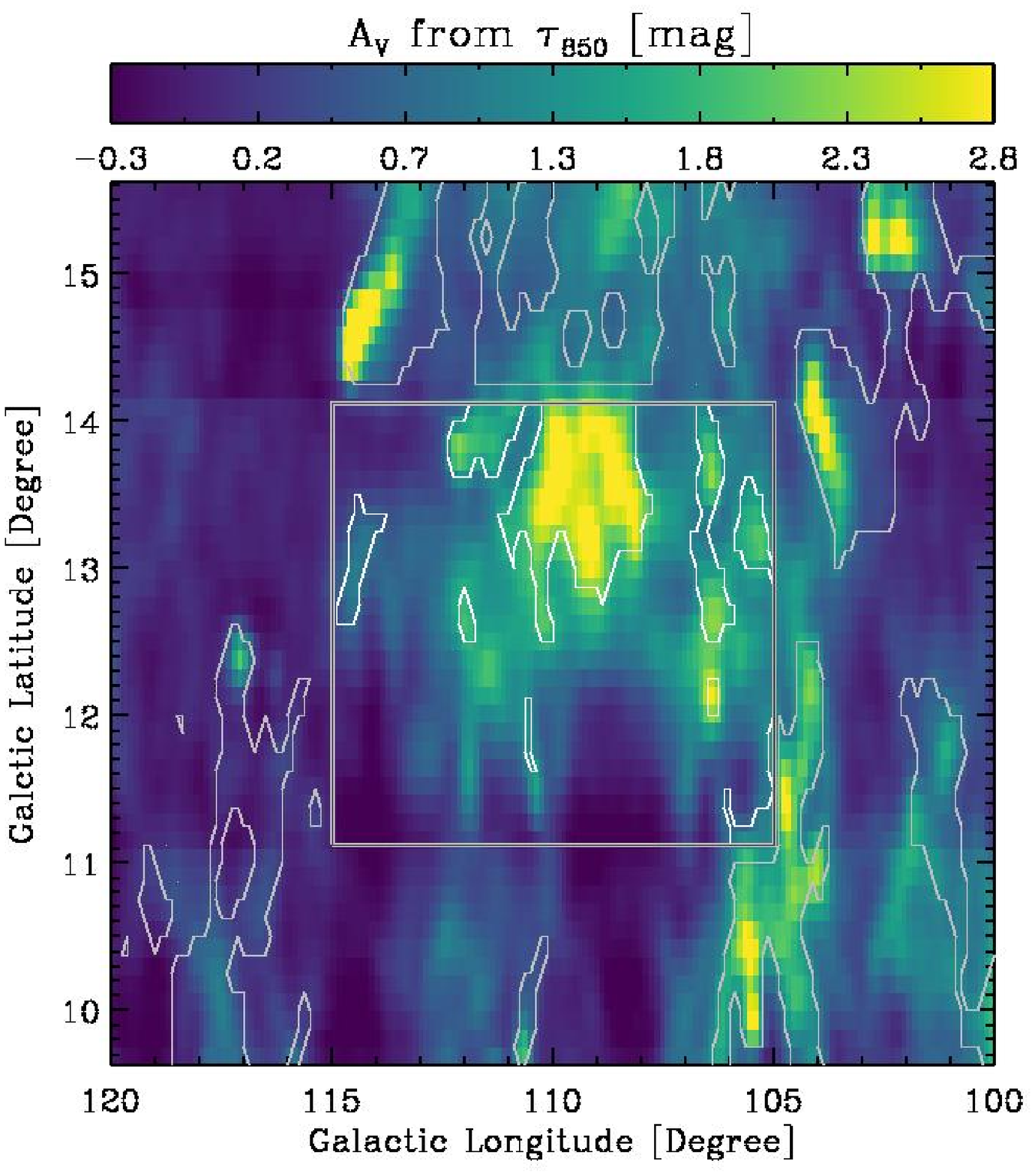}{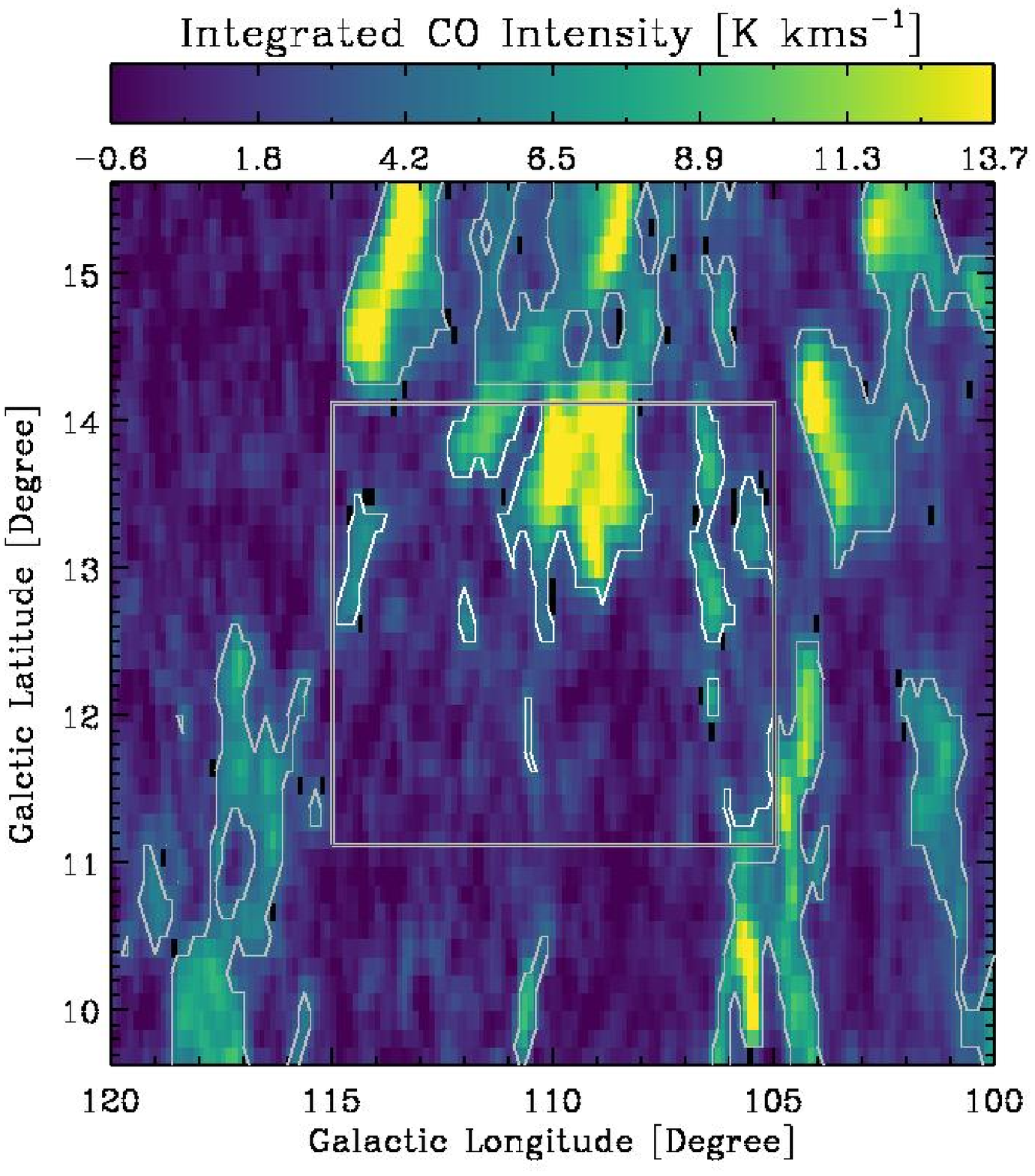}
\plottwo{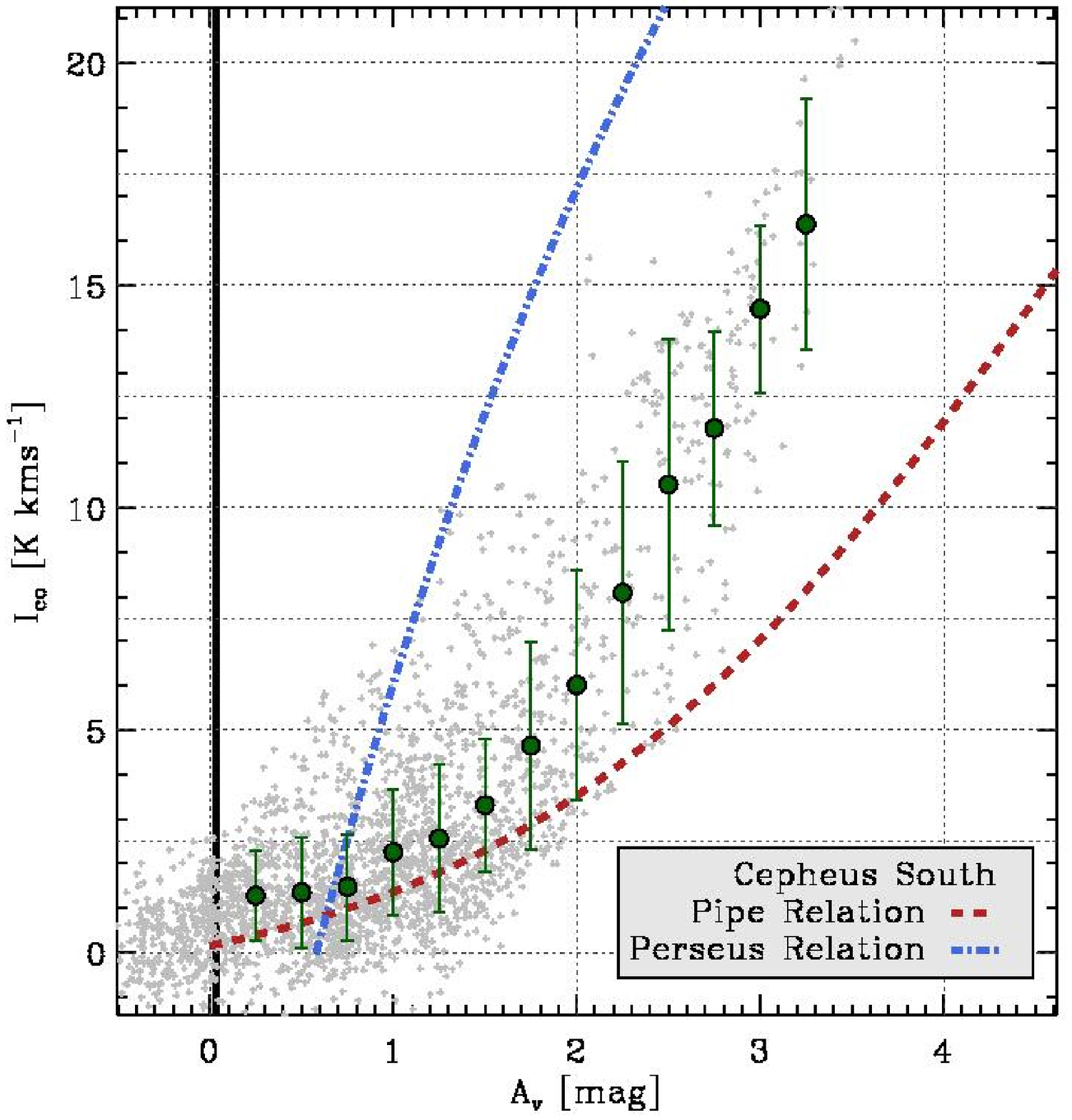}{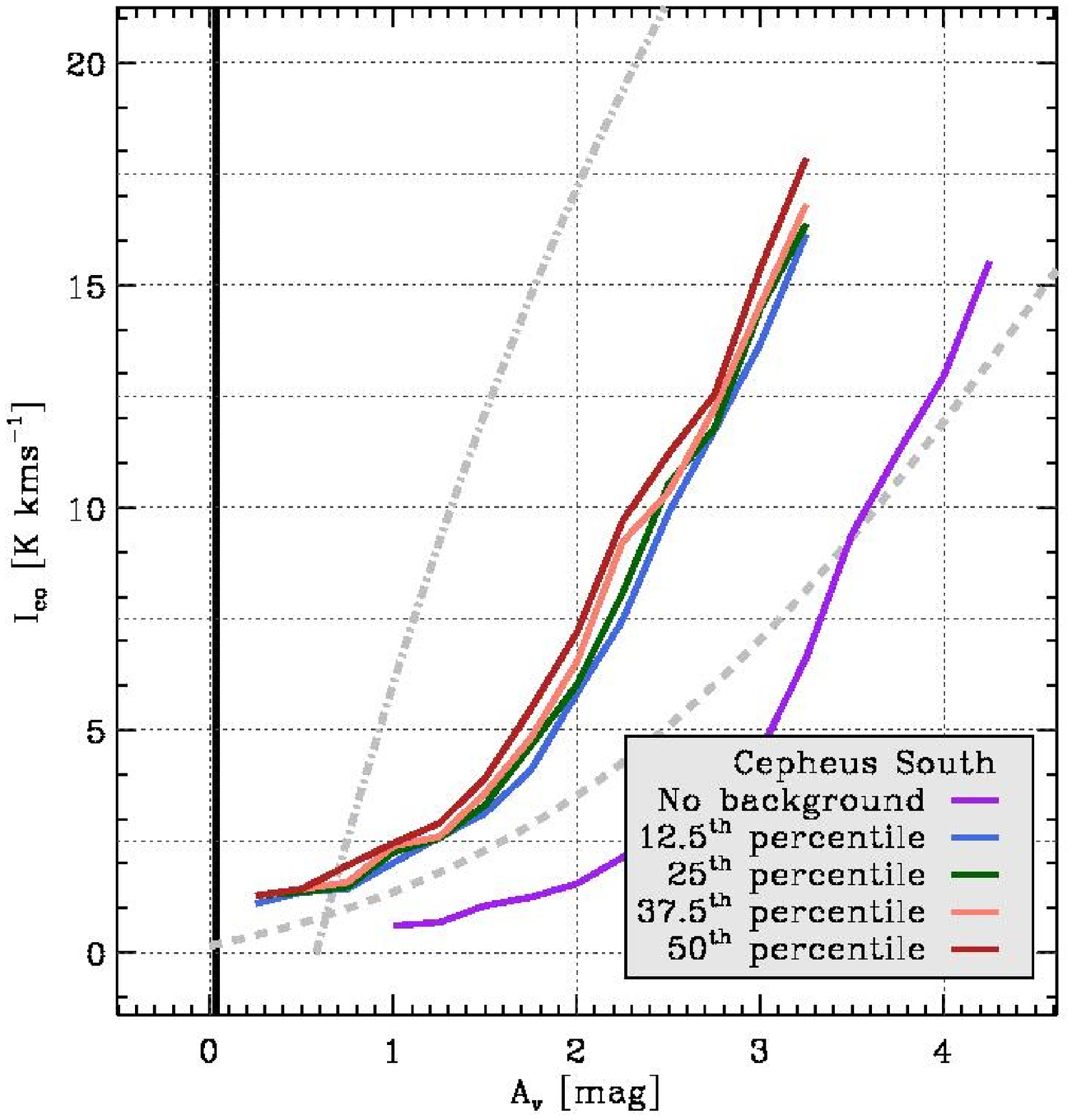}
\caption{Same as Figure~\ref{fig:taurus}, for the case of Cepheus South.}
\end{figure*}

\begin{figure*}
\plottwo{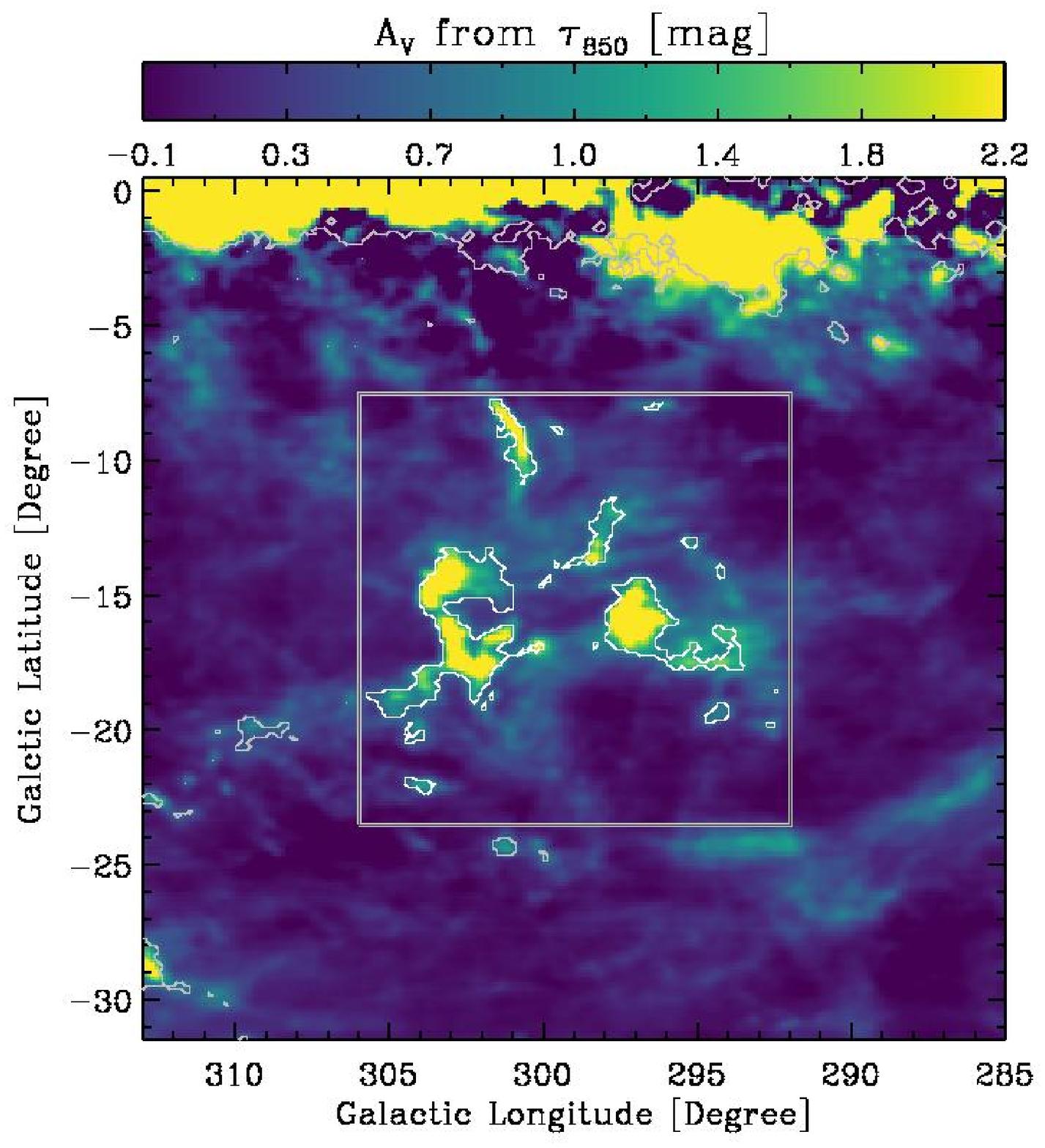}{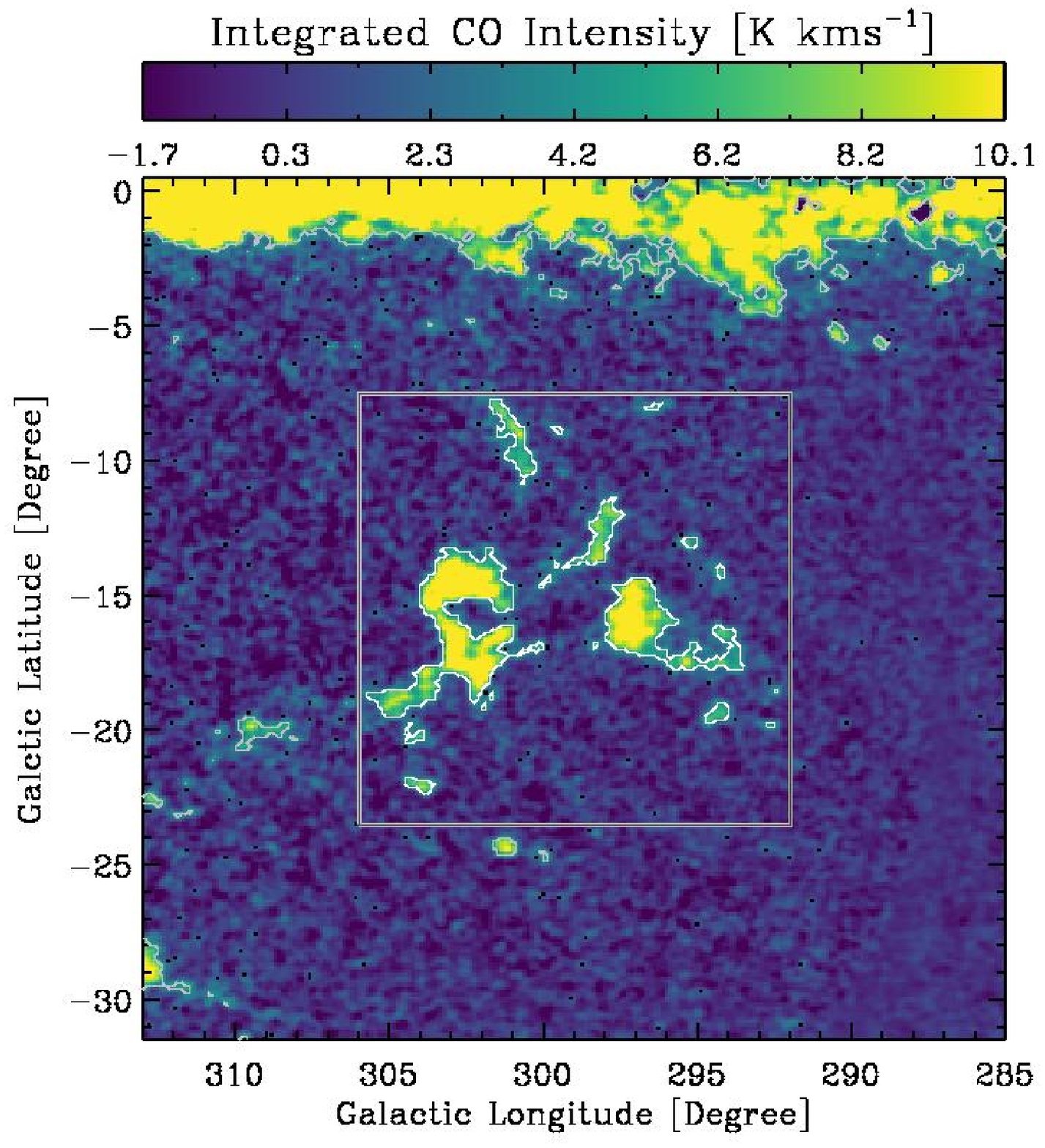}
\plottwo{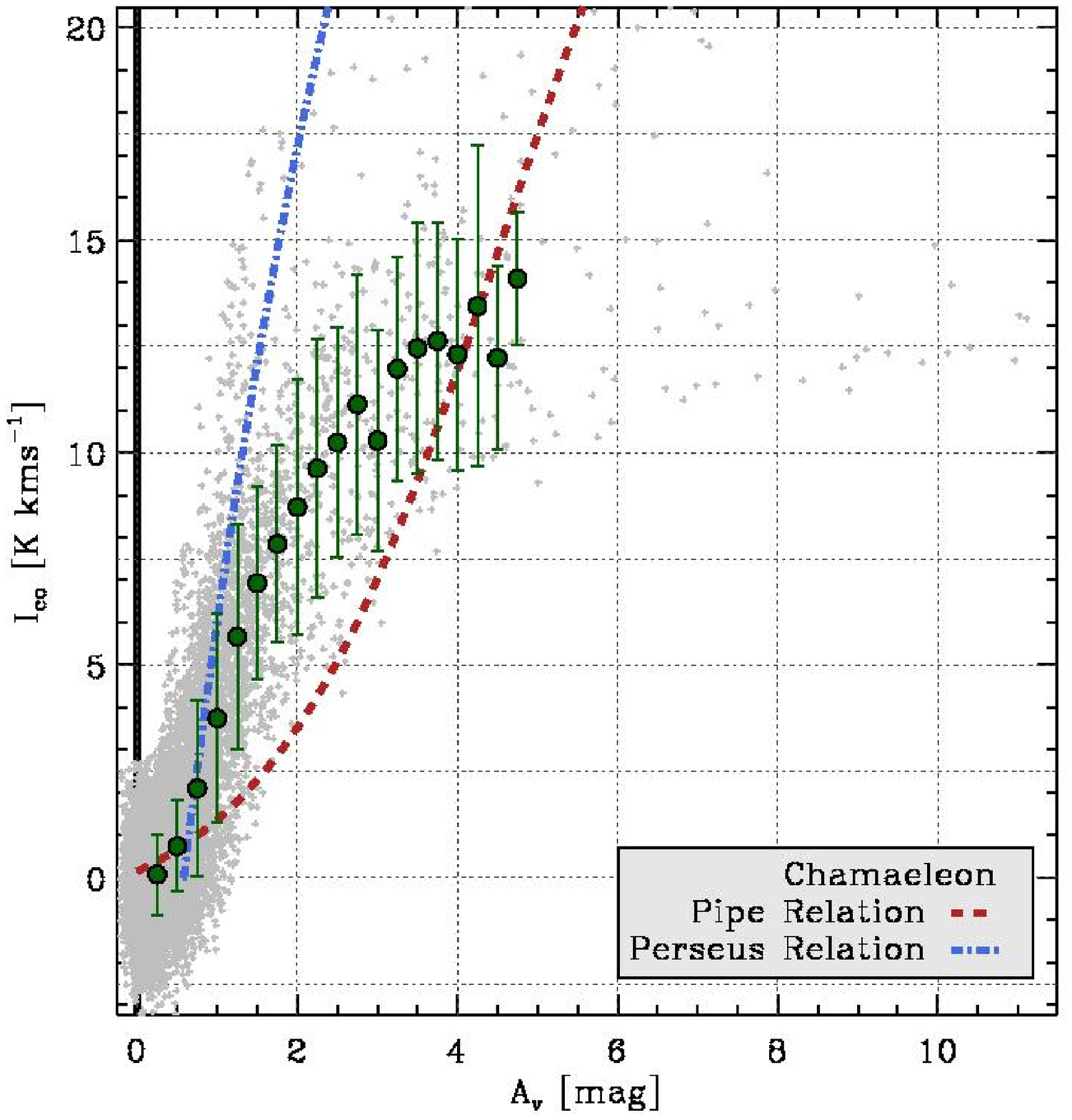}{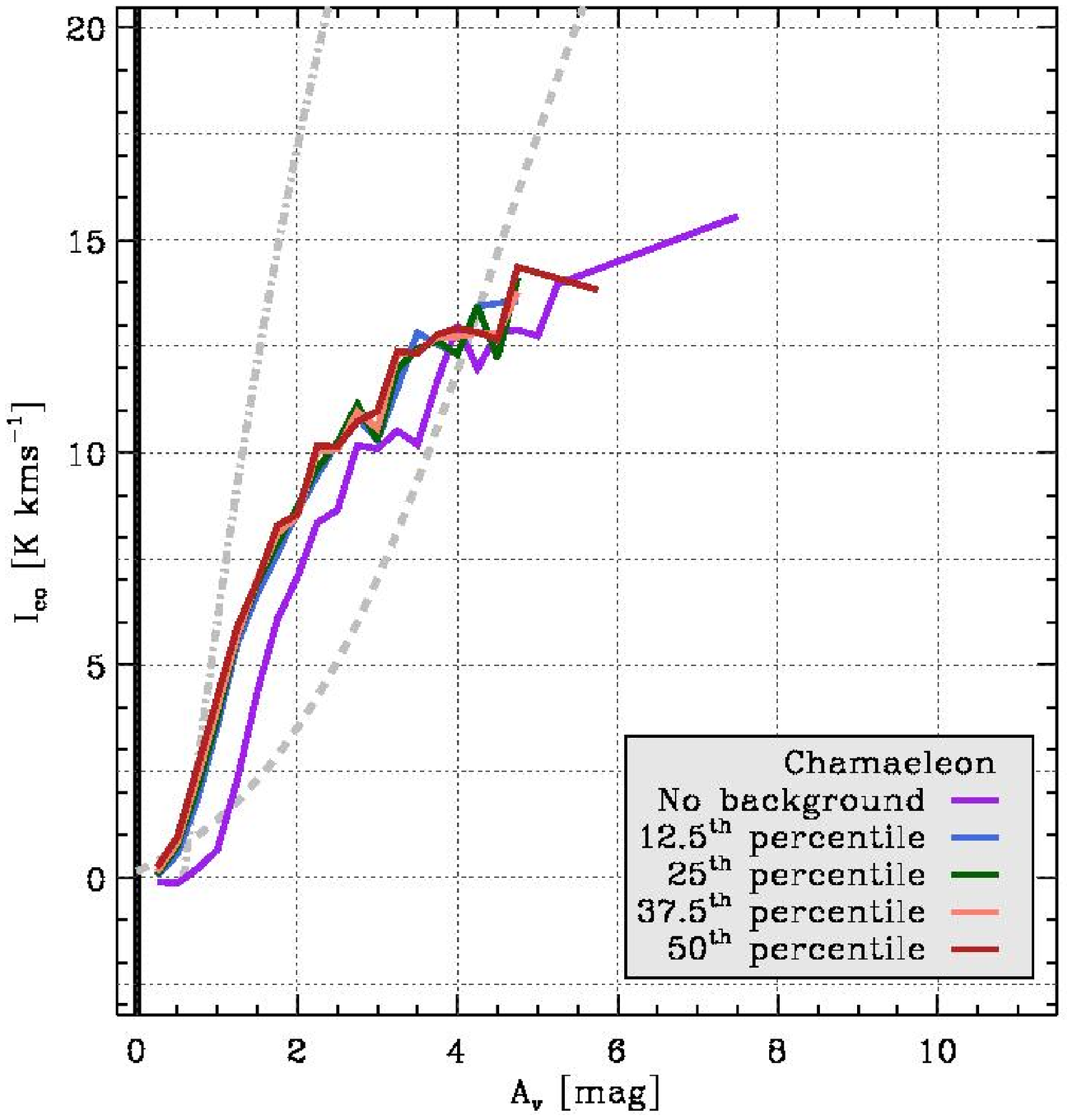}
\caption{Same as Figure~\ref{fig:taurus}, for the case of Chamaeleon.}
\end{figure*}

\begin{figure*}
\plottwo{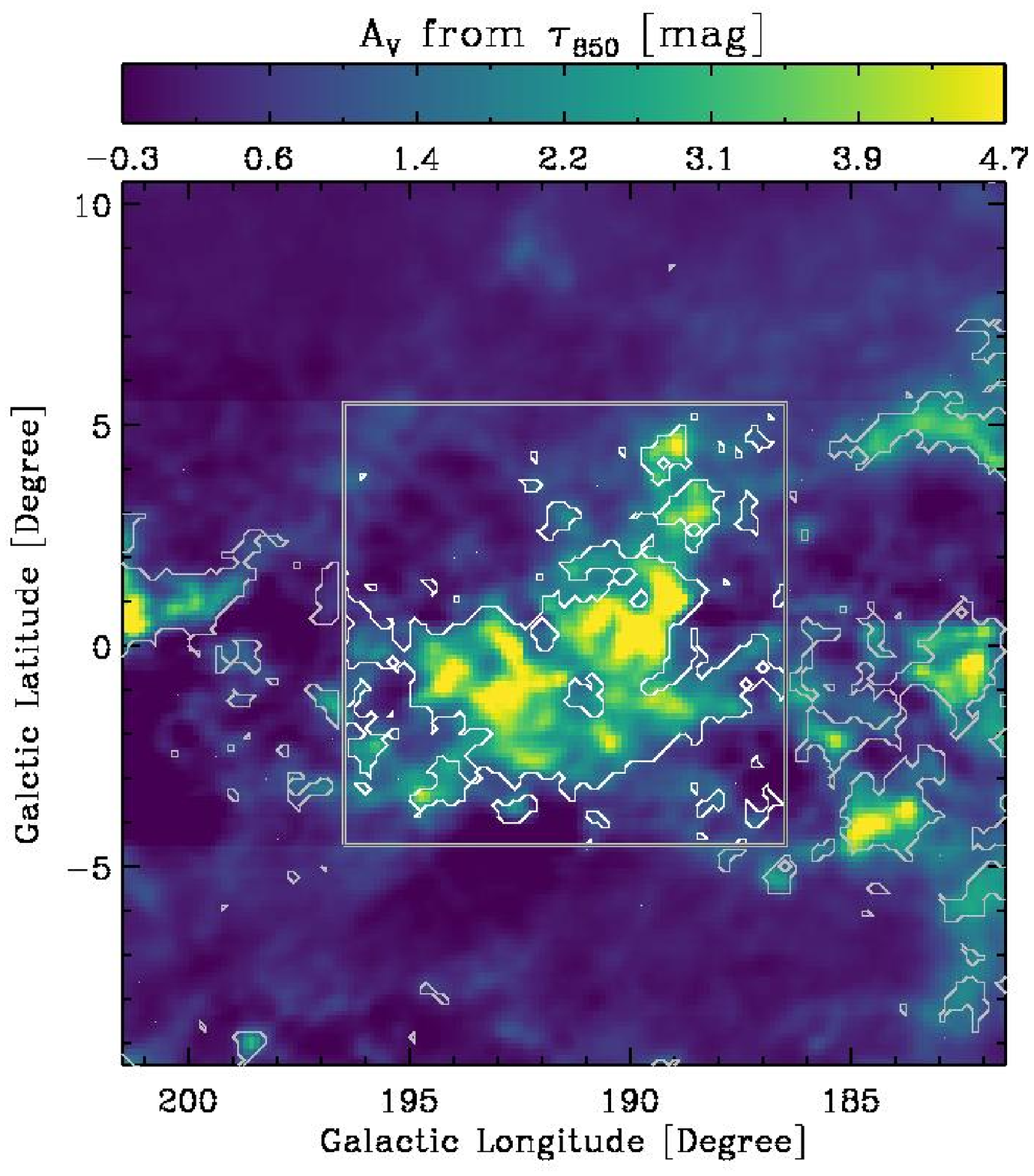}{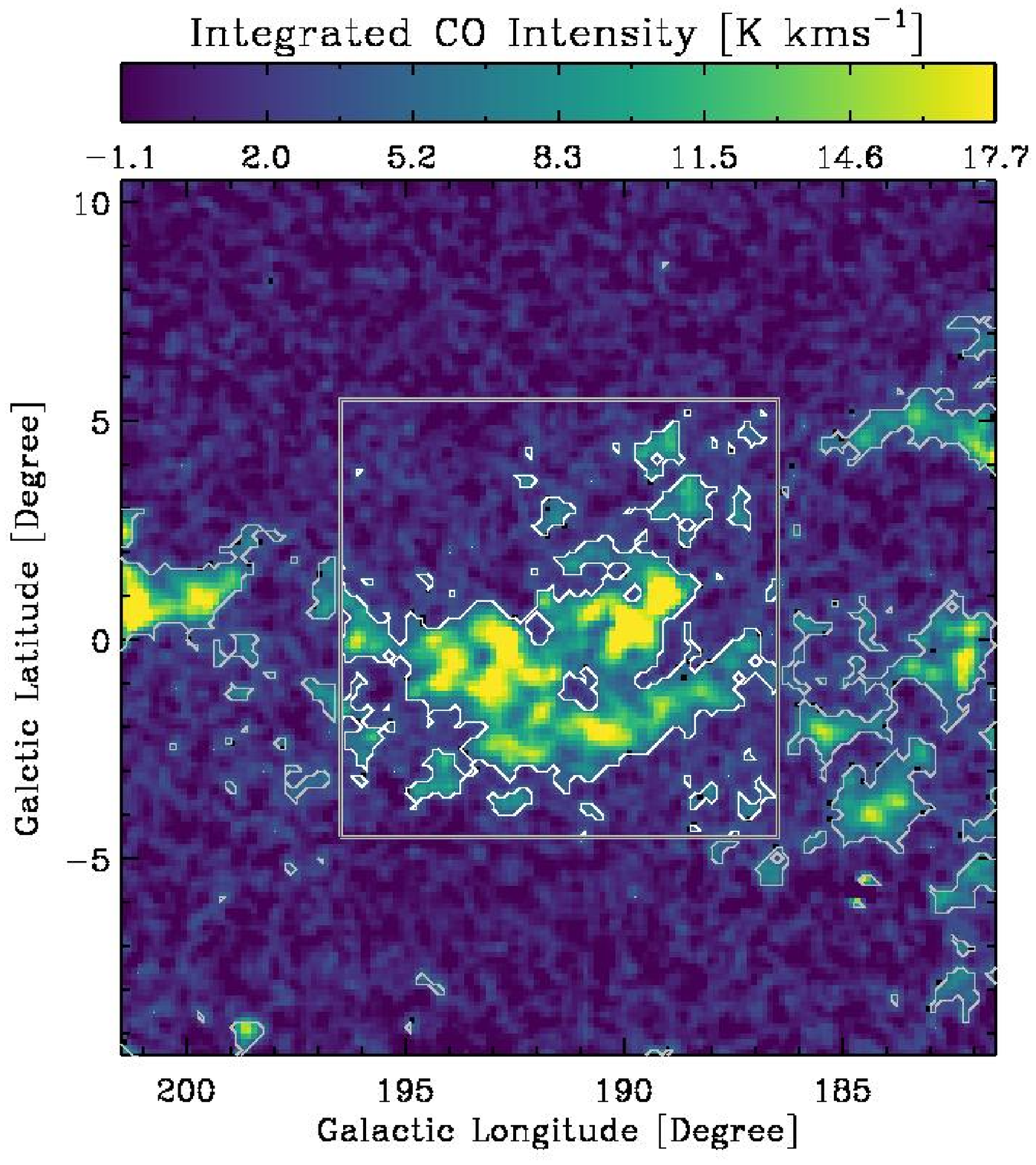}
\plottwo{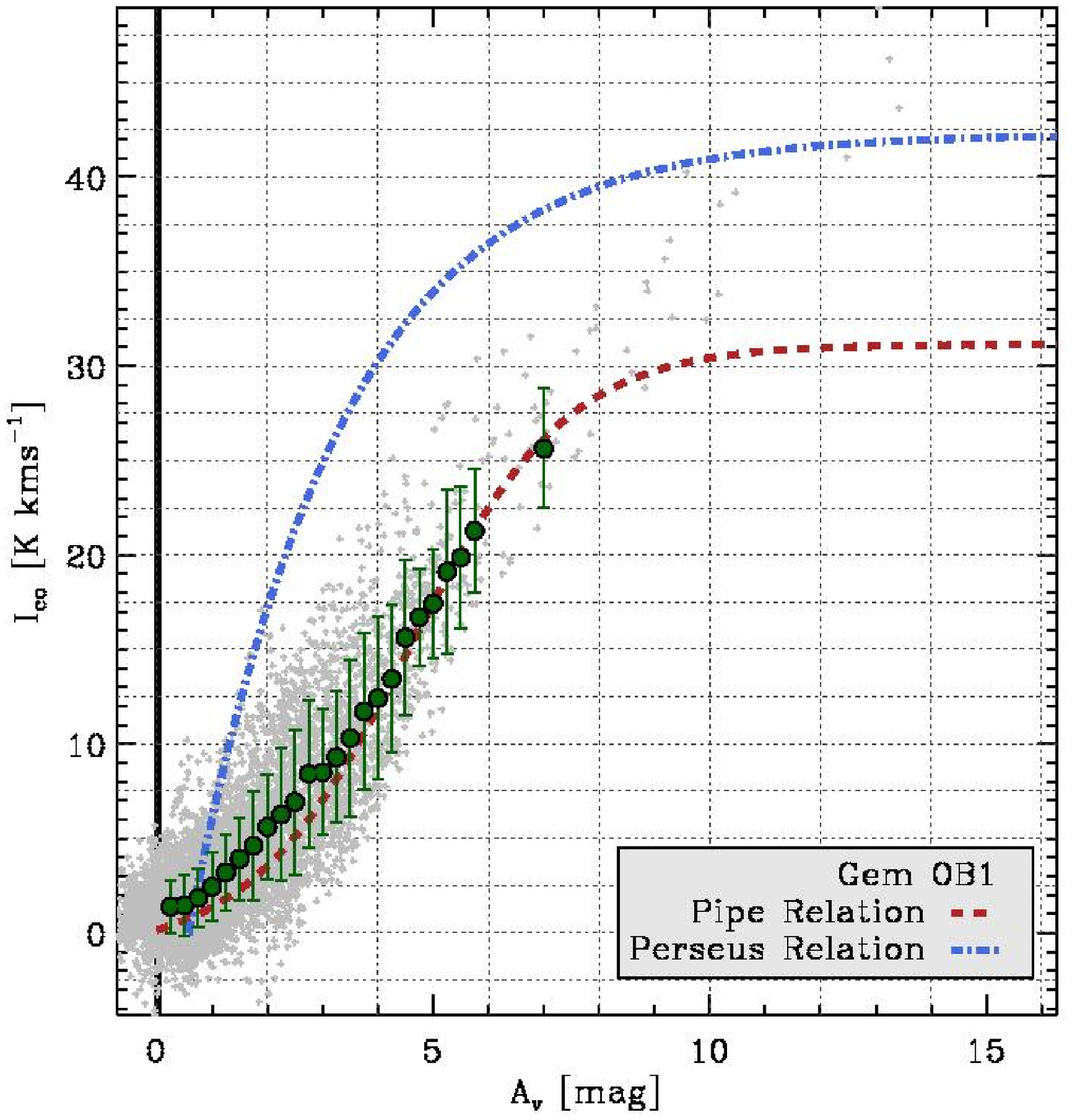}{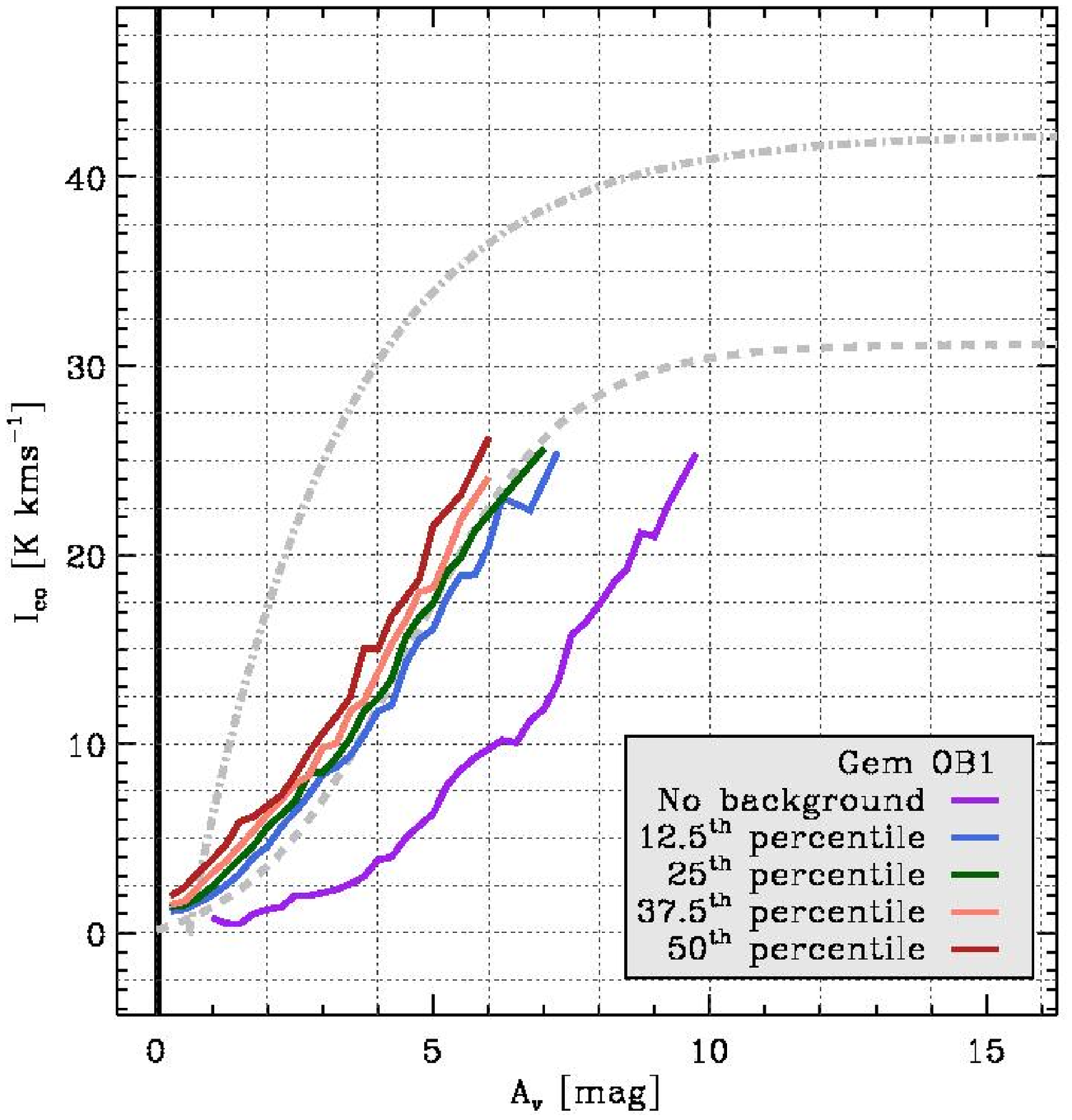}
\caption{Same as Figure~\ref{fig:taurus}, for the case of Gem OB1.}
\end{figure*}

\begin{figure*}
\plottwo{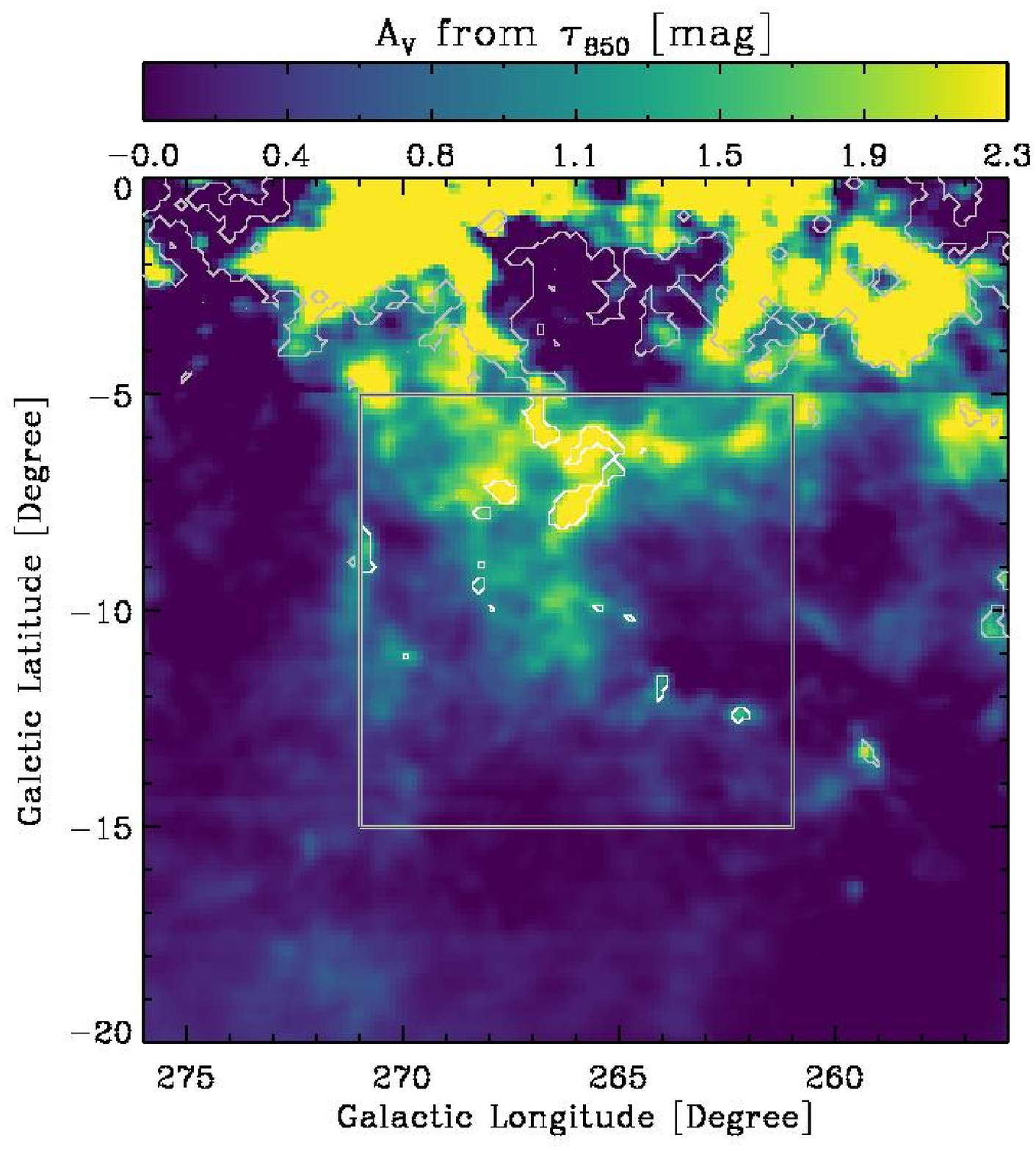}{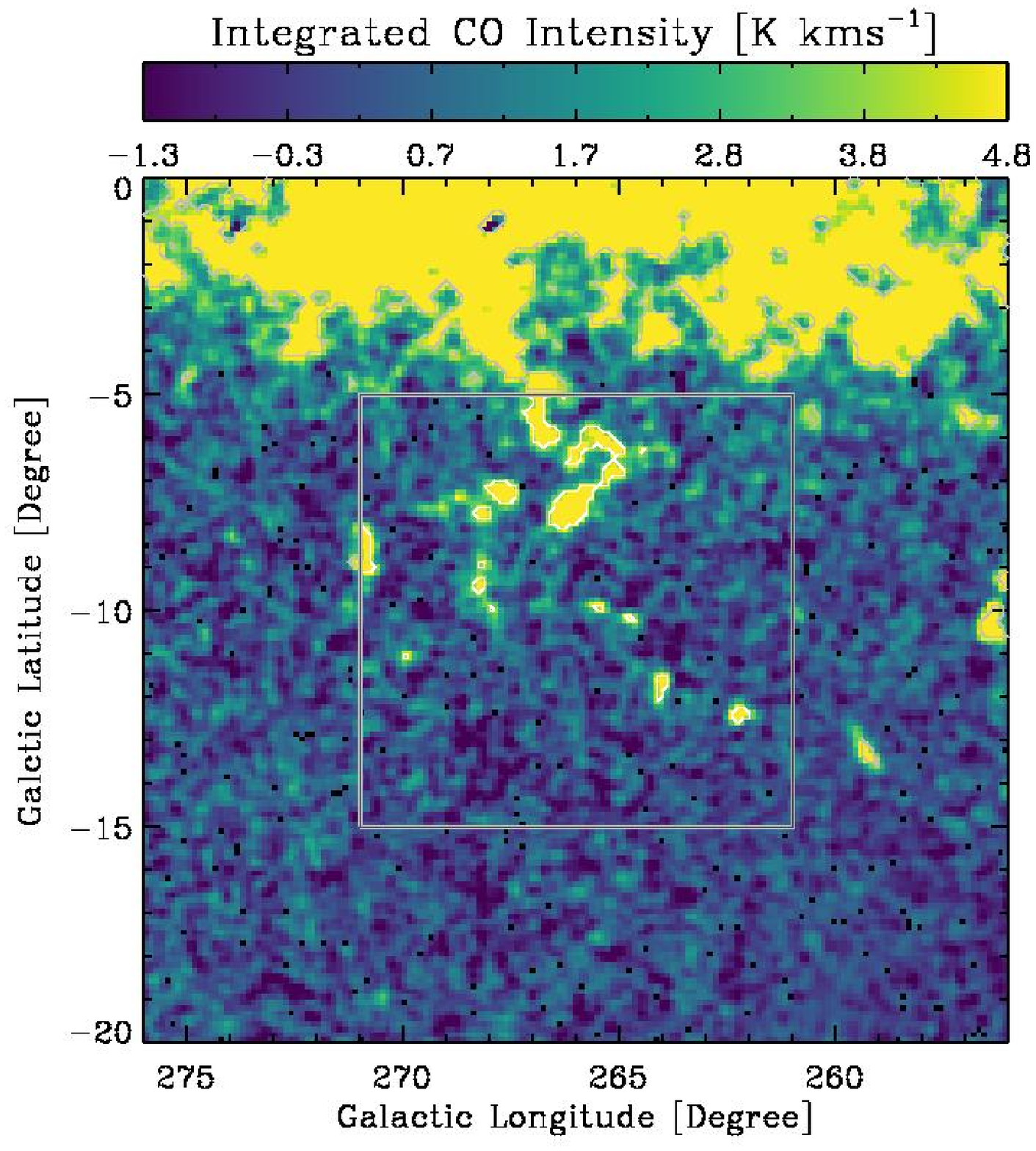}
\plottwo{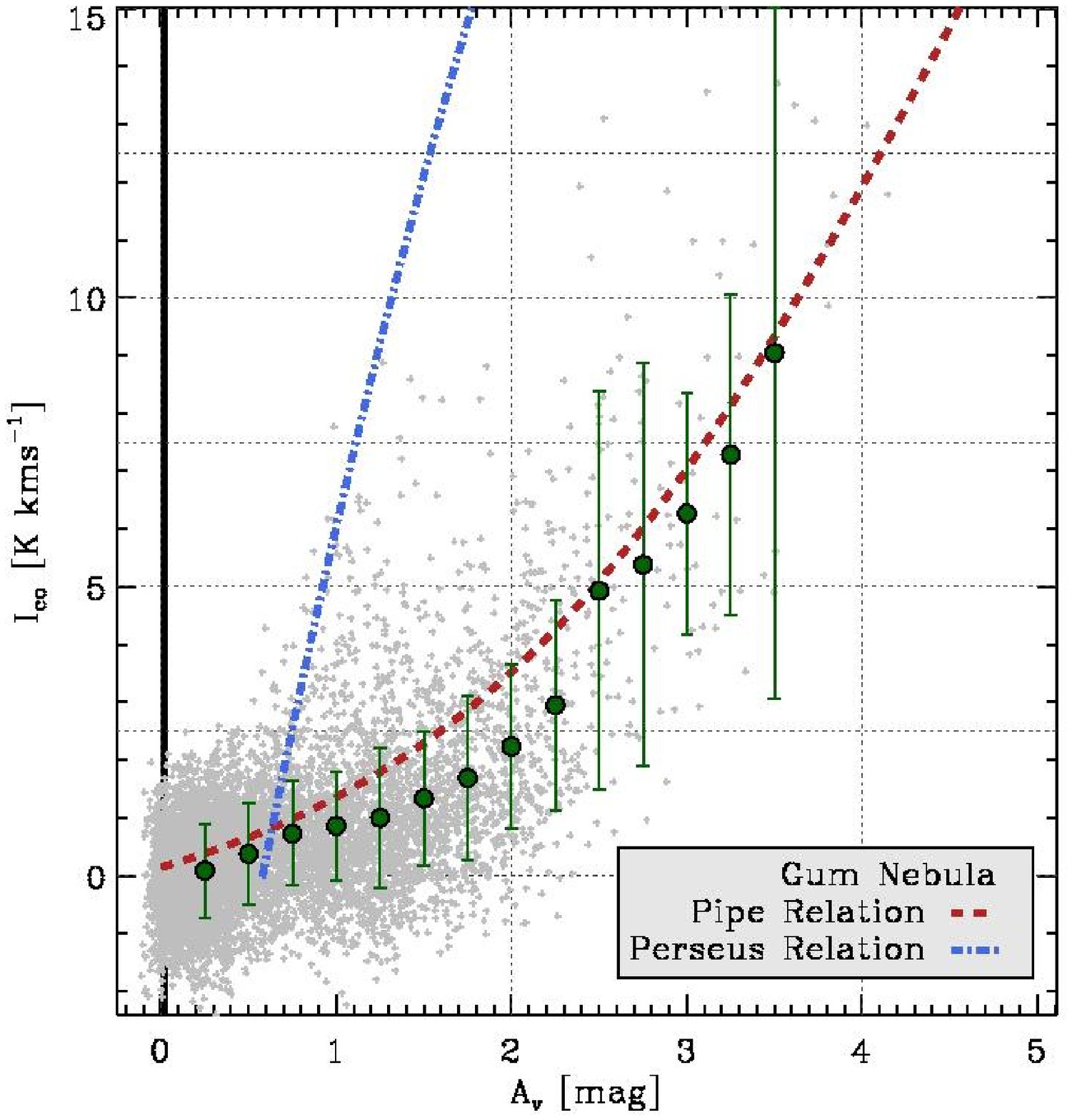}{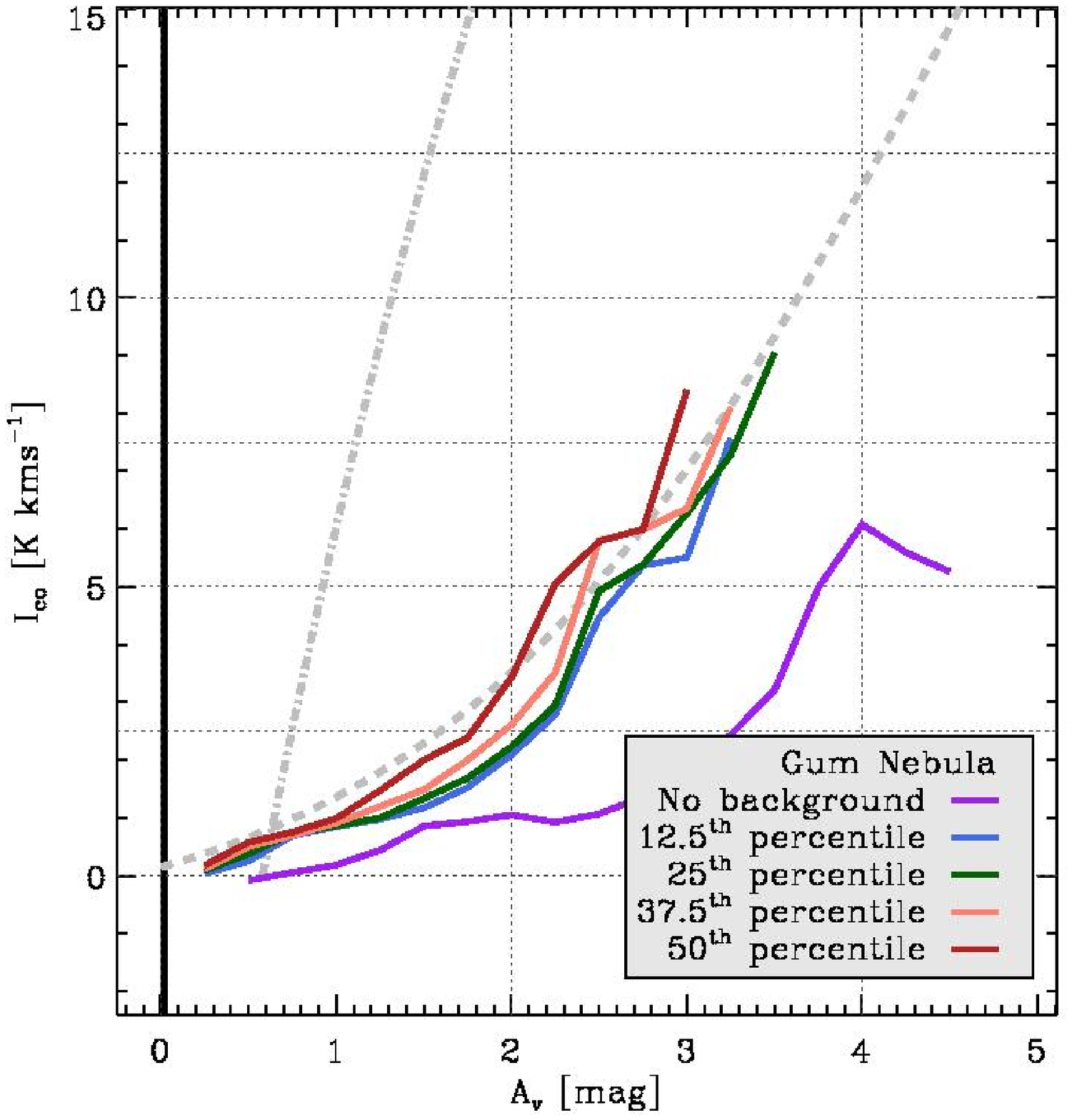}
\caption{Same as Figure~\ref{fig:taurus}, for the case of Gum Nebula.}
\end{figure*}

\begin{figure*}
\plottwo{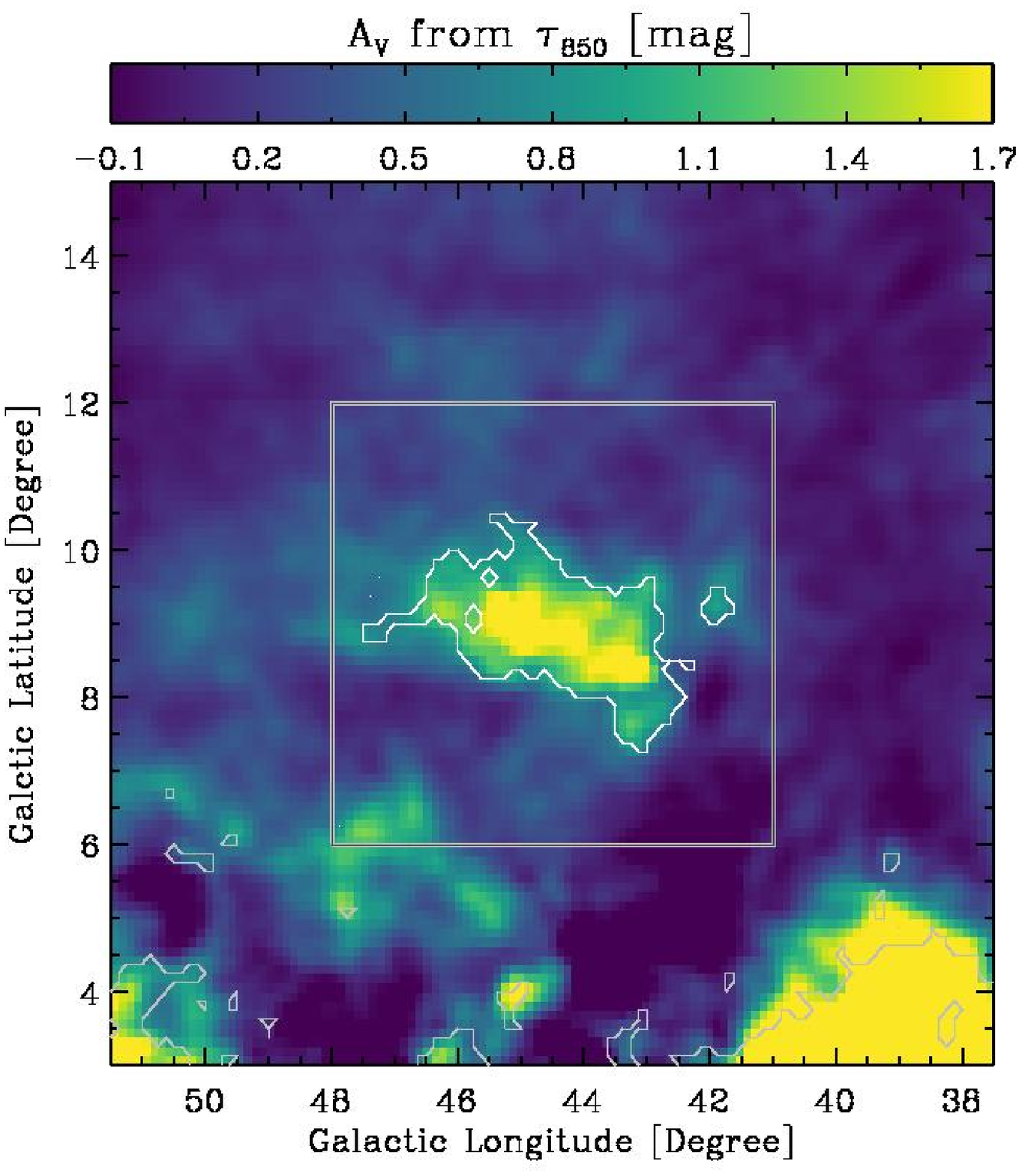}{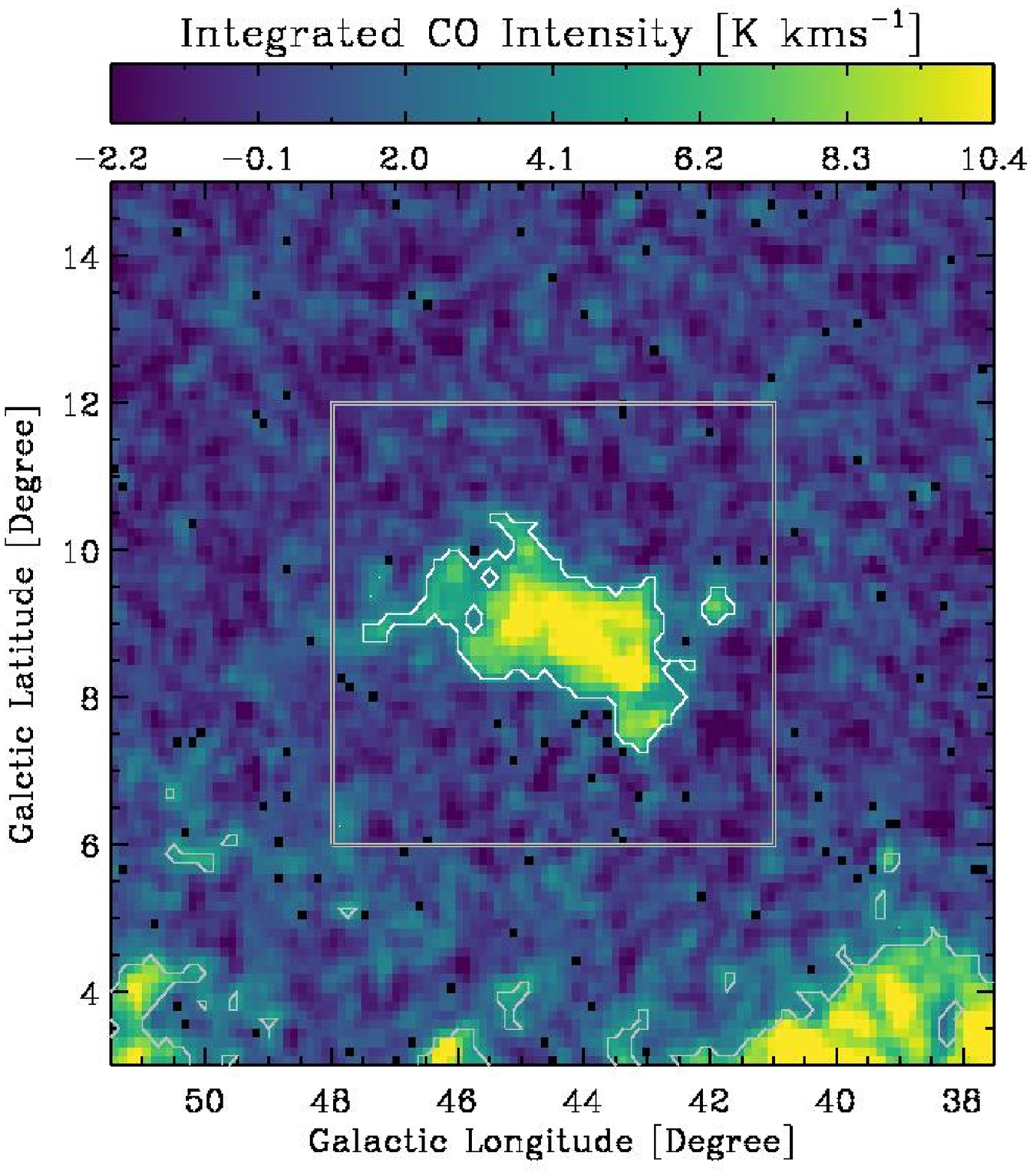}
\plottwo{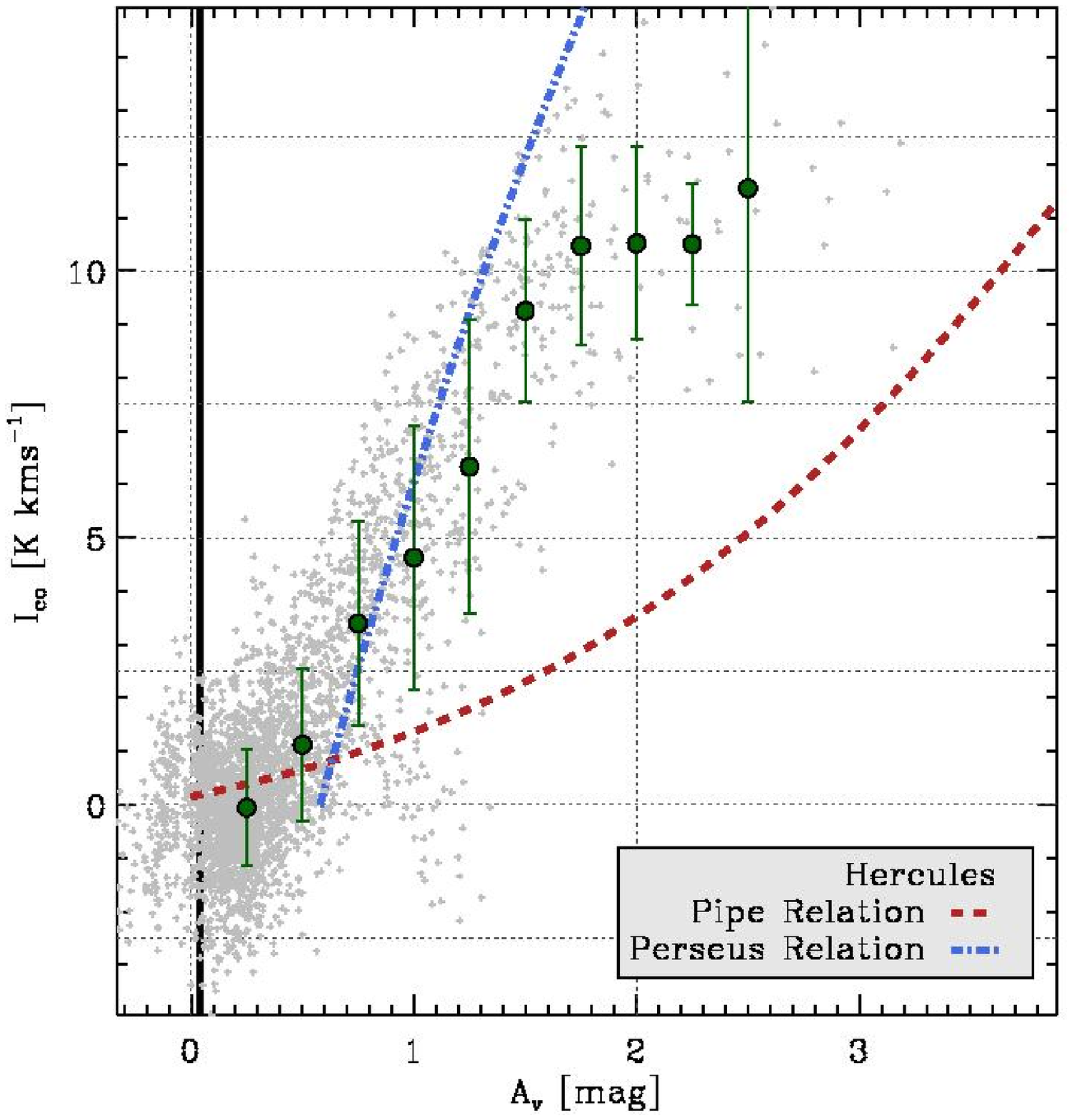}{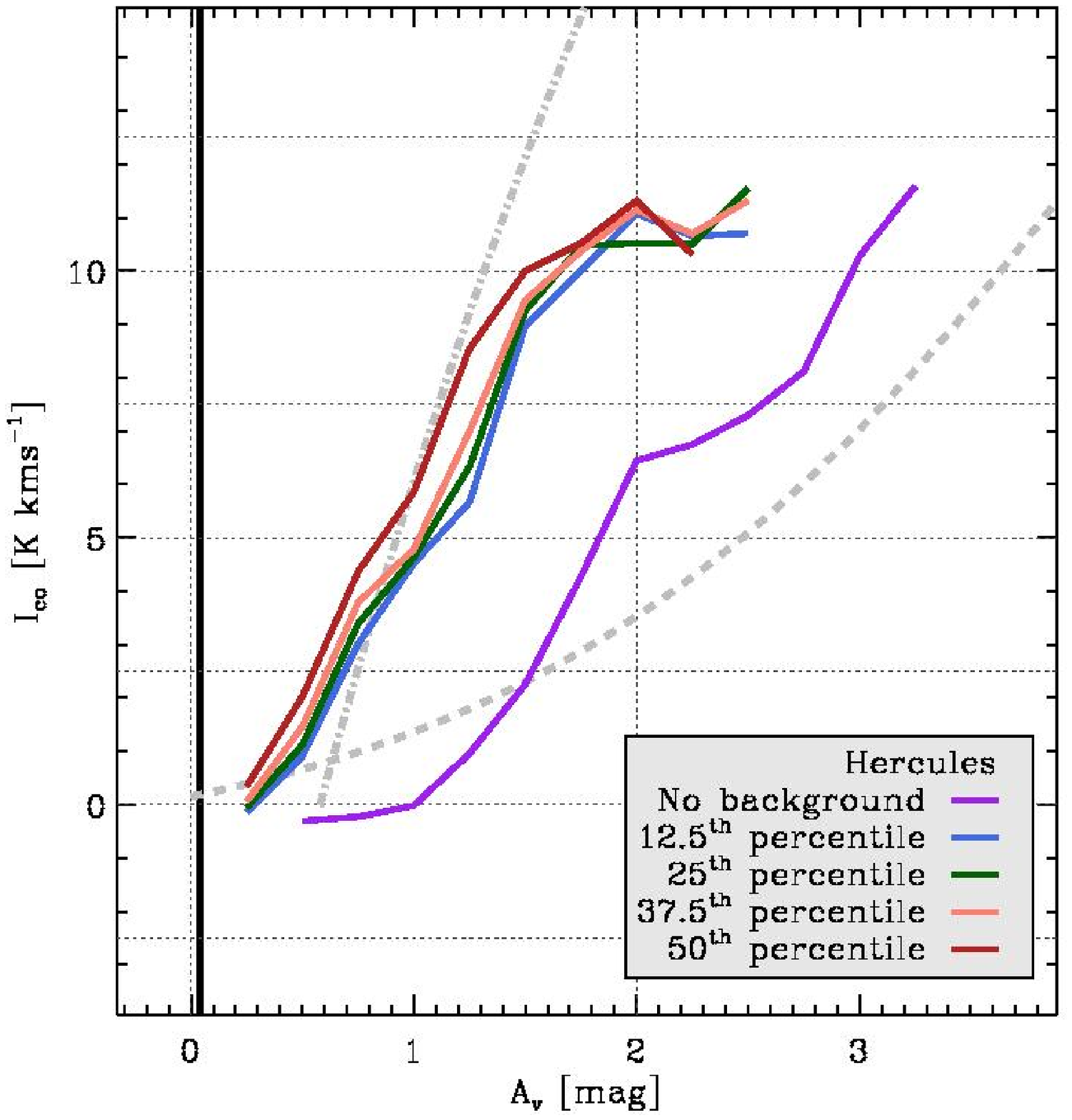}
\caption{Same as Figure~\ref{fig:taurus}, for the case of Hercules.}
\end{figure*}

\begin{figure*}
\plottwo{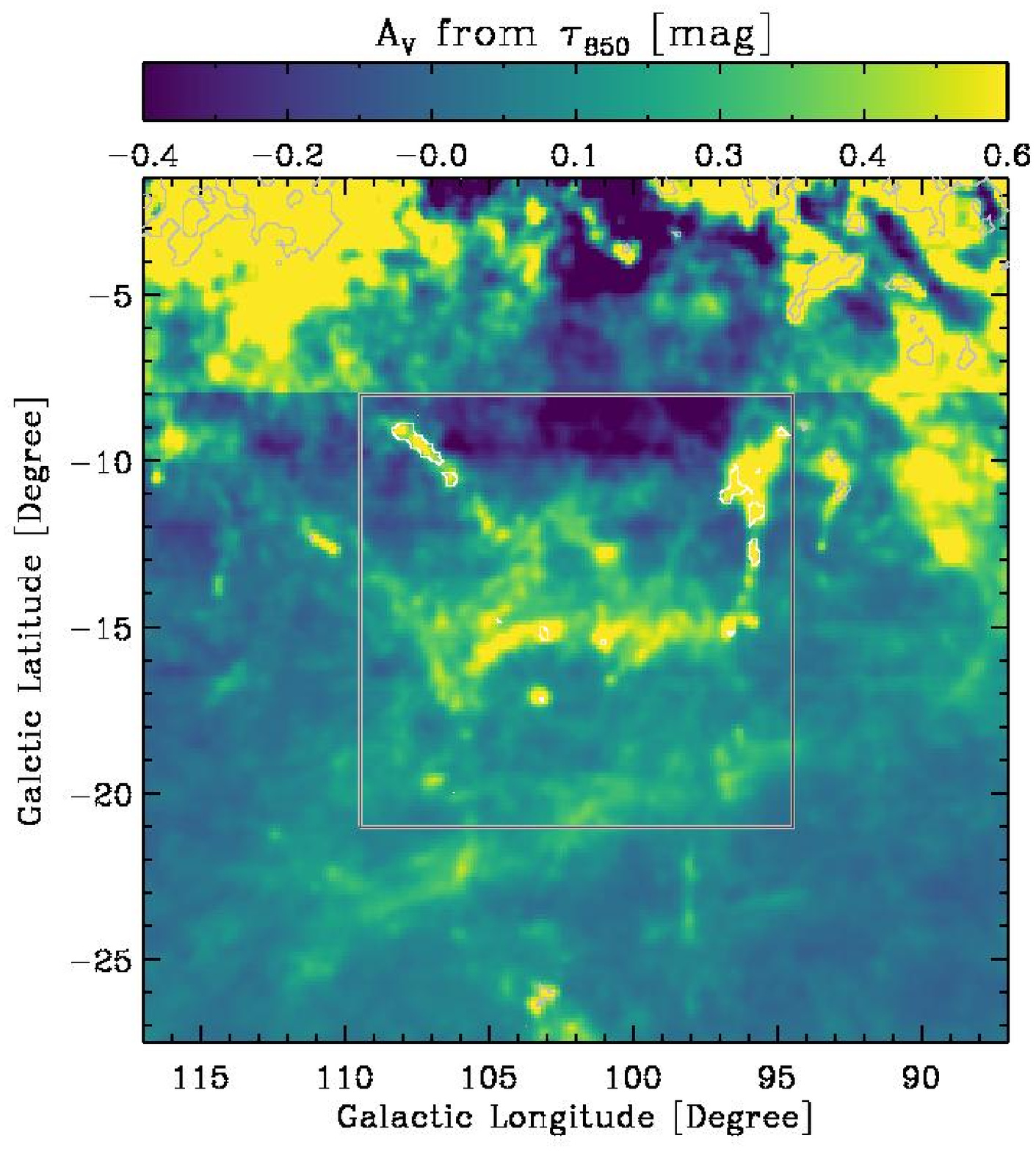}{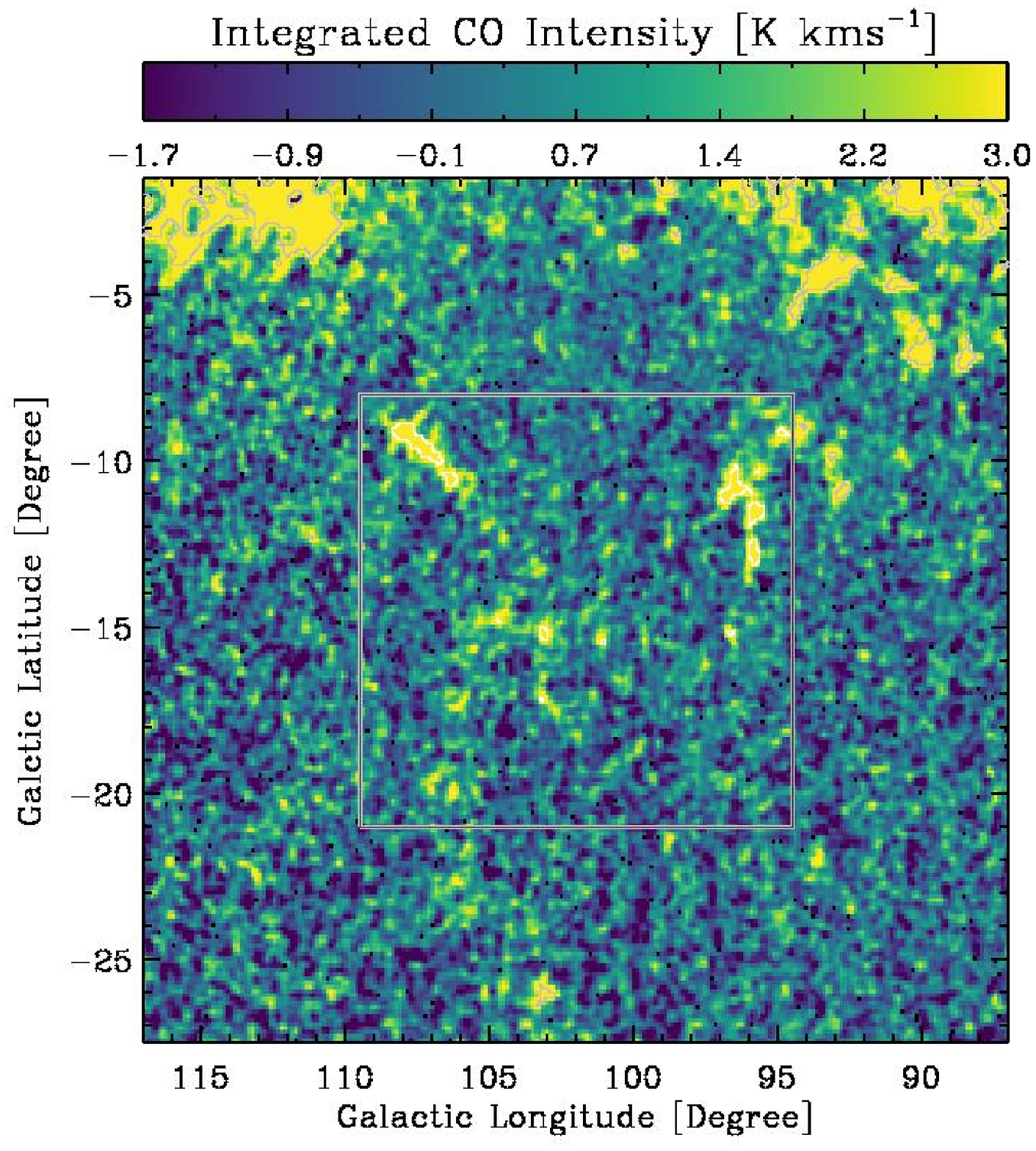}
\plottwo{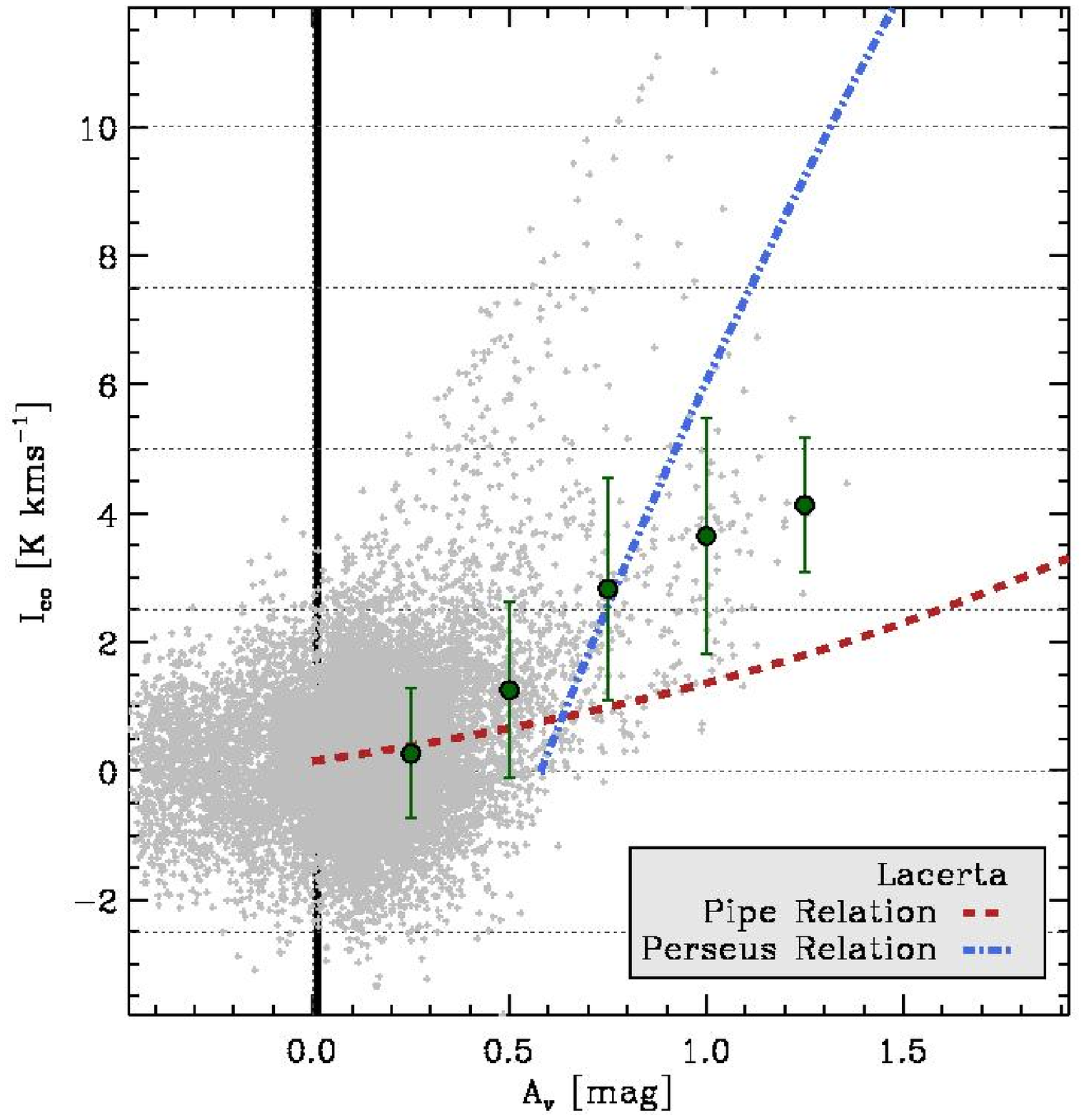}{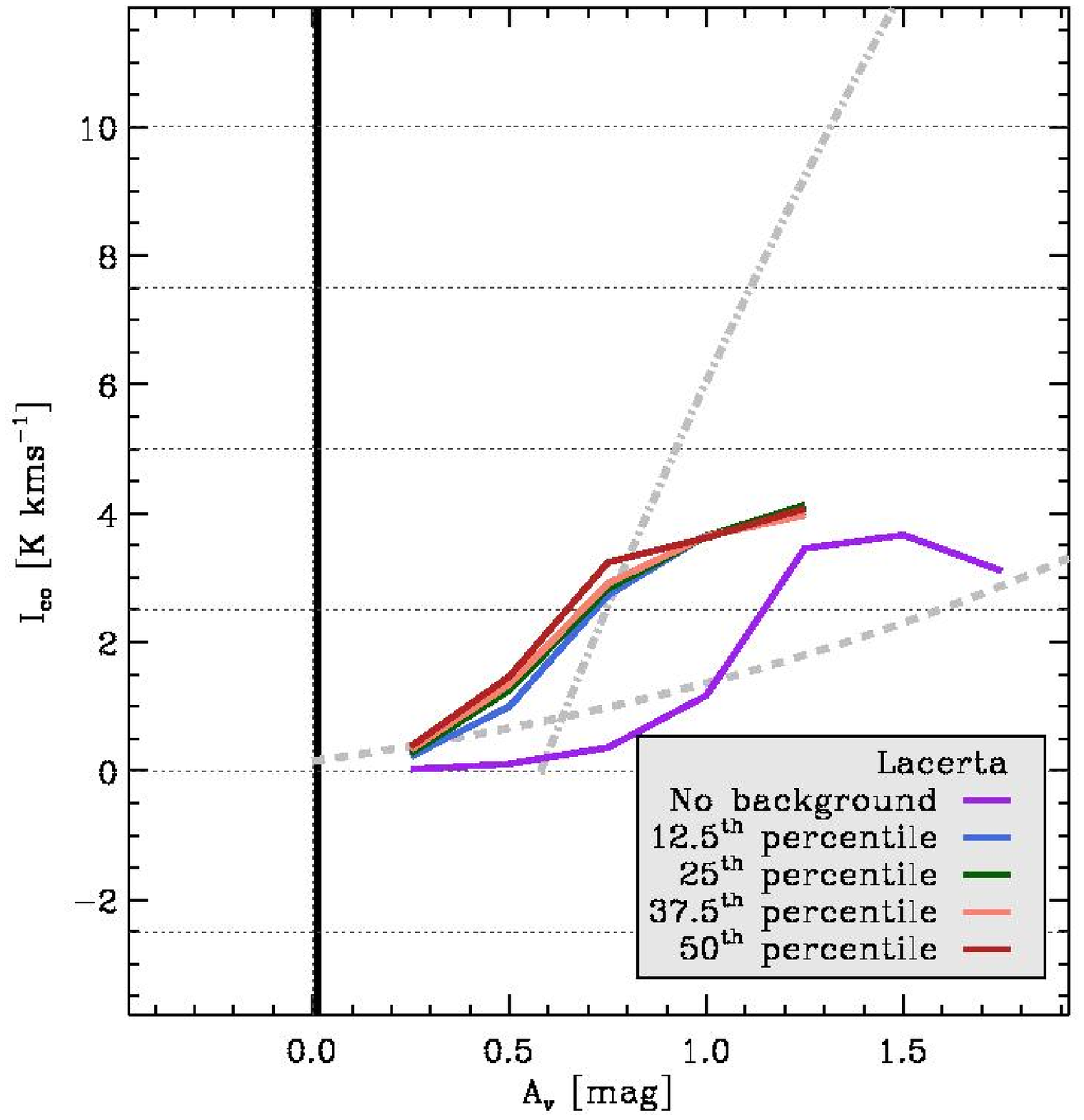}
\caption{Same as Figure~\ref{fig:taurus}, for the case of Lacerta.}
\end{figure*}

\begin{figure*}
\plottwo{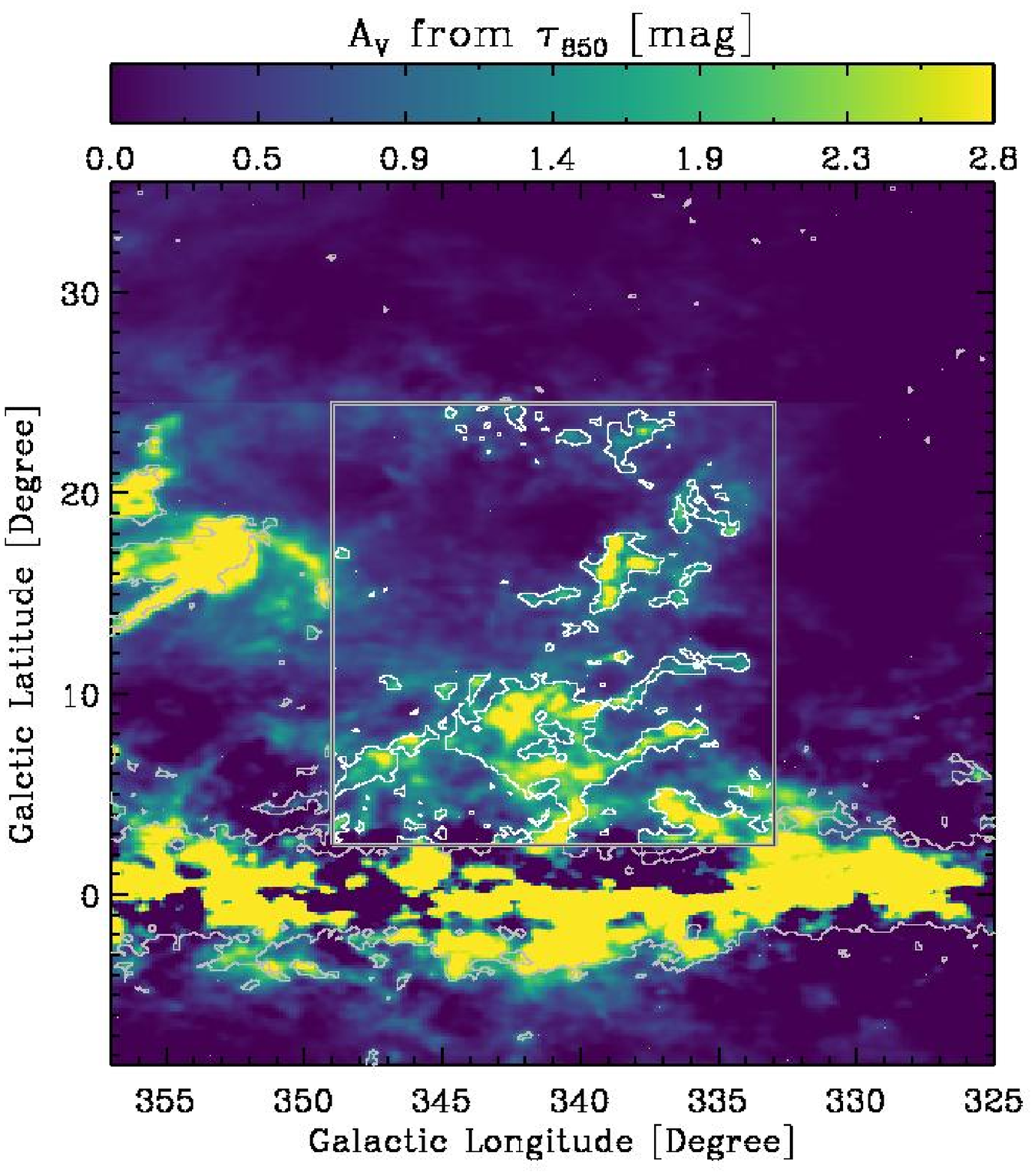}{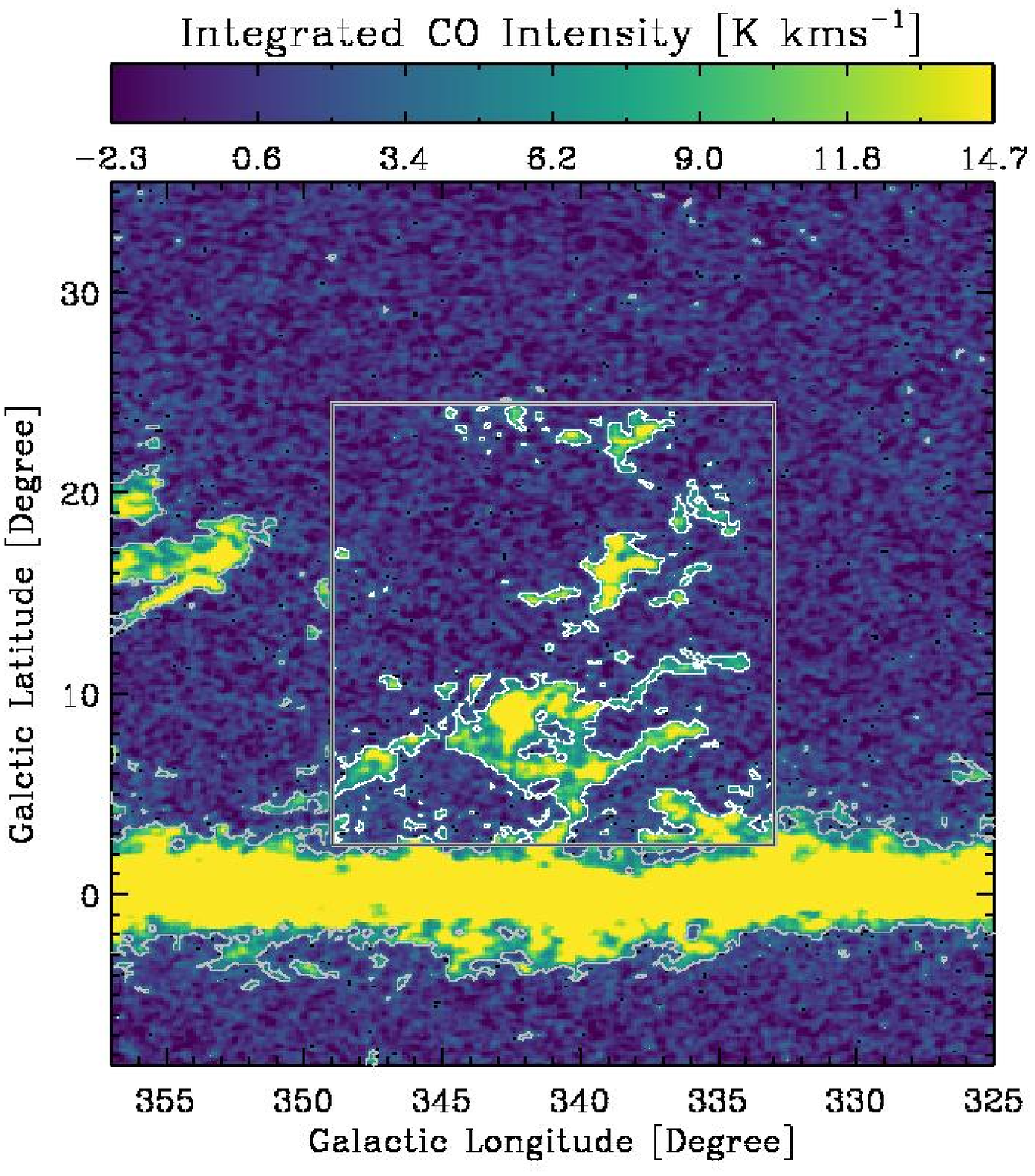}
\plottwo{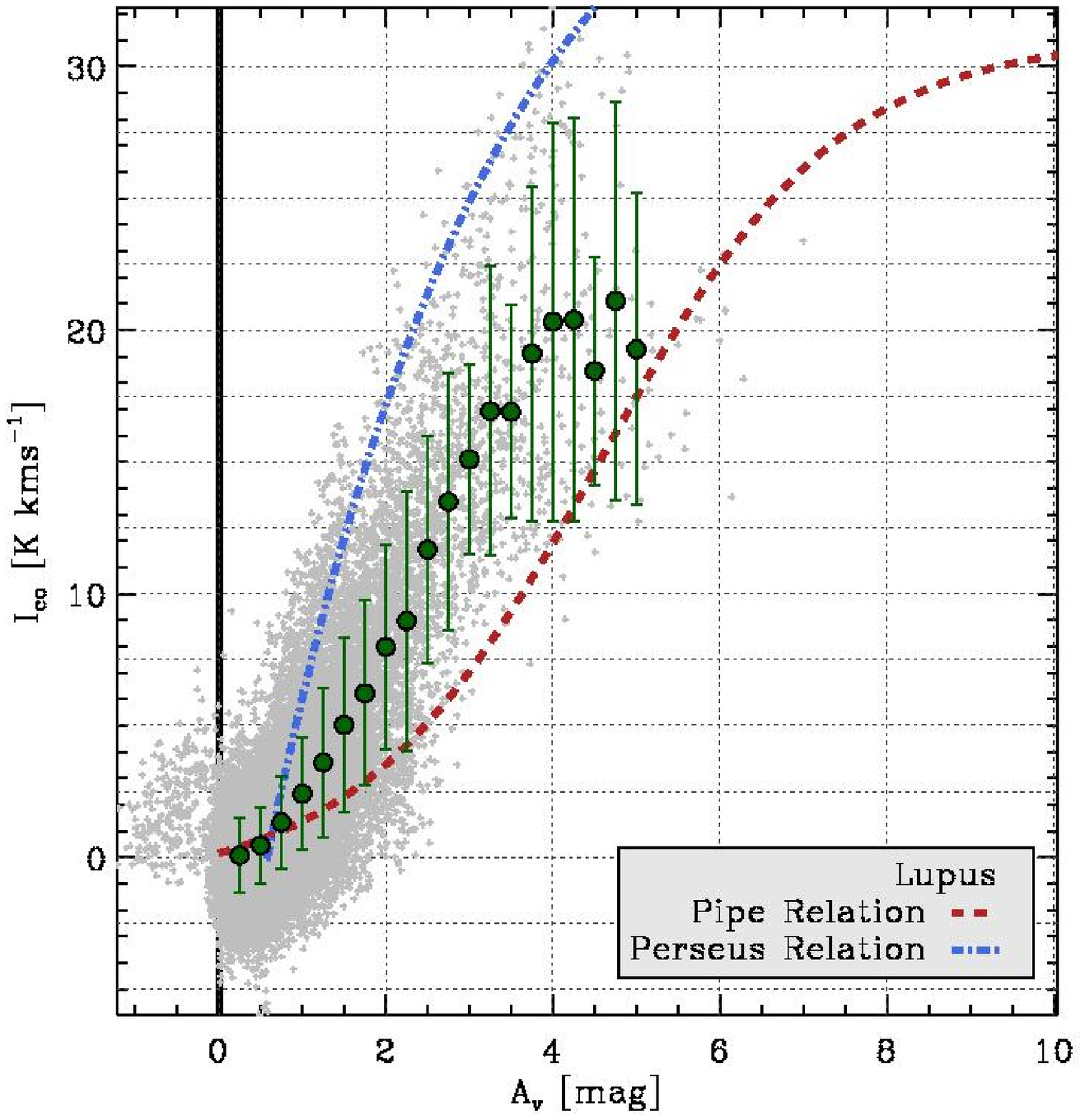}{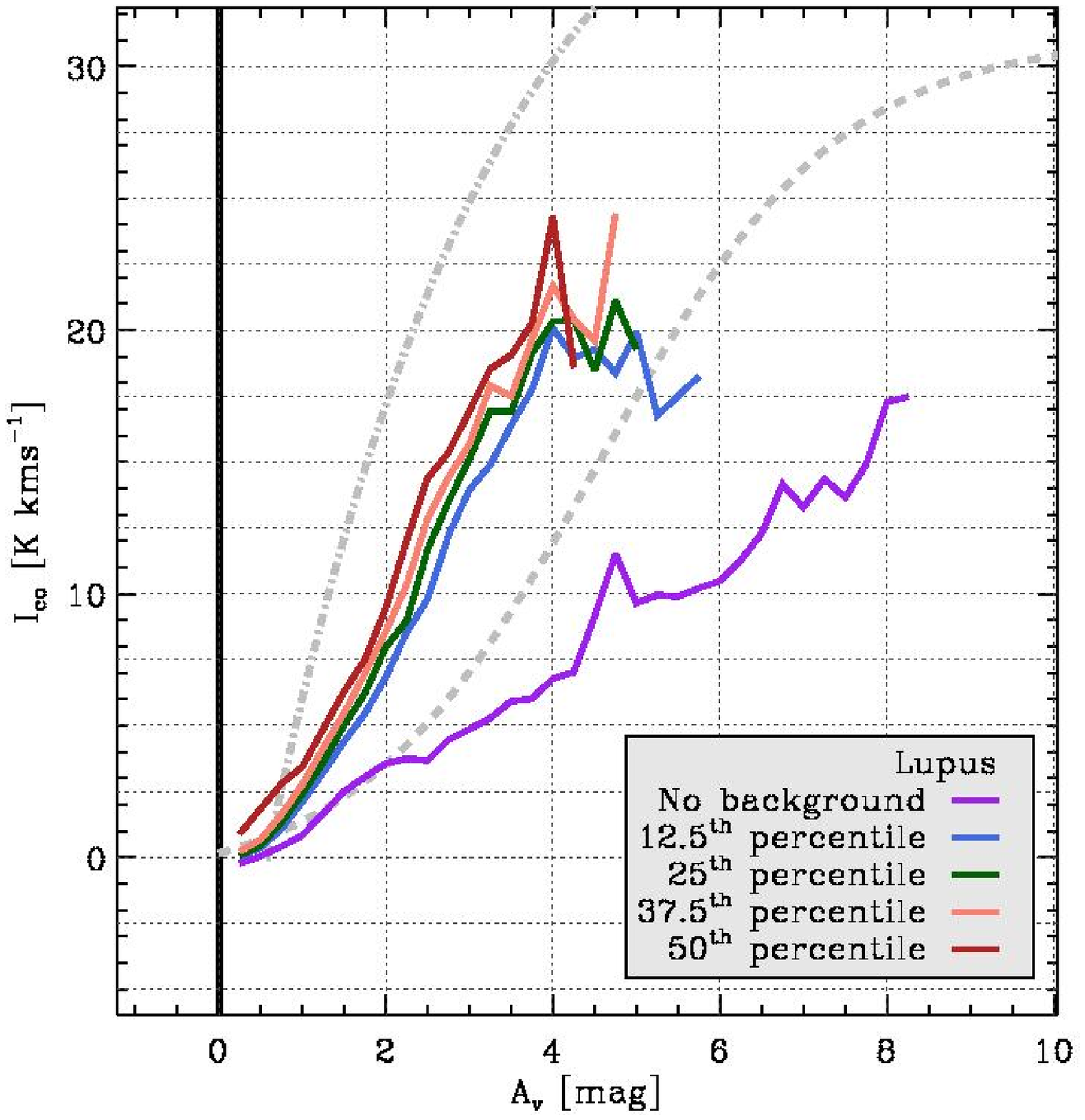}
\caption{Same as Figure~\ref{fig:taurus}, for the case of Lupus.}
\end{figure*}

\clearpage

\begin{figure*}
\plottwo{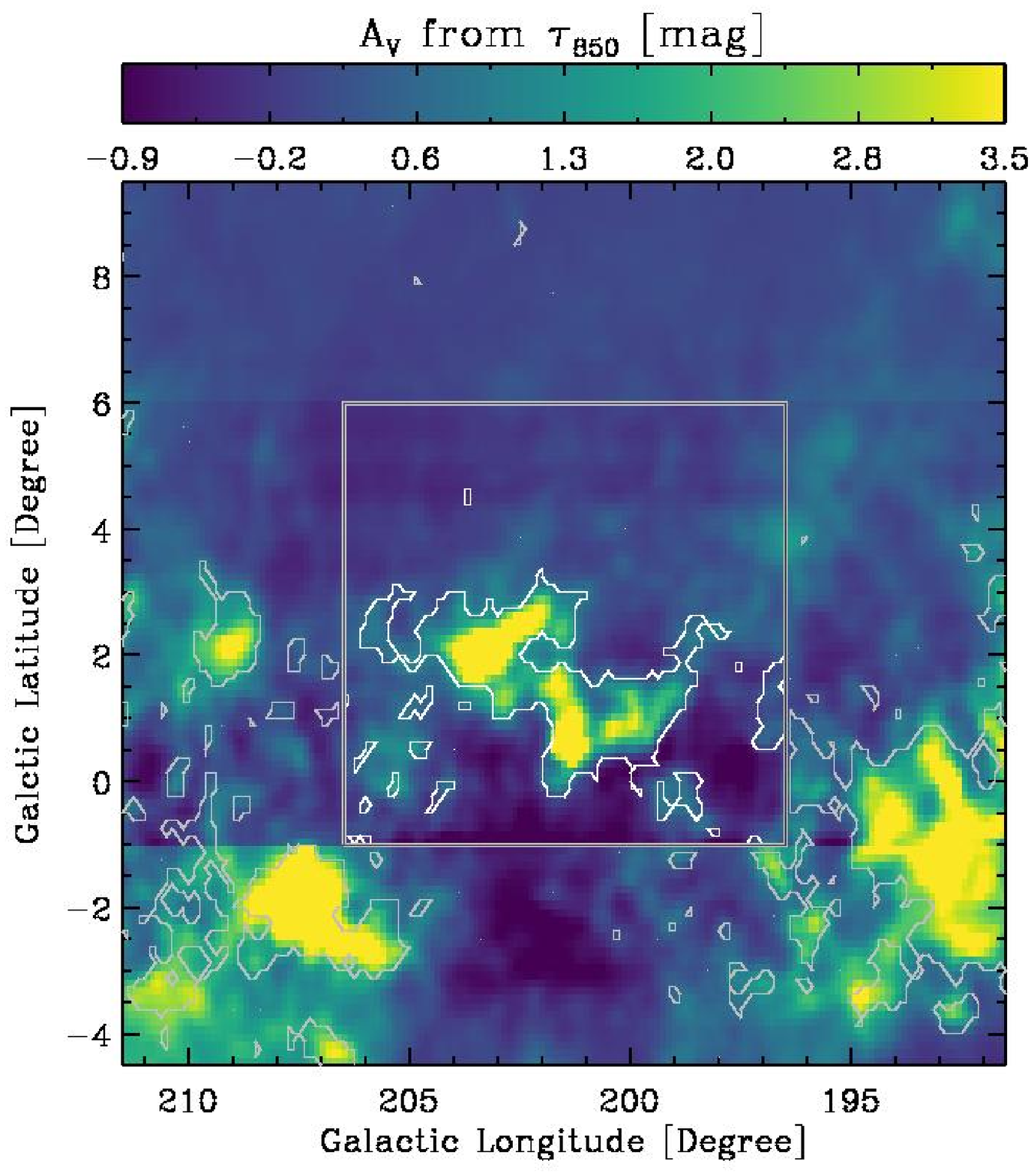}{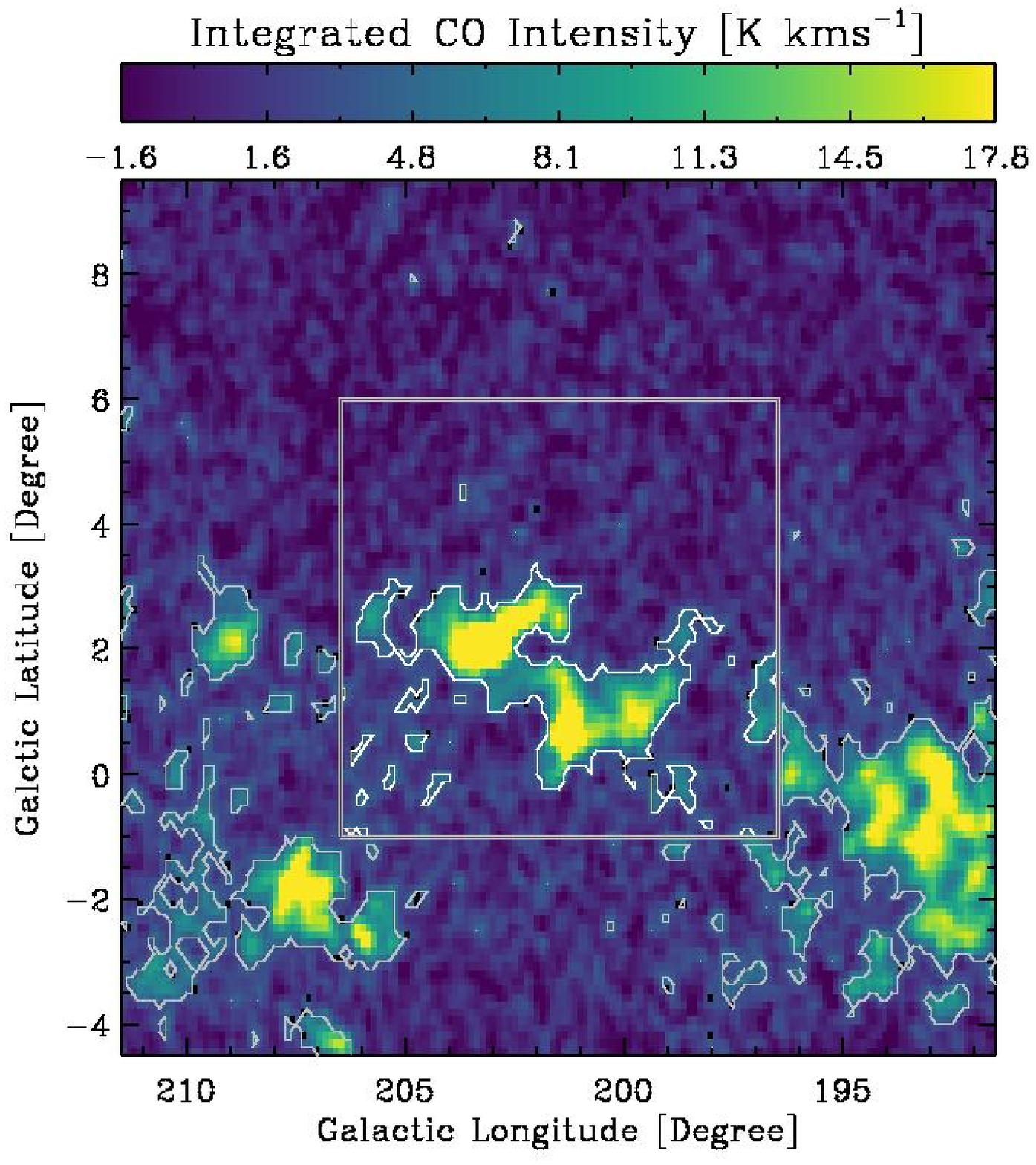}
\plottwo{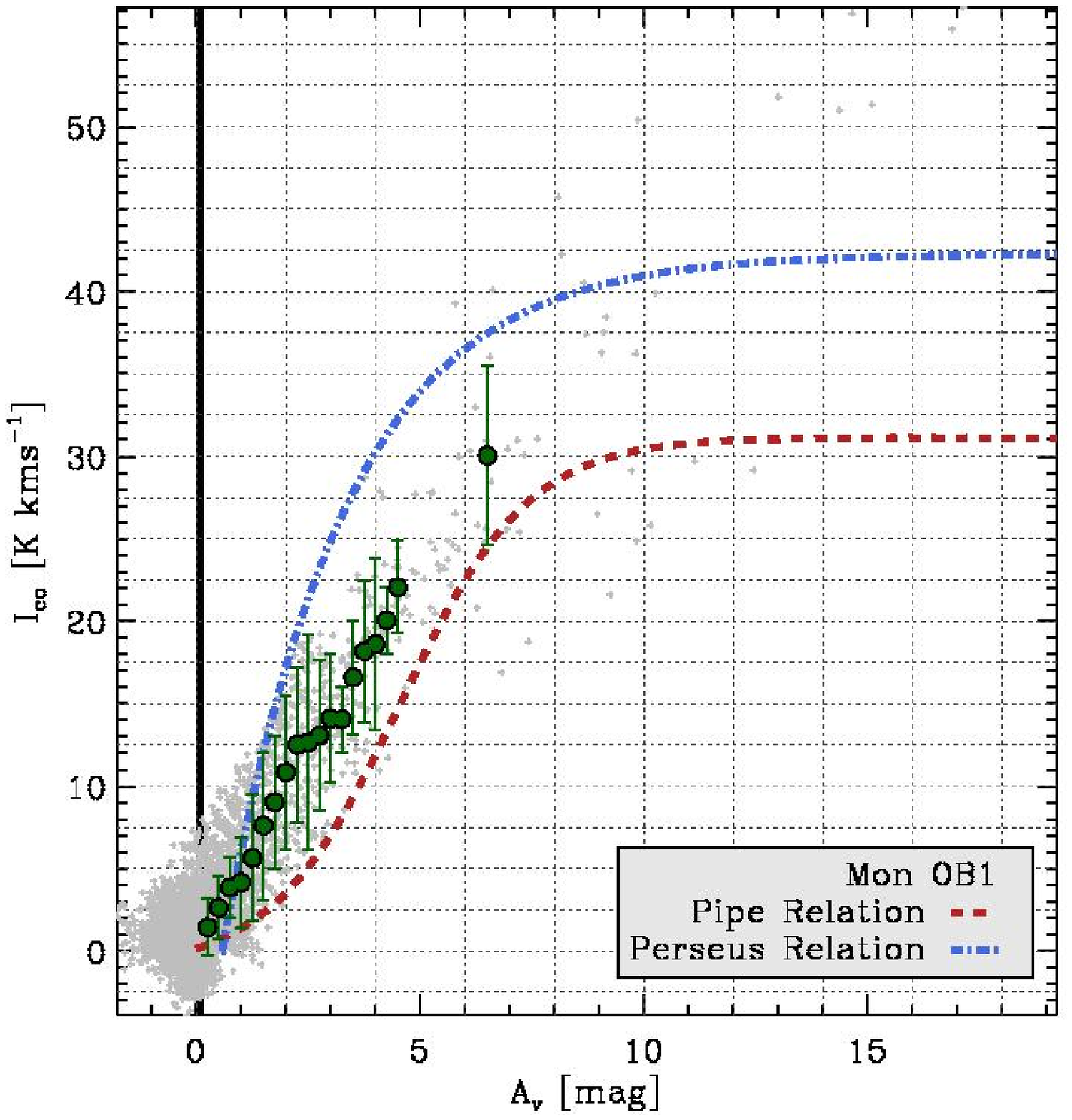}{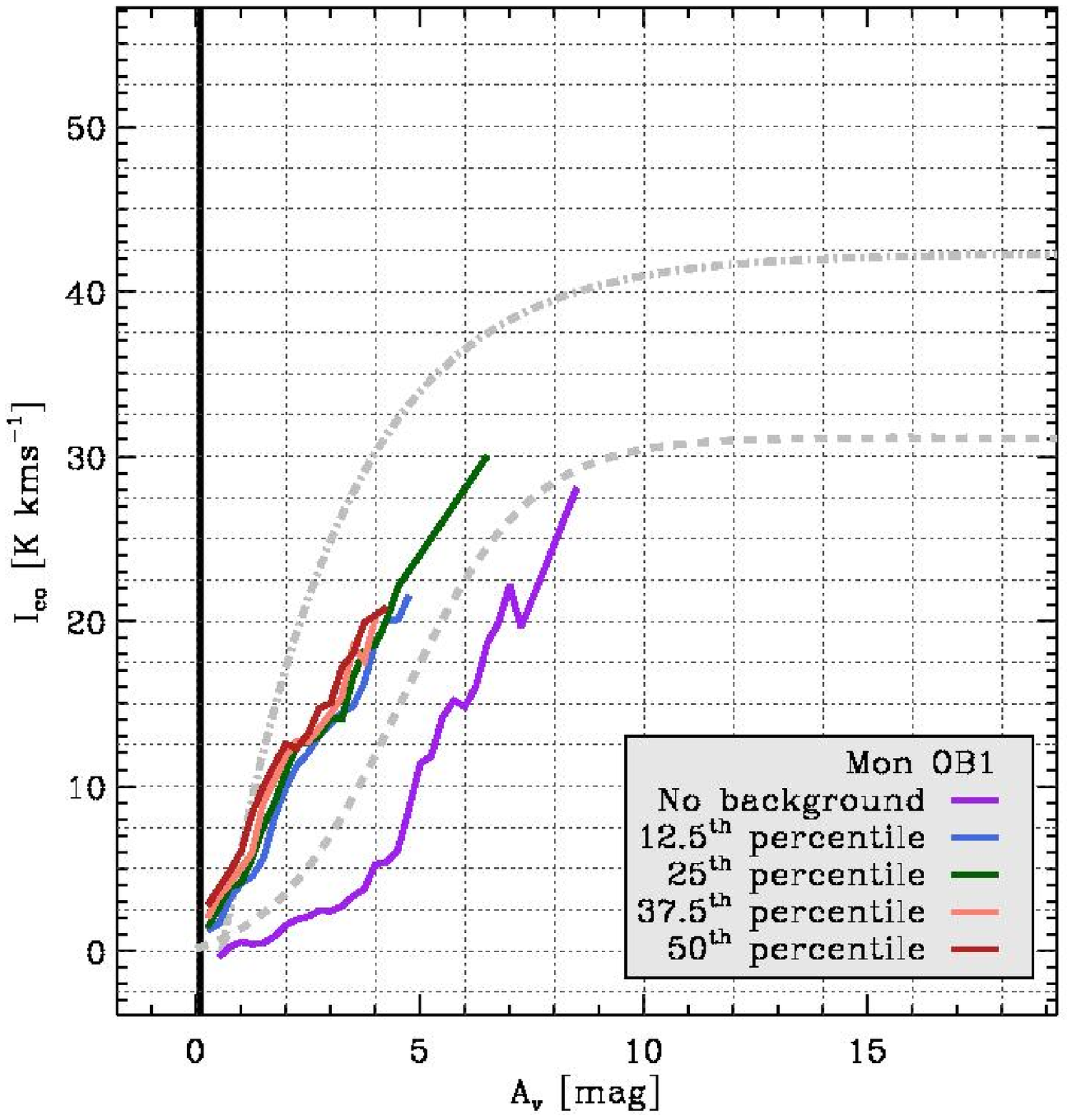}
\caption{Same as Figure~\ref{fig:taurus}, for the case of Mon OB1.}
\end{figure*}

\begin{figure*}
\plottwo{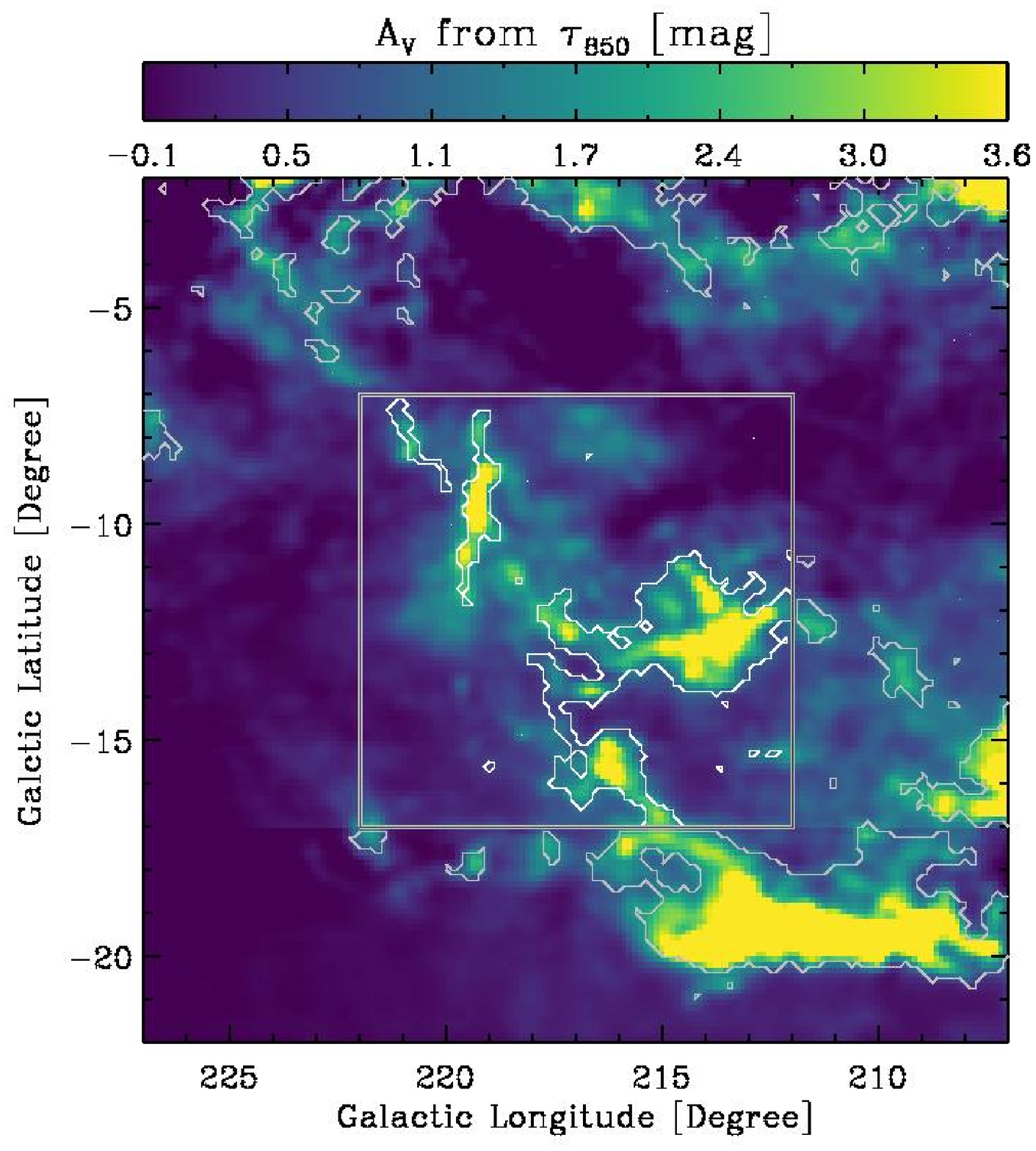}{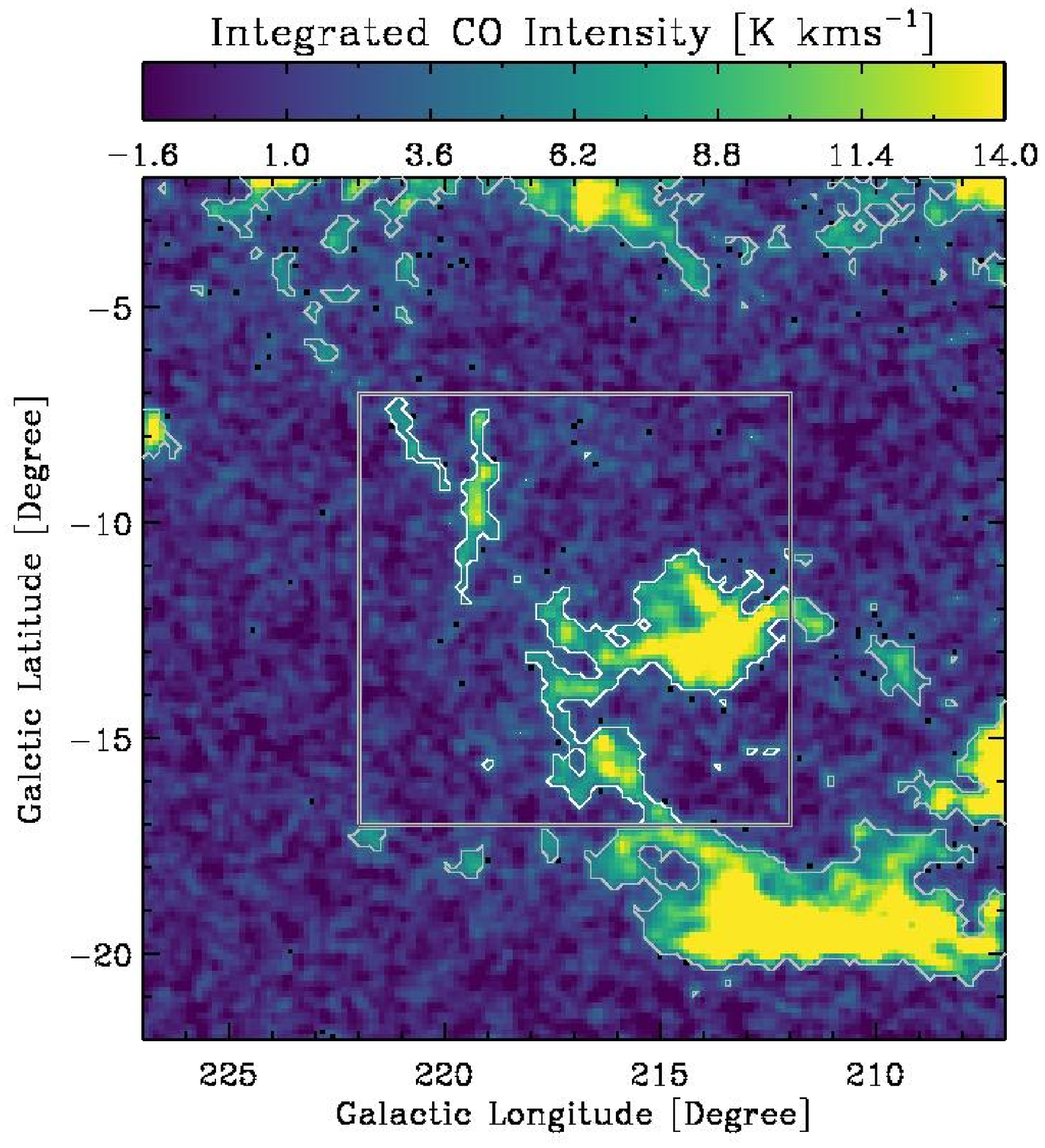}
\plottwo{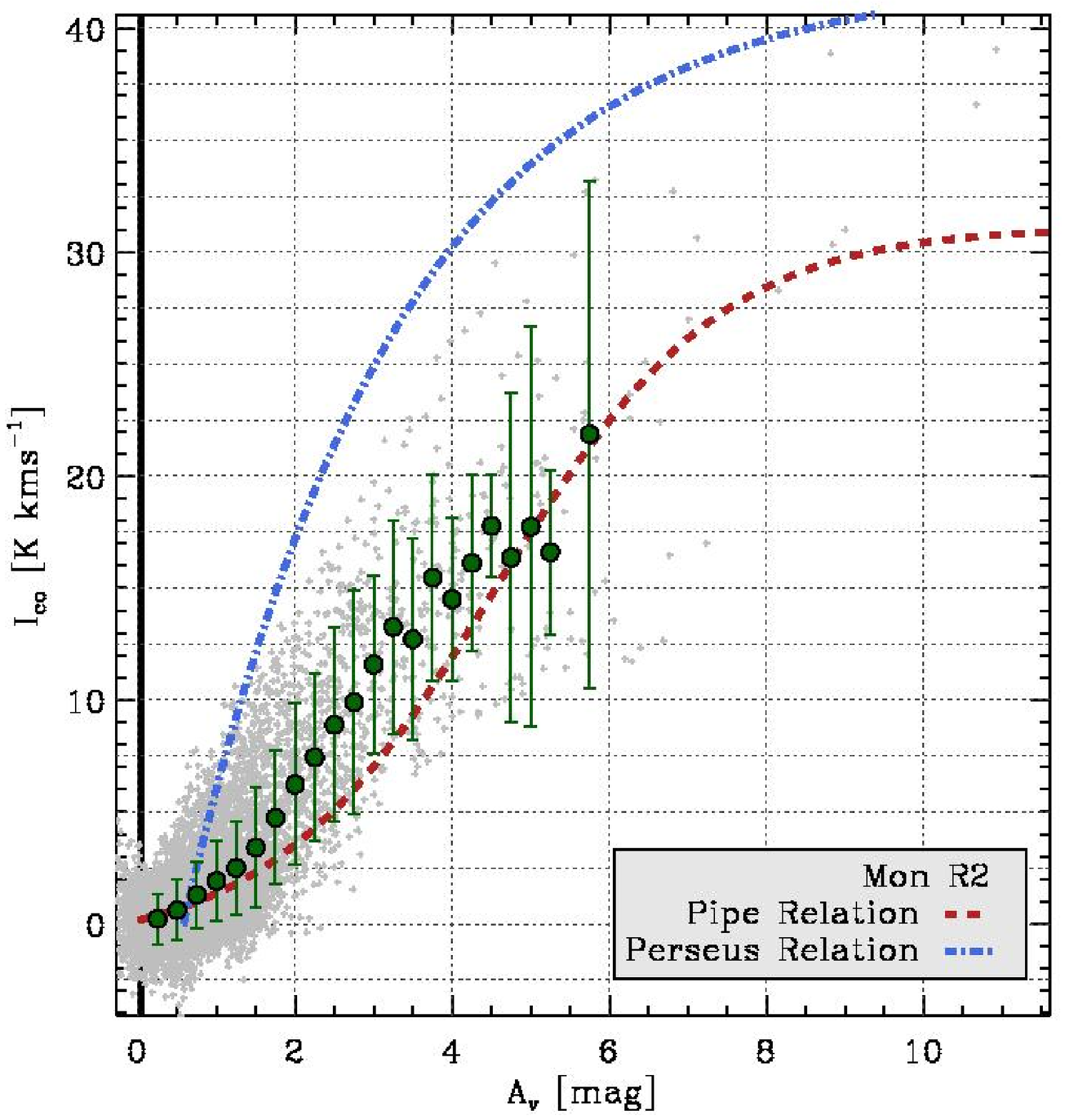}{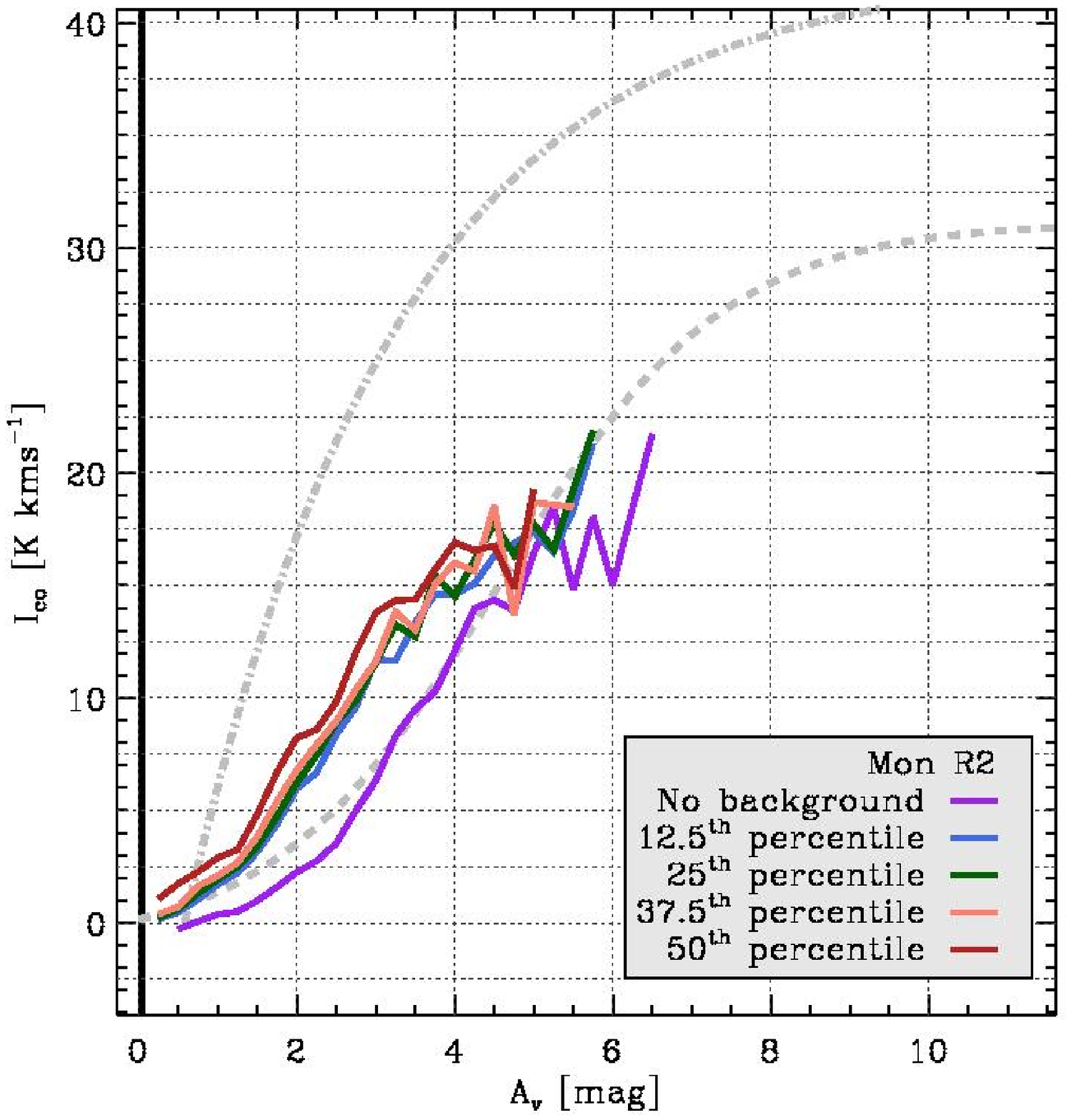}
\caption{Same as Figure~\ref{fig:taurus}, for the case of Mon R2.}
\end{figure*}

\begin{figure*}
\plottwo{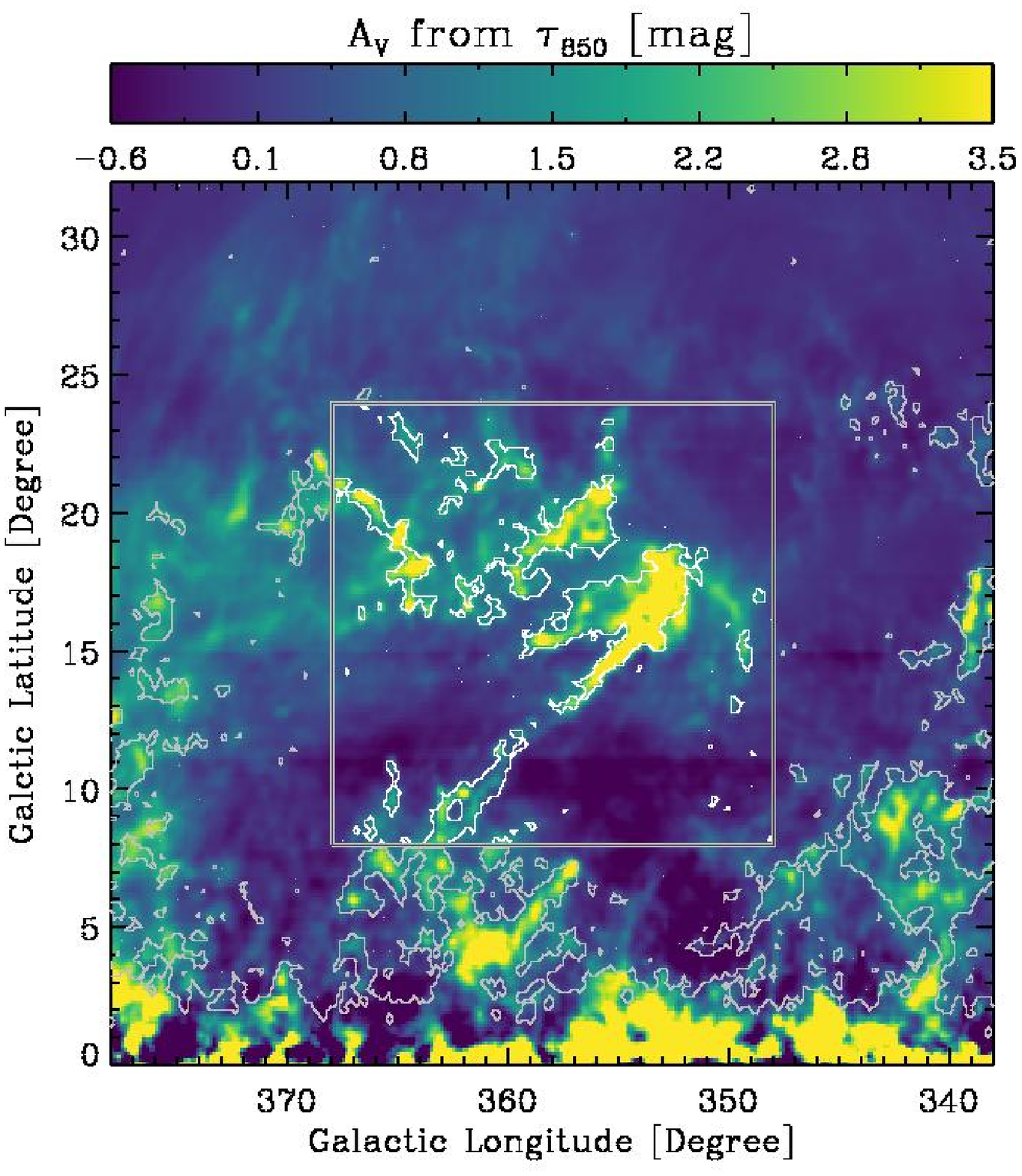}{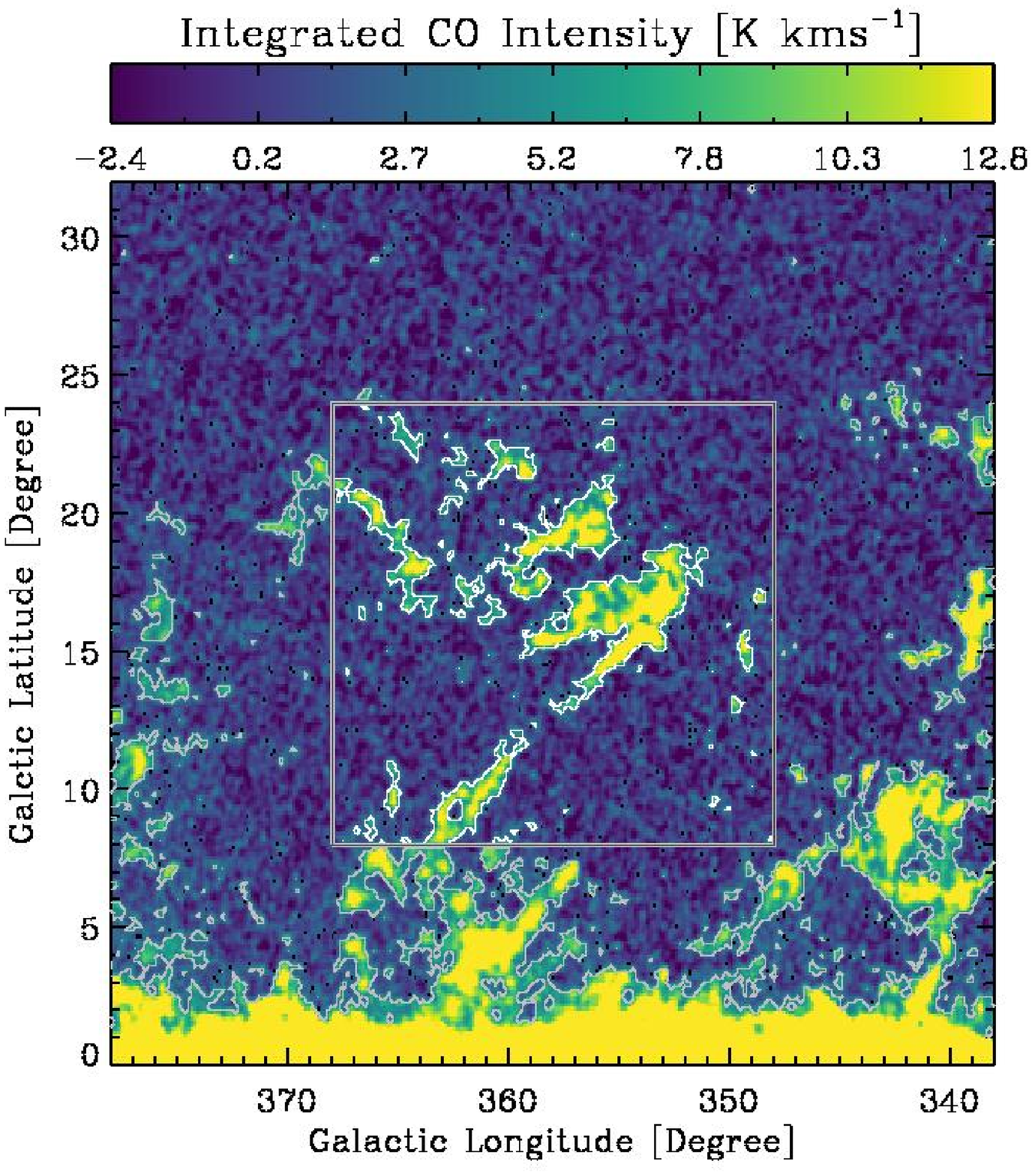}
\plottwo{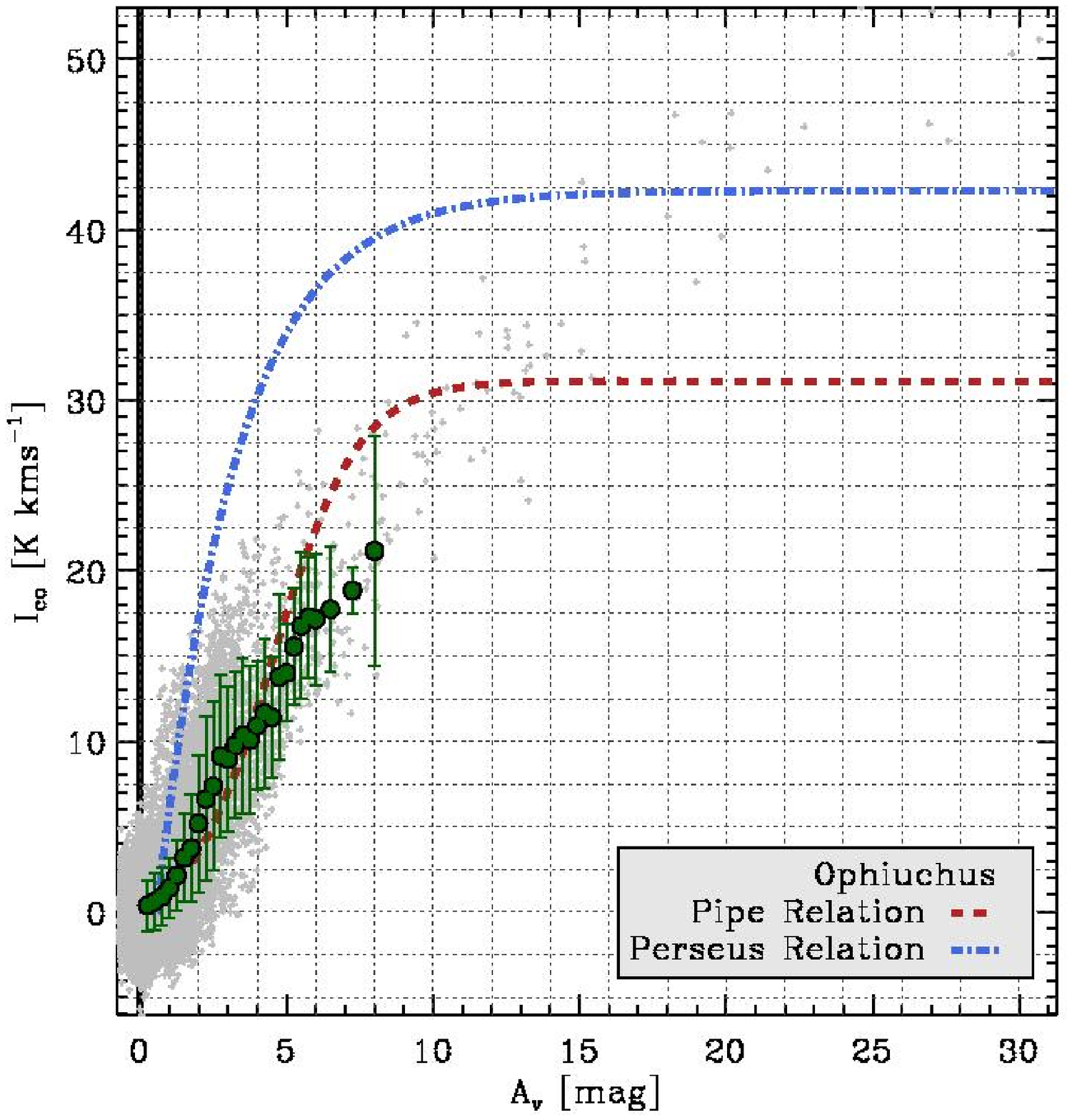}{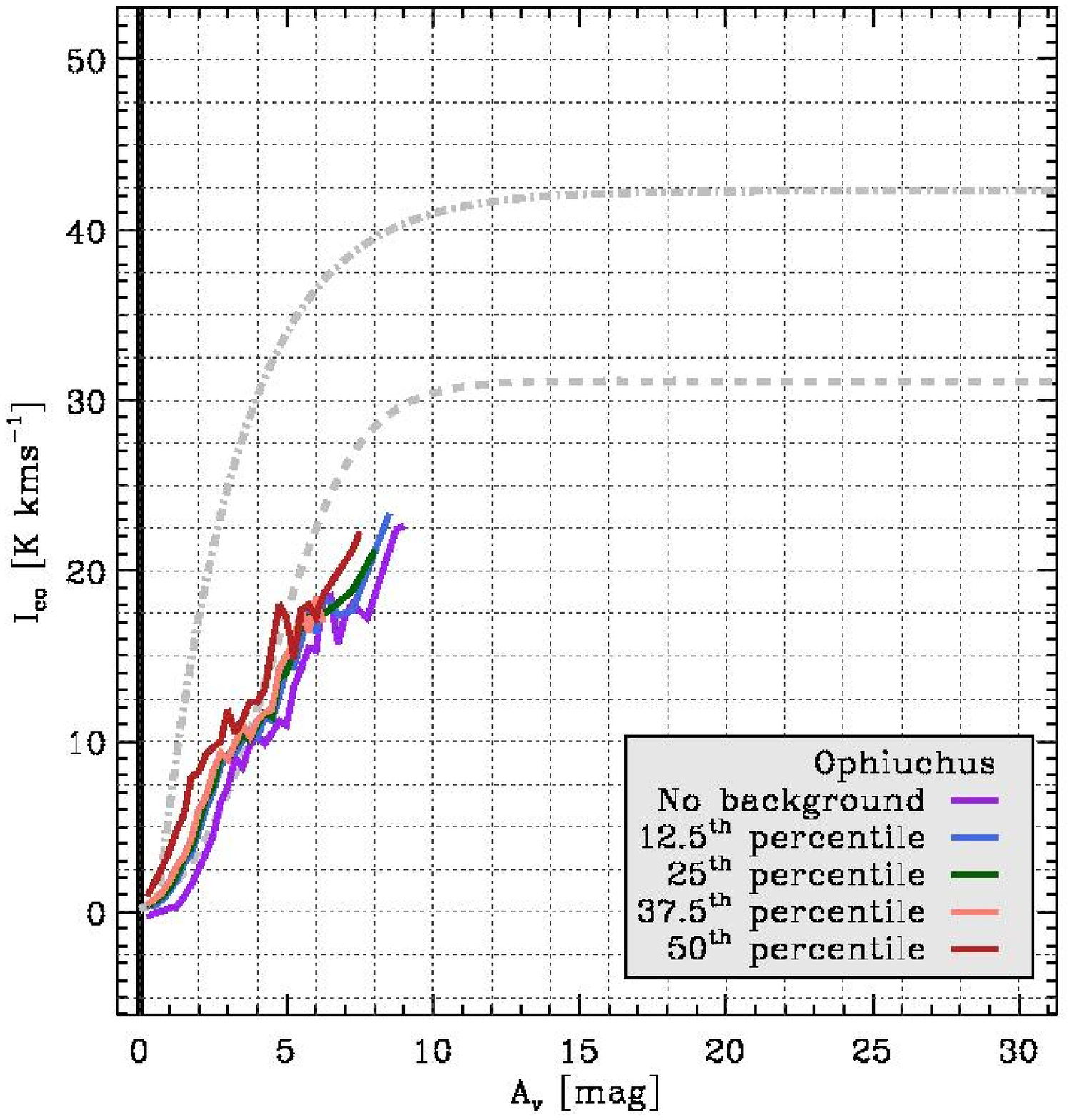}
\caption{Same as Figure~\ref{fig:taurus}, for the case of Ophiuchus.}
\end{figure*}

\begin{figure*}
\plottwo{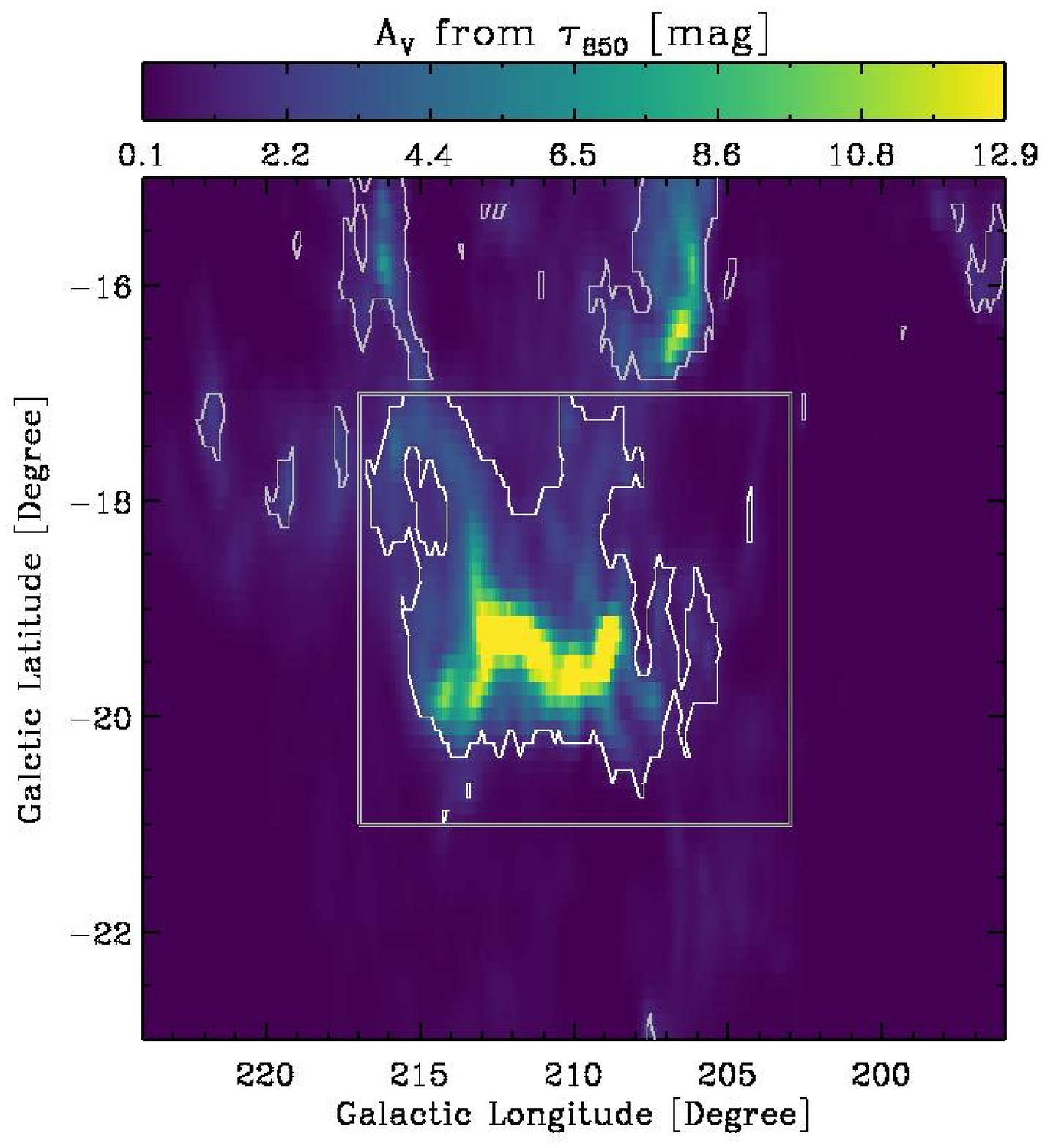}{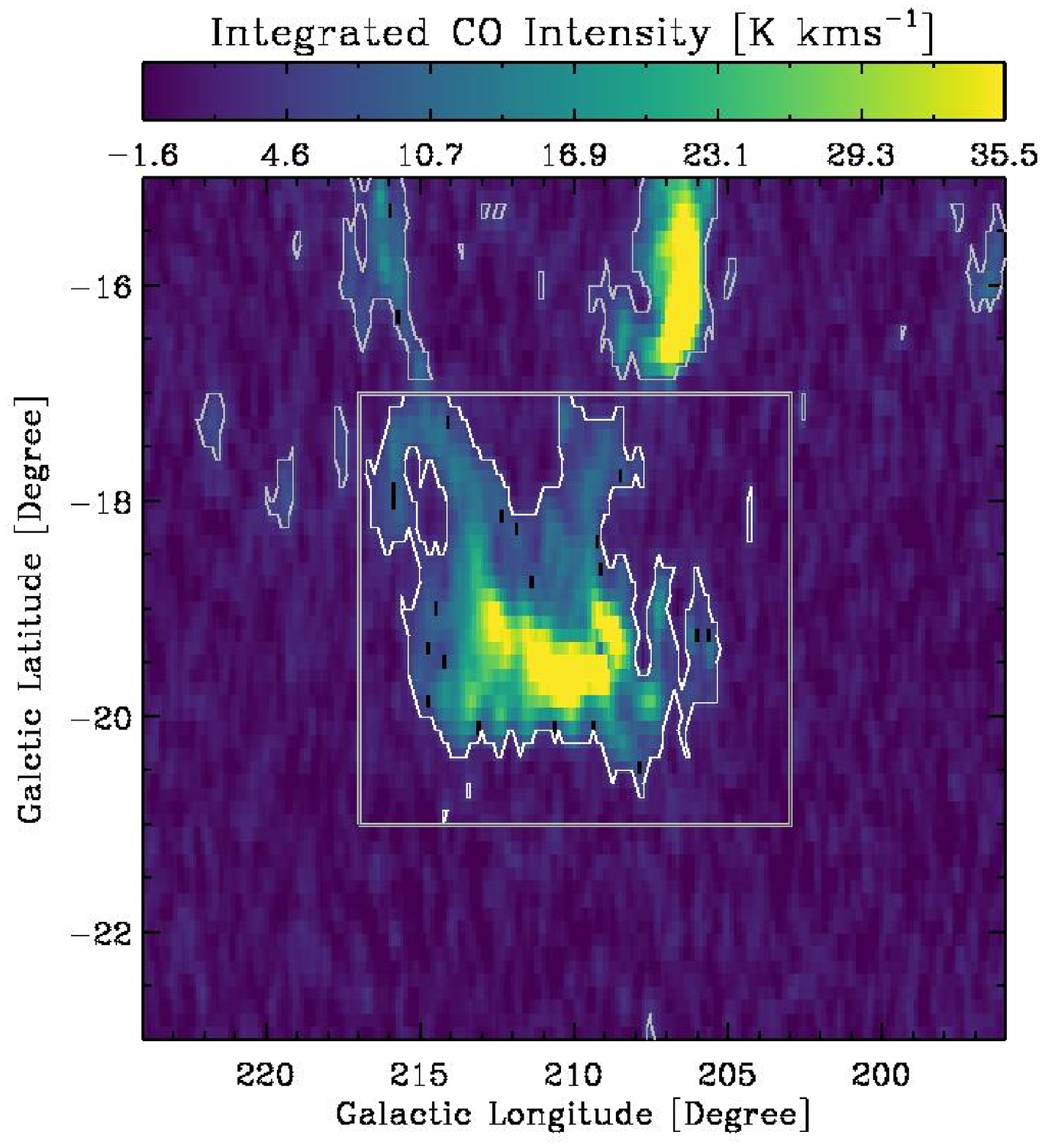}
\plottwo{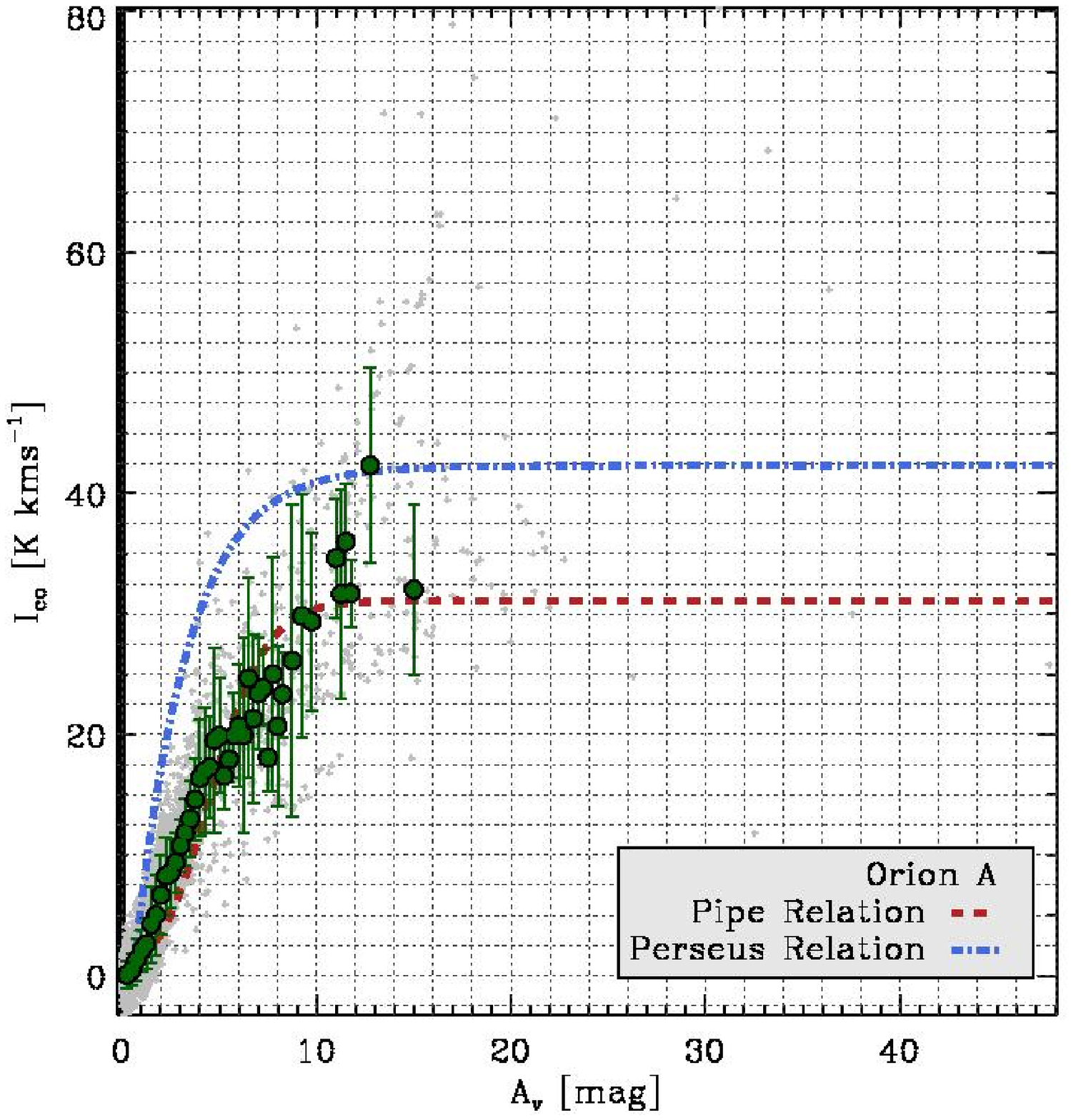}{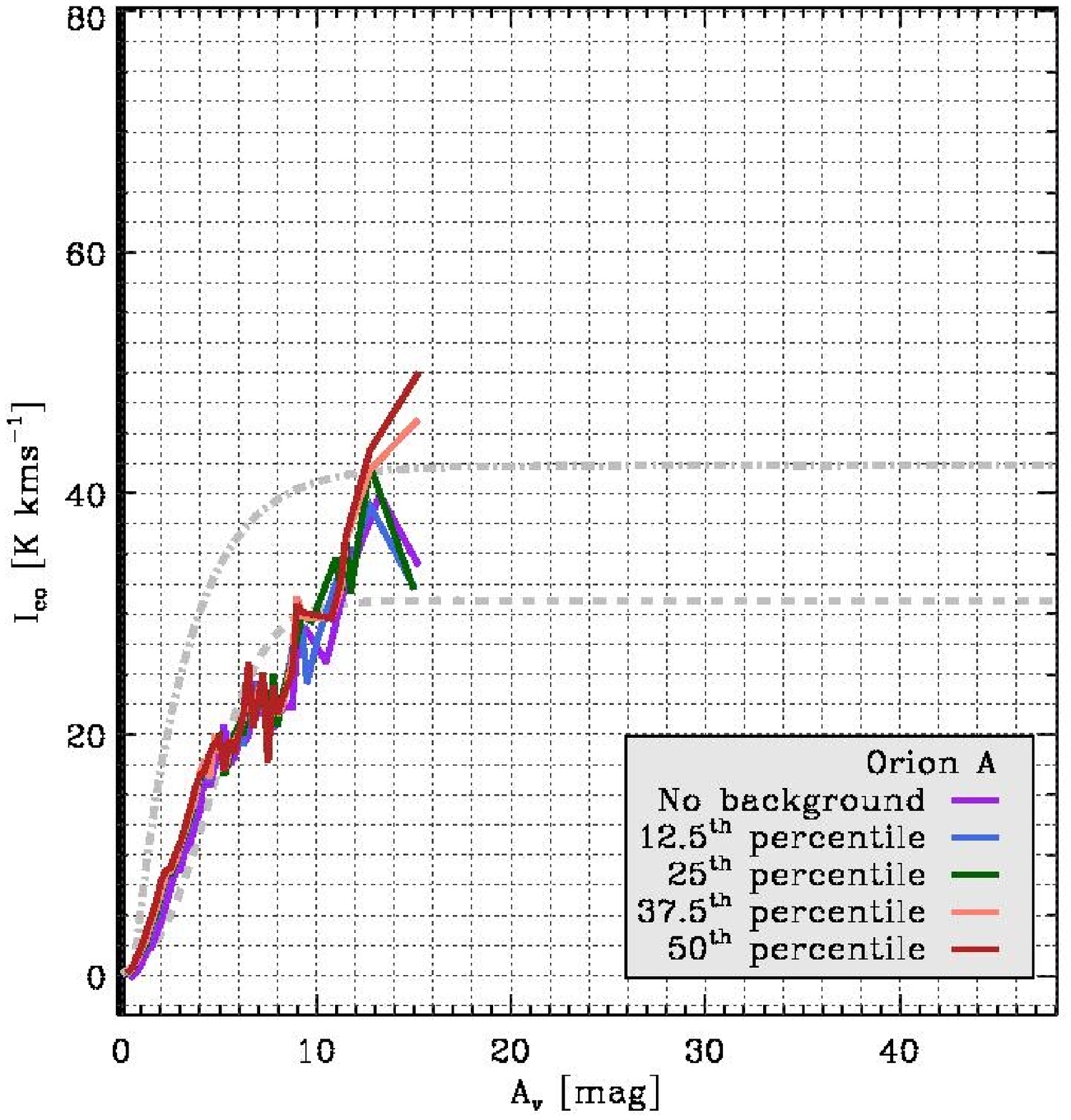}
\caption{Same as Figure~\ref{fig:taurus}, for the case of Orion A.}
\end{figure*}

\begin{figure*}
\plottwo{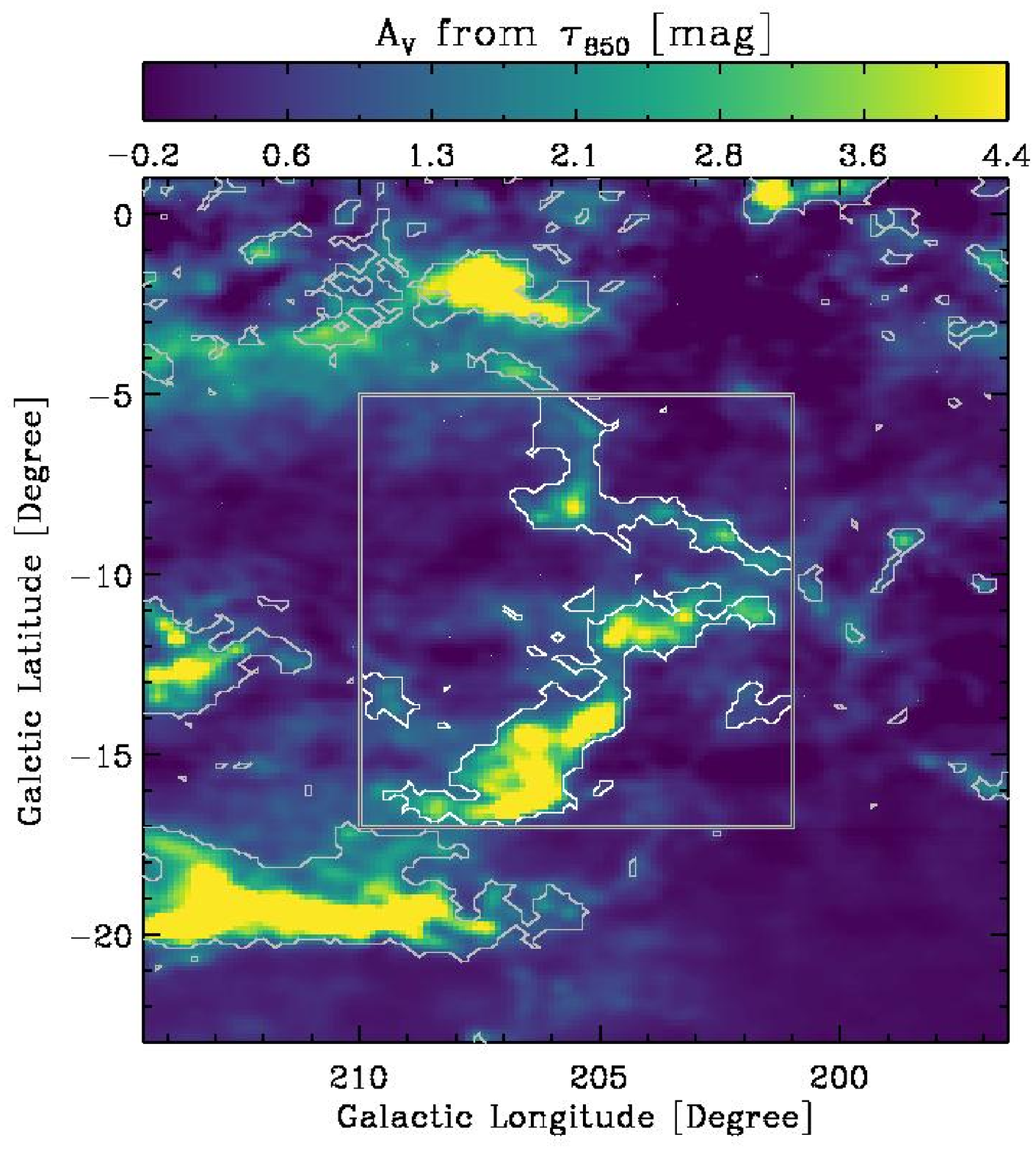}{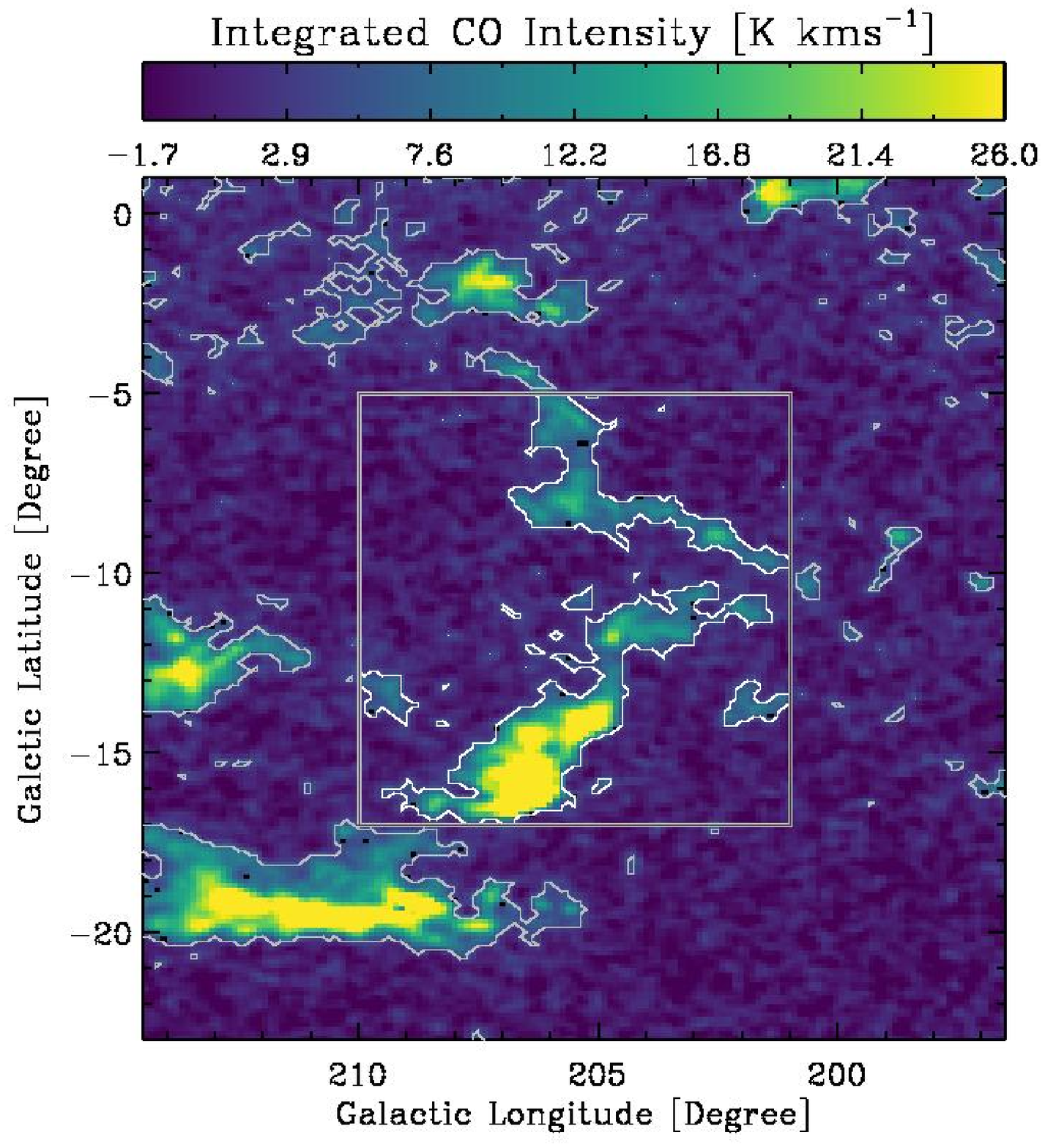}
\plottwo{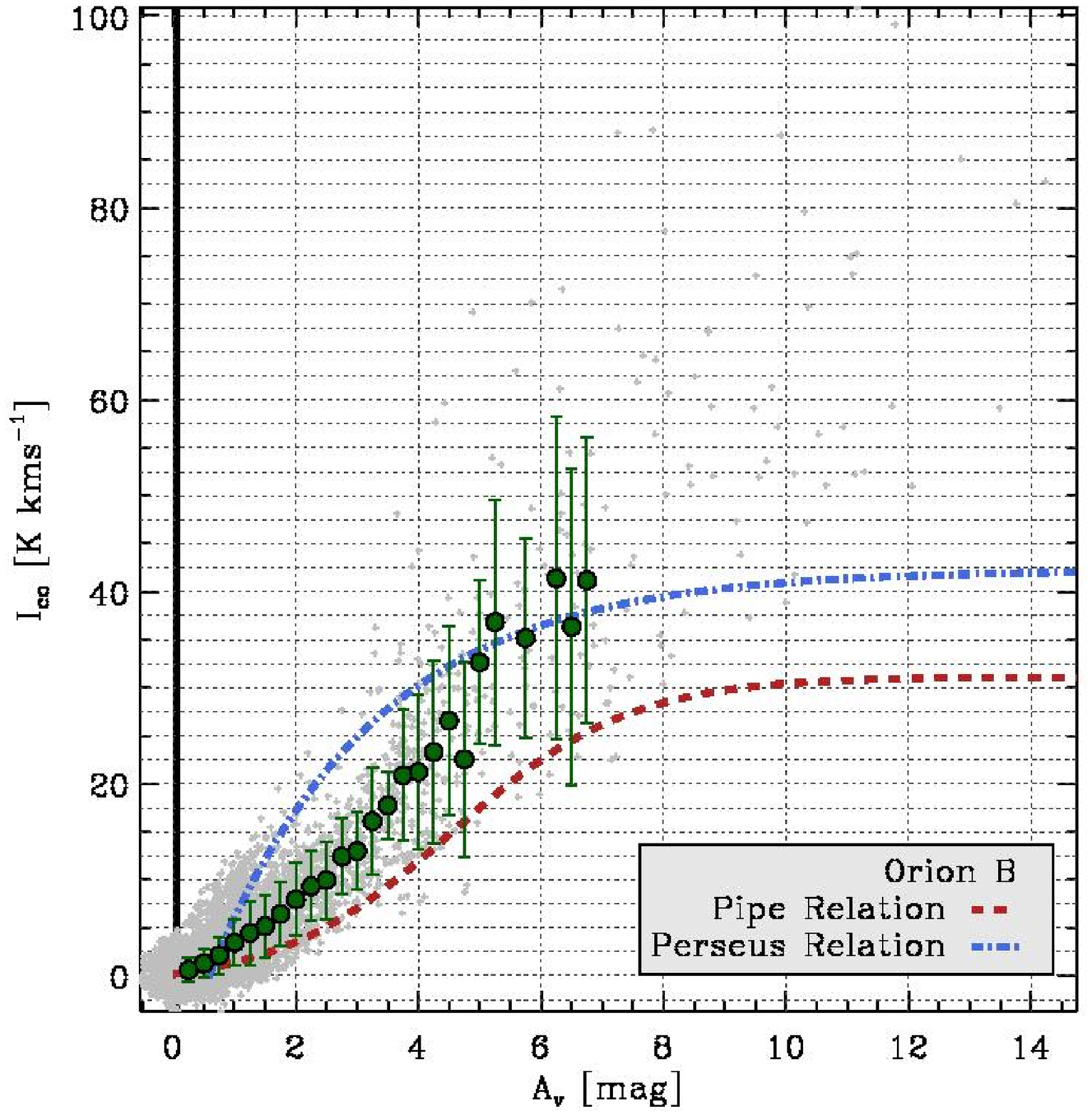}{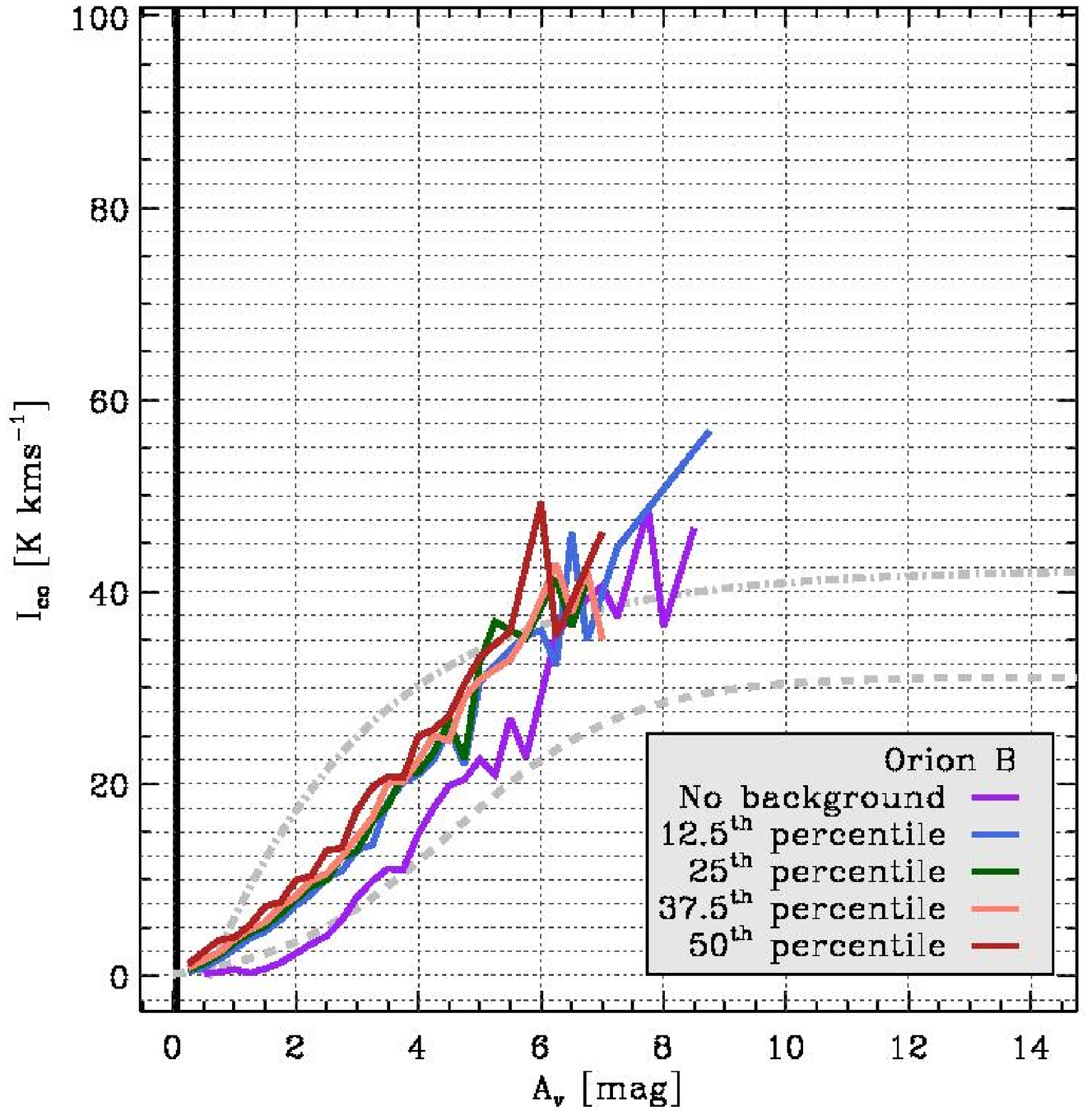}
\caption{Same as Figure~\ref{fig:taurus}, for the case of Orion B.}
\end{figure*}

\begin{figure*}
\plottwo{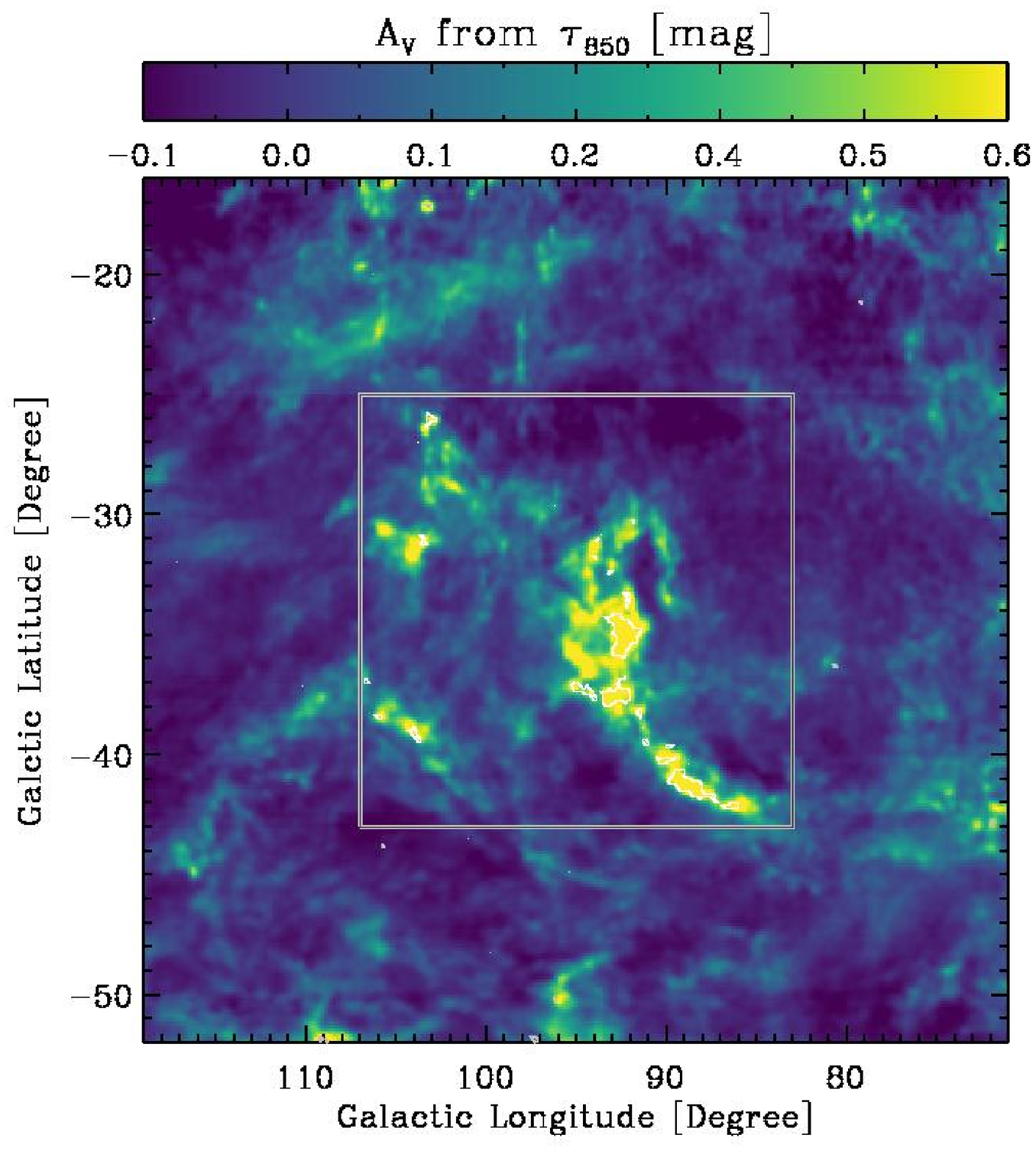}{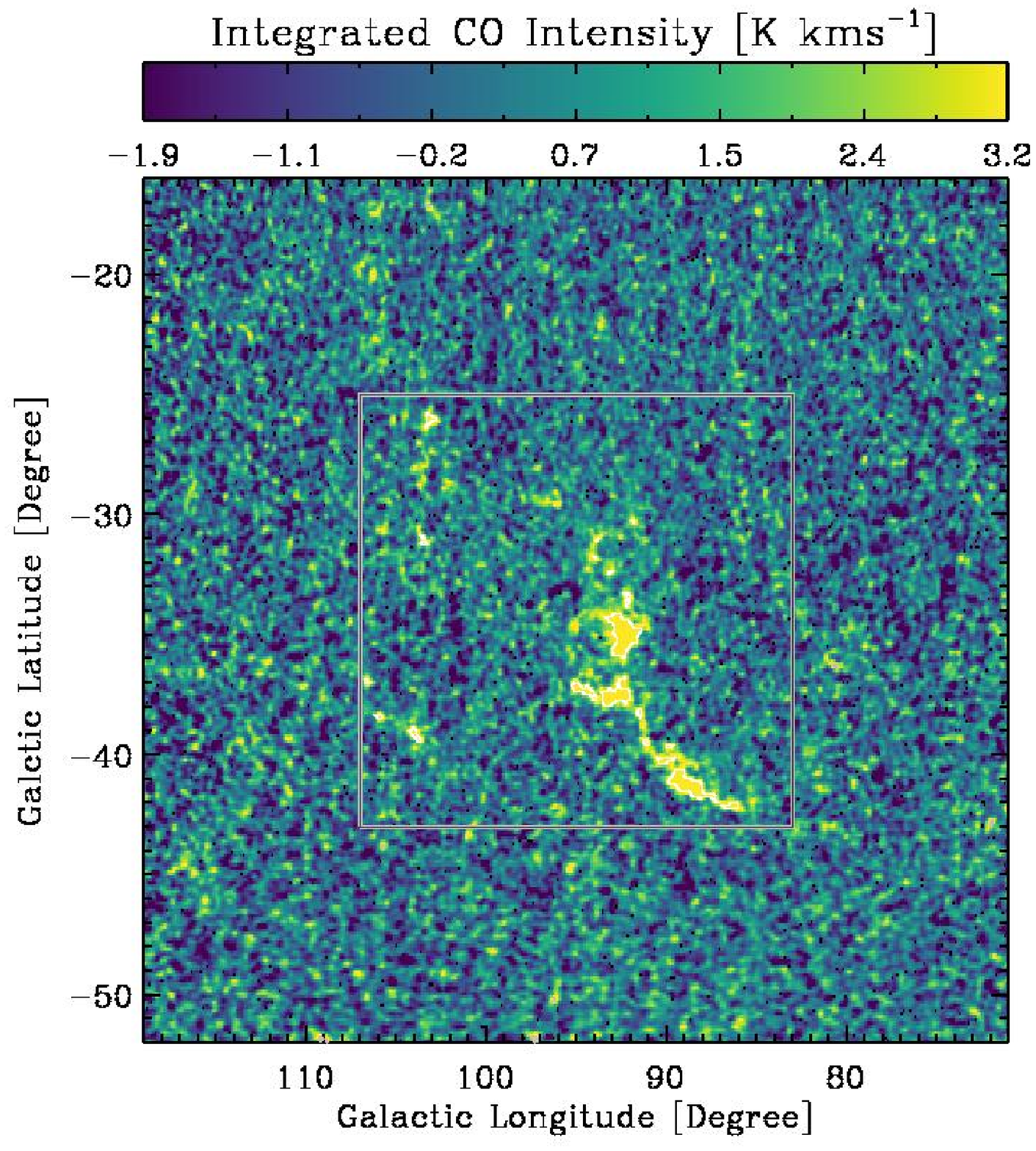}
\plottwo{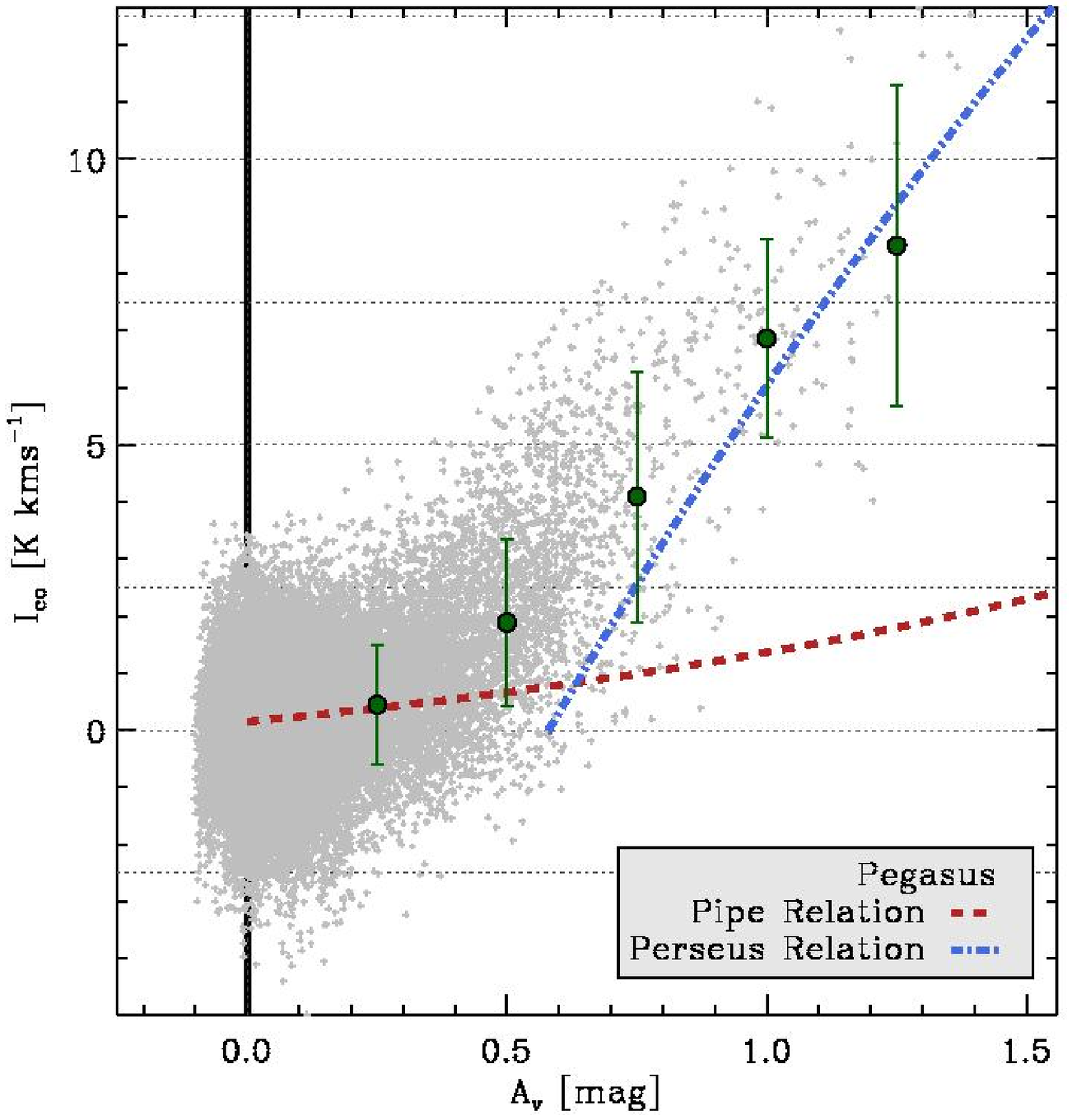}{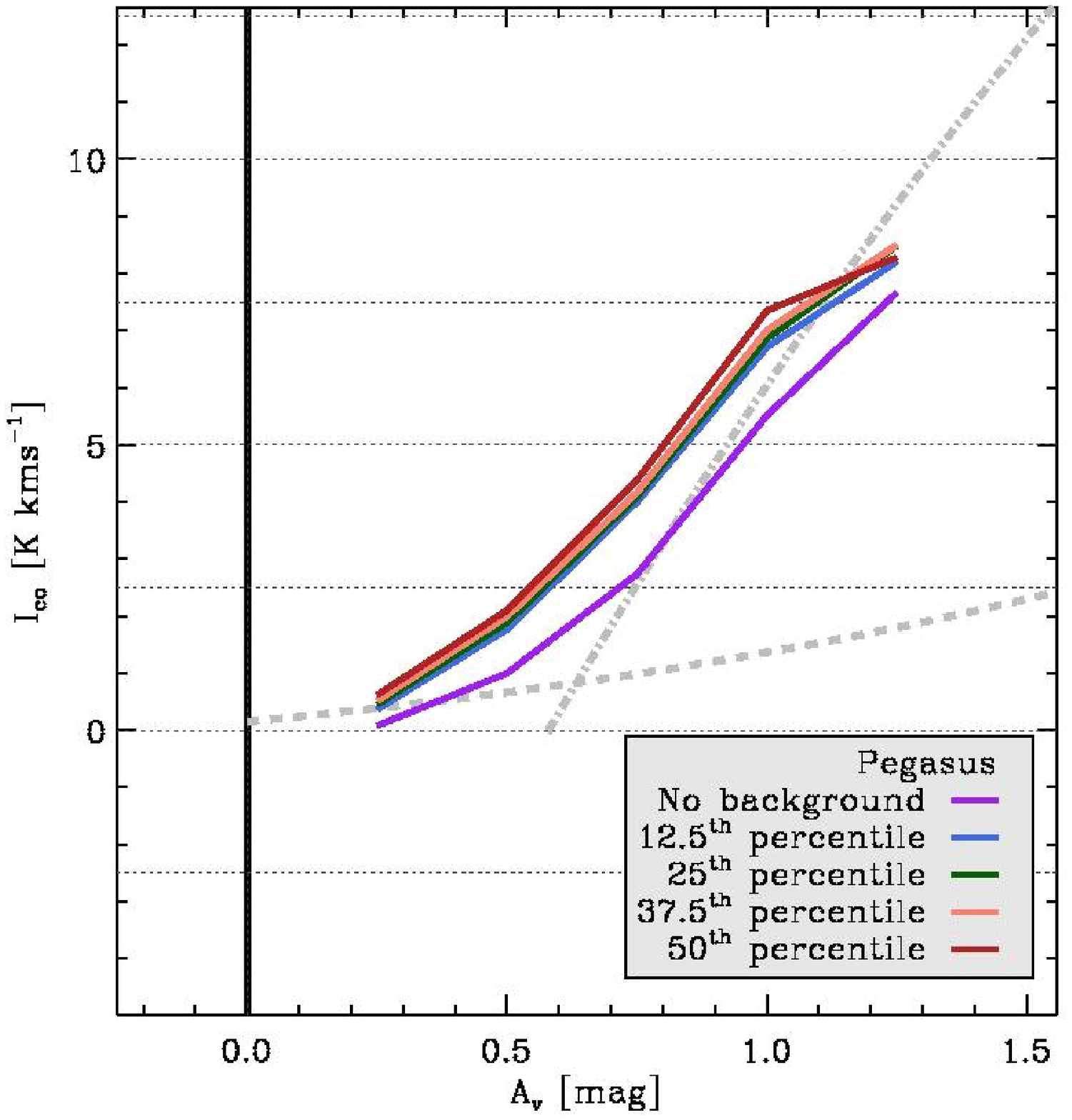}
\caption{Same as Figure~\ref{fig:taurus}, for the case of Pegasus.}
\end{figure*}

\begin{figure*}
\plottwo{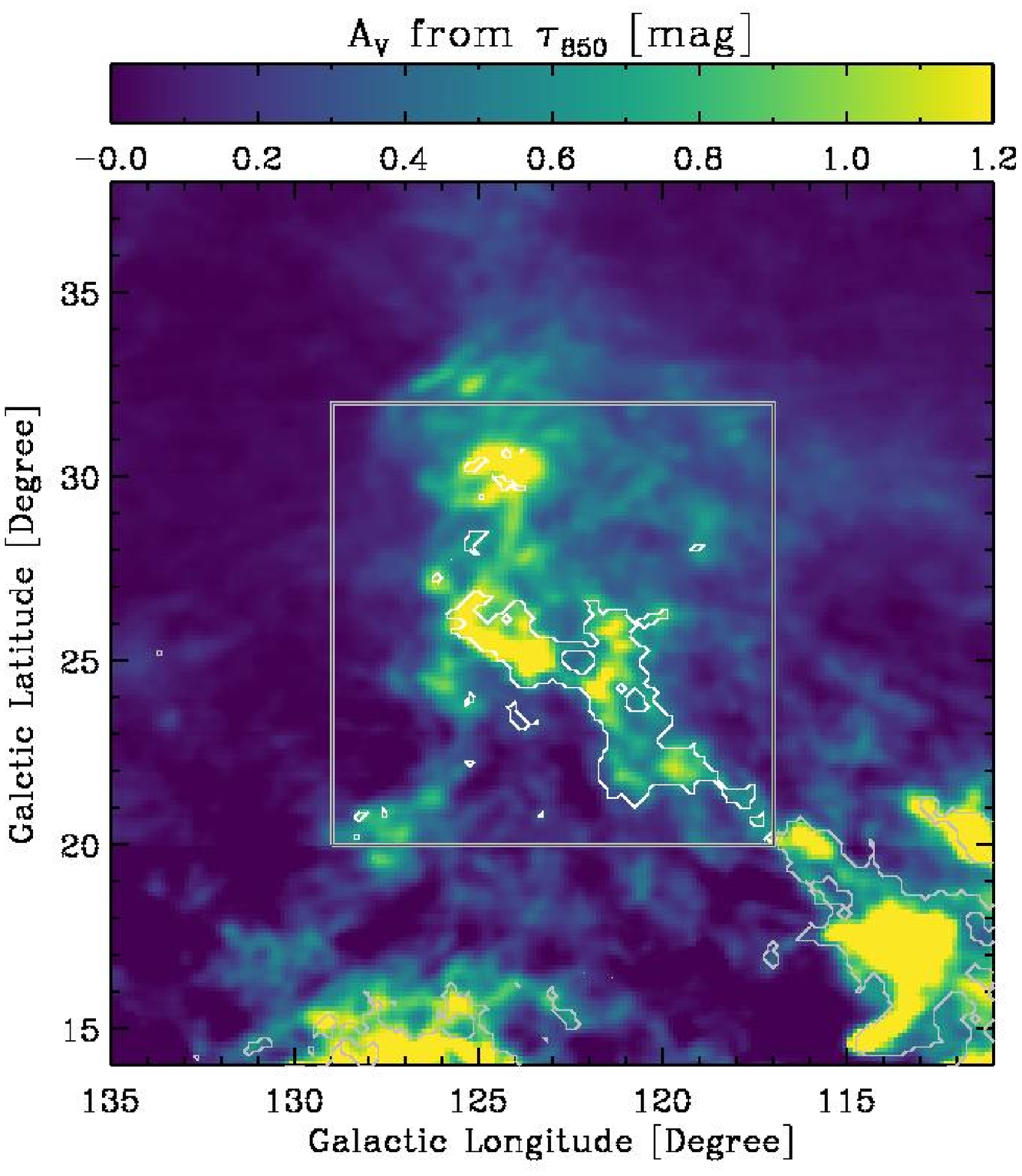}{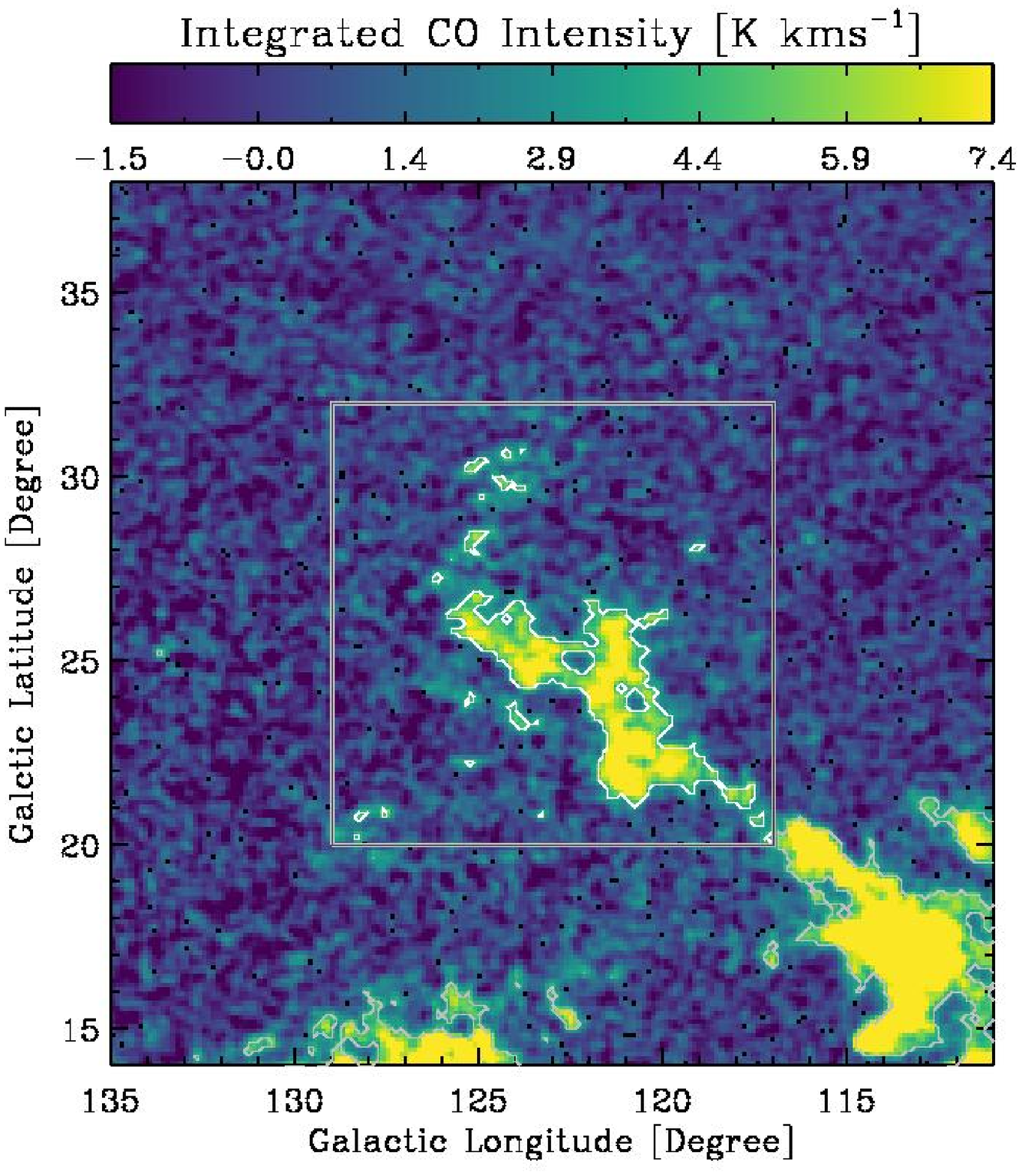}
\plottwo{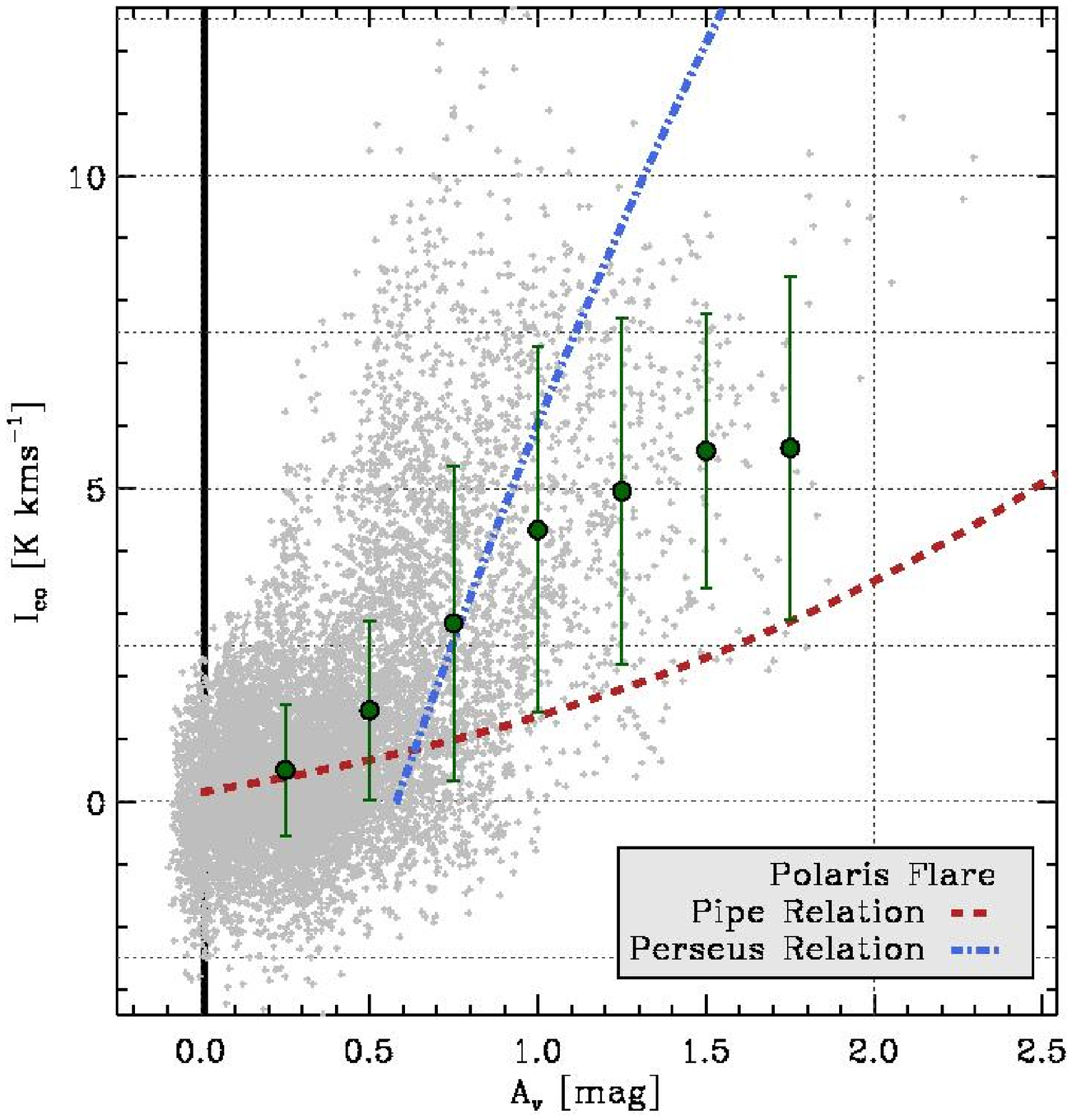}{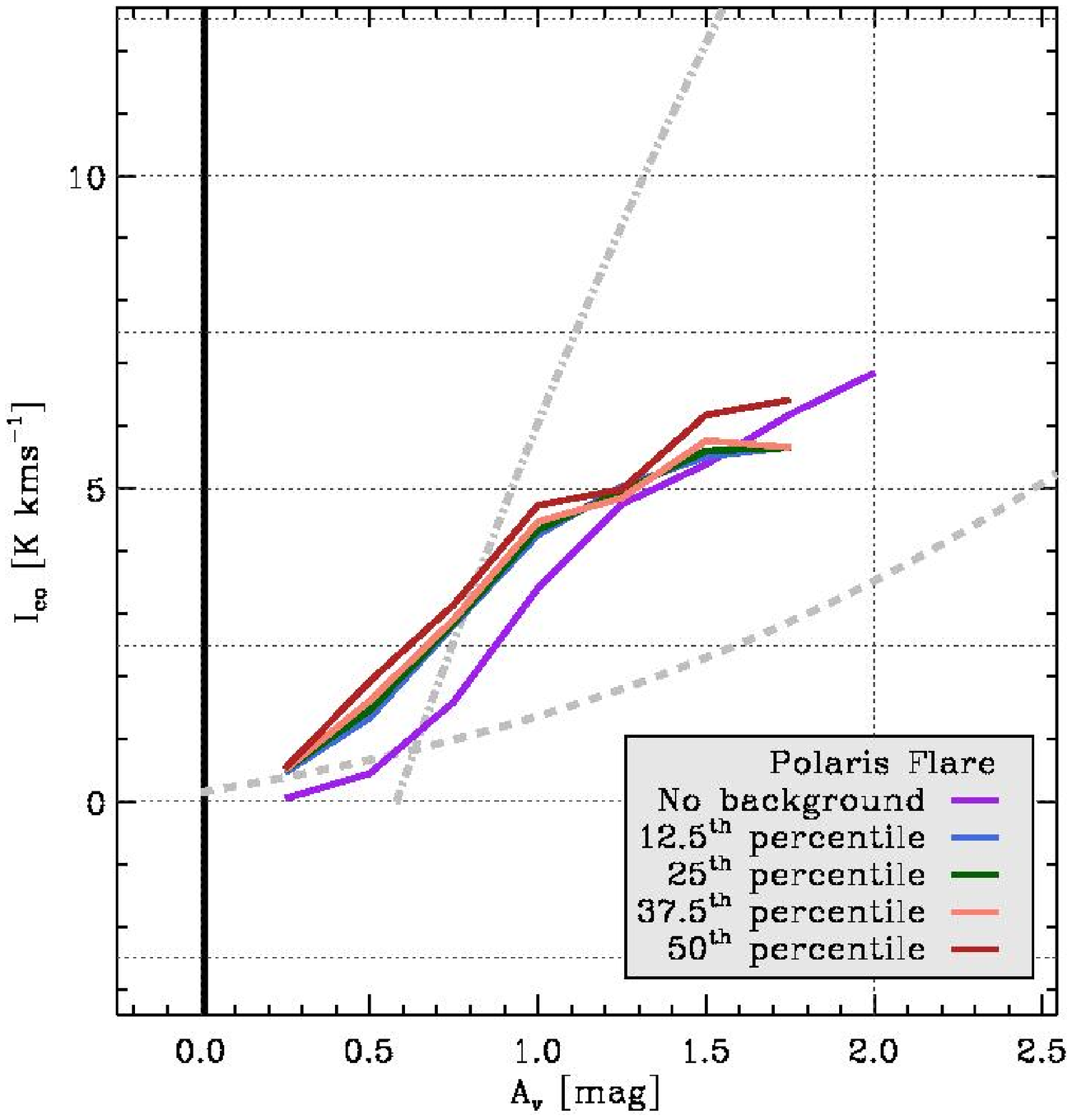}
\caption{Same as Figure~\ref{fig:taurus}, for the case of Polaris Flare.}
\end{figure*}

\begin{figure*}
\plottwo{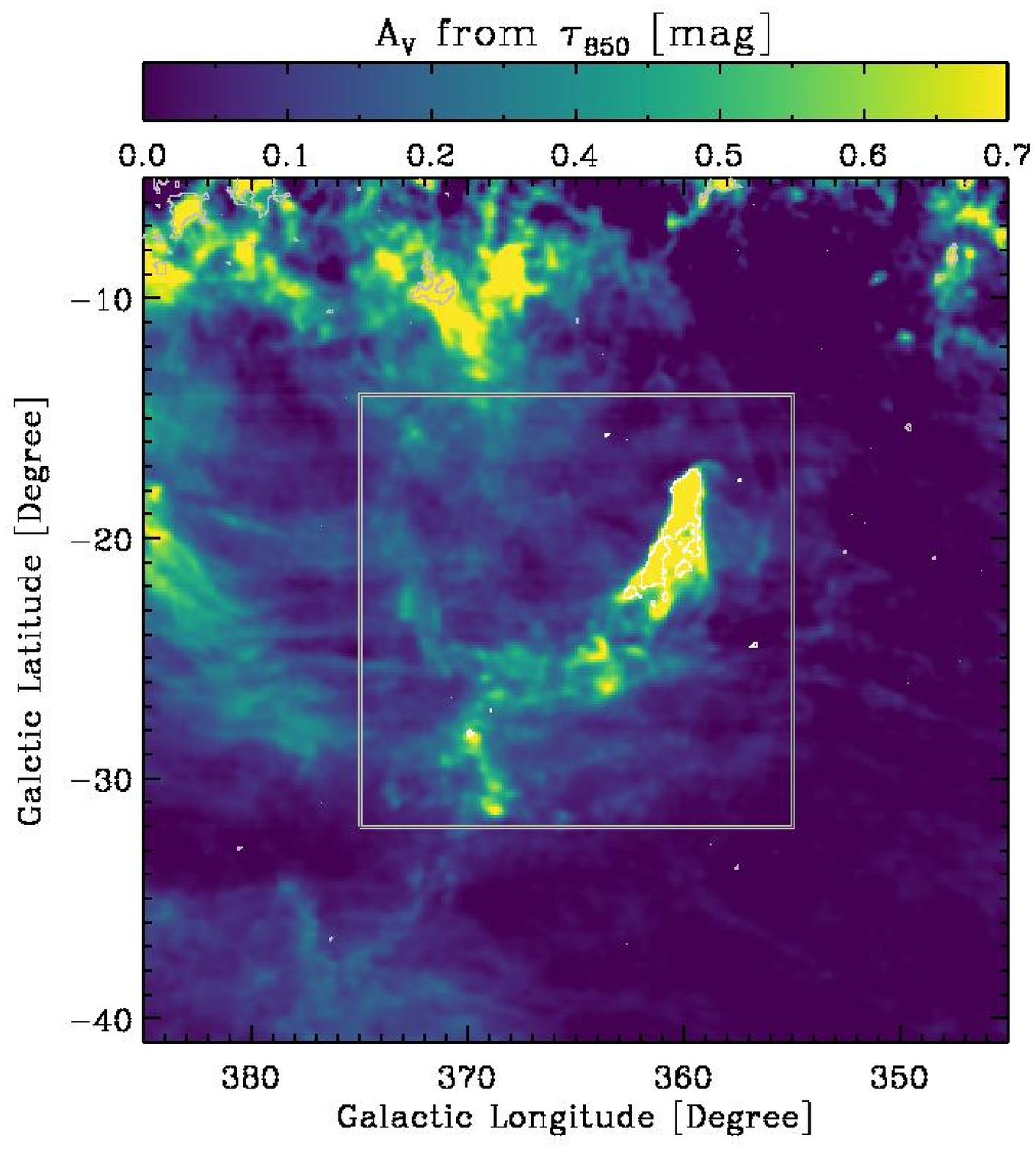}{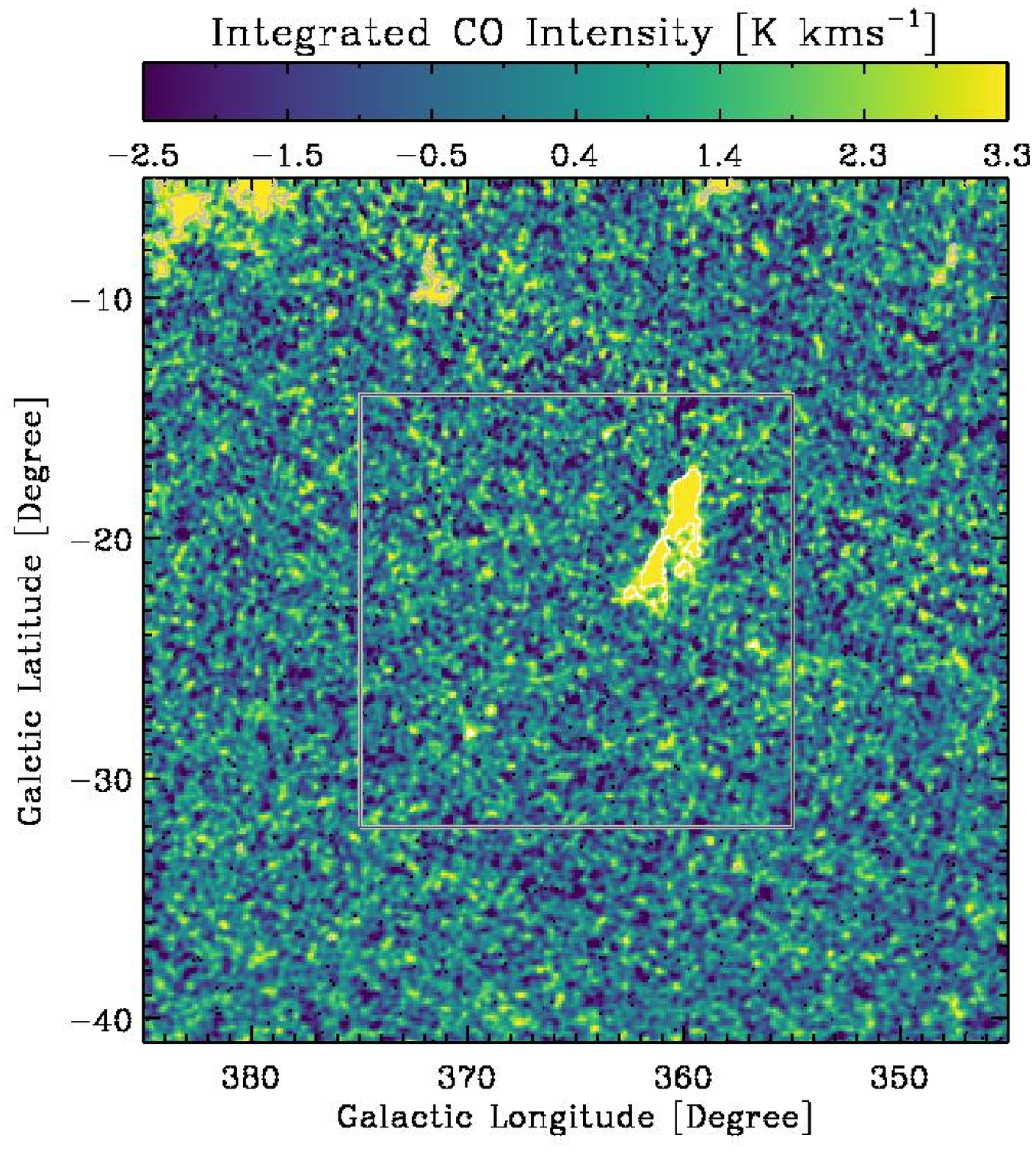}
\plottwo{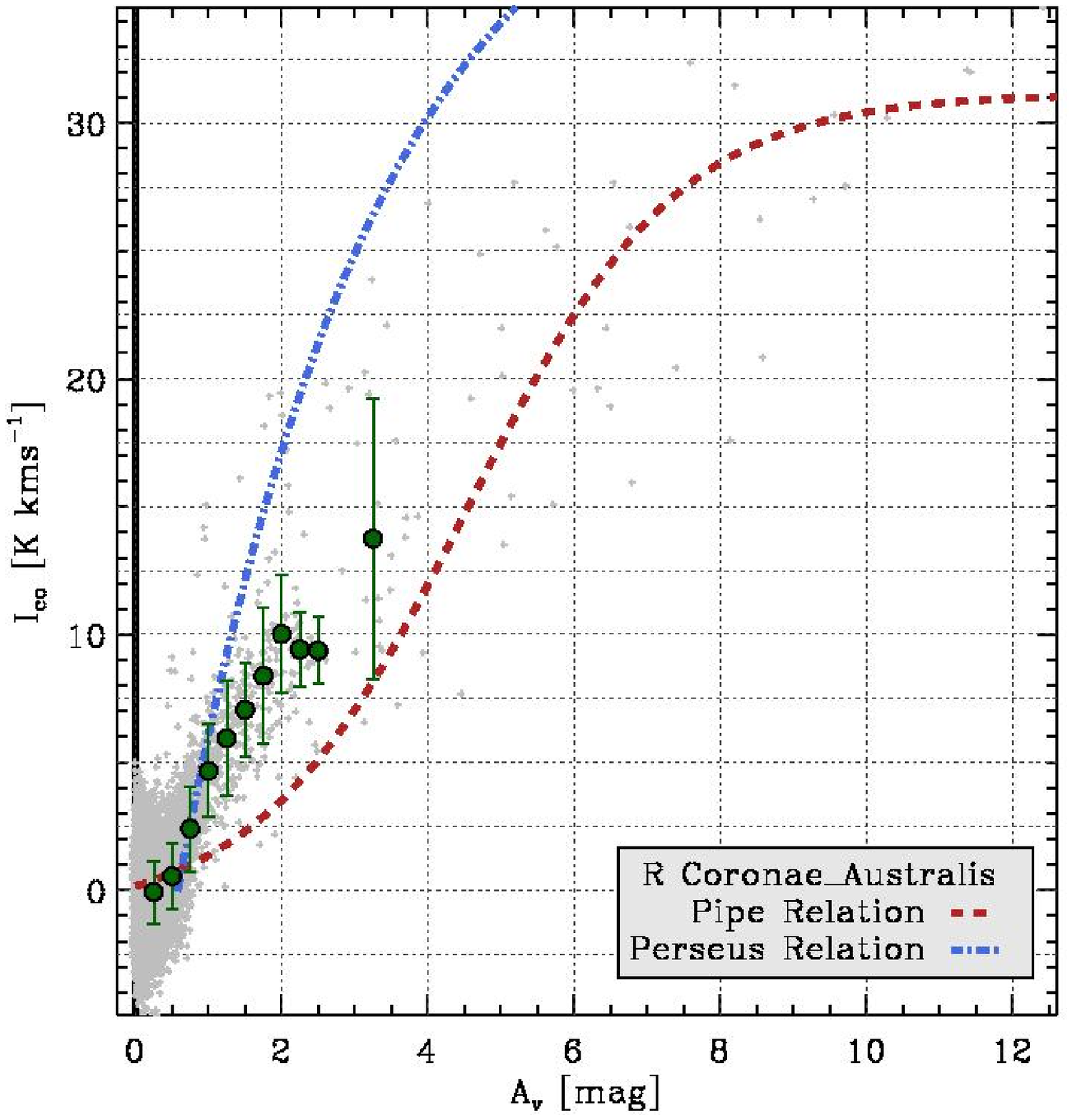}{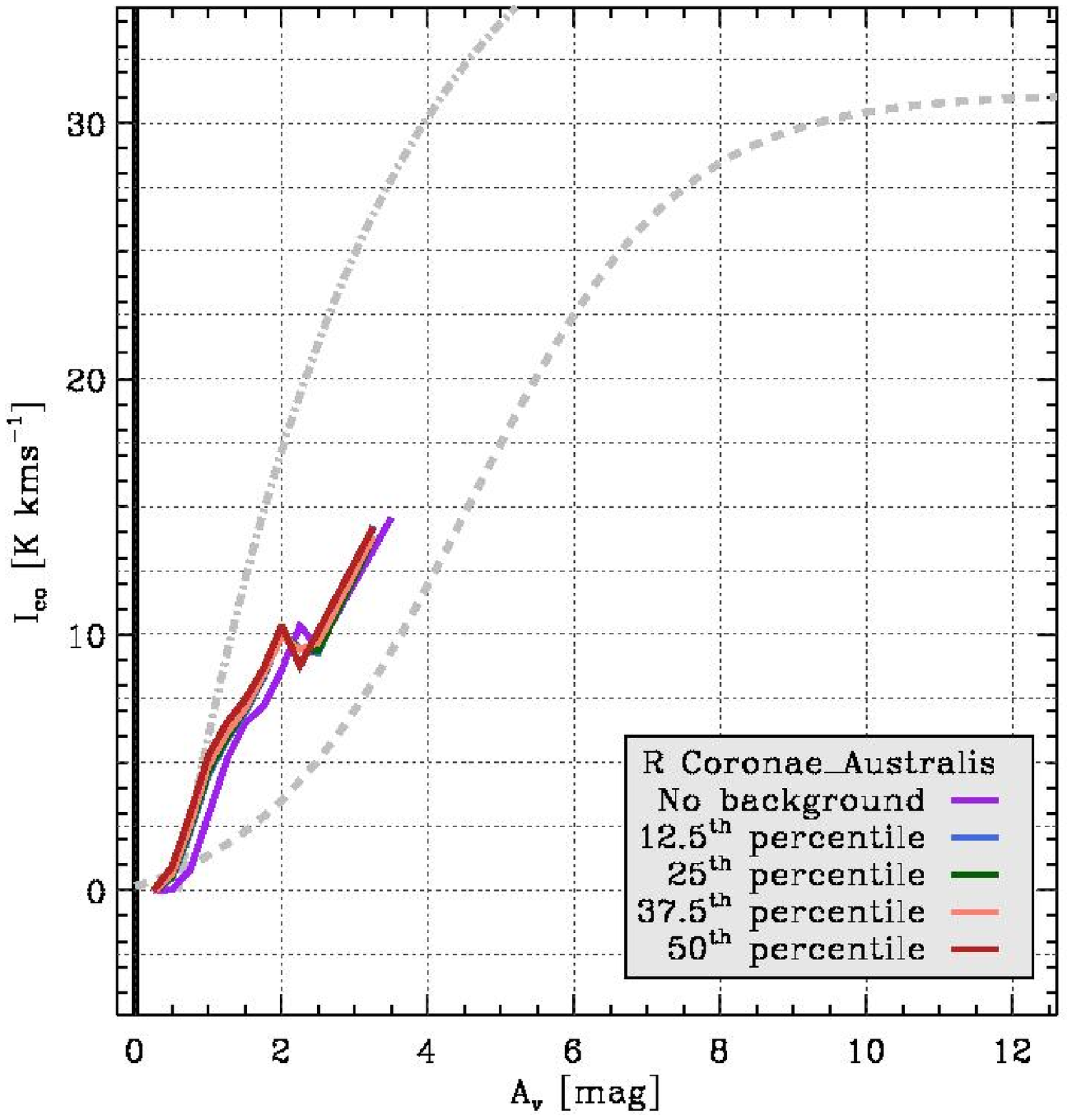}
\caption{Same as Figure~\ref{fig:taurus}, for the case of R Coronae Australis.}
\end{figure*}

\begin{figure*}
\plottwo{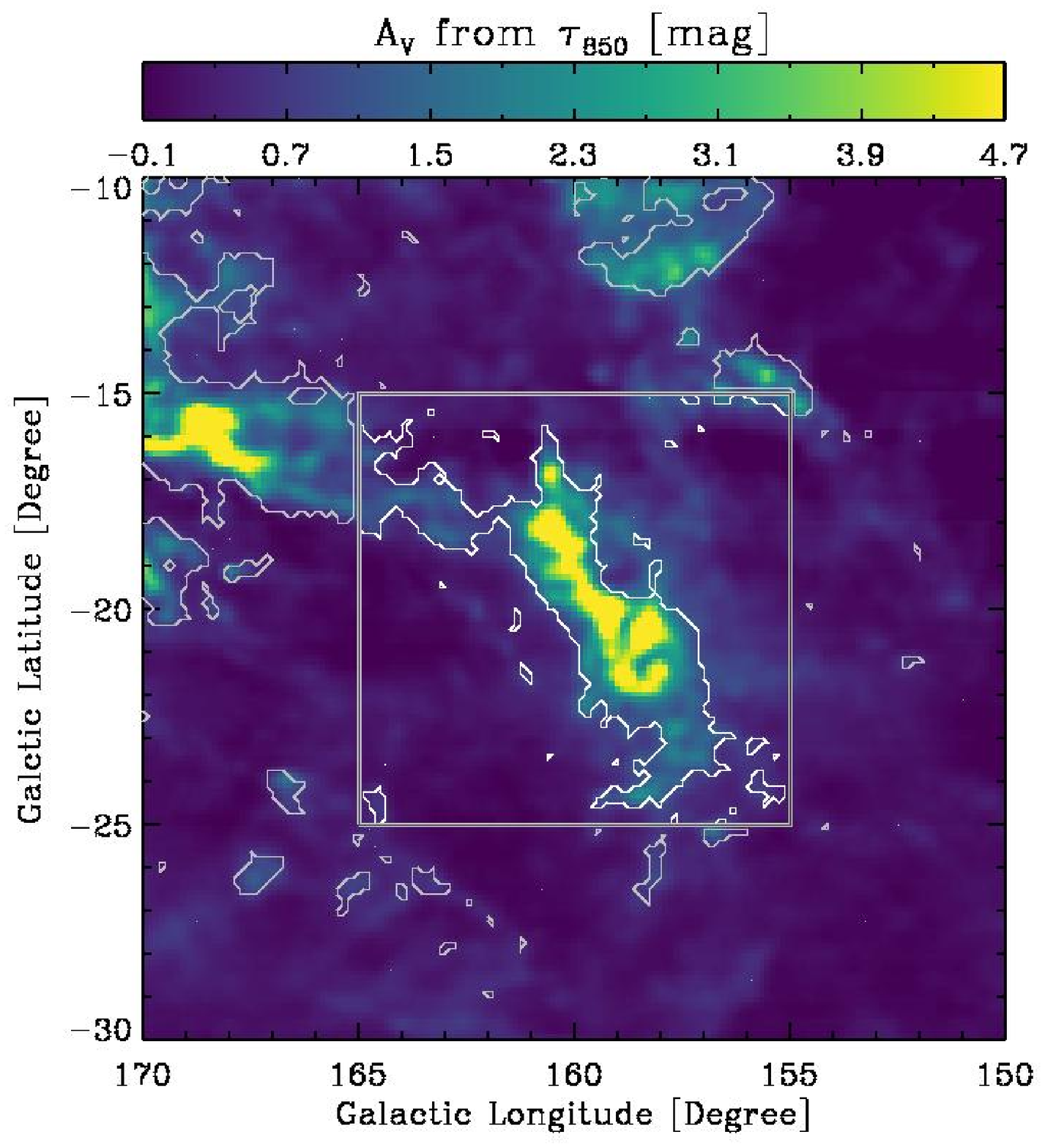}{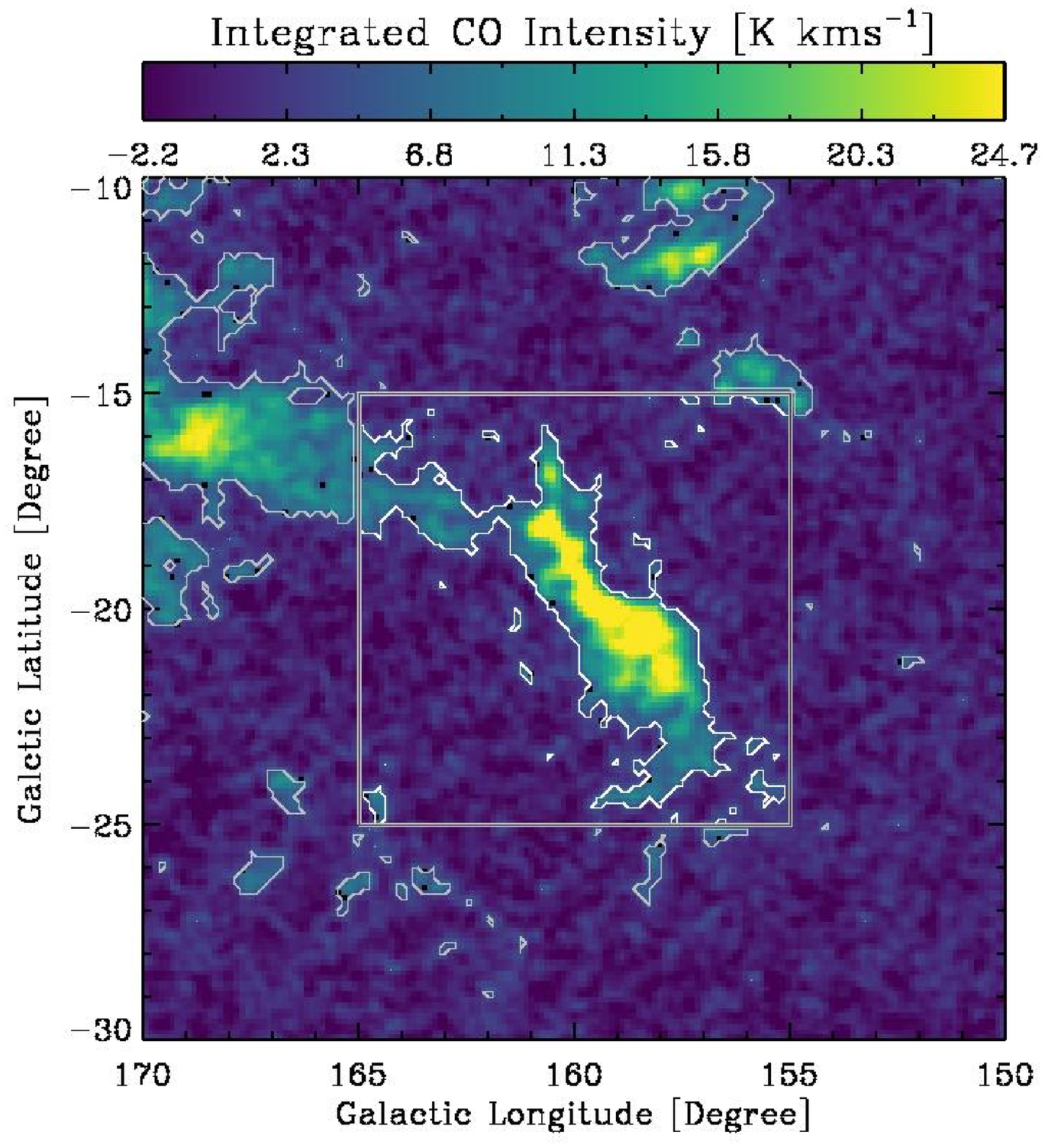}
\plottwo{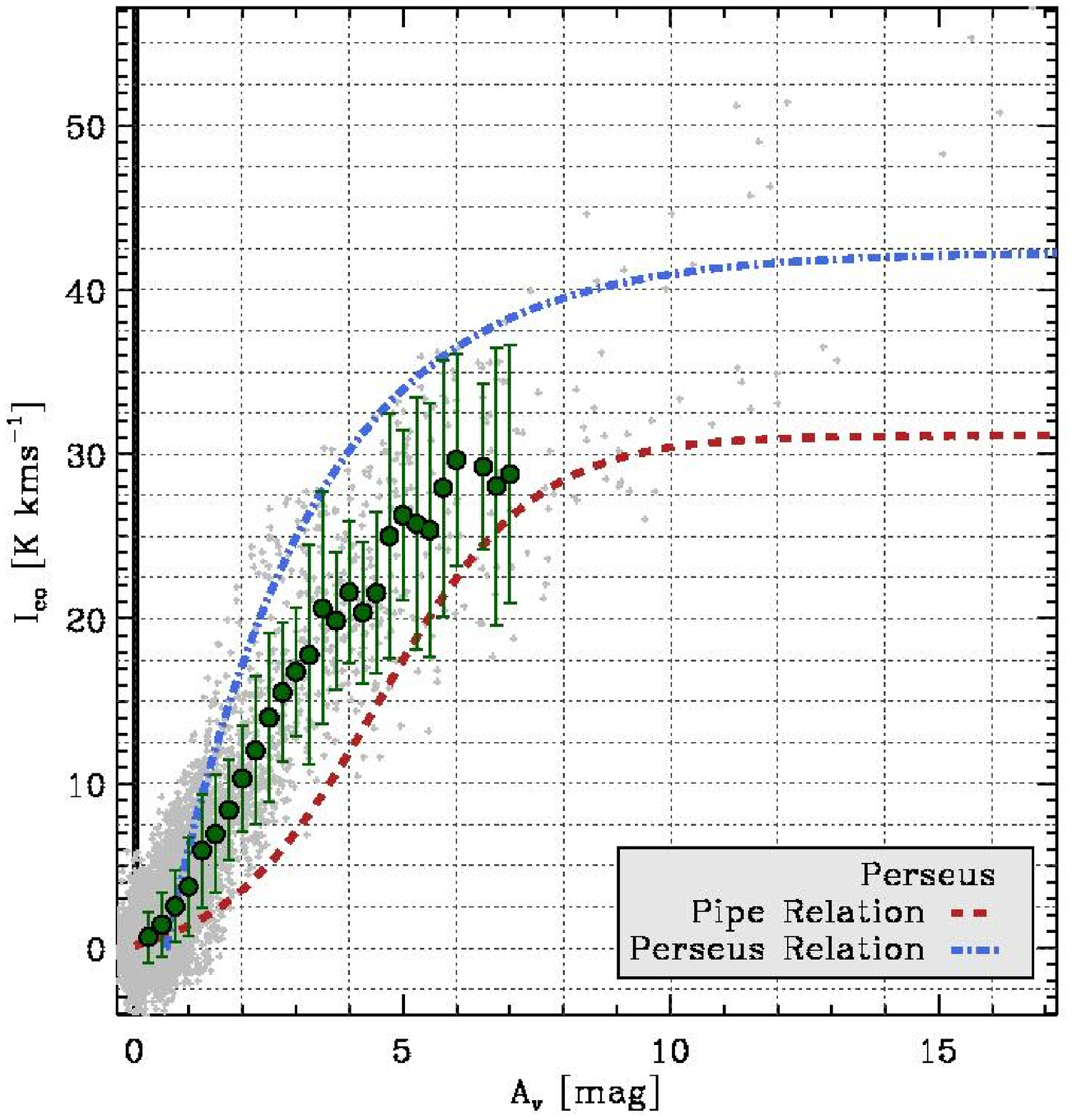}{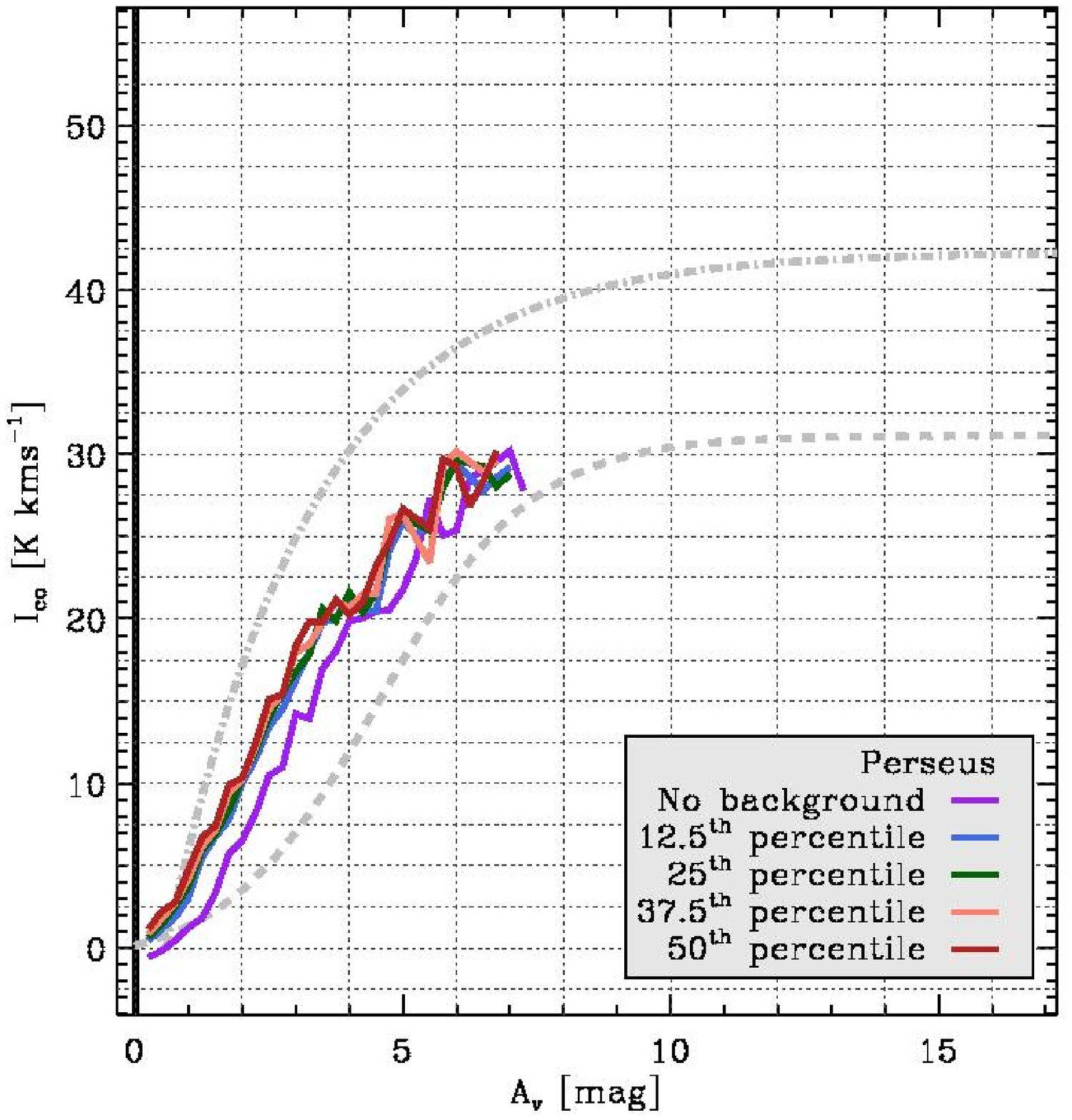}
\caption{Same as Figure~\ref{fig:taurus}, for the case of Perseus.}
\end{figure*}

\begin{figure*}
\plottwo{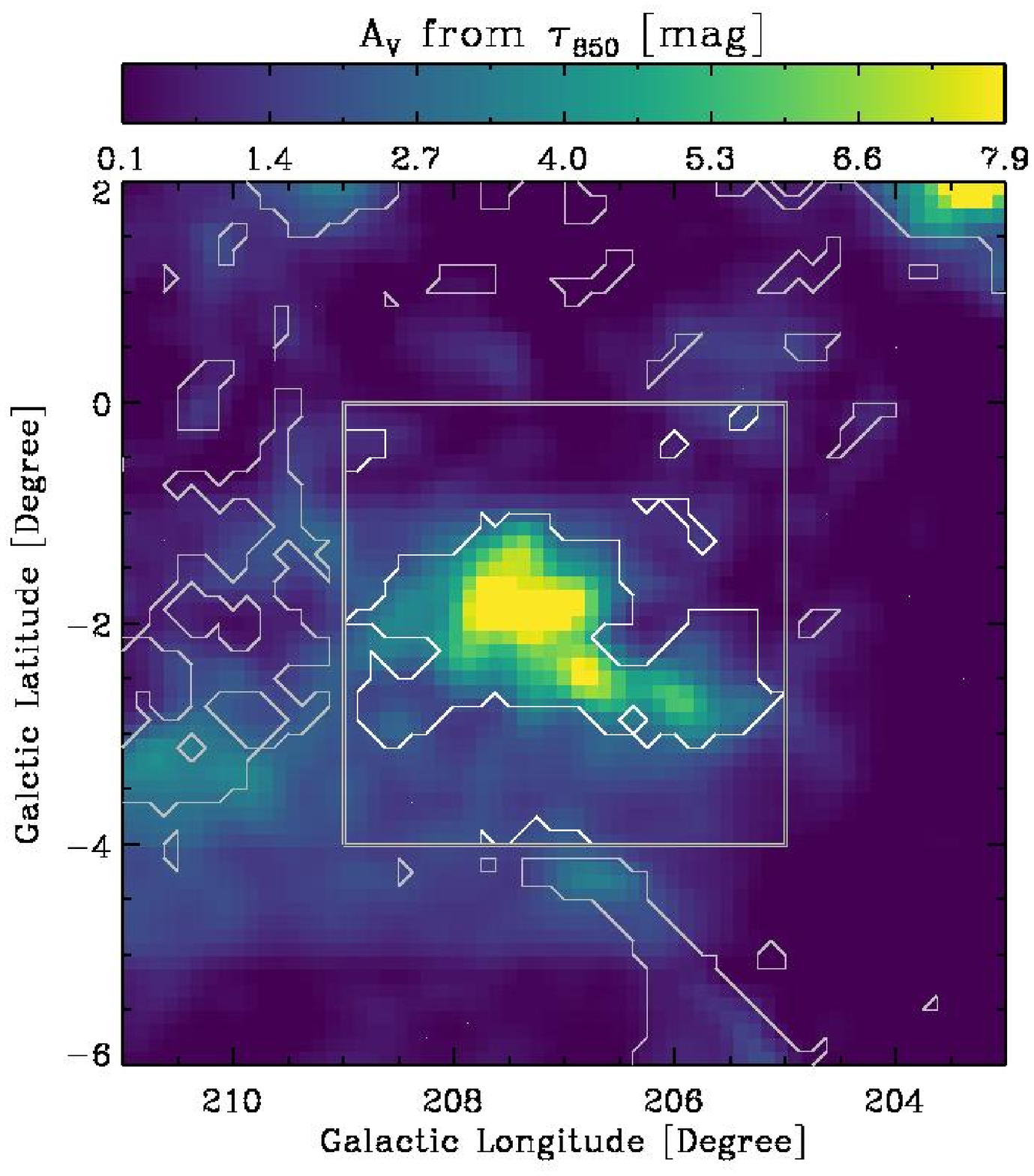}{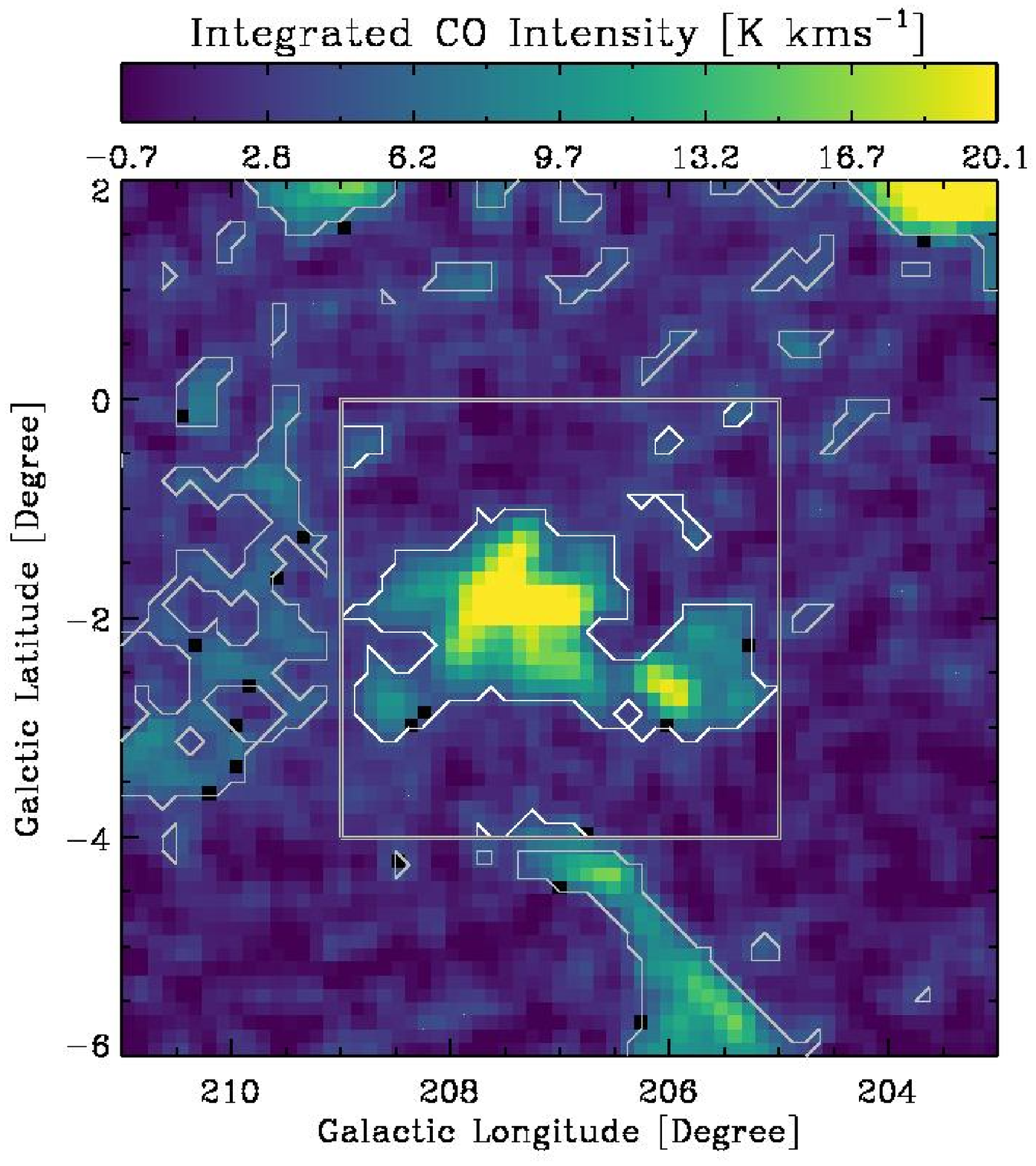}
\plottwo{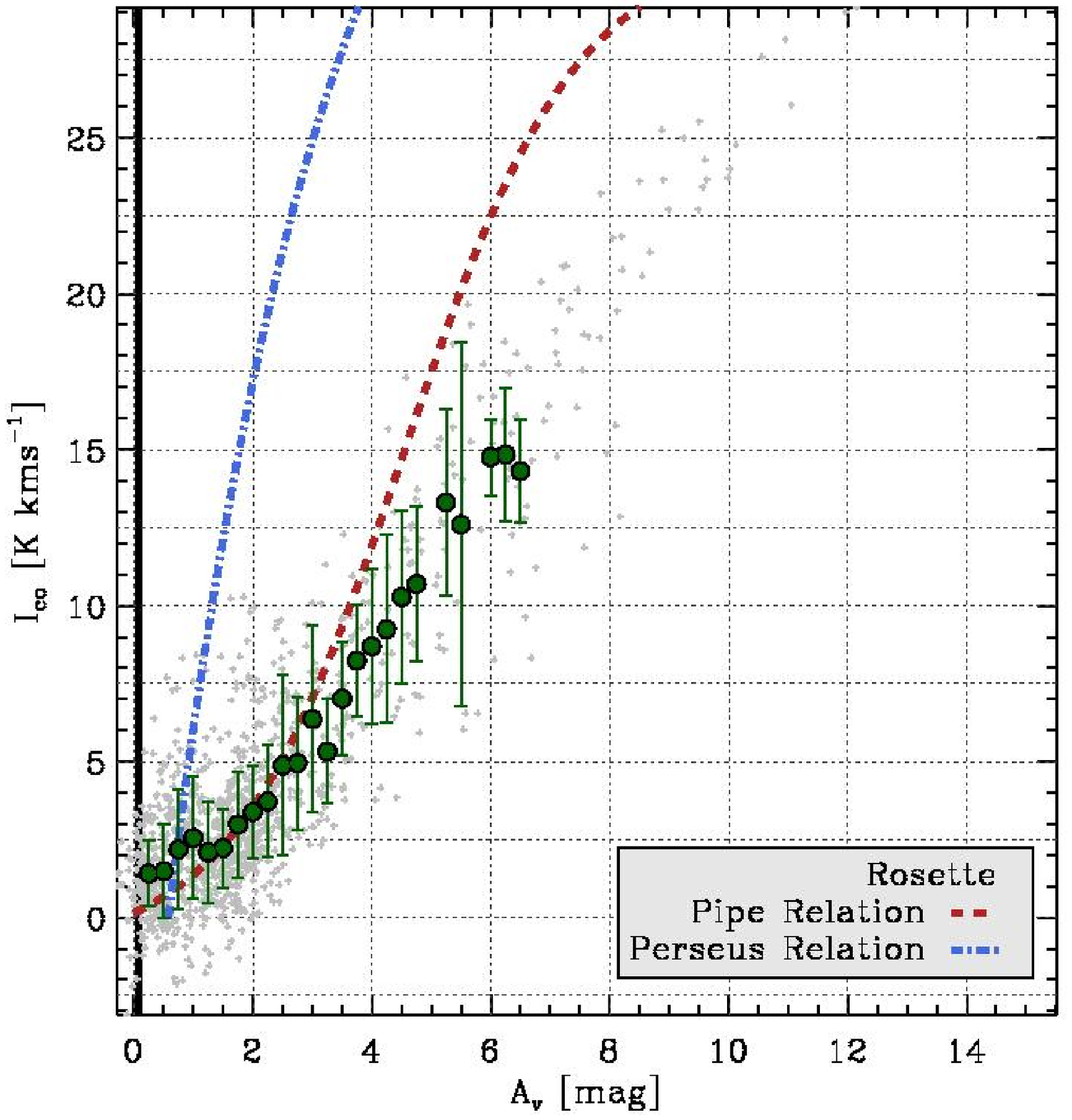}{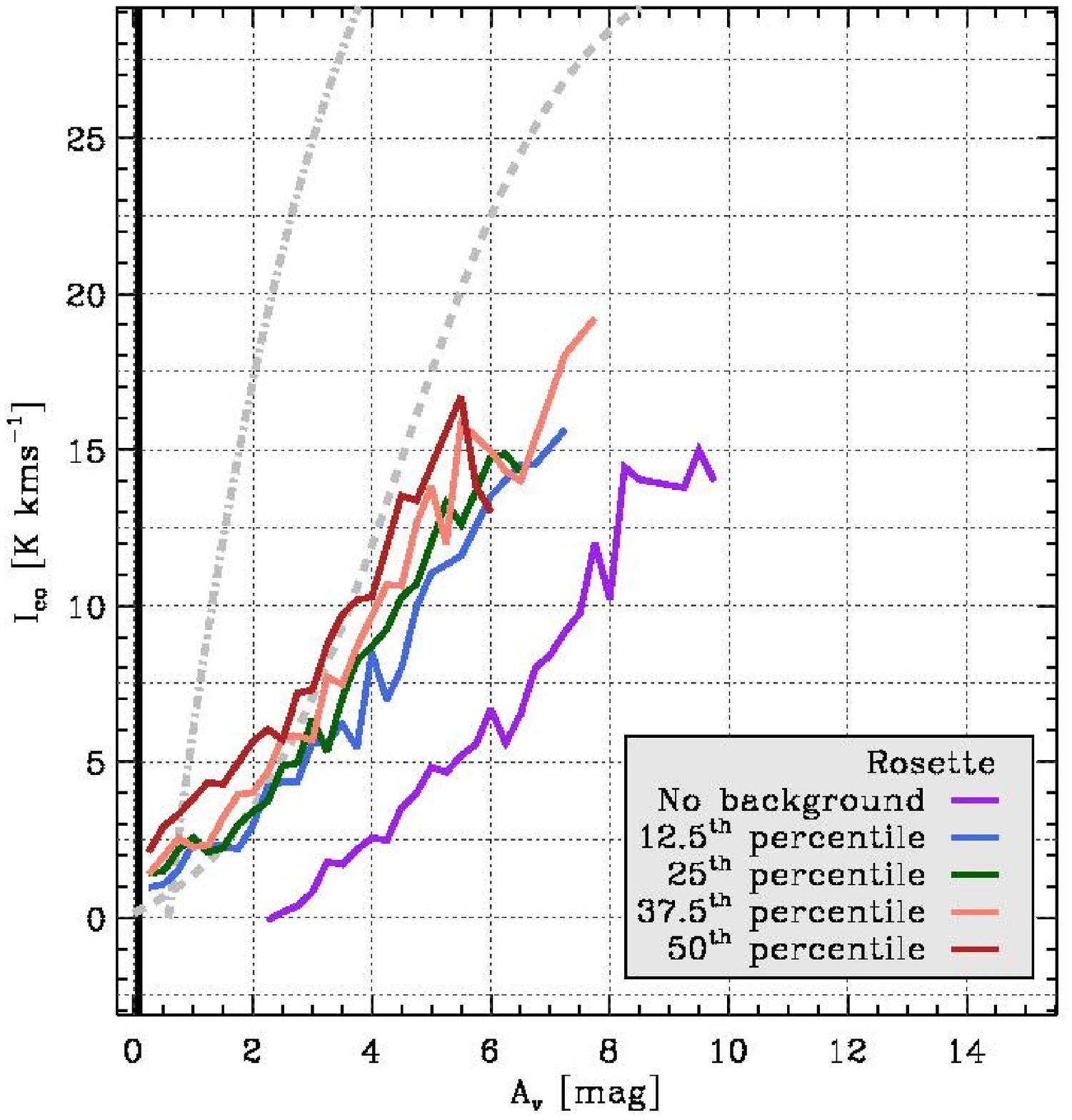}
\caption{Same as Figure~\ref{fig:taurus}, for the case of Rosette.}
\end{figure*}

\bibliographystyle{apj}
\bibliography{sample}
\end{document}